\documentclass[onecolumn,authoryear]{els-mrw} 

\usepackage{amsmath,amssymb,amsfonts,amsthm,makeidx,graphicx}
\usepackage{txfonts}
\usepackage{helvet}

\usepackage{aas_macros}

\begin{document}

\chapter{Cosmic quenching}\label{chap1}

\author[1,2]{Gabriella De Lucia}%
\author[1,2]{Fabio, Fontanot}%
\author[3,1]{Michaela Hirschmann}%
\author[4]{Lizhi Xie}

\address[1]{\orgname{INAF - Osservatorio Astronomico di Trieste}, \orgaddress{via G.B. Tiepolo 11, I-34143 Trieste, Italy}}
\address[2]{\orgname{IFPU – Institute for Fundamental Physics of the Universe}, \orgaddress{Via Beirut 2, 34151 Trieste, Italy}}
\address[3]{\orgname{Institute for Physics, Laboratory for Galaxy Evolution, EPFL, Observatoire de Sauverny}, \orgaddress{Chemin Pegasi 51, 1290 Versoix, Switzerland}}
\address[4]{\orgname{Tianjin Normal University}, \orgaddress{Binshuixidao 393, 300387 Tianjin, PR China}}

\articletag{Chapter Article tagline: update of previous edition,, reprint..}

\maketitle

\begin{abstract}[Abstract]
The quenching of star formation activity represents a critical phase for a non-negligible fraction of the observed galaxy population at all cosmic epochs, marking a transition from an epoch of intense mass growth to an extended period of passive evolution. Over the past years, we have collected a detailed characterization of how the fraction of quenched galaxies evolves as a function of cosmic time (it grows at later cosmic epochs), and correlates with the physical properties of galaxies (more massive galaxies, that also tend to be bulge-dominated, are more likely to be quenched) and with their environment (denser environments host larger fractions of quiescent galaxies). Different physical processes can lead to a suppression of star formation. These include internal processes (processes that do not depend on the environment in which galaxies live, e.g. stellar and AGN feedback mechanisms) and processes that instead depend on the environment (e.g. mergers, gas stripping processes). In this chapter, we summarize the observational and theoretical status of the field, highlighting the most recent results and the questions that are yet to be answered. We also summarize our view on the expected developments in the next few years. 
\end{abstract}

\begin{glossary}[Glossary]
\term{Magnitude} A measure of the brightness of an object, in a defined passband.\\
\term{Color} Magnitude difference between two defined passbands. It is often used in astronomy as a proxy for star formation rates in galaxies, since young stellar populations have blue colors than older stellar populations. \\
\term{Quiescent galaxy} Galaxy that is forming stars at a negligible rate.\\
\term{Quenching} Process that leads to a significant suppression of the star formation activity of a galaxy.\\
\term{Environment} The environment of a galaxy can be broadly defined by the distribution of matter in its surroundings. Depending on the information available, galaxy environments can be defined in many different ways (e.g. mass of the halo in which the galaxy resides, galaxy over-densities on different physical scales, density defined by closest nth neighbors, etc).\\
\term{External (quenching) mechanisms} Processes that regulate the star formation activity of a galaxy, and that are specific of the galaxy environment.\\
\term{Internal (quenching) mechanisms} Processes that regulate the star formation activity of a galaxy and that do not depend on the specific environment in which the galaxy lives.\\
\term{Feedback} Process whereby energy and momentum are fed back to the interstellar medium as a result of astrophysical mechanisms such as star formation and cold gas accretion on super-massive black holes.\\
\term{Merger} Process where two or more galaxies come together due to gravitational interactions, potentially leading to the fusion of the merging systems into a new galaxy.\\
\term{Starvation/strangulation} Removal of the hot halo gas associated with a galaxy infalling into a larger structure.\\
\term{Gas Stripping} Removal of cold gas from the stellar disc of a galaxy moving through a dense intra-cluster or intra-group medium.\\
\end{glossary}

\begin{glossary}[Nomenclature]
\begin{tabular}{@{}lp{34pc}@{}}
AGN & Active Galactic Nuclei \\
IMF & Intial Mass Function \\
IR & Infra-Red \\
ISM & Inter Stellar Medium \\ 
JWST & James Webb Space Telescope \\
NIR & Near Infra-Red \\
SFR & Star formation rate\\
SMBH & Super-Massive Black Hole \\
SN & SuperNova\\
sSFR & specific SFR (SFR/M$_\star$) \\
UV & Ultra-Violet \\
\end{tabular}
\end{glossary}


\section{Objectives}
The layout of the chapter is as follows:
\begin{itemize}
\item Section~\ref{chap1:sec1} gives a general introduction to the subject;
\item Section~\ref{chap1:sec2} provides a detailed overview of available observational information about the abundance of quenched galaxies, at different cosmic epochs, and dependence as a function of galaxy physical properties and/or environment. The section also summarizes uncertainties and limits of different methods to select quiescent galaxies.
\item In Section~\ref{chap1:sec3} we give an overview of the physical processes that are responsible for the suppression of star formation in galaxies, of their impact at different mass scales and in different environments, and of the associated time-scales.
\item In Section~\ref{chap1:sec4}, we summarize our current understanding about the physical processes driving quenching, based on detailed comparisons between observational measurements and predictions from state-of-the-art theoretical models. 
\item Section~\ref{chap1:sec5} provides an outline of what we believe are the future prospects in this field.
\end{itemize}

\section{Introduction}\label{chap1:sec1}

A fraction of the galaxy populations at all cosmic epochs are forming stars at very low or negligible rates. These galaxies are usually referred to as passive, quiescent, or quenched to indicate that some physical process has suppressed their star formation activity at some point in their past. These terms are often used interchangeably in the recent literature, although it should be noted that, depending on the observational information available, it might be difficult to say when in the past a galaxy's star formation has been suppressed, or what is the probability that the same galaxy will start again forming stars at some significant rate in the future.  In this chapter, we will use the terms passive and quiescent interchangeably, with notes of caution where we deem it appropriate. We will use the term quenched each time we refer to a specific physical process that has caused quenching. 

Since the luminosity of stars scales with their mass, and the most massive stars tend to be the shortest lived and the bluest, the starlight from a galaxy that forms stars at a non negligible rate tends to be blue. In contrast, old stellar populations are associated with red colors. However, estimating the age of a stellar population from its colors only is not straightforward because of the well known age-dust-metallicity degeneracy: extinction becomes more important at shorter wavelengths thus making the rest-frame colour redder; increasing metallicities also lead to redder colors because the effective temperatures of stars decrease due to the increasing opacity in the stellar atmosphere. Therefore, an unambiguous identification of passive galaxies requires additional observational information to break the above mentioned degeneracies. As we will discuss in more detail below, this translates into a largely heterogeneous amount of information about the incidence and properties of quiescent galaxies at different cosmic epochs. 

In the local Universe, where star formation can be measured accurately for large samples of galaxies using a variety of indicators, galaxies are found to follow a bimodal distribution \citep[e.g.][]{Blanton_etal_2003,Kauffmann_etal_2003,Baldry_etal_2004a}, with an intermediate region (sometimes dubbed `green valley') that is not well populated, and whose location does not vary significantly when considering galaxies in different environments. The fraction of passive galaxies is found to rise steeply with increasing stellar mass and to depend on the environment in which galaxies reside \citep{Kauffmann_etal_2004, Balogh_etal_2004, Baldry_etal_2006}. At earlier cosmic epochs, accurate measurements of the star formation are more difficult to obtain for statistical samples of galaxies, and a colour-colour selection is usually employed to perform a separation between star forming and quiescent galaxies.  This leads to general trends similar to those observed in the local Universe, with an overall decrease of the fraction of quiescent galaxies at earlier cosmic epochs \citep[e.g.][]{Brammer_etal_2009,Whitaker_etal_2011,Muzzin_etal_2013,Weaver_etal_2023}. The advent of the James Webb Space Telescope (JWST) has opened a new window into the early Universe, revealing a large population of quiescent candidate galaxies during the first two billion years of our Universe, in excess of previous observational inferences \citep{Carnall_etal_2023, Valentino_etal_2023, Nanayakkara_etal_2024}. At the time of writing, many massive (galaxy stellar mass larger than $10^{10.5} \,{\rm M}_{\odot}$) quiescent galaxies have been spectroscopically confirmed \citep{Carnall_etal_2024, deGraaff_etal_2024, Setton_etal_2024}, with the highest reported redshift being $\sim 7.3$ \citep{Weibel_etal_2024}. The very rapid assembly and quenching of such massive galaxies at these early cosmic epochs potentially represent an important challenge for theoretical models of galaxy formation. In Section~\ref{chap1:sec2} below, we will provide a detailed overview of available observational information at different cosmic epochs, summarizing uncertainties and limits of different methods to select quiescent galaxies. 

A plethora of physical processes can efficiently suppress star formation in galaxies. Traditionally, these processes have been often categorized as `internal' processes like e.g. star formation and stellar feedback or `environmental' processes to refer to processes that are at play for galaxies that reside in over-dense regions (e.g. ram-pressure stripping, harassment). The former are associated with what is sometimes referred to as `mass quenching' or `central quenching', while the latter are associated with `environmental quenching' or `satellite quenching'. In Section~\ref{chap1:sec3} we will give an overview of the physical processes that are at play, of their impact at different mass scales and in different environments, and of the relative time-scale associated with the suppression of star formation. One important aspect to bear in mind is that these physical processes do not act in isolation, but are rather entangled in a complex network of actions, back-reactions and self-regulation, which makes it difficult to firmly establish the relative importance of different processes. In addition, the impact of a specific physical process might strongly depend on the environment in which galaxies reside. For example, supernovae (SN) feedback can efficiently move gas from the densest star forming regions towards the galaxy outskirts. If the galaxy resides in an over-dense region, gas can be efficiently removed from the ouskirts of the galaxy by ram-pressure making it easier for SN feedback to drive out larger fractions of gas \citep{Bahe_McCarthy_2015, Bustard_etal_2018}. Finally, it is important to note that it would be in principle possible to separate mass from environment if they are correlated but physically uncoupled. However, in the current standard cosmological paradigm, dark matter collapses in a bottom-up fashion: small haloes form first and progressively merge into larger and larger systems. As structure grows, galaxies are accreted onto more massive haloes therefore experiencing a variety of environments during their lifetimes \citep{DeLucia_etal_2012}. In this context, mass and environment are inevitably and heavily intertwined making it very difficult to separate the effect of internal and external processes.

Over the past years, we have learned much about the physical processes driving galaxy quenching at different scales and in different environments. This requires a careful and detailed comparison with sophisticated theoretical models that attempt to incorporate, simultaneously and in a realistic fashion, both the assembly of the cosmic structures and the different physical processes at play. This can be achieved through direct hydro-dynamical simulations that include an explicit description of gas dynamics in N-body simulations \citep[for recent reviews, see e.g.][]{Somerville_and_Dave_2015,Naab_and_Ostriker_2017,Vogelsberger_etal_2020}. Alternatively, the evolution of the baryonic components of galaxies can be described by coupling dark matter merger trees extracted using high-resolution cosmological dark matter only simulations or based on extended Press-Schechter formalism with physically and/or observationally motivated prescriptions. This approach is typically referred to in the literature as `semi-analytic' \citep[for details see e.g.][and references therein]{Baugh_etal_2006, DeLucia_etal_2019}. The former method comes with relatively high computational costs, in particular for large cosmological volumes and high resolution. In contrast, the latter technique gives access to a large dynamic range in mass and spatial resolution at limited computational costs, allowing an efficient exploration of the parameter space and of the influence of specific physical assumptions. However, assumptions need to be made to compensate for the lack of an explicit treatment of gas dynamics, and for the lack of information about the spatial distribution of the different baryonic components \citep[see][for attempts to model self-consistently the spatial distribution of baryons in semi-analytic models]{Stevens_etal_2016,Yates_etal_2021}. Both methods have to resort to `sub-grid' prescriptions of the relevant physical processes that incorporate a number of free parameters and that are typically calibrated against a subset of observational results in the local Universe. We refer the reader interested in these techniques to the references given above and note that, in the following, we will refer to both approaches as theoretical models distinguishing between the numerical approach used only when necessary. In Section~\ref{chap1:sec4}, we will summarize what we have learned about the physical processes driving quenching, through a detailed comparison of observational measurements and predictions from state-of-the-art theoretical models. 

As mentioned above, new instrumentation has finally allowed spectroscopic confirmation of quiescent galaxies at $z>3$. The combination of deep spectra and photometry has also allowed a first assessment about some physical properties of these galaxies like metallicity and abundance ratios and of their star formation and assembly histories. This is a very rapidly moving subject, with new results appearing on an almost daily basis. In the next sections, we will focus on the evolution of the subject in the last two decades, reserving ample space to the most recent developments, with the caveat that this will soon become out-dated. In Section~\ref{chap1:sec5}, we conclude with an outline of what we believe are the future prospects in this field.  

\section{Observational overview}\label{chap1:sec2}

The first step to identify quiescent galaxies is to measure a galaxy's star formation rate (SFR). This is a fundamental galaxy property that enters many scaling relations, and for which several indicators can be used across the full electromagnetic spectrum \citep[see e.g.][]{Kennicutt_and_Evans_2012,Calzetti_etal_2013}. The goal is to trace emission that comes from newly or just recently formed stars, while avoiding as much as possible the contribution from evolved stellar population or Active Galactic Nuclei (AGN). The emission by young stellar population can be observed directly in the ultraviolet (UV) which, however, is strongly affected by dust attenuation. In addition, a correct calibration of the UV as a star formation indicator depends strongly on the shape of the stellar Initial Mass Function (IMF), on metallicity, and on the adopted modelling for the emission from massive stars that can be complicated by stellar rotation and binaries. At the other end of the electromagnetic spectrum, far infrared emission can be used as a calorimetric measure of the total star formation rate (the IR luminosity depends on the heating rate provided by the stars). There are several caveats to consider also in this case: e.g. not all emission by young stars is absorbed by dust, there might be a contribution from an AGN dust torus, and dust can be heated by old stellar populations. 
The SFR of galaxies can also be obtained combining information at different wavelengths and attempting to fit their full spectral energy distributions \citep[e.g.][and references therein]{Conroy_etal_2013,Chevallard_and_Charlot_2016,Carnall_etal_2018}. This approach requires assumptions about the star formation and metallicity history of the galaxies and an explicit modeling of the stellar populations (this, in turn, includes a treatment of stellar evolution, assumptions about the IMF, etc.). Different studies have shown that SFR estimates based on broad-band photometry are typically associated with large uncertainties (also due to the above mentioned degeneracies) and appear to depend strongly on the templates library adopted and on the assumptions made for the sed-fitting \citep[e.g.][]{Pacifici_etal_2015,Leja_etal_2019}. Finally, star formation can also be estimated by measuring the starlight that has been reprocessed by interstellar gas or dust or employing tracers of the death of massive stars, like optical and near-IR emission lines from ionized gas surrounding massive young stars. H$\alpha$ remains in this case the most widely used SF indicator, for both local and distant galaxies, although other lines have been considered and used ([OII], Paschen and Brackett series lines, [CII], etc.). These estimators are not free from uncertainties either. In particular, there might be large systematic errors due to dust attenuation and/or sensitivity to the massive end of the IMF especially in regions with low absolute values of SFRs. In summary, an accurate measurement of the SFR is very difficult to obtain; the statistical uncertainties of the recovered measurements depend strongly on the quality and kind of information available (e.g. photometric versus spectroscopic); and there might be systematic uncertainties affecting the measurements in specific regimes of the SFR - this is especially true for the regime corresponding to low SFR values, which is the subject of this chapter. 

In the local Universe, large and deep spectroscopic redshift surveys have allowed a statistical and accurate characterization of the galaxy population. E.g. the Sloan Digital Sky Survey (SDSS) has provided spectra for about one million galaxies spread over more than 7000 deg$^2$ of the sky \citep{York_etal_2000,Abazajian_etal_2009}, providing an accurate mapping of the nearby Universe out to $z\sim 0.3$. The Galaxy and Mass Assembly (GAMA; \citealt{Driver_etal_2011,Liske_etal_2015}) survey covers a smaller area than SDSS but provides spectra for galaxies about two magnitudes fainter. At higher redshift, spectroscopic information is available for smaller (and often biased) samples of galaxies and becomes extremely expensive in terms of telescope time for galaxies below the knee of the mass function. Since colors can be more easily computed for statistical samples of galaxies, several broad-band techniques have been designed to efficiently separate different galaxy populations. One selection that has been widely used in the past years is based on the rest-frame $U-V$ versus $V-J$ color-color diagram \citep{Labbe_etal_2005,Williams_etal_2009}: these studies have shown that a separation based on this color space leads to similar results of a separation based on UV+IR estimates of the specific star formation rates at least up to $z\sim 2.5$. At higher redshift, the $UVJ$ selection is still characterized by a high purity (i.e. there are few star-forming interlopers because e.g. of dust), but it becomes increasingly incomplete missing either galaxies that are both old and significantly obscured by dust, or galaxies that are devoid of dust but whose star formation has been abruptly quenched in the last few million years \citep{Merlin_etal_2018,Schreiber_etal_2018}. Alternative selections based on the $NUV-r$ color, instead of $U-V$, have been proposed. This has the advantage of offering a larger dynamical range, a better separation between dusty star forming galaxies and quiescent galaxies, and being still sampled by optical data at $z>2$ \citep{Arnouts_etal_2013,Ilbert_etal_2013}. Rest-frame colours can be estimated from available photometry or extrapolated using templates that fit the observed-band photometry. Therefore, the approach suffers from uncertainties and systematics due to available photometry, as well as from assumptions related to the sed-fitting. In addition, and most importantly, different selections imply different definition of quiescence so that comparisons among different samples and/or at different cosmic epochs should be interpreted with caution.  

\begin{figure*}
    \centering
    \includegraphics[width=0.85\linewidth]{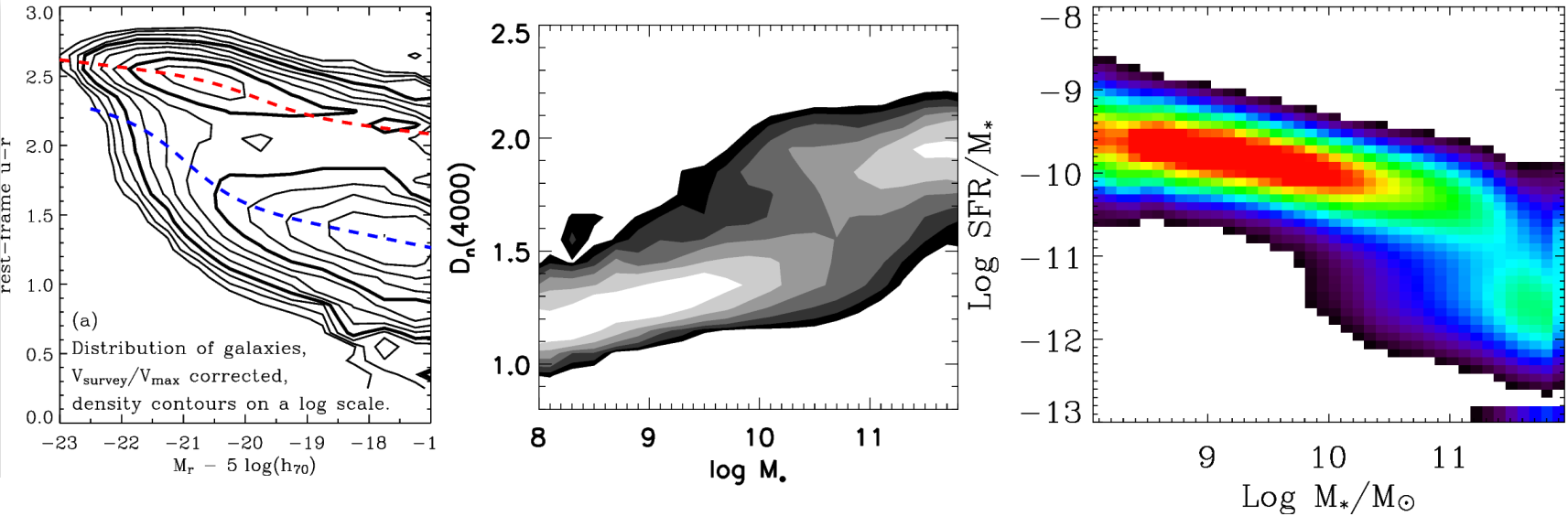}
    \caption{All panels are based on data from SDSS and have been corrected for incompleteness. Left panel: adapted from \citet{Baldry_etal_2004b}, shows the observed colour-magnitude distribution. The contours are on a logarithmic scale in number density, doubling every two levels. The dashed lines represent the colour-magnitude relations of the red and blue sequences obtained by considering the peaks of the colour distributions in bins of luminosity. Middle panel: adapted from \citet{Kauffmann_etal_2003}, shows the distribution of D4000 as a function of stellar mass. Right panel: adapted from \citet{Brinchmann_etal_2004} shows the distribution of SFRs as a function of stellar mass. We refer to the original papers for details.}
    \label{fig:bimod}
\end{figure*}

First results based on SDSS revealed a bimodal distribution in the optical color-luminosity diagram~\citep{Baldry_etal_2004a}, and in the distribution of the SFR and D4000 \footnote{This is a strong discontinuity in the optical spectrum of a galaxy that arises due to the accumulation of many spectral lines in a narrow wavelength region, with the main contribution coming from ionized metals. Since in hot stars elements are multiply ionized, the opacity decreases leading to a smaller break. Thus, small/strong breaks are associated with young/old stellar populations, respectively.} break~\citep{Kauffmann_etal_2003,Brinchmann_etal_2004, Driver_etal_2006}. Figure~\ref{fig:bimod} is a composition of figures from \citet{Kauffmann_etal_2003}, \citet{Baldry_etal_2004b}, and \citet{Brinchmann_etal_2004} showing the (bimodal) distribution of galaxies in the colour-magnitude, D4000-galaxy stellar mass, and SFR-galaxy stellar mass diagrams. The distributions are found to depend strongly on stellar mass, with low mass galaxies dominating the regions occupied by star forming, young, and blue stellar populations and more massive galaxies being predominant among stellar populations dominated by low levels of star formation, old ages and red colors. It has also long been known, and it has been confirmed for statistical samples of galaxies, that galaxies dominated by old stellar populations tend to have an early-type morphology \citep{Yi_etal_2005, Bremer_etal_2018} and lower amounts of neutral gas \citep[][and references therein]{Saintonge_Catinella_2022}. Massive quenched central galaxies also tend to have a denser core in the inner $\sim 1$~kpc \citep{Fang_etal_2013}, and are often associated with relativistic jets visible in the radio band \citep[][and references therein]{Best_etal_2005,Best_etal_2007} that are triggered by accreting super-massive black holes (SMBHs). These jets can play an important role in regulating the star formation histories of the galaxies at the centre of the most massive haloes by reducing or suppressing cooling of the intracluster medium \citep[for reviews, see e.g.][]{McNamara_and_Nulsen_2007, Fabian_2012}.  The relatively tight scaling relation found between the mass of the bulge and that of the central SMBH has been traditionally interpreted as evidence of an important role played by AGN feedback in regulating the star formation of their host galaxies, eventually leading to quenching \citep[e.g.][]{Kormendy_Ho_2013}. In addition to the demographics of radio-loud galaxies mentioned above, this hypothesis is also supported by direct observations of large galactic-scale molecular outflows associated with luminous quasars \citep{Feruglio_etal_2010,Bischetti_etal_2019} that can potentially remove all molecular gas from massive systems \citep{Fiore_etal_2017}.

\begin{figure*}
    \centering
    \includegraphics[width=0.9\linewidth]{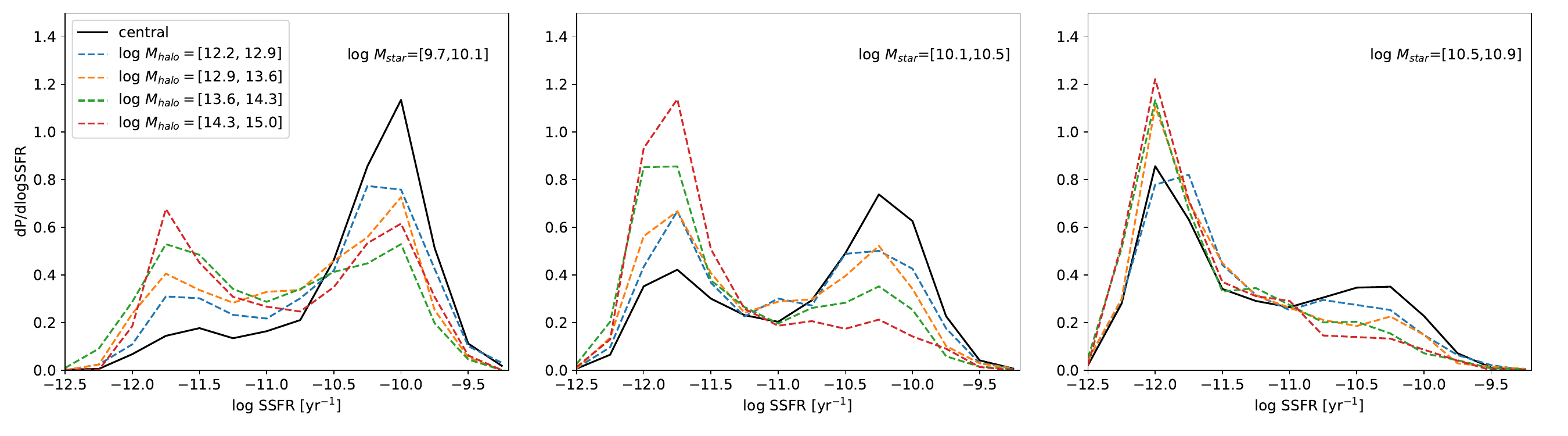}
    \caption{Adapted from Fig.~1 in \citet{Wetzel_etal_2012}. The figure shows the specific SFR distributions for central galaxies (solid black lines) and satellite galaxies (dashed coloured lines) in three bins of stellar masses (different panels) and four bins of halo mass (different colours). For all galaxies, independently of their hierarchy and halo mass, the distribution is bimodal with a similar peak at large values of the sSFR and a similar fraction of galaxies at the break of the distribution, around $10^{-11}\,{\rm yr}^{-1}$. The figure has been remade using the SDSS DR7 catalogue \citep[model C of][]{Yang_etal_2007, Yang_etal_2008, Yang_etal_2009, Yang_etal_2012} publicly available at https://gax.sjtu.edu.cn/data/Group.html.}
    \label{fig:sSFR_distribution}
\end{figure*}

The location of the intermediate break of the sSFR distributions, around $10^{11}\ {\rm yr}^{-1}$, does not vary significantly as a function of stellar mass or halo mass \citep{Wetzel_etal_2012}, while the overall fraction of quiescent galaxies increases with halo mass and local environmental density \citep{Kauffmann_etal_2004,Balogh_etal_2004,Baldry_etal_2006}. Figure~\ref{fig:sSFR_distribution}, adapted from \citet{Wetzel_etal_2012}, shows the sSFR distribution for three bins of stellar mass (different columns) and for satellites residing in haloes of different mass (dashed lines of different colours, as indicated in the legend). The thick black lines in each panel show the corresponding distributions for central galaxies only. As noted in the original paper by \citet{Wetzel_etal_2012}, all galaxies exhibit a similar high sSFR peak and there is no specific environment that exhibits an excess of transition galaxies. These trends suggest that the rate at which galaxies convert gas into galaxies is largely unaffected by the environment for several Gyrs and that, when quenching starts, it must occur on a very rapid time-scale. These findings, that as we will see below have long represented a challenge for theoretical model of galaxy formation, led to the formulation of a  `delayed-then-rapid' quenching scenario for satellite galaxies \citep[see also][]{Wetzel_etal_2013}.

\begin{figure*}
    \centering
    \includegraphics[width=0.9\linewidth]{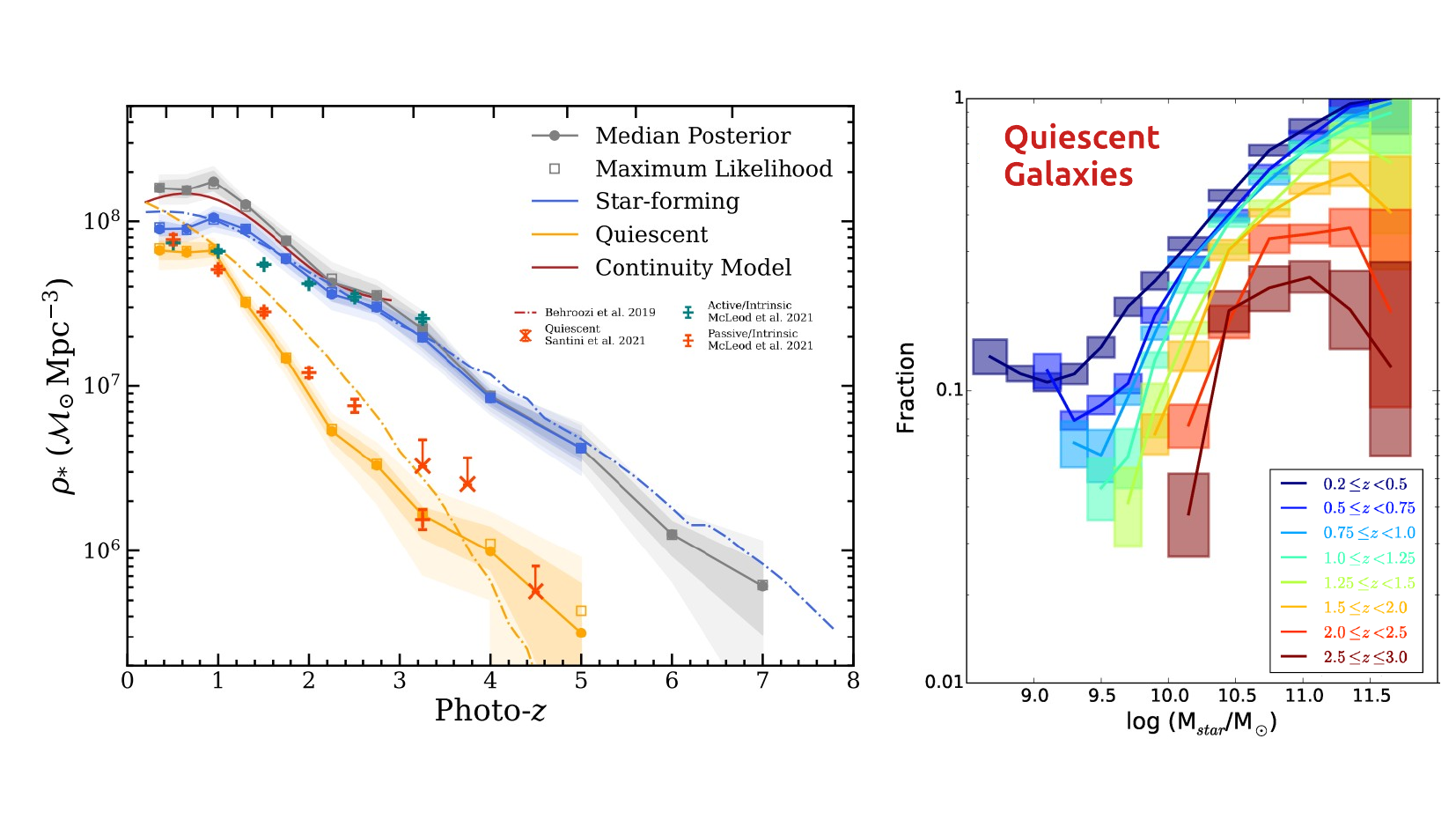}
    \caption{The left panel, adapted from \citet{Weaver_etal_2023}, shows the evolution of the cosmic stellar mass density for all galaxies with stellar mass larger than $10^8\,{\rm M}_{\odot}$ (in gray), only star forming galaxies (in light blue), and quiescent galaxies (in orange). Observational estimates presented in the original paper are compared with other literature measurements, as indicated in the legend. The right panel, adapted from \citet{Martis_etal_2016}, shows the fraction of quiescent galaxies as a function of stellar mass and at different redshifts, as indicated in the legend.}
    \label{fig:quiesclu}
\end{figure*}

Large photometric and spectroscopic surveys such as COMBO-17 \citep{Wolf_etal_2003}, GOODS \citep{Giavalisco_etal_2004}, DEEP \citep{Weiner_etal_2005}, VVDS \citep{LeFevre_etal_2005}, COSMOS \citep{Scoville_etal_2007} and zCOSMOS \citep{Lilly_etal_2007} have allowed the extension of these studies to higher redshifts. The color-bimodality persists out to $z\sim 2$ and potentially even at higher redshift \citep{Bell_etal_2004,Williams_etal_2009,Brammer_etal_2011,Muzzin_etal_2013,Weaver_etal_2023} with the mass density of star forming galaxies remaining approximately flat and decreasing below $z\sim 1$ and the mass density of quiescent galaxies with $M_{\rm star} > 3\times 10^{10}\, {\rm M}_{\odot}$ increasing by a factor $\sim 10$ below $z\sim 2$. The left panel of Figure~\ref{fig:quiesclu}, adapted from \citet{Weaver_etal_2023}, shows the evolution of the cosmic stellar mass densities for both quiescent and star forming galaxies from a few recent studies. These densities have been computed integrating the observed galaxy stellar mass functions down to $\sim 10^8\,{\rm M}_{\odot}$ (it should be noted that observational samples are not complete down to this limit so these estimates are based on extrapolations). The fraction of quiescent galaxies decreases at earlier cosmic epochs and is a strong function of galaxy stellar mass at all epochs, as shown in the right panel of Figure~\ref{fig:quiesclu} (adapted from \citealt{Martis_etal_2016}). Galaxy growth appears to be mass-dependent, with the most massive galaxies assembling most of their mass at high-redshift and low-mass galaxies growing more slowly - a trend that is usually referred to as `downsizing' or `anti-hierarchical' \citep[for a discussion, see][]{Fontanot_etal_2009, Hirschmann_etal_2016}. Red galaxies are found to populate a rather tight colour-magnitude relation that was originally identified in galaxy clusters \citep[][and references therein]{Bower_etal_1998}, but that is found to exist also when averaging over all environments. Red galaxies are also found to be more strongly clustered than those actively forming stars \citep{Li_etal_2006, Hartley_etal_2010, Mostek_etal_2013, Coil_etal_2017}, which is typically taken as an indication that red galaxies reside in more massive dark matter haloes than their bluer counter-parts of similar stellar mass. 

\begin{figure*}
    \centering
    \includegraphics[width=1.05\linewidth]{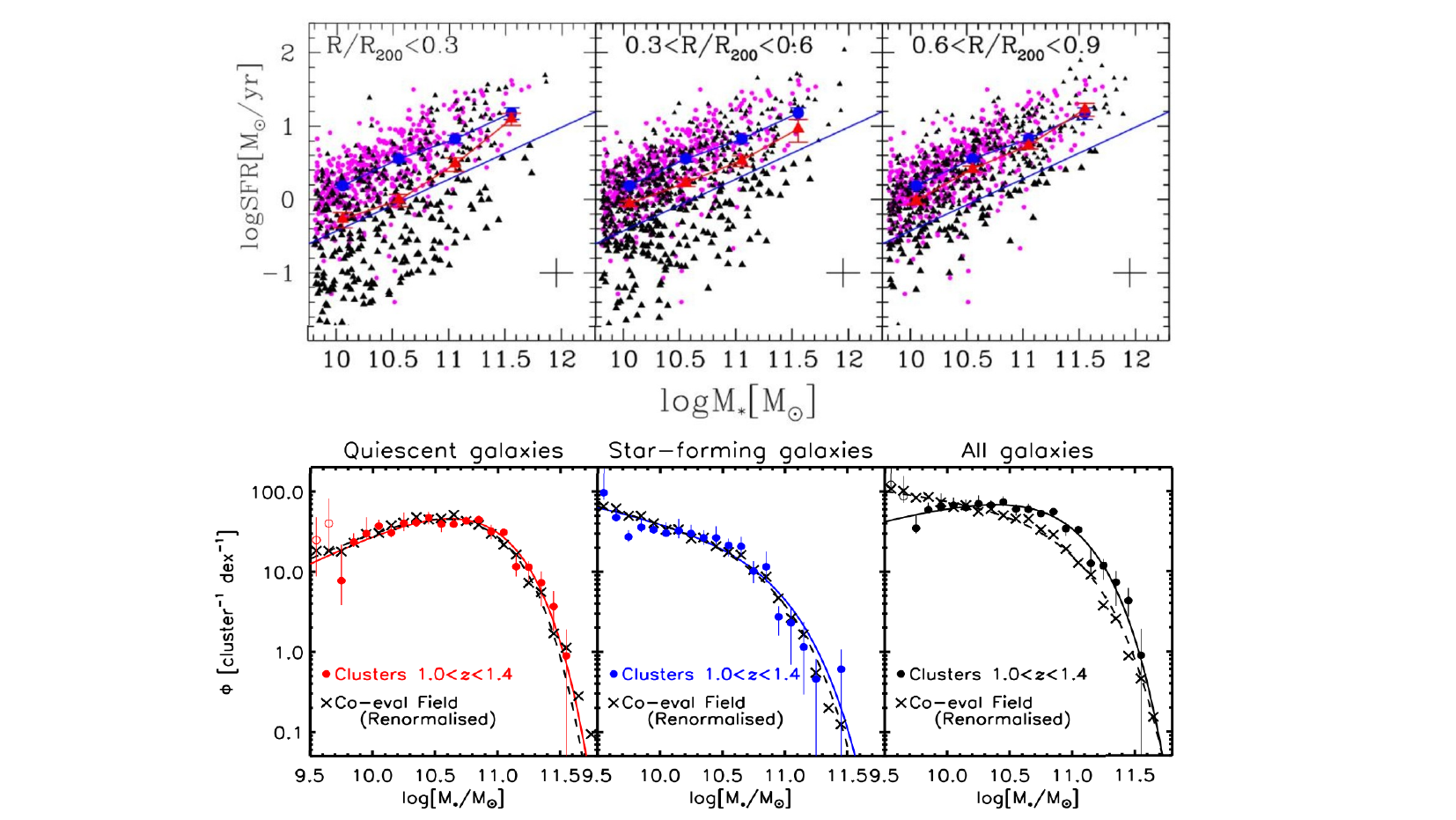}
    \caption{The top panel, adapted from \citet{Paccagnella_etal_2016}, shows the distribution in the SFR-galaxy mass plane of galaxies in three bins of cluster-centric distance. Black triangles and magenta circles show galaxies in a sample of nearby clusters and in the field, respectively. The bottom panel, from \citet{vanderBurg_etal_2020}, shows the galaxy stellar mass function from the GOGREEN cluster survey, compared with the coeval field measurements. The three panels show the galaxy stellar mass function for all galaxies cluster and filed galaxies (right), and for the quiescent and star forming population (left and middle, respectively).}
    \label{fig:radclu}
\end{figure*}

Taking a purely empirical approach, \citet{Peng_etal_2010} argued that the effect of mass quenching and environmental quenching are `separable' at least out to $z\sim 0.6$. This conclusion was driven by the finding that, when considering over-densities on scales that correspond to about one comoving Mpc for typical galaxies, the average SFR depends only on galaxy mass, with no residual dependence on the environment. This is in agreement with the results by \citet{Wetzel_etal_2012} discussed above. However,  there is no consensus on this topic: a number of studies have shown that even star-forming galaxies exhibit an environmental dependence of the SFR \citep[e.g.][]{vonderLinden_etal_2010, Vulcani_etal_2010, Tran_etal_2015, Erfanianfar_etal_2016} and that there is a non-negligible population of galaxies in clusters with reduced SFR with respect to the field. The top panel of Figure~\ref{fig:radclu}, adapted from \citet{Paccagnella_etal_2016}, shows the distribution in the SFR-galaxy mass plane of galaxies in three bins of cluster-centric distance. Black triangles and magenta circles show galaxies in clusters and in the field, respectively. The study is based on a large sample of nearby galaxy clusters and measurements of SFR are obtained by fitting available spectroscopy with a spectrophotometric model and accounting for all main spectrophotometric features (we refer to the original paper and references therein for details). The figure shows that galaxies with reduced SFR are more abundant in the central region of the clusters. For these galaxies, the transition from star forming to passive must occur on a sufficiently {\it long} time-scale for us to detect them in the process of being quenched. One possible reason for the apparently contradictory results on this matter could be related to different possible definition of `environment' (this can be estimated e.g. through some evaluation of the local galaxy density, parent halo mass, distance from the halo centre in massive clusters, etc.). In fact, the description proposed by \citet{Peng_etal_2010} fails in very dense environments \citep[e.g.][]{Darvish_etal_2016,Pintos-Castro_etal_2019, vanderBurg_etal_2020}. The bottom panel of Figure~\ref{fig:radclu} shows the galaxy stellar mass function from the GOGREEN cluster survey \citep{Balogh_etal_2017,Balogh_etal_2021}, compared with the coeval field measurements. The figure shows that the galaxy stellar mass functions of both quiescent and star forming galaxies in clusters are identical to those in the (coeval) field while the total galaxy stellar mass function in the clusters is radically different from that in the field. This is due to the different fractions of quiescent galaxies in the two environments. As discussed in detail in \citet{vanderBurg_etal_2020}, this cannot be reproduced in the framework of a mass-independent quenching scenario for infalling galaxies. Another important aspect to consider about environmental processes is that they likely happen in a `probabilistic way': the time-scale of star formation suppression will depend strongly on the orbital distribution of accreted satellites with more radial orbits likely leading to shorter quenching time-scales due to the stronger tidal and ram-pressure stripping in the inner regions of groups/clusters, although this expected behavior can be complicated significantly by rapid encounters with other group/cluster members \citep{Villalobos_etal_2012,Villalobos_etal_2014,Lotz_etal_2019}.

Historically, observational (and theoretical) studies trying to assess the role of environment on galaxy evolution have been mainly focused on galaxy clusters. The primary reason for this is the practical advantage of having many galaxies in a relatively small region of the sky and all approximately at the same redshift. This allows efficient observations to be carried out, even with modest fields of view and modest amount of telescope time. Since the seminal work by \citet{Dressler_1980}, it is well established that galaxies in clusters are dominated by early-type, quiescent galaxies. As mentioned above, systematic differences are measured also in the late-type population, with spiral galaxies in clusters having on average a lower atomic and molecular gas content than their counter-parts in the average field \citep{Cayatte_etal_1990,Fumagalli_etal_2009,Catinella_etal_2013,Boselli_etal_2014}. As a consequence, their star formation activity is also reduced, especially in galaxies that appear to be deficient of their molecular gas (the fuel for star formation). As we will see in the next section, there are several mechanisms that are expected to be effective in over-dense regions. Disentangling the processes responsible for the observed correlations has proved difficult, and it remains unclear to what extent the observed relations are imprinted during formation or by physical processes at work in dense environments \citep{DeLucia_2011,Boselli_etal_2022}. Over the past decades, the characterization of the population of cluster galaxies has been extended to higher and higher redshift showing significant evolution in the morphological mix \citep{Dressler_etal_1997,Postman_etal_2005} and overall fraction of blue/star-forming galaxies \citep{Butcher_and_Oemler_1984,Loh_etal_2008,Poggianti_etal_1999,Poggianti_etal_2006}. Taken at face value, the observed trends suggest that high-redshift gas-rich and star forming galaxies have had their star formation suppressed at the time they have been accreted onto the cluster environment. The extent to which this suppression is important, and how this is associated to morphological transformation is still not well understood. 

In recent years, multi-wavelength surveys have progressively extended the discovery space for galaxy clusters to higher and higher redshift, enabling the study of the cluster galaxy population out to $z\sim 1.5$. At these early cosmic epochs, observational studies are plagued by small and heterogeneous samples with a poorly defined selection function, and with data often being unable to robustly estimate cluster mass or virialization status.  Current evidence suggests that the fraction of quiescent galaxies is very high among massive galaxies in clusters up to $z\sim 1$ \citep{Mei_etal_2009,vanderBurg_etal_2020}, and that these are already more evolved than their field counterparts \citep{Nantais_etal_2017,Lee-Brown_etal_2017}. The epoch around $z\sim 1.5-2$ appears to mark a critical transition for galaxy clusters: quiescent galaxies are still found in the cores of massive systems up to $z\sim 2$ \citep[e.g.][]{Strazzullo_etal_2010,Andreon_etal_2014,Strazzullo_etal_2019,Willis_etal_2020}, although with a large cluster-to-cluster variation, while significant star formation activity is often observed in massive cluster galaxies at $z>1.3$, sometimes even within the cluster cores \citep[e.g.][]{Brodwin_etal_2013,Tran_etal_2015,Alberts_etal_2016,Klutse_etal_2024}. The epoch at $z>2$, is largely populated by `proto-clusters', very extended over-densities that we believe are in the process of collapse (and will eventually collapse into a massive cluster by present time), and that host substantial star formation (for a review, see \citealt{Overzier_2016} and \citealt{Alberts_and_Noble_2022}). Unfortunately, there is no good operational definition of proto-clusters. This is in part due to observational challenges posed by the nature of these objects: their ICM might not be heated sufficiently to be detected via X-ray or SZ \citep[but see][]{DiMascolo_etal_2023}, and they extend over large portions of the sky (up to 10-30') so that identifying them through galaxy over-densities requires wide and very deep surveys and good spectroscopic information. To circumvent these demanding requests, proto-clusters have often been identified using rare populations (e.g. dusty star forming galaxies, or high-z radio galaxies and quasars) as biased tracers of the underlying matter distribution. Although the field is progressing rapidly, studies of proto-clusters are still based on limited and heterogeneous samples, preventing us from obtaining a clear view of how the environment affects galaxy properties in these early stages of cluster formation. 

\begin{figure*}
    \centering
    \includegraphics[width=0.85\linewidth]{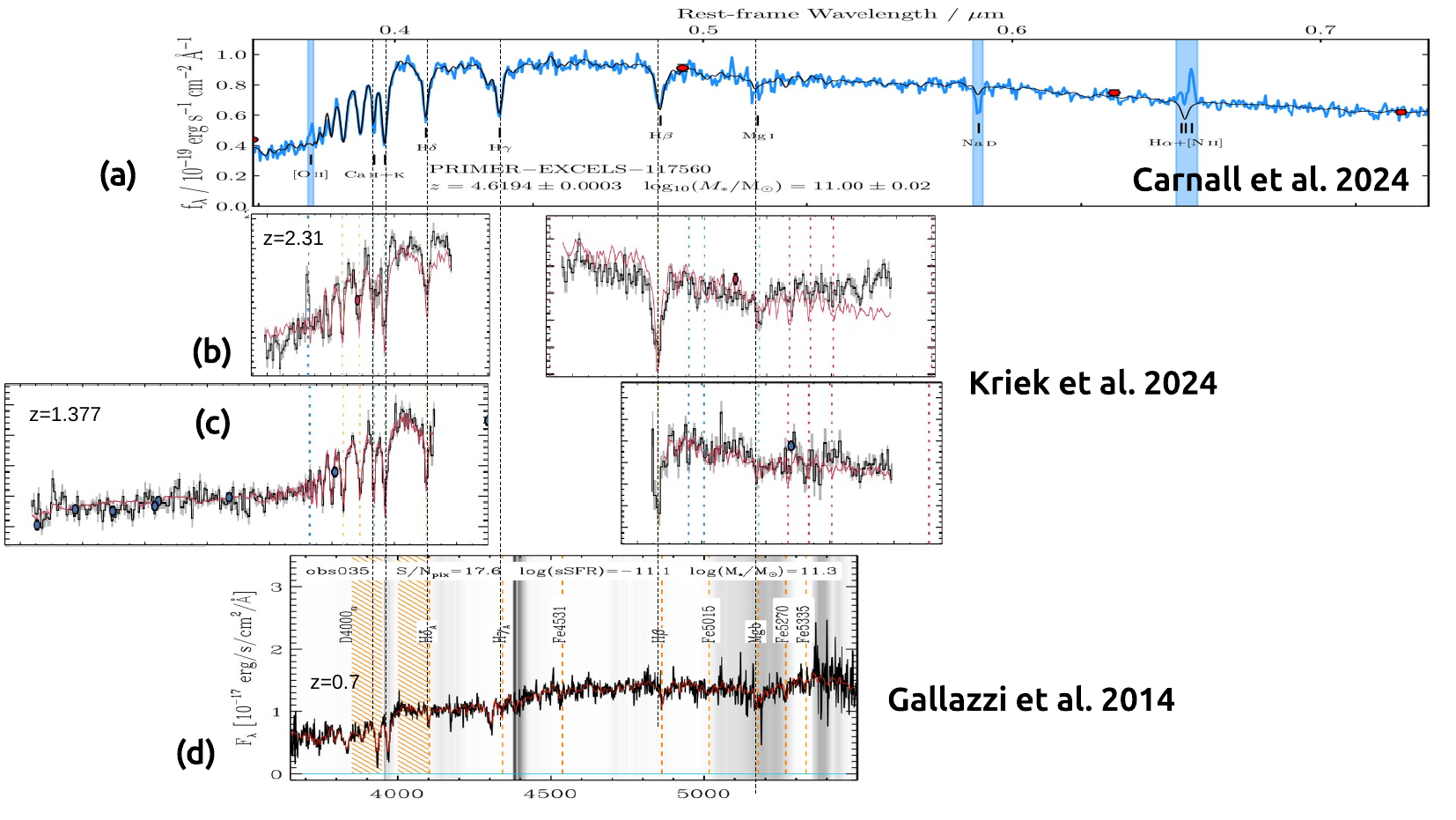}
    \caption{From top to bottom (a) NIRSpec observations of a massive quiescent galaxy at $z \sim 4.6$ (adapted from \citealt{Carnall_etal_2024});  (b) MOSFIRE J-band (12-14 hr) and H-band spectra (16-17 hr) of a massive quiescent galaxy at $z \sim 2.3$ (adapted from \citealt{Kriek_etal_2024}); (c) LRIS (4-6 hr) and MOSFIRE J band spectra (12 hr) of a quiescent galaxy at $z \sim 1.4$ (adapted from \citealt{Kriek_etal_2024}); (d) IMACS@Magellan 10 hr spectrum of a massive quiescent galaxy at $z \sim 0.7$ (adapted from \citealt{Gallazzi14}).}
    \label{fig:spectral_comparison}
\end{figure*}

The advent of JWST has allowed us to push studies of quenched galaxies to unprecedented epochs: observations from space are not affected from the atmospheric cutoffs in the NIR and high thermal backgrounds that limit ground based observations of high redshift galaxies. In particular, JWST allows the exploration of the spectral range redward of 0.6~$\mu$ that corresponds to the restframe optical-to-ultraviolet at $z>4$. Early results from JWST revealed a large population of massive and apparently quiescent galaxies at $z > 3$, in excess of previous observational inferences although the reported number densities have a large scatter due to different selection criteria and cosmic variance \citep{Carnall_etal_2023,Valentino_etal_2023,Weaver_etal_2023,Nanayakkara_etal_2024}. The number of quiescent candidates being spectroscopically confirmed has been increasing rapidly in the past few years. Available spectroscopy provides a first assessment the intrinsic properties of these galaxies, that appear to have experienced short periods of intense star formation, leading to a rapid mass growth in the first $1-2$ billion years of the Universe, and then to have suffered a rapid decline of their star formation within a few tens of million years. Figure~\ref{fig:spectral_comparison} shows one example of an ultra-deep, medium-resolution spectrum of a massive quiescent galaxy at $z\sim 4.6$ in the top panel (adapted from \citealt{Carnall_etal_2024}). The spectrum has been obtained as part of the EXCELS survey, using NIRSpec onboard of JWST and targeting $120$ galaxies before cosmic noon (we refer to the original paper for more details about the survey). As a comparison, we show in the other panels ground based spectra of quiescent galaxies of similar mass ranging from $z\sim 0.7$ (panel d, adapted from \citealt{Gallazzi14}) to $z\sim 2.3$ (panel c, adapted from \citealt{Kriek_etal_2024}). The spectra in panel (b) and (c) come from the Heavy Metal Survey conducted with MOSFIRE and LRIS on the Keck I telescope. This sample of 21 massive quiescent galaxies at $1.3 < z < 2.3$ required integration times of up to $16$ hours per band per galaxy against the total integration times within each grating (3 gratings in total) and each pointing (four in total) of the EXCELS survey of $\sim 4-5.5$~h. 

A fraction of the massive quenched galaxies at high-z are found to host luminous AGN \citep{Park_etal_2024, Baker_etal_2024}, suggesting that this physical process might play an important role in galaxy quenching at early cosmic epochs. Some quenched galaxies are found in overdense environments \citep[e.g.][]{Jin_etal_2024, Kakimoto_etal_2024}, suggesting that environmental quenching may also contribute as early as $z\sim 4$. However, much work remains to be done to firmly establish the contribution from AGN and/or environment in quenching galaxies at these early epochs. The (likely related) questions of (i) how such massive systems assemble at such early cosmic epochs and (ii) which physical mechanisms are responsible for the (abrupt) end of their star formation activity represent important challenges for modern theoretical models of galaxy formation, that struggle to reproduce the estimated large number densities of massive high-z passive galaxies and/or their very short formation times \citep{Lagos_etal_2024b}. We will come back to this in Section~\ref{chap1:sec4}.

\section{Physical processes}\label{chap1:sec3}

From the theoretical viewpoint, many different physical mechanisms could lead to galaxies experiencing a decline in their star formation.  As mentioned in section ~\ref{chap1:sec1}, we can broadly consider two different groups: internal physical processes that act in galaxies independently of the environment in which they reside, and external processes that are expected to be efficient only in specific (over-dense) environments. Star formation takes place in giant condensations of cold ($T<10^2$~K) molecular gas that form after the gas that is accreted from cosmological filaments onto dark matter haloes condenses and cools towards their center. The balance between gas accretion and gas consumption leads to a relatively tight correlation between galaxy stellar mass and SFRs observed out to high redshift \citep[][and references therein]{Popesso_etal_2023}. Any disruption of this balance leads to a suppression of star formation and therefore to galaxy quenching. We can consider four main routes to quenching \citep[][see their Fig.~1 for a schematic view]{Man_and_Belli_2018}:  (i) infalling of gas onto dark matter haloes is prevented or hot gas in dark matter haloes is removed by some physical process. Galaxies can still form stars from the available reservoir of cold gas (if any) and from gas produced by stellar evolution, but this is not sufficient to maintain high level of star formation for long time leading to galaxy quenching.  (ii) Gas is being accreted onto the halo, but for whatever reason is not condensing or is not forming stars at non-negligible rates; (iii) cold gas is consumed very efficiently, on a time-scale that is shorter than that of its replenishment;  or (iv) cold gas is removed from the galaxy due to some feedback processes or interactions with surrounding environment and therefore cannot be used for star formation. Below, we will explain the physical processes that can lead to these different quenching routes in more detail. 

While the above division is convenient to illustrate how quenching can happen, we stress that in reality several physical processes are in place simultaneously and that this entanglement can significantly affect the relevant time-scales. In addition, there is no easy one-to-one correspondence between a given physical process and a specific quenching route. For example, AGN feedback can both remove gas from galaxies (thus `actively' suppressing or even terminating the star formation) or suppress significantly the cooling flow at the center of massive haloes (in this sense, it is a maintenance quenching mode as it prevents further star formation that would otherwise take place). To complicate the picture even further, AGN can also lead to a `positive' feedback by compressing the surrounding gas and triggering star formation \citep{Cresci_etal_2015,Nesvadba_etal_2020,Shin_etal_2019}. Therefore, {\it it is the combination and interplay between different physical processes that explain the variety of the observed galaxy properties and the observed trends.}

\subsection{Internal physical processes}

Internal processes that regulate the rate at which gas is converted into stars, at different stages of the star formation process, include mechanisms of energy release into the interstellar medium (ISM) from various evolutionary stages of massive stars (in this case one typically refers to stellar feedback), as well as from accreting supermassive black holes (AGN feedback). 

Stellar feedback owes to a combination of ionizing radiation, radiation pressure, stellar winds, supernovae explosions and possibly cosmic rays  whose collective effect leads to inefficient star formation in giant molecular clouds \citep[e.g.][]{Gatto_etal_2017,Krumholz_etal_2019,Grudic_etal_2022}. As soon as massive stars leave their main evolutionary sequence, they start ejecting large quantities of mass in winds -- high-speed outflows of particles -- that inject energy and momentum into the surrounding gas, dispersing it and creating bubbles or cavities in the ISM. These stellar winds are believed to contribute to driving turbulence in the ISM \citep[e.g.][]{MacLow_and_Klessen_2004}, which helps regulate future star formation by dispersing dense clouds or mixing gas over larger scales. Additionally, young, massive O and B stars emit large amounts of high-energy radiation, particularly in the ultraviolet. This radiation can ionize nearby gas, heating it to temperatures around $10^4$~K thereby increasing the thermal pressure in star-forming clouds. This pressure counteracts gravitational collapse, slowing or even preventing further star formation in the immediate vicinity of massive stars. Moreover, this stellar radiation exerts pressure on surrounding gas and dust, pushing it away from dense, star-forming regions, thereby regulating the star formation rate at galaxy scales. When massive stars (typically ${\rm M} > 8 \,{\rm M}_{\odot}$, depending on the stellar evolution model adopted) reach the end of their lives, they explode as supernovae, releasing vast amounts of energy \citep[approximately $10^{51}$~erg per event, e.g.][]{Kasen_Woosley_2009} in a short time and injecting both energy and momentum into the ISM. It is worth reminding that stellar feedback also pollutes the ISM with the end-products of stellar evolution (i.e. metals), which has an additional impact on the thermodynamical properties of the gas. 

\begin{figure*}
    \centering
    \includegraphics[width=0.85\linewidth]{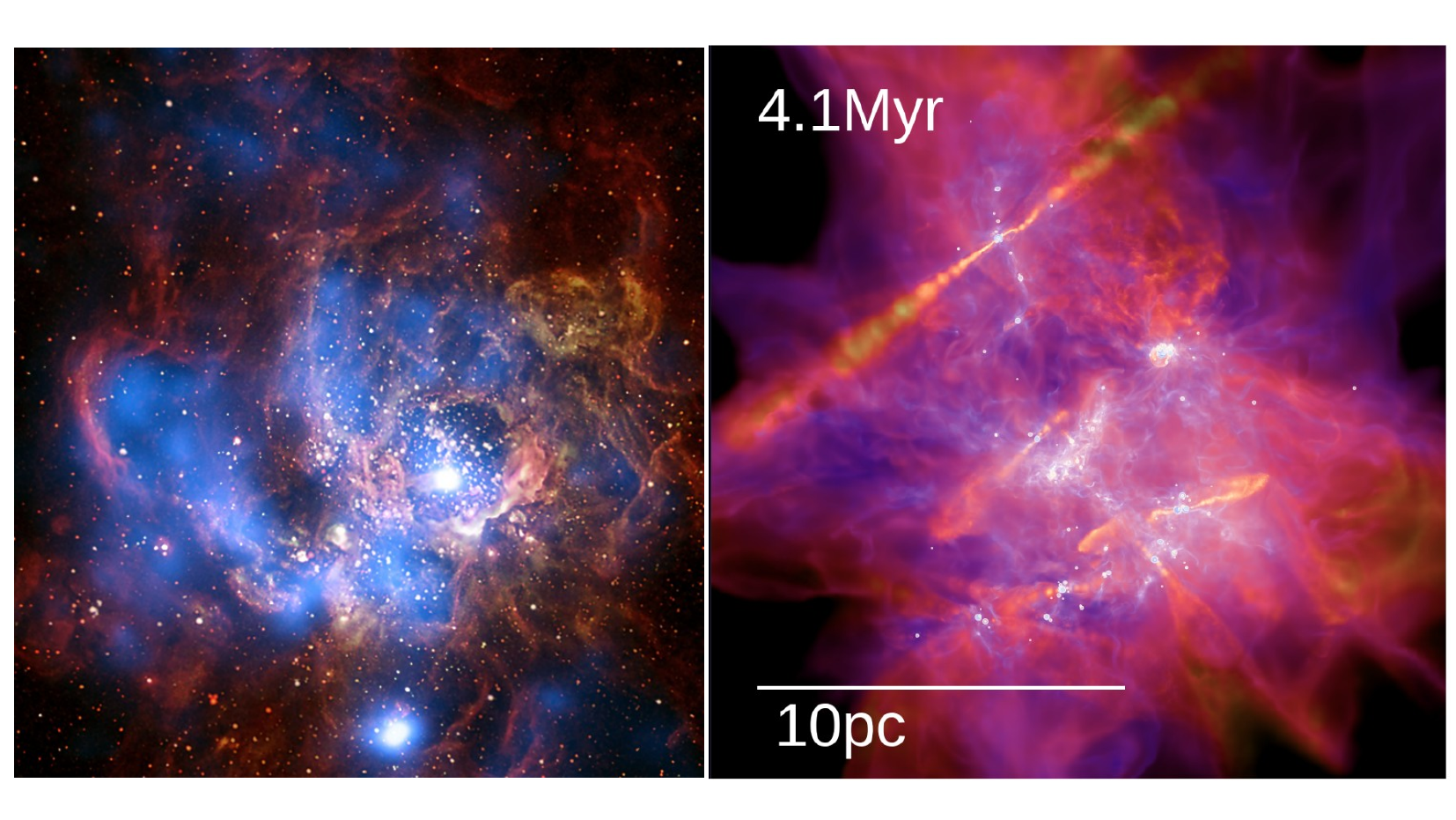}
    \caption{Left panel: color composite image of the star forming region NGC604, also known as the Tarantula Nebula in the Large Magellanic Cloud. The image has been obtained combining optical data from the Hubble Space Telescope with X-ray data from the Chandra Observatory \citep{Tuellemann_etal_2008}. Right panel: hydro-dynamical simulation of a $2\times 10^4\,{\rm M}_{\odot}$ giant molecular cloud including the combined effect of proto-stellar jets, stellar wind, radiation, and supernova feedback (from the STARFORGE collaboration - \citealt{Grudic21}).}
    \label{fig:stellarfeedback}
\end{figure*}

Sophisticated simulations including gravity, magneto-hydrodynamics, radiative transfer, cooling, chemical physics, stellar evolution and feedback in the form of radiation, stellar winds, proto-stellar jets, and core-collapse supernovae have been developed in the past few years. These simulations confirm that the radiation and winds released by massive stars are capable of expanding bubbles that can completely disrupt star forming clouds inducing a significant suppression of the SFR. The simulations also successfully reproduce the observed IMF and star formation efficiency \citep[][and references therein]{Grudic_etal_2022}. Figure~\ref{fig:stellarfeedback} shows, in the right panel, a visualization of a simulated $2\times 10^4\,{\rm M}_{\odot}$ giant molecular cloud including the combined effect of proto-stellar jets, stellar wind, radiation, and supernova feedback (from the STARFORGE collaboration - \citealt{Grudic21}). The simulation has a resolution of $10^{-3}\,{\rm M}_{\odot}$; color coding provides information about the 1D line-of-sight velocity dispersion (purple is $\sim 0.1{\rm kms}^{-1}$, orange is $\sim 10\,{\rm kms}^{-1}$), and is modulated with gas surface density. The left panel shows a composite image of the Tarantula Nebula, a star forming region in the Large Magellanic Cloud. The image shows bubbles and cavities that are filled with a hot and tenous X-ray emitting gas. On the western side of the nebula, measurements indicate that the material is likely heated by winds from young massive stars. On the eastern side, cavities are older, likely created by supernovae explosions from the end evolutionary phases of massive stars.  

While this work is crucially needed to understand and model star formation, it offers only limited guidance to galactic-scale simulations that remain unable to include all relevant physical processes in a proper cosmological framework, due to the large range of physical scales involved and the necessity to make additional assumptions about the evolution of a more complex system. Significant progress has been achieved with faithfully modelling stellar feedback in high-resolution, cosmological zoom simulations of low-mass galaxies \cite[LYRA, EDGE-INFERNO simulations of][]{Gutcke22, Andersson24}, via sampling individual massive stars and numerically resolving blast waves driven by supernovae explosions. These simulations indicate that gas ejection is primarily driven by supernovae, while ionising radiation and stellar winds are crucial for reducing the gas density at sites of the SN explosions. In large cosmological boxes, such high accuracy is not achievable, and stellar feedback is modelled using theoretically and/or observationally motivated prescriptions (one typically refers to these as sub-grid physics) carrying a number of parameters that can be calibrated by trying to reproduce a specific (sub)set of observational measurements \citep[][]{Schaye_etal_2015,Hirschmann_etal_2016}. In fact, this translates in quite different prescriptions being adopted in different models (see e.g. \citealt{Valentini_etal_2017} and \citealt{Hang_etal_2024} for two independent analyses of the impact of a few different prescriptions for stellar feedback on the properties of disc galaxies). Given our still limited understanding of the flow of energy and mass associated with stellar feedback and the strong self-regulation with star formation, it is currently very difficult to use observed statistical properties of galaxies to discriminate among different models. As for its relevance in the framework of galaxy quenching, it is generally recognized that stellar feedback is unable to quench star formation over long time-scales. However, it is an important regulator for star formation in low-mass galaxies. Moreover, stellar feedback turned out to be crucial for cosmological simulations to obtain realistically looking spiral galaxies with thin disks, thanks to late re-accretion of high-angular momentum gas, which was ejected by stellar feedback at earlier cosmic times \cite[e.g.,][and references therein]{Aumer13, Naab_and_Ostriker_2017}. As we will discuss more in detail below, stellar feedback does play an important role in quenching channels (iii) and (iv) introduced above. 

\begin{figure*}
    \centering
    \includegraphics[width=1.\linewidth]{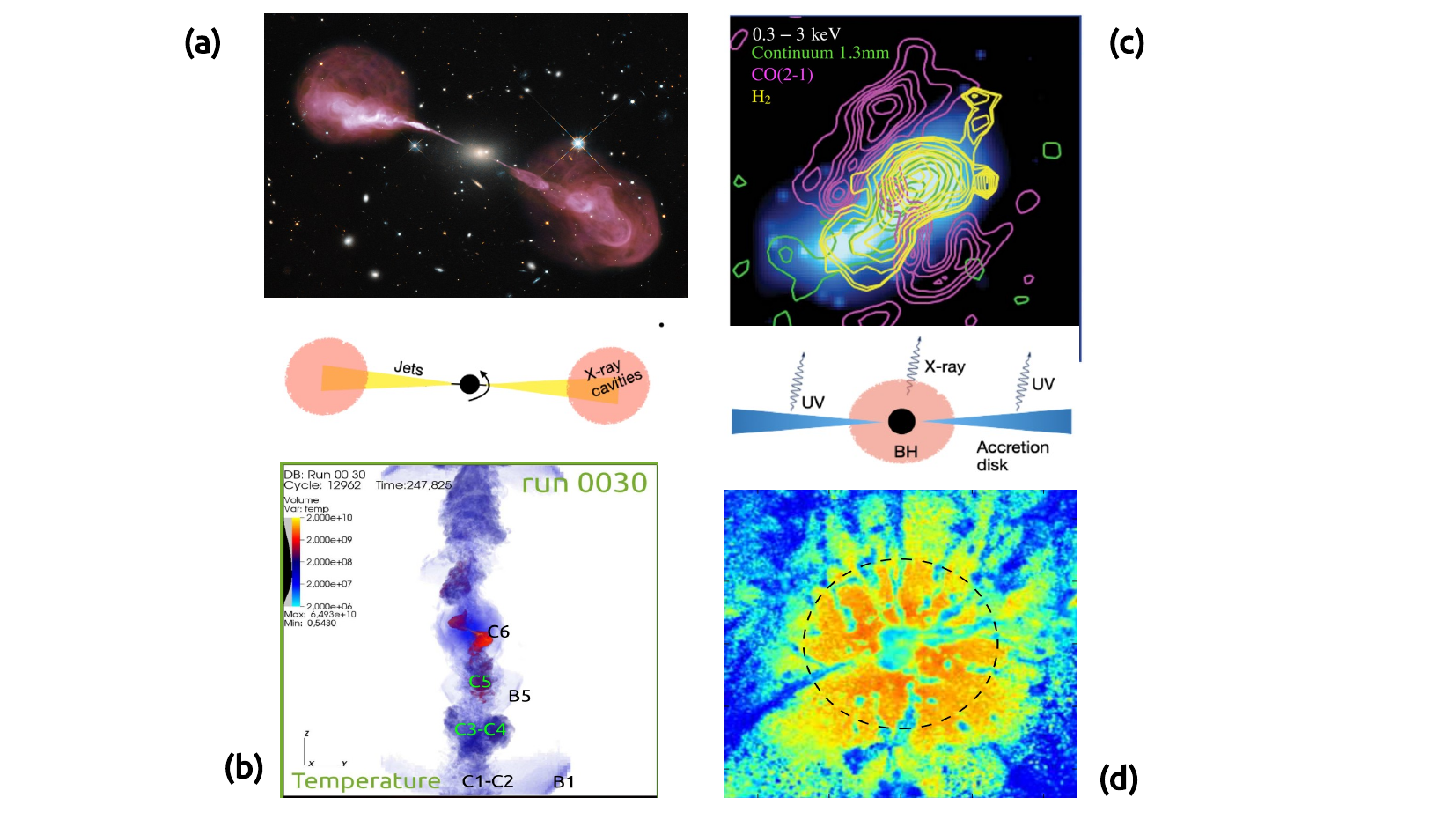}
    \caption{Different modes of AGN Feedback. Left panels: collimated jets of relativistic particles streams from radiatively inefficient accretion flows onto central SMBHs. (a) Composite image (optical image from the Hubble Space Telescope superposed with a radio image from the Very Large Array) of the jetted galaxy Hercules A. (b) Three-dimensional rendering of gas temperature from a hydrodynamical simulation of jetted AGN \citep{Cielo18}. Right panels: Energetic radiation from accretion disk around radiatively efficient accretion flows onto central SMBHs: (c) Composite zoom-in image of the nuclear region of the Seyfert galaxy ESO428–G14: blue colored contours correspond to soft X-ray emission from Chandra; green and magenta contours represent $1.3$~mm continuum and velocity-integrated CO(2–1) line emission detected by ALMA; yellow contours trace the warm molecular gas (H$_2$ 2.12 $\mu$m line) from SINFONI/VLT data \citep{Feruglio20}. (d) Projected gas temperature in a (500 $h^{-1}$ kpc)$^3$ comoving volume around the a massive galaxy at $z=6$ (the black dashed circle shows the galaxy virial radius), from a zoom-in hydrodynamical simulation \citep{Barai18}. The cartoons in the middle provide a visual impression of the two different modes of feedback.}
    \label{fig:AGNfeedback}
\end{figure*}

For more massive galaxies, stellar feedback cannot significantly impact star formation due to their larger potential wells requiring larger velocities for gas to be expelled. At this mass scale, AGN feedback is generally considered as the main quenching mechanism. It is useful to distinguish between energy release processes related to the radiatively efficient and inefficient regimes characterized, respectively, by large accretion rates close to the Eddington limit and very low accretion rates significantly below the Eddington limit. In the radiatively efficient regime, intense radiation (in the ultraviolet and X-ray) and relativistic particle winds are produced by different processes (e.g, magneto-rotational instabilities, viscous forces etc.) in a geometrically thin and optically thick BH accretion disk and its surrounding hot gaseous corona. The radiation can heat and ionise the ambient gas, raising its thermal pressure. Acting on the nearby gas and dust, this pressure can potentially drive large-scale outflows that expel gas from the central regions of the galaxy and/or prevent the inflow of fresh gas needed for star formation. AGN-driven winds can also be driven by mechanical energy from the accretion disk (fast, relativistic particle flows), which can accelerate gas to high velocities, creating shock waves in the ISM and surrounding halo. Therefore, radiatively efficient AGN feedback can lead to long-lasting quenching of star formation by efficiently ejecting or heating the cold gas reservoir available for star formation \citep{Choi15, Choi17}. These physical processes are linked to channel (iv) introduced above.

In the radiatively inefficient regime of BH accretion, feedback is classically thought to be driven by relativistic jets of particle streams launched from the interaction of the BH spin and magnetic field lines. Such jets may extend far beyond the galaxy’s central regions so that they may be particularly effective in heating the hot gas within a halo or cluster-like environment. The jets inflate bubbles of hot plasma (also called “cavities”), which can prevent the hot halo gas from cooling and being accreted onto the galaxy. By preventing cooling, this feedback mode is thought to stabilise the hot gaseous halo and reduce star formation over long timescales (it is also referred to as maintenance mode). This process is linked to channel (ii) introduced above. While it has been repeatedly shown that some form of additional energy (e.g. from a central accreting BH) is needed to sufficiently suppress star formation in massive galaxies, the {\it relative} impact of radiative and jet-mode feedback remains debated. Recent simulations, such as IllustrisTNG \citep{Pillepich_etal_2018} or Simba \citep{Dave_etal_2019}, point towards a dominant relevance of jet-mode feedback in quenching star formation in {\it low-redshift} massive galaxies. A general conclusion from recent theoretical work is that if AGN feedback is implemented in a kinetic form (direct momentum injection into the ambient gas elements), it leads to effective star formation quenching in massive galaxies over several Gyrs. Figure~\ref{fig:AGNfeedback} shows observational examples of the two modes of feedback discussed (top row), together with a visual impression of them. In the bottom row, we show examples from two independent groups of hydrodynamical simulations including these modes of feedback. 

In addition to stellar and AGN feedback, another process that has been discussed in the literature as a potentially important actor for the suppression of star formation is `morphological quenching' \citep{Martig_etal_2009}. This refers to the fact that the presence of a stable, massive stellar bulge or of a central concentration of stars can suppress or slow down the formation of new stars despite the presence of a gaseous disk, by stabilizing it against fragmentation into bound, star-forming clumps.
The process, that is related to channel (ii) discussed above, would provide a simple and natural explanation to the fact that colour evolution of a galaxy is tightly linked to its morphological evolution \citep{Bluck_etal_2014} and can be effective at all mass scales. The relative importance of morphological quenching compared to stellar and AGN feedback remains very debated \citep[e.g.][]{Fang_etal_2013,Su_etal_2019}.

\subsection{External physical processes}

A plethora of physical processes are at play in over-dense regions (massive groups and clusters) that can be effective in suppressing star formation (and affecting the morphology of cluster galaxies). Broadly speaking, these can be grouped into two big families: (i) interactions with other group/cluster members and/or with the halo potential; (ii) interactions with the hot gas that permeates massive galaxy systems. In the following, we briefly review the relevant physical processes, commenting on the associated time-scales of quenching. 

{\bf Galaxy mergers}: galaxy mergers and more generally strong galaxy-galaxy interactions are commonly viewed as a rarity in massive clusters because of the large velocity dispersions of the systems. However, merger events are relatively common in group-like environments and may still represent an important `pre-processing' element in the evolution of cluster galaxies. Numerical simulations \citep[e.g.][and references therein]{Cox_etal_2008, DiMatteo_etal_2008, Hayward_etal_2014} have shown that close interactions can lead to a strong internal dynamical response driving the formation of spiral arms and, in some cases, of strong bar modes. Sufficiently close encounters can completely destroy the disc leaving a kinematically hot mergers with photometric and structural properties that resemble those of elliptical galaxies. 
When merging galaxies are gas-rich and have comparable stellar masses, the merger can drive compressive gas motions and effective angular momentum loss, triggering an intense starburst and eventually enhancing central AGN activity. This accelerated gas depletion, that is linked to quenching scenario (iii), may subsequently reduce star formation, leading to what is known as a post-starburst phase \citep[e.g.][]{Snyder_etal_2011, Pawlik_etal_2019}. This reduced star formation can represent just a temporary phase of the galaxy evolution if the remnant galaxy is able to accrete large amount of gas at later times. Cosmological galaxy formation simulations that only include this physical process have generally failed to maintain the star formation suppressed over extended timescales, and to explain the observed number densities of quiescent galaxies over cosmic time. 

{\bf Starvation}: Current theories of galaxy formation assume that, when a galaxy is accreted onto a larger structure, the gas supply can no longer be replenished by cooling that is suppressed by the removal of the hot halo gas associated with the infalling galaxy \citep{Larson_etal_1980}. This process is sometimes referred to as `strangulation' and is linked to the quenching channels (i) introduced above. It is common to read in discussions related to environmental quenching that starvation is expected to affect star formation of cluster galaxies on relatively long time-scales. However, if this is combined with an efficient removal of the gas due to e.g. stellar feedback, tidal stripping, and ram pressure stripping, the time-scale of quenching can become very short (see Section~\ref{chap1:sec4} and discussion about over-quenching of satellites). Therefore, the effective quenching time-scales depend on how efficient stellar feedback is and what is the fate of the gas that is reheated or ejected by stellar feedback. Numerical simulations have shown that the stripping of the hot halo associated with infalling satellites should not happen instantaneously \citep{McCarthy_etal_2008} potentially making the quenching time-scales longer, with an obvious and expected dependence on the galaxy orbit. The relevance of this process in quenching galaxies remains debated, with some studies claiming this could represent the primary process responsible for the environmental dependence of the quiescent galaxy fractions, particularly for low-mass galaxies \citep{Stevens_etal_2017, Cora_etal_2018, Xie_etal_2020}.

\begin{figure*}
    \centering
    \begin{minipage}{0.335\textwidth}
        \centering
        \includegraphics[width=\linewidth]{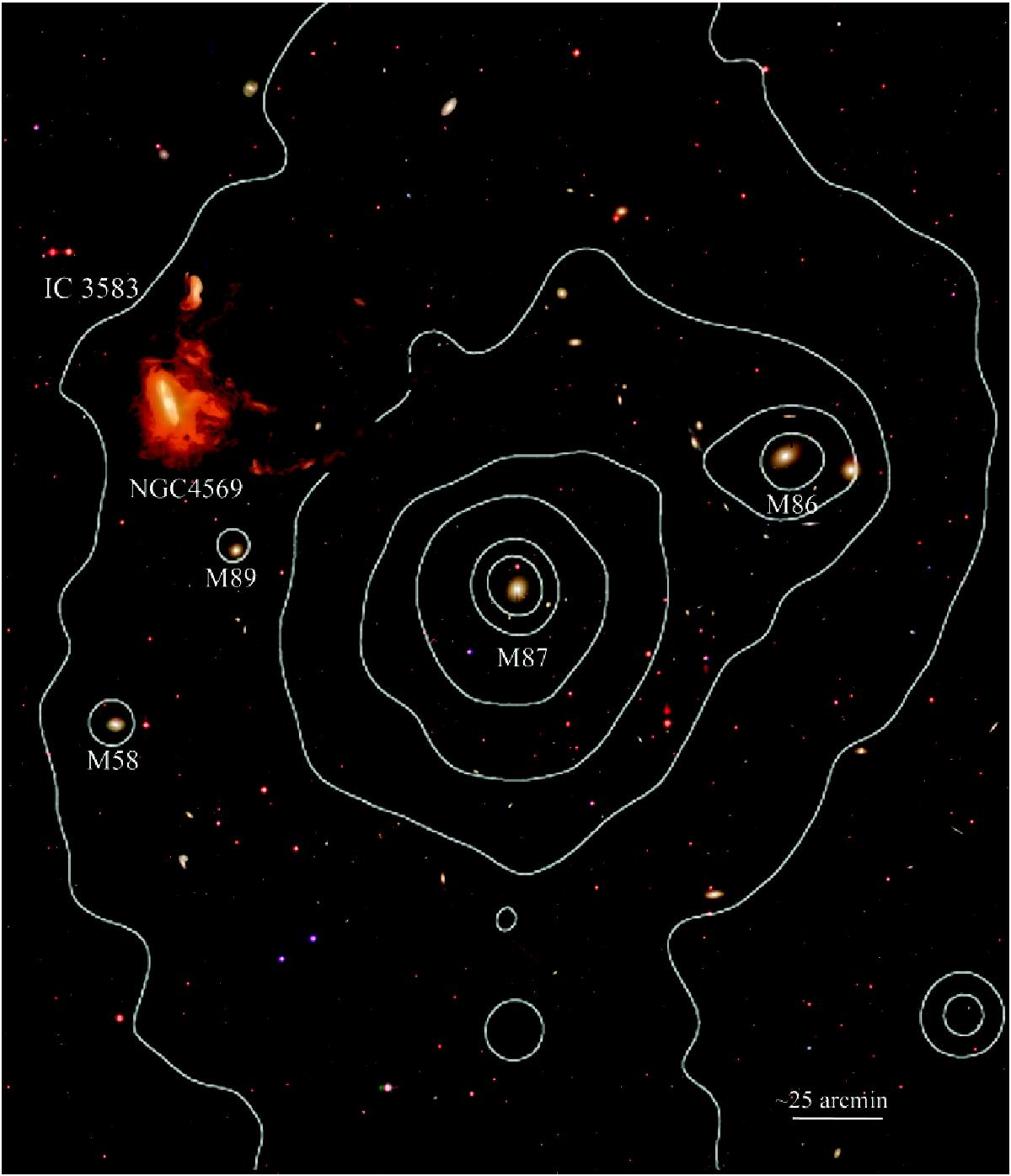}
    \end{minipage}
    \begin{minipage}{0.515\textwidth}
        \centering
        \includegraphics[width=\linewidth]{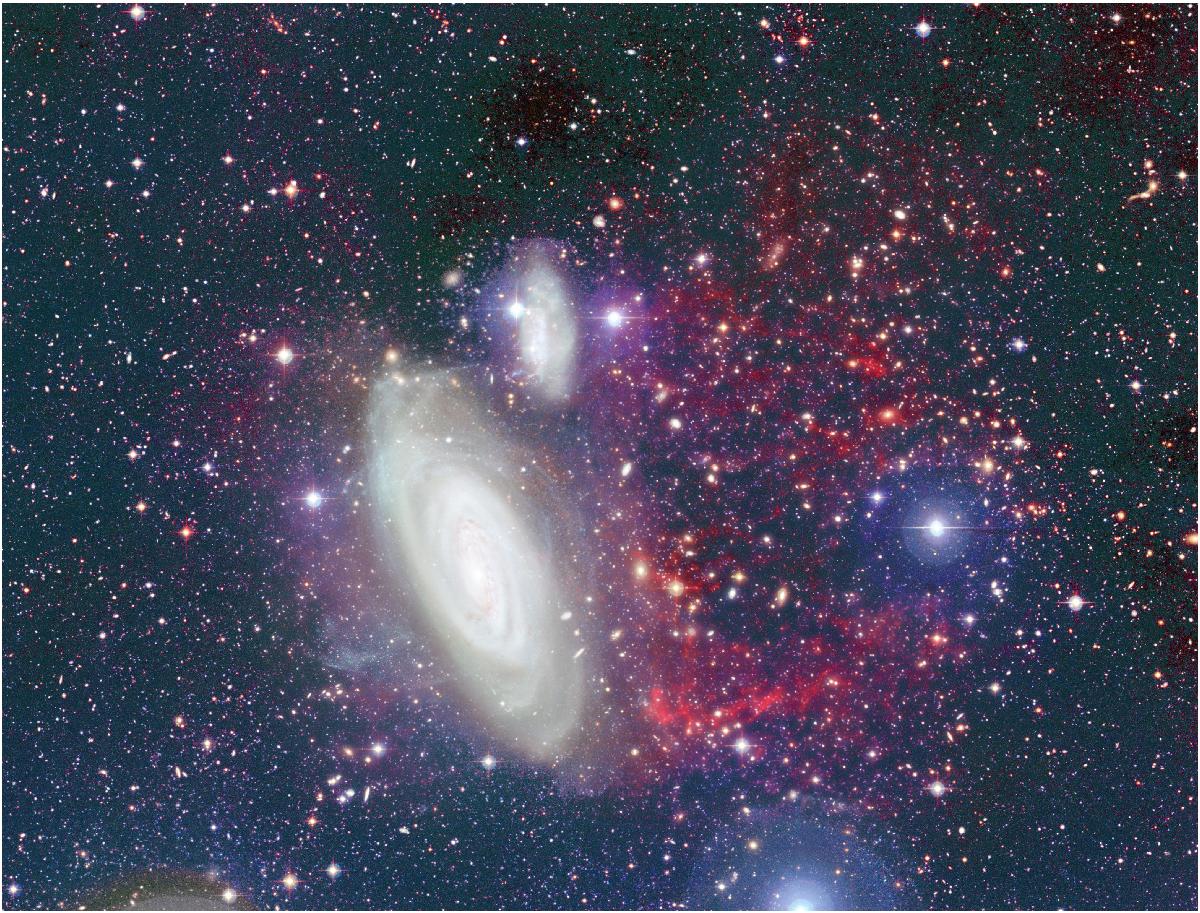}
    \end{minipage}
    \begin{minipage}{0.85\textwidth}
        \includegraphics[width=\linewidth]{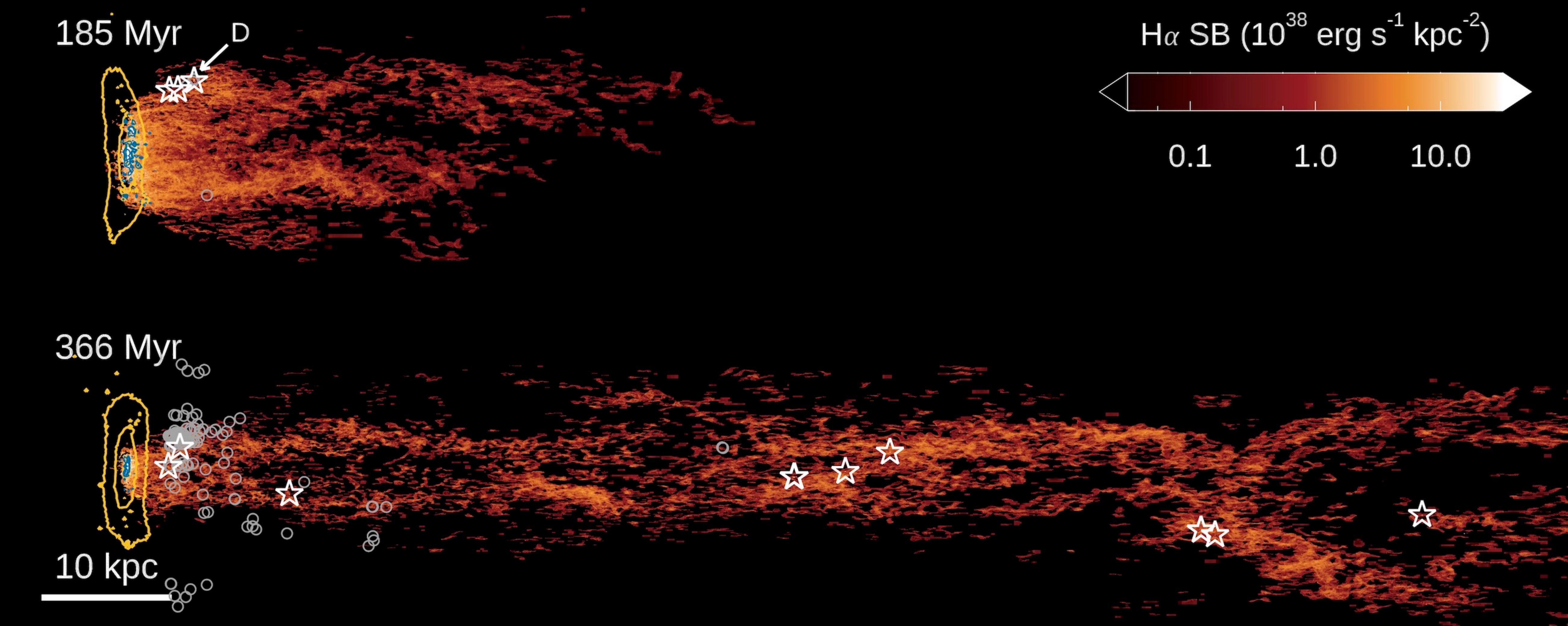}
    \end{minipage}
    \caption{Galaxies undergoing ram pressure stripping. Upper left panel: location within the Virgo cluster of two galaxies (NGC 4569 and IC 3583) undergoing ram-pressure stripping. The white contours show the distribution of the X-ray emitting gas. Upper right panel:  pseudo-color image of the same galaxies, obtained combining optical g (blue) and i (green) images with H$_{\alpha}$+[NII] narrow-band image (red). Both figures are from \citet{Boselli_etal_2016a}. Bottom panel: dust-obscured  H${\alpha}$ map of a simulated galaxy at $t =$ 185 and 366 Myr from \citet{Lee_Kimm_etal_2022}. Yellow and blue contours show the distribution of all stars and the stars younger than 20 Myr in the galactic disk, respectively. Stars symbols show the position of stellar particles younger than 20~Myr in the ram-pressure tail. Open circles show the position of older stellar particles formed in the tail between $100$~Myr and $344$~Myr.}
    \label{fig:rampressure}
\end{figure*}

{\bf Cold gas stripping}: galaxies travelling through a dense intra-cluster or intra-group medium suffer a strong ram-pressure stripping that can sweep cold gas out of the stellar disc \cite{Gunn_and_Gott_1972}. Depending on the binding energy of the gas in the galaxy, the medium can either blow through the galaxy removing some of the diffuse inter-stellar medium, or can be forced to flow around the galaxy \citep{Cowie_and_Songaila_1977,Nulsen_etal_1982}. This physical process is related to the quenching scenario (iv) introduced above, and is expected to be important at the centre of massive systems because of the large relative velocities and densities of the intra-cluster medium. When a galaxy loses most of its interstellar gas, its potential for future star formation is significantly diminished. Therefore, this effect could explain why dense environments, such as clusters, exhibit a marked deficit of gas-rich, star-forming galaxies. By considering the distribution and ‘history of ram-pressure’ experienced by galaxies in clusters, \citet{Brueggen_and_DeLucia_2008} estimated that strong episodes of ram-pressure are indeed expected to dominate in the inner core of clusters. However, according to their calculations, virtually all cluster galaxies suffered weaker episodes of ram-pressure, suggesting that this physical process might have a significant role in shaping the observed properties of the entire cluster galaxy population. In addition, they found that ram-pressure fluctuates strongly so that episodes of strong ram-pressure alternate to episodes of weaker ram pressure, possibly allowing the gas reservoir to be replenished and intermittent episodes of star formation to occur. Although spectacular examples of ram-pressure stripping caught in the act are available  \citep[see e.g.][]{Poggianti17, Jachym_etal_2019, Vulcani21} and a large body of numerical work has focused on the impact of this process \citep{Steinhauser_etal_2016, Kulier_etal_2023, Roediger_Bruggen_2006, Tonnesen_2019}, the extent to which gas stripping contributes to star formation suppression remains debated and the complex interplay between different gas phases has yet to be understood. Since it is more loosely bound to the galaxy and has lower densities, material at the disk outskirts is expected to be stripped first. The remaining gas may actually be compressed by the ram pressure, potentially enhancing star formation within the disk (this is another example of `positive' feedback). In some cases, the stripped gas might stay gravitationally bound to the galaxy, eventually falling back and re-initiating star formation. In a few cases, detections of significant amounts of molecular gas in the tail of ram-stripped galaxies have been reported \citep[e.g.][and references therein]{Moretti_etal_2018,Jachym_etal_2019}. This gas maybe formed {\it in-situ} and can lead to the formation of new stars. Figure~\ref{fig:rampressure} shows, in the top right panel, a pseudo-color image of two galaxies (NGC 4569 and IC 3583) orbiting the nearby Virgo cluster (their location within the cluster is shown in the top left panel of the same figure), and undergoing ram-pressure stripping of their cold gas. The diffuse red emission escaping from the galaxy in the western direction is ionised gas that extends up to $\sim 150$~kpc from the stellar disc. The bottom panel of Figure~\ref{fig:rampressure} shows two dust-obscured H${\alpha}$ maps from a radiation-hydrodynamic simulation of a gas-rich dwarf galaxy with a multiphase interstellar medium, exposed to a wind with velocity $\sim 1000\,{\rm km}\,{\rm s}^{-1}$ that starts influencing the galaxy at $t\sim 135$~Myr and is run up to $t=366$~Myr. Yellow and blue contours show the distributions of all stars and stars younger than 20~Myr in the galactic disk. Stars symbols show the position of stellar particles younger than 20~Myr in the ram-pressure tail. Open circles show the position of older stellar particles formed in the tail between $100$~Myr and $344$~Myr.    

\subsection{Other physical processes}

Other processes that are not easily classifiable as purely internal or external processes can play a role in galaxy quenching. These include virial shock heating and gravitational heating. 

In the current picture of galaxy formation, gas being trapped into the potential wells of dark matter halos is compressed until a virial shock is created that transfers kinetic energy that has been built during the collapse into internal gas energy just behind the shock \citep{Rees_and_Ostriker_1977,Binney_1977}. If cooling times are long compared to dynamical times, then the shock is expected to form at a radius that is comparable or slightly larger than the virial radius. If instead cooling times are short, the shock is expected to form at much smaller radii, close to the central galaxy \citep[see e.g.][and references therein]{Benson_and_Bower_2011}.  The transition between these two cooling regimes, that are often referred to as slow and rapid respectively, is rather well defined and occurs at $\sim 10^{12} \,{\rm M}_{\odot}$. Haloes where shocks form close to the virial radius are expected to contain a quasi-hydrostatic atmosphere of hot gas. Although this hot gas may stay in equilibrium within the halo potential for some time, virial shock heating alone cannot quench star formation, because the hot halo gas will later cool and condense towards the halo center. As we have discussed above, AGN feedback can efficiently contrast cooling. 

Another process that has been discussed in the literature in this context is gravitational heating: gravitational interactions inject kinetic energy into the halo gas, increasing its thermal energy and leading to a rise in temperature. This can come in different forms: e.g. collisions between the infalling gas and the ambient gas lead to shock waves; when galaxies merge or when smaller satellite galaxies are accreted onto larger ones, dynamical friction causes the gravitational potential energy of the interacting galaxies to be converted into thermal energy. Some studies \citep[e.g.][]{Johannson_etal_2009} have argued that this physical process might be responsible for the formation of passive galaxies (with a late history that is dominated by accretion and minor mergers) by $z\sim 1$ even in absence of stellar and AGN feedback. However, cosmological hydrodynamic simulations automatically include these heating processes self-consistently, and by no means these are sufficient to overcome the problem of excessive gas cooling at the centre of massive clusters if no treatment for AGN feedback is included.

\begin{figure*}
    \centering
    \includegraphics[width=0.85\linewidth]{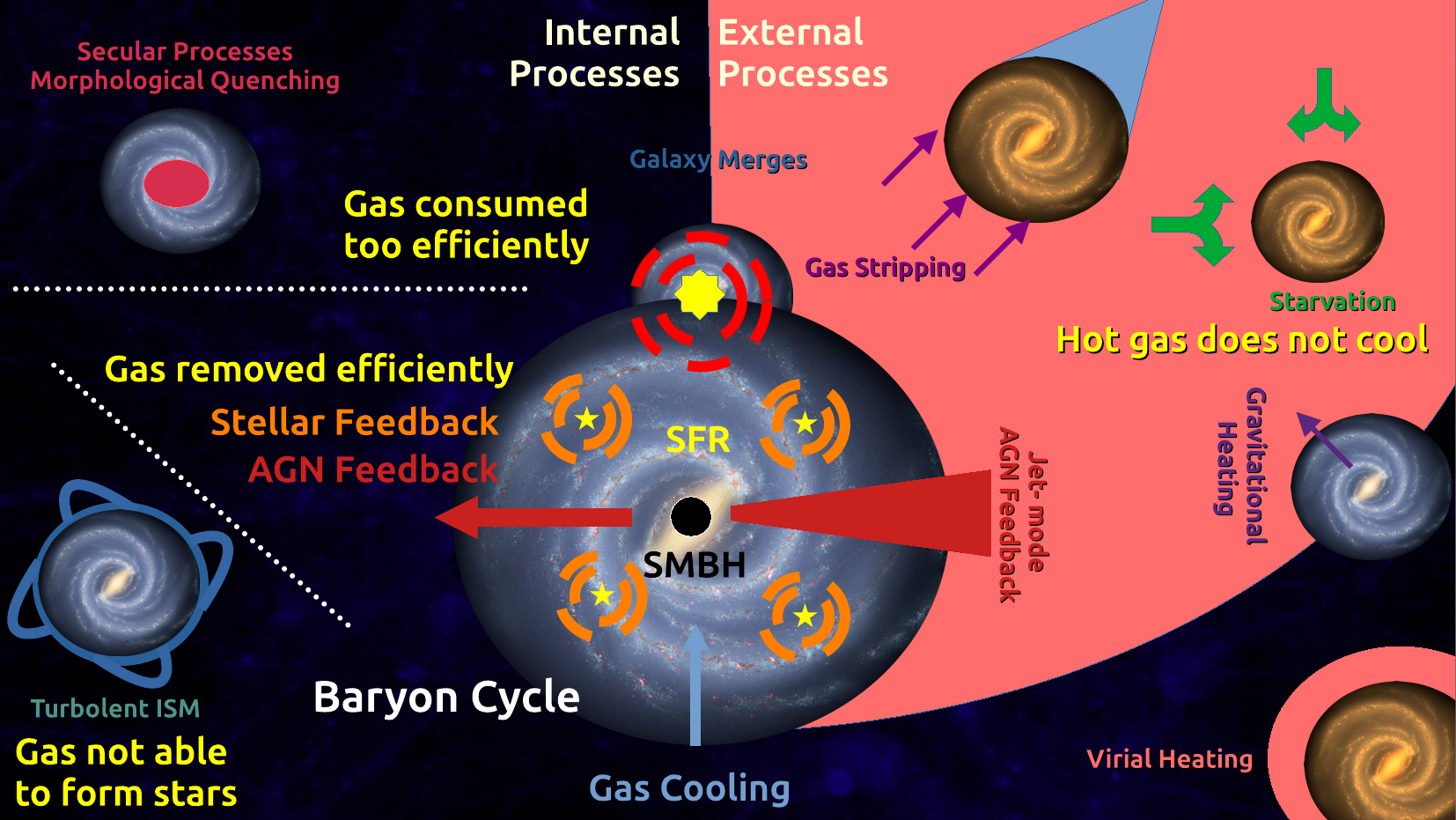}
    \caption{A schematic summary of the physical processes described in this section. Internal processes are shown on the left, and external processes on the right. AGN feedback is on both sides to account for the different phenomenology and physics associated with different accretion rate regimes. (The cartoon includes an artist visual impression of the Milky Way, credits: NASA/JPL-Caltech/ESO/R. Hurt)}
    \label{fig:cartoon}
\end{figure*}

Figure~\ref{fig:cartoon} gives a schematic summary of the physical processes described in this section. Internal and external processes are shown on the left and the right of the figure, respectively. AGN feedback is sitting in the middle of the figure to account for the different phenomenology and physics associated with different accretion rate regimes. The figure is by no means complete (e.g. it does not include the positive AGN feedback effect discussed above), but it already shows how complicated is the network of processes that one should account for when trying to understand galaxy evolution (and quenching). 

\section{What we have learned so far}\label{chap1:sec4}

Since early renditions of theoretical models of galaxy formation, it was clear that some physical process should prevent an efficient conversion into stars of the baryonic gas falling into the potential wells of dark matter haloes. At low stellar masses, where galaxies have shallow potential wells and the fraction of satellite galaxies is larger, a combination of stellar feedback and environmental processes is expected to play an important role in regulating star formation. 

\begin{figure*}
    \centering
    \includegraphics[width=0.85\linewidth]{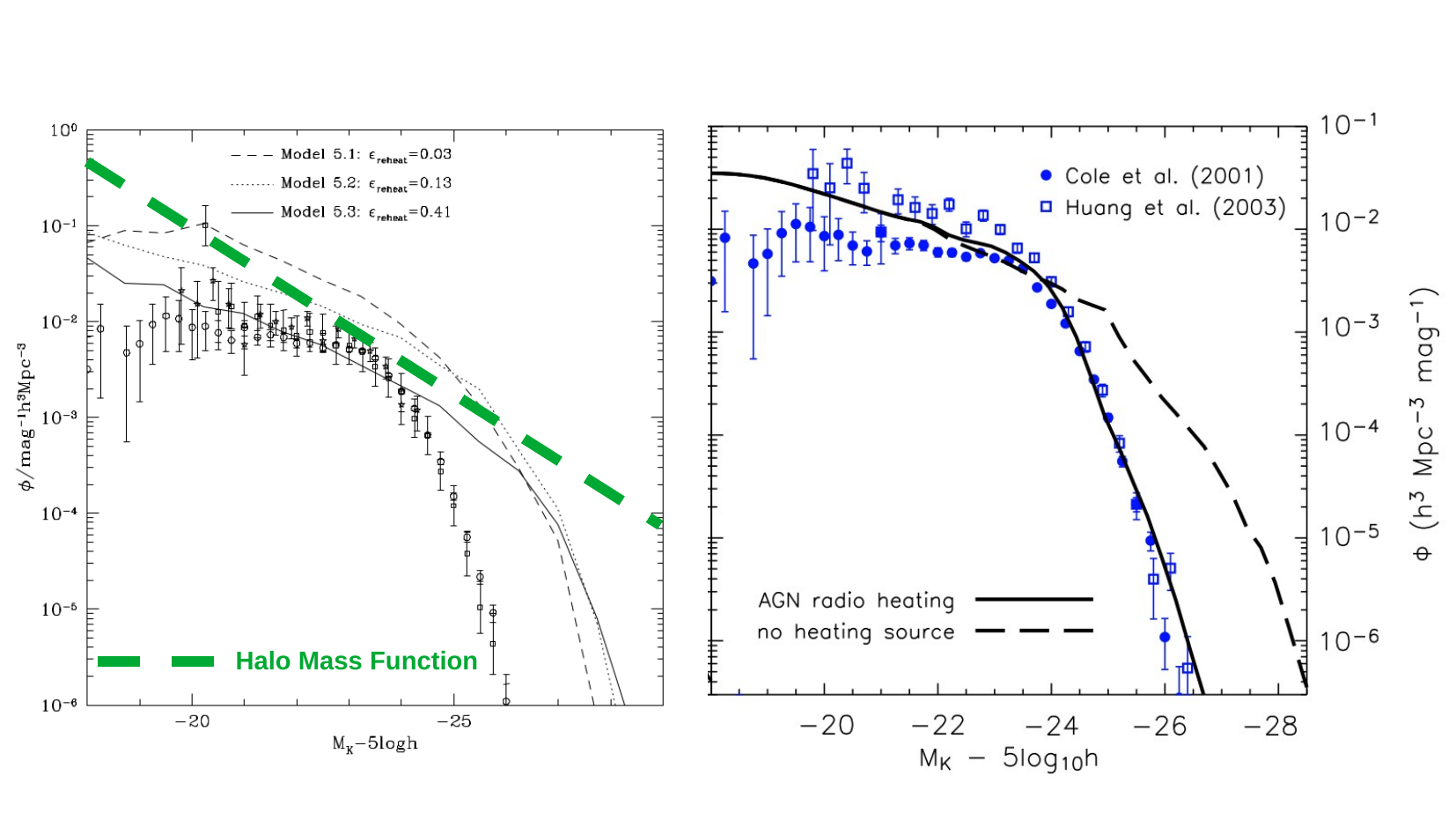}
    \caption{Left panel: adapted from \citet{Benson_etal_2003}. Points show observational determinations of the observed K-band luminosity function; lines show predictions from models with increasing efficiency of stellar feedback as indicated in the legend. The dashed green line shows the number densities obtained by scaling the halo mass function so as to match the observed number densities at the knee. Right panel: adapted from \citet{Croton_etal_2006}. Points are again different observational measurements of the observed K-band luminosity function, while the solid and dashed lines show model predictions from two different runs where radio-mode feedback has been switched on and off, respectively (both runs also include a treatment for stellar feedback).}
    \label{fig:smf}
\end{figure*}

The left panel of Figure~\ref{fig:smf}, adapted from \citet{Benson_etal_2003}, shows model predictions for the K-band luminosity function, compared with observational estimates (symbols with error bars). Lines with different styles show model predictions obtained for increasing efficiency of stellar feedback, as indicated in the legend. As a reference, we have added a dashed colored line showing the expected number densities obtained by simply scaling the halo mass function by a constant factor so as to match the observations at the knee. Such a simple model, as noted in many early studies, dramatically over-predicts the number densities of both low-luminosity and luminous galaxies. Adopting a relatively strong feedback, the agreement with the observational data improves significantly at low luminosities, but a significant excess with respect to observational predictions remains at the most luminous end. It is also interesting to note that increasing the efficiency of stellar feedback comes at the expenses of exacerbating the excess of luminous bright galaxies. This is due to the fact that, in the models considered, the material reheated and/or ejected by low-mass galaxies ends up in the hot gas that is associated with the corresponding central galaxies. At later times, this material cools efficiently onto the central galaxies increasing their luminosities and star formation rates, at odds with observational data. 

Matching simultaneously both the faint and the bright end of the luminosity function has proven difficult: massive galaxies (M$_\star >$10$^{10.5}$ M$_\odot$) typically reside at the center of relatively massive dark matter haloes in which stellar feedback is neither effective at regulating star formation rate, nor at preventing efficient gas cooling. The necessity of introducing a physical process to suppress the condensation of gas at the centre of massive haloes and the hypothesis that this might be due to feedback from AGN was noted in early work \citep[e.g.][]{Kauffmann_etal_1999,DeLucia_etal_2004}. In later years, a reasonable `success' was achieved in models based on a semi-analytic approach through the inclusion of physically motivated prescriptions of a relatively strong `radio-mode' feedback \citep{Bower_etal_2006, Croton_etal_2006}. The right panel of Figure~\ref{fig:smf}, adapted from \citet{Croton_etal_2006}, shows predictions for the K-band luminosity function from two model versions where this process has been switched on and off (solid and dashed lines; both runs include a treatment for stellar feedback). The figure shows that these two physical processes are crucial to reproduce, simultaneously, the number densities of galaxies above and below the knee of the galaxy luminosity function. 

It is worth commenting that the main reason for the success of the radio-mode feedback is related to the fact that it does not require star formation to be effective. As a consequence, this mode of feedback allows suppressing the luminosity of massive galaxies while, at the same time, keeping their stellar population old, in qualitative agreement with observational data. These models also seem to reproduce, at least qualitatively, the observed trend for more massive galaxies to have shorter star formation time-scales \citep[][but see also \citealt{DeLucia_etal_2017}]{DeLucia_etal_2006}. As discussed in the previous section, AGN feedback may also drive fast and large-scale outflows due to radiation pressure or mechanical energy. Cosmological hydro-dynamical simulations that attempt to model the injection of momentum into the gas, due to AGN feedback, have shown that this can act simultaneously as a preventative and ejective mode, i.e. the strong outflow rates associated with AGN feedback also prevent inflow of new and recycling gas onto the halo \citep{Brennan_etal_2018}. Considering the large amounts of energy that can be released by an accreting SMBH, AGN feedback provides an appealing explanation for numerous observational results including the origin of various BH scaling relation and galaxy quenching \citep[e.g.][]{Silk_and_Rees_1998, DiMatteo_etal_2003, Monaco_and_Fontanot_2005, Springel_etal_2005, Shankar_etal_2020}. Current hydro-dynamical simulations of moderately large cosmological boxes that successfully reproduce the main observed trends for the global galaxy populations, do so by implementing rather different prescriptions for AGN feedback. For example, in the TNG simulation suite, Simba, and HorizonAGN \citep{Pillepich_etal_2018, Dave_etal_2019, Dubois_etal_2016}, AGN feedback is modelled as kinetic jet-mode feedback for the radiatively inefficient regime. In the radiatively efficient regime, AGN feedback is either modelled via thermal energy input (e.g. TNG, HorizonAGN) or also via a kinetic scheme (Simba). On the other hand, EAGLE assumes a stochastic thermal feedback model independently of the accretion regime \citep{Schaye_etal_2015}. To make progress on how the released energy from the central BH may couple to the ambient medium, idealised simulations are being conducted, e.g. studying Blandford-Znajek jets \citep{Talbot22} or radiation-driven outflows \citep{Costa17, Bieri17} in isolated galaxies. A fully self-consistent description of gas accretion onto BHs, their spin evolution and their related energy release, would require general-relativistic MHD simulations of gas accretion disks to be connected to galaxy-scale simulations, which bears a number of complications \citep{Gaspari17}.

Reproducing the observed fractions of quiescent galaxies over the entire range of galaxy stellar masses, cosmic epochs end environments sampled by observational studies has represented a challenge for theoretical models. One long standing problem that still affects a large number of recently published models is that of the `over-quenching' of satellite galaxies \citep{Weinmann_etal_2006,Wang_etal_2007}, i.e. the finding that low-mass galaxies\footnote{This population is dominated by satellite galaxies but also includes low-mass galaxies that are unaffected by environmental processes.} tend to be too old and passive compared with observational measurements. Early attempts to address this problem focused on the simplified treatment of satellite galaxies adopted in most models based on a semi-analytic approach published in those years: these models assumed that, when a galaxy is accreted onto a larger structure, its reservoir of hot gas is stripped instantaneously which, combined with the efficient supernovae feedback typically adopted, induces a very rapid decline of the star formation histories of satellite galaxies. Later studies assuming a more gradual stripping of the hot gas reservoir showed an improved but still far from satisfactory agreement with observational measurements \citep{Font_etal_2008,Kang_and_vandenBosch_2008,Weinmann_etal_2010,Guo_etal_2011,Hirschmann_etal_2014}. It should be noted that the problem has not been limited to semi-analytic models and has been identified also in state-of-the-art hydrodynamical simulations that should, by construction, include an explicit and consistent treatment of the gas stripping processes \citep[e.g.][]{Bahe_etal_2017,Donnari_etal_2021}. In fact, large-scale hydro-dynamical simulations do not trace the multi-phase inter-stellar medium and resort to `effective' equations of state to describe dense gas that is actually warm or even hot rather than cold \citep{Hu_etal_2016}. Resolution might also play an important role: due to limited resolution, galaxies might artificially and prematurely lose significant fractions of their gas reservoir. These limitations can have an important impact on the treatment of gas stripping in cosmological hydro-dynamical simulations, and can explain at least in part the disagreement with observational results. 

In recent years, the comparison between theoretical predictions and observed quenched fractions has been pushed to higher redshifts showing that the the problem of satellite over-quenching appears to be exacerbated at earlier cosmic epochs and in the high-density environments of galaxy clusters \citep{DeLucia_etal_2019,Lustig_etal_2023,Kukstas_etal_2023}. One element that has been often overlooked in discussions focused on quenching of satellite galaxies is the impact of internal physical processes that are at play {\it before} the galaxies are accreted onto larger systems and become satellites. In fact, previous work has shown that a good agreement with observational measurements of the passive fractions in the local Universe and for stellar mass larger than $\sim 10^{10}\, {\rm M}_{\odot}$, can be obtained in models that still assume an instantaneous stripping of the hot gas reservoir associated with infalling satellites \citep{DeLucia_etal_2019}. Taken at face value, these results suggest that the abundance of passive galaxies in this stellar mass range is primarily determined by the self-regulation between star formation and stellar feedback, with environmental processes playing a more marginal role. 

\begin{figure*}
    \centering
    \includegraphics[width=0.85\linewidth]{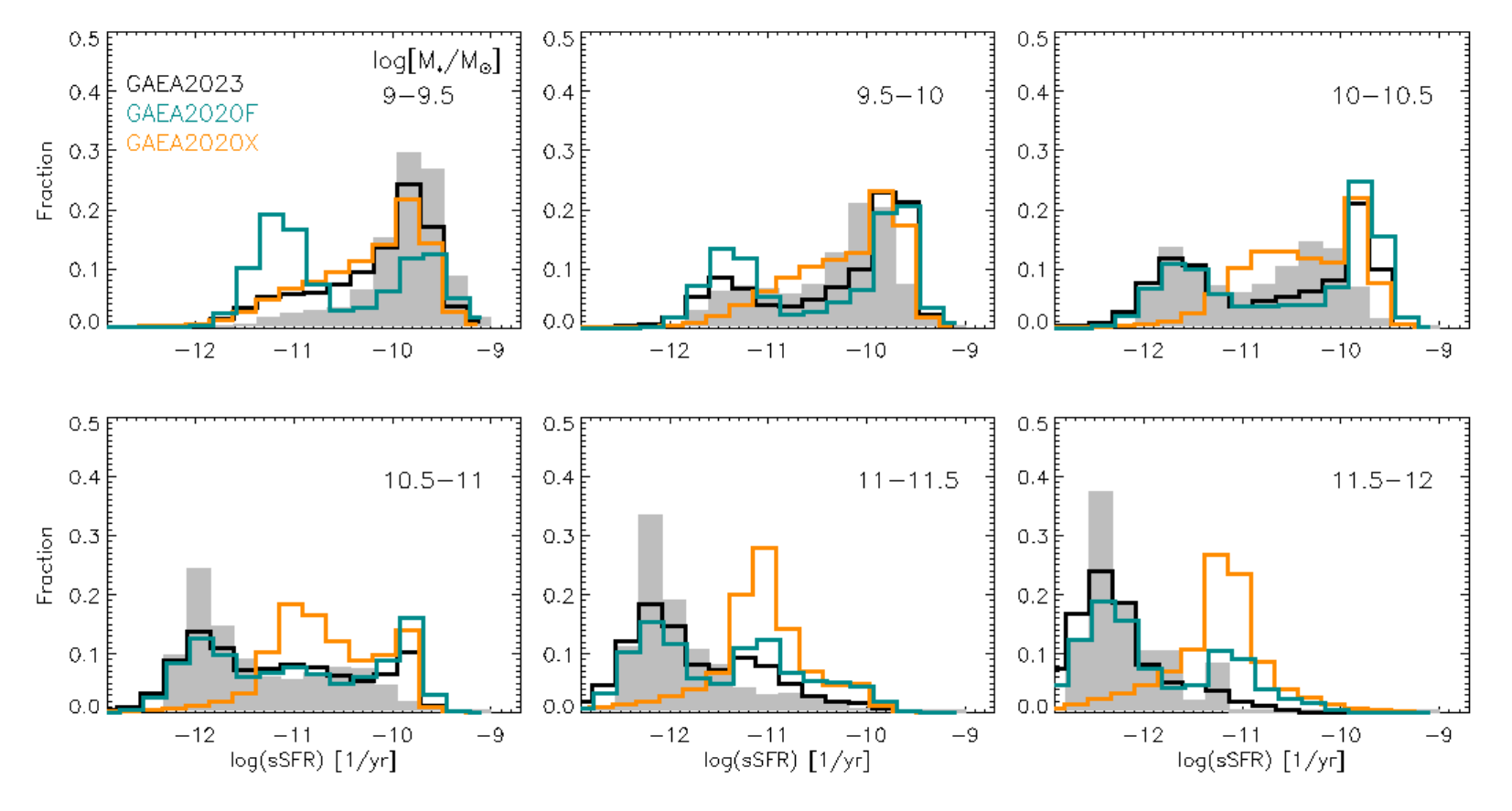}
    \caption{From \citet{DeLucia_etal_2024}. Specific star formation rate (sSFR) distributions as predicted by three different versions of the GAEA model, and compared to observational estimates based on SDSS  (grey shaded histograms). Each panel corresponds to a different bin in galaxy stellar mass, as indicated in the top-right legend. Both the GAEA2020X and GAEA2020F model versions are based on the model presented in \citet{Hirschmann_etal_2016}. The GAEA2020X version (orange lines - described in detail in \citealt{Xie_etal_2020}) include an explicit partition of the cold gas in its atomic and molecular components and an updated treatment of satellite galaxies featuring a gradual stripping of the hot gas reservoir and an explicit treatment for ram pressure stripping. The GAEA2020F model (dark cyan lines - described in detail in \citealt{Fontanot_etal_2020}) includes an updated treatment of AGN feedback featuring an explicit treatment of quasar driven winds. Finally, the GAEA2023 model combines all improvements presented in the two previous model versions.}
    \label{fig:delucia24}
\end{figure*}

In recent years, some success has been achieved in reproducing the observed statistics of quiescent and star forming galaxies at late cosmic epochs ($z<2$) within the GAlaxy Evolution and Assembly (GAEA) semi-analytic model \citep[][and references therein]{DeLucia_etal_2014,Hirschmann_etal_2016,DeLucia_etal_2024}. Figure~\ref{fig:delucia24}, from \citet{DeLucia_etal_2024}, shows the sSFR distributions as predicted by three different versions of the GAEA model and compared to observational estimates based on SDSS. Both the GAEA2020X and GAEA2020F models are based on the model presented in \citet{Hirschmann_etal_2016} but include the following improvements: the GAEA2020X version (described in detail in \citealt{Xie_etal_2020}) includes an explicit partition of the cold gas in its atomic and molecular components and an updated treatment of satellite galaxies featuring a gradual stripping of the hot gas reservoir and an explicit treatment for ram pressure stripping. The GAEA2020F model (described in detail in \citealt{Fontanot_etal_2020}) includes an updated treatment of AGN feedback featuring an explicit treatment of quasar driven winds. The latest rendition of the model, GAEA2023 in the figure, combines the improvements of the two previous model versions. Figure~\ref{fig:delucia24} shows that both the GAEA2020X and GAEA2020F versions have problems in reproducing the observed sSFR distributions, while the GAEA2023 model captures well the observed trends, for all galaxy mass bins considered. Comparison between the different model versions show that: at low stellar masses, the better agreement with observational data is driven by the improved treatment of the environmental processes; the inclusion of powerful winds associated with the bright QSO phase, triggered by high-efficiency gas accretion events like galaxy mergers and disk instabilities, leads to a rapid suppression of the star formation that manifests as a clear bimodality for galaxies with stellar masses between $10^{9.5}$ and $10^{11} \,{\rm M}_\odot$. AGN feedback also has a strong impact in regulating the star formation activity in massive galaxies. However, this model ingredient alone is unable to remove the residual excess activity that is found, in this mass range, with respect to observational estimates. As discussed in \citet{DeLucia_etal_2024},  including an explicit partition of the cold gas in its atomic and molecular components and relating star formation to the surface density of molecular gas leads to a more physical treatment of the star formation threshold improving the agreement with observational data. In this model, the interplay between stellar-driven and AGN-driven outflows emerges as a key mechanism governing the amount of cold gas available in galaxies, balancing the relative role of star formation and reheating \citep[see also][]{Monaco_and_Fontanot_2005}. At the same time, environmental processes represent key actors in the the baryon cycle of low-mass galaxies, either by removing the circum-galactic gas and/or by regulating the amount of new gas able to cool down and accrete onto galaxies.

\begin{figure*}
    \centering
    \begin{minipage}{0.45\textwidth}
    \includegraphics[width=\linewidth]{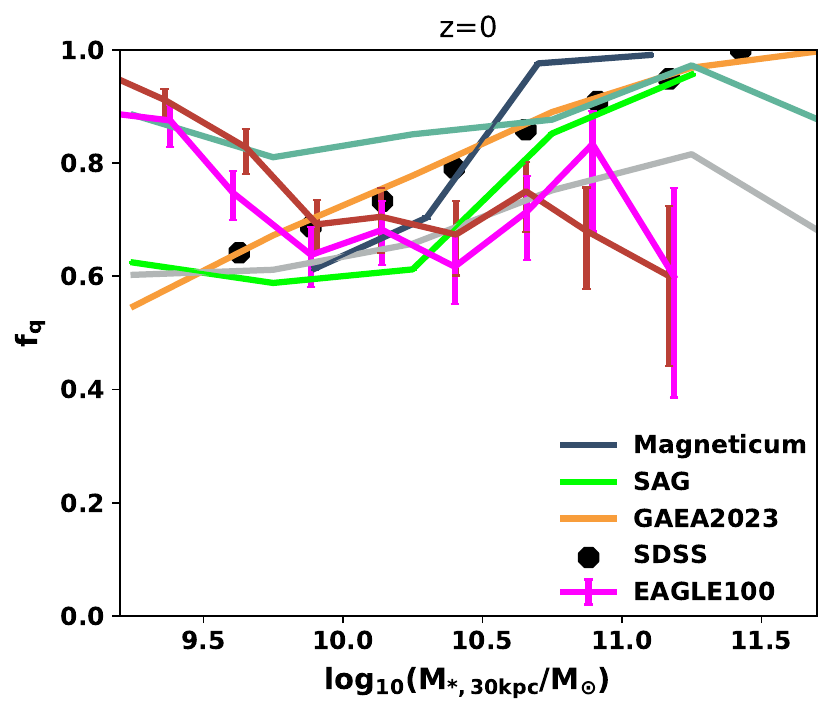}
    \end{minipage}
    \begin{minipage}{0.45\textwidth}
    \includegraphics[width=\linewidth]{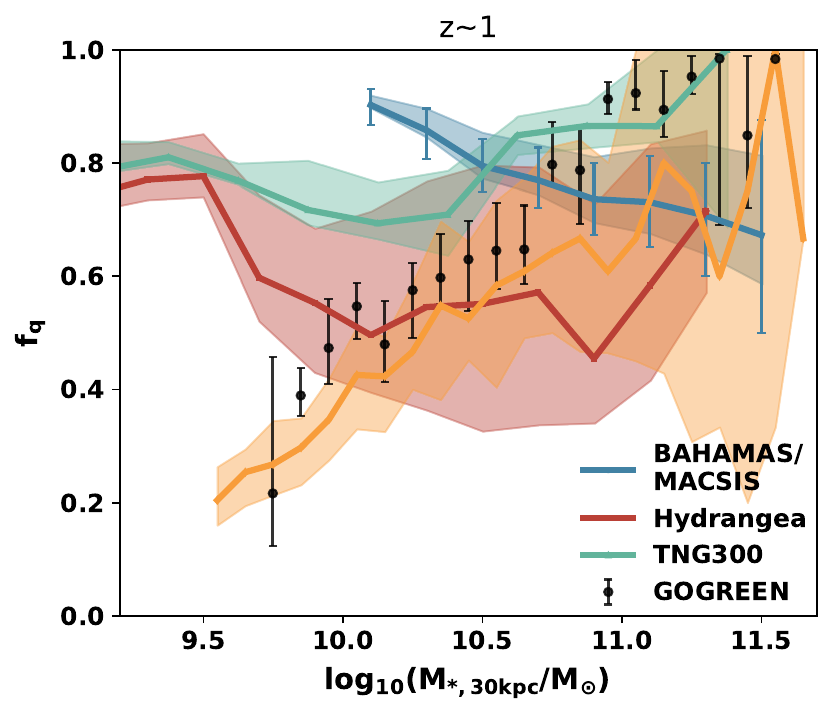}
    \end{minipage}
    \caption{Passive fractions of satellite galaxies in cluster halos ($10^{14} M_{\odot} < M_h  < 10^{14.5} M_{\odot}$ ) at $z=0$ (left) and $z\sim 1$ (right). The left panel is adapted from \citet{Xie_etal_2020}. The right figure is adapted from \citet{Kukstas_etal_2023} and \citet{DeLucia_etal_2024}. }
    \label{fig:qf_comparemodel}
\end{figure*}

At the scale of massive galaxy clusters, dominated by satellite galaxies, theoretical models and hydro-dynamical simulations in particular, still struggle with the over-quenching problem. This can be appreciated in Figure~\ref{fig:qf_comparemodel} that shows a comparison between several state-of-the-art theoretical models and observational measurements at $z=0$ in the left panel and $z\sim 1$ on the right. Most of hydrodynamical simulations predict quenched fractions that are significantly larger than observational measurements for low-mass satellites. As mentioned above, limited resolution might be in part responsible for these disagreements. At low-z, it has been shown that the excess of passive satellites translates into an excess of HI-poor satellite galaxies in hydro-dynamical simulations of cluster haloes, especially in the central regions that are dominated by satellites accreted at earlier times \citep{Diemer_etal_2019, Chen_etal_2024}. Albeit more successful, theoretical models based on semi-analytic approach also struggle to reproduce the different observational measurements available for cluster galaxies. In particular, they tend to under-predict the number densities of quenched galaxies around the knee of the galaxy stellar mass function \citep{DeLucia_etal_2024} and to under-predict the HI content of satellite galaxies \citep{Brown_etal_2017, Xie_etal_2020}. Therefore, modelling the quenching of satellite galaxies in dense environments remains a challenge for modern theoretical models.

\begin{figure*}
    \centering
    \includegraphics[width=0.45\linewidth]{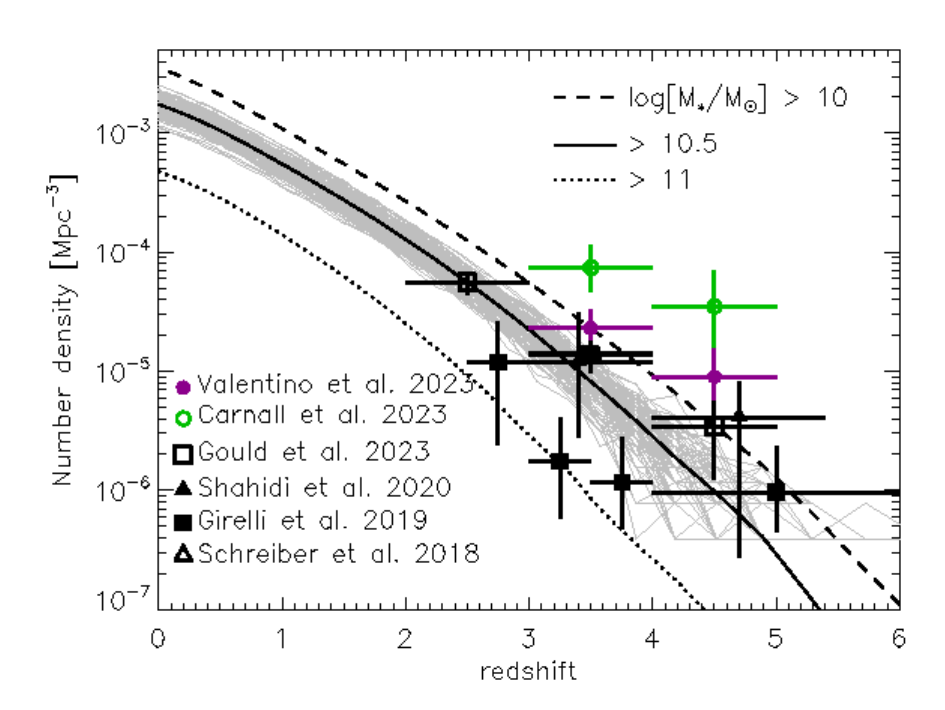}
    \includegraphics[width=0.45\linewidth]{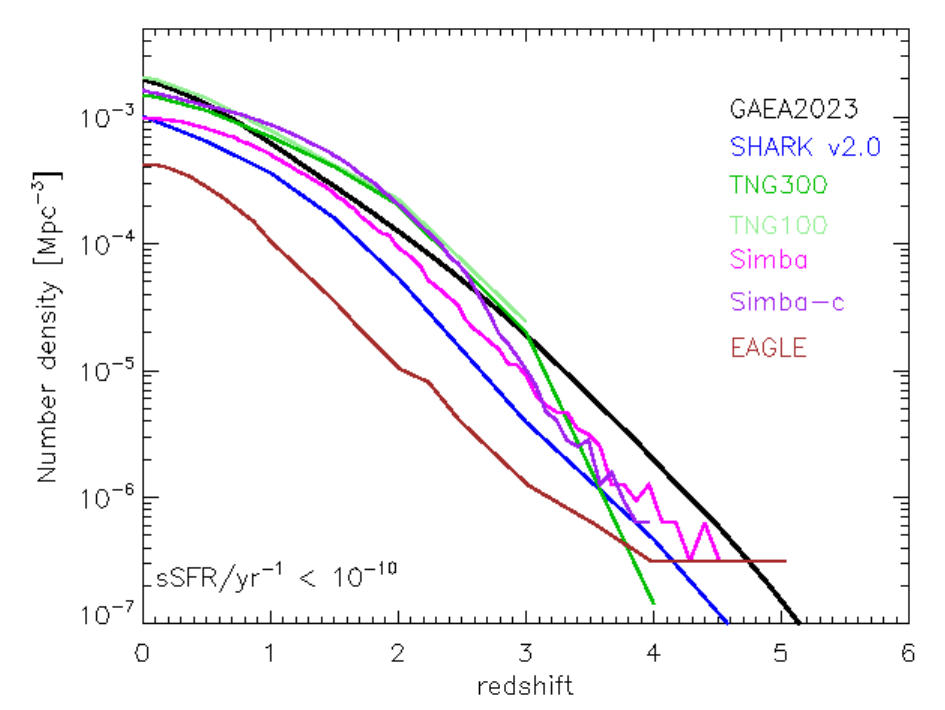}
    \caption{Cumulative number density of massive quiescent galaxies as a function of redshift, adapted from \citep{DeLucia_etal_2024}. The left panel shows a compilation of recent observational measurements compared with predictions from the GAEA model. The solid black line shows the predicted number densities for a mass cut similar to that of the observational measurements. The gray lines show predictions obtained considering 125 independent sub-volumes of $\sim 140$~Mpc on a side. The right panel shows the predicted number densities of galaxies more massive than $10^{10.5}\,{\rm M}_{\odot}$ from a compilation of recently published theoretical models.}
    \label{fig:numdens}
\end{figure*}

As discussed in Section~\ref{chap1:sec2}, the assembly of massive quiescent galaxies in a short time ($\sim 1.5-2$ billion years when focusing on observations at $z>3$) places strong constraints on galaxy formation models. Indeed, despite the large uncertainties in the determination of stellar mass and SFR at these early epochs, most modern galaxy evolution models have difficulties reproducing the measured space densities for this population \citep[e.g.][and references therein]{Lagos_etal_2024a, DeLucia_etal_2024}. The left panel of Figure~\ref{fig:numdens} shows a compilation of recent measurements compared with predictions from the GAEA model introduced above. The solid black line shows the predicted number densities for a mass cut similar to that of the observational measurements. The gray lines show predictions obtained considering 125 independent sub-volumes of $\sim 140$~Mpc on a side, approximately equal to the total effective volume analysed in \citet{Valentino_etal_2023}. The right panel of Figure~\ref{fig:numdens} shows the predicted number densities of galaxies more massive than $10^{10.5}\,{\rm M}_{\odot}$ from a compilation of recently published theoretical models. The figure shows that the expected cosmic variance is large, and can easily accommodate at least some of the recently published measurements. Characterizing the environment of high-z quiescent galaxies and what is the impact of different selections for quiescent galaxies is the next obvious step both from the theoretical and the observational point of view. 

\begin{figure*}
    \centering
    \includegraphics[width=0.85\linewidth]{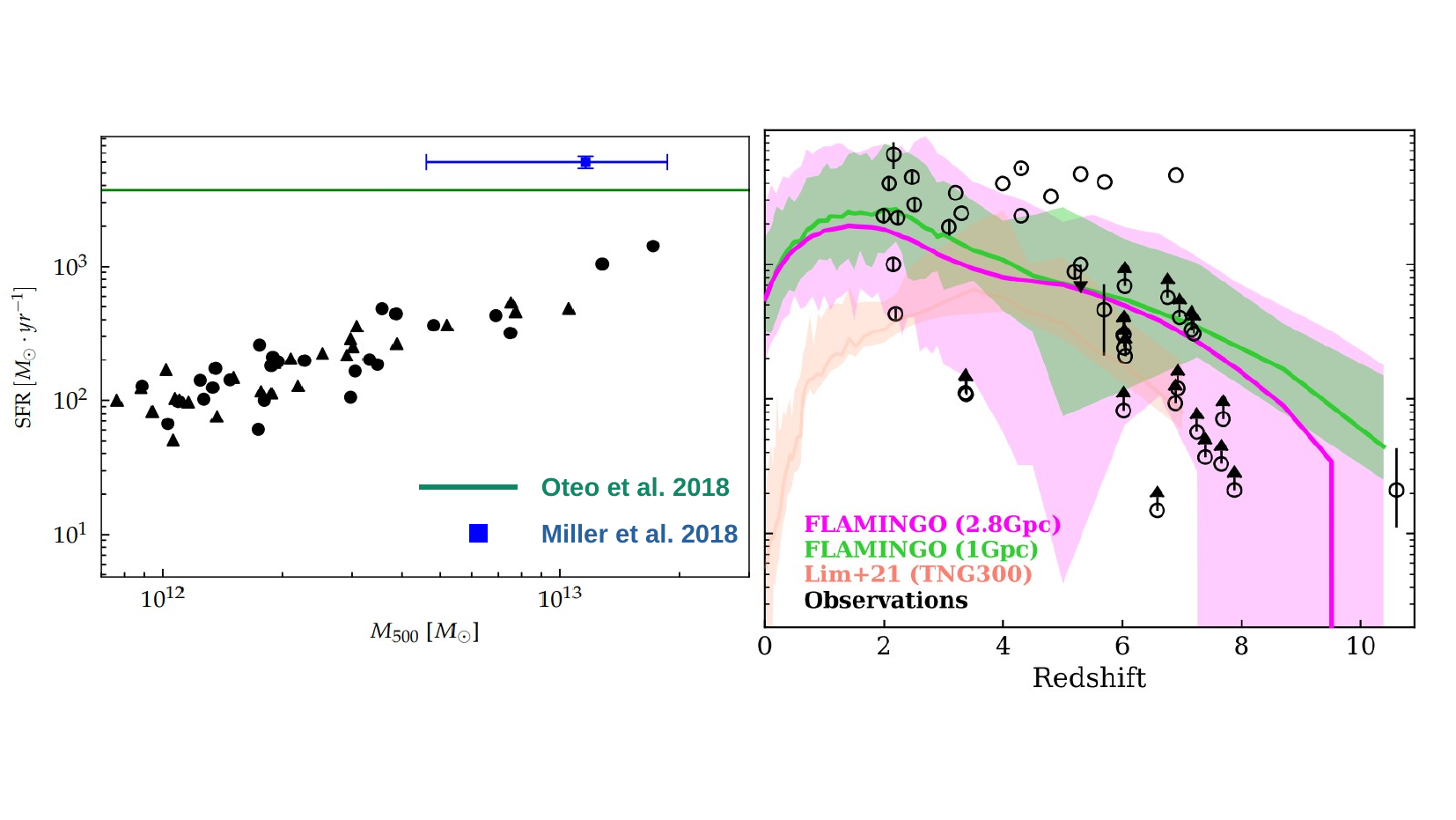}
    \caption{The left panel is from \citet{Bassini_etal_2020} and shows the star formation rate as a function of $M_{500}$ at $z\sim 4.3$.  The blue square and the green line show observational estimates; black symbols are for simulated proto-cluster regions from the DIANOGA simulations. The right panel, from \citet{Lim_etal_2024}, shows the cosmic star formation history corresponding to proto-cluster regions extracted from two different suites of large hydro-dynamical simulations compared to observational estimates. FLAMINGO provides a better agreement with observational estimates at $z>2$, but overpredicts significantly observational measurements at low-z (not shown in the figure).}
    \label{fig:clusfr}
\end{figure*}

A problem that is possibly related to tension between predicted and observed number densities of high-z quiescent galaxies is the {\it lack} of star formation activity with respect to observational estimates in proto-cluster regions. Figure~\ref{fig:clusfr} shows a comparison between observational estimates of SFRs in proto-cluster regions at $z\sim 4$ and results obtained for simulated proto-cluster regions at the same redshift \citep{Bassini_etal_2020}.  The right panel shows the cosmic star formation history corresponding to proto-cluster regions extracted from two suites of hydro-dynamical simulations (lines of different colours) and observational estimates for cluster and proto-cluster regions at $z>2$. This deficit of star formation in proto-cluster regions is a reflection of a general trend seen in virtually all models recently published that tend to under-predict the normalization of the relation between galaxy stellar mass and star formation (sometimes called the `main sequence') at $z\sim 2$. The origin of this discrepancy is unclear, with both biases and selection effects in observational measurements and systematic differences between theoretical and observational measurements likely playing a role. 

Significant theoretical work has recently focused on the physical processes that lead to early galaxy quenching. In recent theoretical studies, this is generally attributed to AGN feedback although predictions can vary significantly from one model to another depending on the detailed implementation of this specific process. For example, as mentioned above,  AGN feedback is modelled as kinetic jet-mode feedback for the radiatively inefficient regime in the simulation suite TNG. There is a strong and clear correlation between AGN feedback and galaxy quenching that happens when the radiatively inefficient AGN feedback mode is triggered \citep{Kurinchi-Vendhan_etal_2024}. At $z\sim 3$ this happens when the BH reaches a mass of $10^{8.6}\,{\rm M}_{\odot}$ \citep[see Fig.~12 in][]{Lagos_etal_2024b}. In the most recent rendition of the GAEA model mentioned above, high-z quenched galaxies experience large black hole accretion rates and suffer radiatively efficient AGN winds right before quenching \citep{Xie_etal_2024}. These large accretion rates onto the central black holes are driven by galaxy mergers for massive galaxies and by disk instabilities for low-to-intermediate mass galaxies. In this model, AGN feedback can efficiently suppress star formation at lower BH masses than TNG. We refer to \citet{Lagos_etal_2024b} for a detailed description of how AGN feedback shapes model predictions for high-z quenching in different recent theoretical models \citep[see also][for analyses based on specific theoretical models]{Kimmig_etal_2023,Szpila_etal_2024}.  The strong correlation between high-z quenching and AGN found in the models appears consistent with the large fraction of active AGN that is found in JWST surveys \citep[][]{Kocevski_etal_2023}. The fraction of AGN is even larger when computed just among quenched galaxies, although these findings are based on small samples and bright AGN. As the observational samples of high redshift quiescent galaxies grow in size and quality of the data, they will represent an important and new testbed against which we can test current implementations of AGN feedback.

\section{Outlook}\label{chap1:sec5}

As discussed in previous sections, the endeavour of understanding cosmic quenching (like many aspects of galaxy evolution in general) is complicated by the interplay between different physical mechanisms, and by the intricate connection between the evolution of galaxies and that of the environment in which they reside. In the previous sections, we have summarized progress that has been made on this subject both from the observational and the theoretical view points, and we have touched upon several questions that remain open as to the time of writing (November 2024). Below we briefly summarize our view on how the field will evolve in the next few years, and what are the main steps to be taken.  

From the observational point of view, we are truly living in a golden age in terms of accumulation of high-quality data that will aid in shedding light on galaxy evolution in general and star-formation quenching in particular. As discussed above, the successful launch of JWST has represented a watershed moment for the study of quiescent galaxies at high-z. However, while the telescope is continuing accumulating data, the identification of statistical samples of high-z massive quiescent galaxies requires large areas to beat their low number densities, and this makes even JWST inefficient given the small field of view. Other space facilities that will aid to overcome these limitations are rapidly coming into operation. As an example, Euclid was successfully launched in July 2023 and the first data demonstrated its potential to gather in one shot sharp images of an unprecedentedly large patch of the sky. In $\sim 6$ years, Euclid will carry out a wide area imaging and spectroscopic survey (the Euclid Wide Survey; \citealt{Scaramella_etal_2022}) in one visible and three near-IR bands, of $\sim 14,000 \, {\rm deg}^2$ down to ${\rm m}_{\rm AB} = 24$ in the near IR, and of $\sim 50\, {\rm deg^2}$ down to two magnitudes deeper (the Euclid Deep Survey). While the mission has been optimized for cosmology, it will yield a treasure trove with unique legacy value for galaxy evolution studies: the vast volume explored will allow us to characterize, for the first time simultaneously, a wide range of environments, ranging from voids to filaments and groups, up to the most extreme over-densities of clusters and proto-clusters.  Euclid data, combined with ground-based imaging, will allow us to estimate photometric redshifts, stellar masses, star formation rates, presence and relevance of central active nuclei for billions of galaxies out to $z\sim 2$ ($\sim 3$ in the deep fields). Colour selections and novel machine-learning methodologies can be used to efficiently identify passive galaxies \citep{Bisigello_etal_2020, Humphrey_etal_2023}. In addition, spectroscopy in the deep fields will efficiently identify the rarest, most massive passive galaxies ($>10^{11}\,{\rm M}_{\odot}$) at $z>1.5$ and up to $z\sim 3$ via detection of the D4000 break \citep{Mellier_etal_2024}. Wide-field surveys currently under definition\footnote{https://roman.gsfc.nasa.gov/science/core$\_$community$\_$survey$\_$definition.html} with the Nancy Grace Roman Space Telescope, scheduled to launch 2027, complemented with optical and IR data will also play a crucial role for statistical studies of rare these rare systems. Some important limitations of surveys that are optimized for cosmology are the relatively poor spectral resolution (the majority of the spectra will have modest signal-to-noise, often with just a single emission line to grant the measurement of the redshift), biases in the tracers used that are not necessarily ideal for studies focusing on quenching (e.g. Euclid will primarily trace H$\alpha$ and, therefore, star-forming galaxies), and non-uniform sampling rate in different environments (this is related to the use of biased tracers). 

On the ground, new precious information will be collected through upcoming optical/near-IR multi-object spectroscopy at 8-10m class telescopes like e.g. MOONS (the Multi-Object Optical and Near-infrared Spectrograph) at the Very Large Telescope (VLT) in Chile, and PSF (the Prime Focus Spectrograph) at the Subaru telescope in Hawaii. The large multiplex and wavelength coverage of these new facilities will provide high-quality spectra for very large number of galaxies at cosmic noon ($1 < z < 3$), effectively matching the range of volume, stellar masses, and environments sampled by SDSS in the local Universe \citep{Maiolino_etal_2020,Greene_et_al_2022}. Redshift surveys conducted at these new facilities will allow detailed simultaneous characterizations of the galaxy physical properties and of their local and larger scale environments, shedding new light on the physical processes leading to cosmic quenching at an epoch when galaxies formed about half of their current stellar mass.  On a longer time scale, there are proposals to bring these capabilities to a new level with spectroscopic survey facilities on dedicated wide field of view 10-12m class telescopes \citep[][]{MSE_2019, Mainieri_etal_2024}. However, data at these facilities will be inevitably affected by the atmosphere limiting the redshift window of the observations, the signal-to-noise of the spectrum, and the capability to observe the continuum for faint systems. As we have learned in the past decades, it is to be expected that real progress in the next future will come from combining observations at different wavelengths, taken at different facilities, as well as from taking advantage of resolved spatial information. Both existing state-of-the-art instrumentation (including e.g. ALMA that will continue mapping the cold molecular gas reservoirs out of which stars form, and thermal emission from interstellar dust being, and MUSE that enables resolved kinematic and stellar population studies of nearby galaxies as well as relating gas traced by absorbers to galaxies) and future facilities (like e.g. SKA that will allow us to trace atomic gas content of galaxies to $z\sim 2$) will play a major role. 

From the theoretical viewpoint, decades of work have clarified which are the physical processes that can lead to quenching of star formation. As discussed in Section~\ref{chap1:sec3}, a large body of work has also led to a quite detailed understanding of the impact of these different physical mechanisms at different mass scales and cosmic epochs, {\it when these processes are considered in isolation}. In order to achieve a better and {\it quantitative} understanding of the physical origin of cosmic quenching, detailed and careful comparisons are needed between observational measurements and predictions from theoretical models that attempt to incorporate both realistic treatments of different physical processes and of their connection with the environmental history of galaxies in a cosmological framework. As discussed above, progress is being made on several fronts, ranging from the the construction of extremely high-resolution and sophisticated hydro-dynamical simulations that are needed to better understand and characterize the physics and impact of specific processes (see Section~\ref{chap1:sec3}), to simulations of ever larger volumes that are highly needed to account for the full range of mass and spatial scales accessible by observations as well as the full range of environments in which galaxies reside \citep[e.g. the new FLAMINGO project][]{Schaye_etal_2023}.  When focusing on cosmological simulations, significant effort is going into implementing a better (more direct) modeling of the cold gaseous phase ($T < 10^4$~K). This requires incorporating different processes like the formation of molecules and dust as well as their effect on gas cooling and in general on the gas thermodynamical properties. Given the high density of the cold gas, attempts to improve its modeling leads to highly non-linear gas dynamics and very short time-steps resulting in greater computational expense. As discussed earlier, theoretical models based on a semi-analytic approach are also becoming more and more sophisticated, including an explicit treatment for the partition of the cold gas in its atomic and molecular components \citep{Lagos_etal_2011,Somerville_etal_2015, Xie_etal_2017} and an explicit treatment for the formation, growth and destruction of dust \citep{Popping_etal_2017, Vijayan_etal_2019, Triani_etal_2021, Parente_etal_2023}. While all these studies are very much needed, it is clear that the combination of volumes and resolution that is already accessible today and that will dramatically increase in the next future also calls for the necessity to find new ways of combining different theoretical techniques to overcome their specific limitations (i.e. either small volumes but very high resolution, or very large volumes but lack of internal spatial resolution or lack of an explicit treatment of the hydro-dynamics). Novel machine learning tools and artificial intelligence are likely going to play an important role in this field in the coming decades. 

Another important aspect that requires attention is related to the adequacy of the models that we have in hands to the new regimes that we can now probe observationally. In fact, most if not all prescriptions that are employed to model the evolution of the baryons in theoretical models of galaxy formation are based on observations in the `local' Universe. These might not adequately describe the physical conditions at different cosmic epochs. For example, it has been suggested that a very high efficiency of conversion of accreted gas into stars is expected in the most massive haloes at high redshift ($z>10$), as a consequence of the high densities and low metallicities expected at this epoch: these lead to free-fall times that are shorter than the delay between a starbust and the onset of stellar winds and supernovae feedback leading to a `feedback-free' starburst phase \citep{Dekel_etal_2023}. In addition, at these early epochs, one would expect a more top-heavy (i.e. with a larger fraction of massive stars) stellar Initial Mass Function due both to the higher CMB temperature and lower gas metallicity. An evolving IMF can have a strong impact on galaxy properties as it affects chemical enrichment patters, the efficiency of stellar feedback, and the overall fraction of baryons locked in long-lived stars \citep[e.g.][and references therein]{Fontanot_etal_2017}. The impact of these effects on the observable properties of galaxies, at different cosmic epochs, has not yet been tested fully and self-consistently. Finally, similar considerations apply to the tools that we routinely use to infer physical properties from observational data like e.g. the star formation histories or IMF adopted in sed-fitting that can have a strong impact on the inferred stellar masses and star formation rates, or the lack of well established and throughly tested $\alpha$-enhanced stellar population models for young stellar populations that are required for meaningful measurements of stellar metallicities of high-z quiescent galaxies. 




\bibliographystyle{Harvard}
\bibliography{reference}

\begin{thebibliography*}{246}
\providecommand{\bibtype}[1]{}
\providecommand{\natexlab}[1]{#1}
{\catcode`\|=0\catcode`\#=12\catcode`\@=11\catcode`\\=12
|immediate|write|@auxout{\expandafter\ifx\csname
  natexlab\endcsname\relax\gdef\natexlab#1{#1}\fi}}
\renewcommand{\url}[1]{{\tt #1}}
\providecommand{\urlprefix}{URL }
\expandafter\ifx\csname urlstyle\endcsname\relax
  \providecommand{\doi}[1]{doi:\discretionary{}{}{}#1}\else
  \providecommand{\doi}{doi:\discretionary{}{}{}\begingroup
  \urlstyle{rm}\Url}\fi
\providecommand{\bibinfo}[2]{#2}
\providecommand{\eprint}[2][]{\url{#2}}

\bibtype{Article}%
\bibitem[{Abazajian} and {et al.}(2009)]{Abazajian_etal_2009}
\bibinfo{author}{{Abazajian} KN} and  \bibinfo{author}{{et al.}}
  (\bibinfo{year}{2009}), \bibinfo{month}{Jun.}
\bibinfo{title}{{The Seventh Data Release of the Sloan Digital Sky Survey}}.
\bibinfo{journal}{{\em \apjs}} \bibinfo{volume}{182} (\bibinfo{number}{2}):
  \bibinfo{pages}{543--558}. \bibinfo{doi}{\doi{10.1088/0067-0049/182/2/543}}.
\eprint{0812.0649}.

\bibtype{Article}%
\bibitem[{Alberts} and {Noble}(2022)]{Alberts_and_Noble_2022}
\bibinfo{author}{{Alberts} S} and  \bibinfo{author}{{Noble} A}
  (\bibinfo{year}{2022}), \bibinfo{month}{Oct.}
\bibinfo{title}{{From Clusters to Proto-Clusters: The Infrared Perspective on
  Environmental Galaxy Evolution}}.
\bibinfo{journal}{{\em Universe}} \bibinfo{volume}{8} (\bibinfo{number}{11}),
  \bibinfo{eid}{554}. \bibinfo{doi}{\doi{10.3390/universe8110554}}.
\eprint{2209.02781}.

\bibtype{Article}%
\bibitem[{Alberts} et al.(2016)]{Alberts_etal_2016}
\bibinfo{author}{{Alberts} S}, \bibinfo{author}{{Pope} A},
  \bibinfo{author}{{Brodwin} M}, \bibinfo{author}{{Chung} SM},
  \bibinfo{author}{{Cybulski} R}, \bibinfo{author}{{Dey} A},
  \bibinfo{author}{{Eisenhardt} PRM}, \bibinfo{author}{{Galametz} A},
  \bibinfo{author}{{Gonzalez} AH}, \bibinfo{author}{{Jannuzi} BT},
  \bibinfo{author}{{Stanford} SA}, \bibinfo{author}{{Snyder} GF},
  \bibinfo{author}{{Stern} D} and  \bibinfo{author}{{Zeimann} GR}
  (\bibinfo{year}{2016}), \bibinfo{month}{Jul.}
\bibinfo{title}{{Star Formation and AGN Activity in Galaxy Clusters from z=1-2:
  a Multi-Wavelength Analysis Featuring Herschel/PACS}}.
\bibinfo{journal}{{\em \apj}} \bibinfo{volume}{825} (\bibinfo{number}{1}),
  \bibinfo{eid}{72}. \bibinfo{doi}{\doi{10.3847/0004-637X/825/1/72}}.
\eprint{1604.03564}.

\bibtype{Article}%
\bibitem[{Andersson} et al.(2024)]{Andersson24}
\bibinfo{author}{{Andersson} EP}, \bibinfo{author}{{Rey} MP},
  \bibinfo{author}{{Pontzen} A}, \bibinfo{author}{{Cadiou} C},
  \bibinfo{author}{{Agertz} O}, \bibinfo{author}{{Read} JI} and
  \bibinfo{author}{{Martin} NF} (\bibinfo{year}{2024}), \bibinfo{month}{Sep.}
\bibinfo{title}{{EDGE-INFERNO: Simulating every observable star in faint dwarf
  galaxies and their consequences for resolved-star photometric surveys}}.
\bibinfo{journal}{{\em arXiv e-prints}} ,
  \bibinfo{eid}{arXiv:2409.08073}\bibinfo{doi}{\doi{10.48550/arXiv.2409.08073}}.
\eprint{2409.08073}.

\bibtype{Article}%
\bibitem[{Andreon} et al.(2014)]{Andreon_etal_2014}
\bibinfo{author}{{Andreon} S}, \bibinfo{author}{{Newman} AB},
  \bibinfo{author}{{Trinchieri} G}, \bibinfo{author}{{Raichoor} A},
  \bibinfo{author}{{Ellis} RS} and  \bibinfo{author}{{Treu} T}
  (\bibinfo{year}{2014}), \bibinfo{month}{May}.
\bibinfo{title}{{JKCS 041: a Coma cluster progenitor at z = 1.803}}.
\bibinfo{journal}{{\em \aap}} \bibinfo{volume}{565}, \bibinfo{eid}{A120}.
  \bibinfo{doi}{\doi{10.1051/0004-6361/201323077}}.
\eprint{1311.4361}.

\bibtype{Article}%
\bibitem[{Arnouts} et al.(2013)]{Arnouts_etal_2013}
\bibinfo{author}{{Arnouts} S}, \bibinfo{author}{{Le Floc'h} E},
  \bibinfo{author}{{Chevallard} J}, \bibinfo{author}{{Johnson} BD},
  \bibinfo{author}{{Ilbert} O}, \bibinfo{author}{{Treyer} M},
  \bibinfo{author}{{Aussel} H}, \bibinfo{author}{{Capak} P},
  \bibinfo{author}{{Sanders} DB}, \bibinfo{author}{{Scoville} N},
  \bibinfo{author}{{McCracken} HJ}, \bibinfo{author}{{Milliard} B},
  \bibinfo{author}{{Pozzetti} L} and  \bibinfo{author}{{Salvato} M}
  (\bibinfo{year}{2013}), \bibinfo{month}{Oct.}
\bibinfo{title}{{Encoding of the infrared excess in the NUVrK color diagram for
  star-forming galaxies}}.
\bibinfo{journal}{{\em \aap}} \bibinfo{volume}{558}, \bibinfo{eid}{A67}.
  \bibinfo{doi}{\doi{10.1051/0004-6361/201321768}}.
\eprint{1309.0008}.

\bibtype{Article}%
\bibitem[{Aumer} et al.(2013)]{Aumer13}
\bibinfo{author}{{Aumer} M}, \bibinfo{author}{{White} SDM},
  \bibinfo{author}{{Naab} T} and  \bibinfo{author}{{Scannapieco} C}
  (\bibinfo{year}{2013}), \bibinfo{month}{Oct.}
\bibinfo{title}{{Towards a more realistic population of bright spiral galaxies
  in cosmological simulations}}.
\bibinfo{journal}{{\em \mnras}} \bibinfo{volume}{434}:
  \bibinfo{pages}{3142--3164}. \bibinfo{doi}{\doi{10.1093/mnras/stt1230}}.
\eprint{1304.1559}.

\bibtype{Article}%
\bibitem[{Bah{\'e}} and {McCarthy}(2015)]{Bahe_McCarthy_2015}
\bibinfo{author}{{Bah{\'e}} YM} and  \bibinfo{author}{{McCarthy} IG}
  (\bibinfo{year}{2015}), \bibinfo{month}{Feb.}
\bibinfo{title}{{Star formation quenching in simulated group and cluster
  galaxies: when, how, and why?}}
\bibinfo{journal}{{\em \mnras}} \bibinfo{volume}{447} (\bibinfo{number}{1}):
  \bibinfo{pages}{969--992}. \bibinfo{doi}{\doi{10.1093/mnras/stu2293}}.
\eprint{1410.8161}.

\bibtype{Article}%
\bibitem[{Bah{\'e}} et al.(2017)]{Bahe_etal_2017}
\bibinfo{author}{{Bah{\'e}} YM}, \bibinfo{author}{{Barnes} DJ},
  \bibinfo{author}{{Dalla Vecchia} C}, \bibinfo{author}{{Kay} ST},
  \bibinfo{author}{{White} SDM}, \bibinfo{author}{{McCarthy} IG},
  \bibinfo{author}{{Schaye} J}, \bibinfo{author}{{Bower} RG},
  \bibinfo{author}{{Crain} RA}, \bibinfo{author}{{Theuns} T},
  \bibinfo{author}{{Jenkins} A}, \bibinfo{author}{{McGee} SL},
  \bibinfo{author}{{Schaller} M}, \bibinfo{author}{{Thomas} PA} and
  \bibinfo{author}{{Trayford} JW} (\bibinfo{year}{2017}), \bibinfo{month}{Oct.}
\bibinfo{title}{{The Hydrangea simulations: galaxy formation in and around
  massive clusters}}.
\bibinfo{journal}{{\em \mnras}} \bibinfo{volume}{470} (\bibinfo{number}{4}):
  \bibinfo{pages}{4186--4208}. \bibinfo{doi}{\doi{10.1093/mnras/stx1403}}.
\eprint{1703.10610}.

\bibtype{Article}%
\bibitem[{Baker} et al.(2024)]{Baker_etal_2024}
\bibinfo{author}{{Baker} WM}, \bibinfo{author}{{Lim} S},
  \bibinfo{author}{{D'Eugenio} F}, \bibinfo{author}{{Maiolino} R},
  \bibinfo{author}{{Ji} Z}, \bibinfo{author}{{Arribas} S},
  \bibinfo{author}{{Bunker} AJ}, \bibinfo{author}{{Carniani} S},
  \bibinfo{author}{{Charlot} S}, \bibinfo{author}{{de Graaff} A},
  \bibinfo{author}{{Hainline} K}, \bibinfo{author}{{Looser} TJ},
  \bibinfo{author}{{Lyu} J}, \bibinfo{author}{{Rinaldi} P},
  \bibinfo{author}{{Robertson} B}, \bibinfo{author}{{Schaller} M},
  \bibinfo{author}{{Schaye} J}, \bibinfo{author}{{Scholtz} J},
  \bibinfo{author}{{Ubler} H}, \bibinfo{author}{{Williams} CC},
  \bibinfo{author}{{Willmer} CNA}, \bibinfo{author}{{Willott} C} and
  \bibinfo{author}{{Zhu} Y} (\bibinfo{year}{2024}), \bibinfo{month}{Oct.}
\bibinfo{title}{{The abundance and nature of high-redshift quiescent galaxies
  from JADES spectroscopy and the FLAMINGO simulations}}.
\bibinfo{journal}{{\em arXiv e-prints}} ,
  \bibinfo{eid}{arXiv:2410.14773}\bibinfo{doi}{\doi{10.48550/arXiv.2410.14773}}.
\eprint{2410.14773}.

\bibtype{Inproceedings}%
\bibitem[{Baldry} et al.(2004{\natexlab{a}})]{Baldry_etal_2004b}
\bibinfo{author}{{Baldry} IK}, \bibinfo{author}{{Balogh} ML},
  \bibinfo{author}{{Bower} R}, \bibinfo{author}{{Glazebrook} K} and
  \bibinfo{author}{{Nichol} RC} (\bibinfo{year}{2004}{\natexlab{a}}),
  \bibinfo{month}{Dec.}, \bibinfo{title}{{Color bimodality: Implications for
  galaxy evolution}}, \bibinfo{editor}{{Allen} RE},
  \bibinfo{editor}{{Nanopoulos} DV} and  \bibinfo{editor}{{Pope} CN}, (Eds.),
  \bibinfo{booktitle}{The New Cosmology: Conference on Strings and Cosmology},
  \bibinfo{series}{American Institute of Physics Conference Series},
  \bibinfo{volume}{743}, \bibinfo{publisher}{AIP},  \bibinfo{pages}{106--119},
  \eprint{astro-ph/0410603}.

\bibtype{Article}%
\bibitem[{Baldry} et al.(2004{\natexlab{b}})]{Baldry_etal_2004a}
\bibinfo{author}{{Baldry} IK}, \bibinfo{author}{{Glazebrook} K},
  \bibinfo{author}{{Brinkmann} J}, \bibinfo{author}{{Ivezi{\'c}} {\v{Z}}},
  \bibinfo{author}{{Lupton} RH}, \bibinfo{author}{{Nichol} RC} and
  \bibinfo{author}{{Szalay} AS} (\bibinfo{year}{2004}{\natexlab{b}}),
  \bibinfo{month}{Jan.}
\bibinfo{title}{{Quantifying the Bimodal Color-Magnitude Distribution of
  Galaxies}}.
\bibinfo{journal}{{\em \apj}} \bibinfo{volume}{600} (\bibinfo{number}{2}):
  \bibinfo{pages}{681--694}. \bibinfo{doi}{\doi{10.1086/380092}}.
\eprint{astro-ph/0309710}.

\bibtype{Article}%
\bibitem[{Baldry} et al.(2006)]{Baldry_etal_2006}
\bibinfo{author}{{Baldry} IK}, \bibinfo{author}{{Balogh} ML},
  \bibinfo{author}{{Bower} RG}, \bibinfo{author}{{Glazebrook} K},
  \bibinfo{author}{{Nichol} RC}, \bibinfo{author}{{Bamford} SP} and
  \bibinfo{author}{{Budavari} T} (\bibinfo{year}{2006}), \bibinfo{month}{Dec.}
\bibinfo{title}{{Galaxy bimodality versus stellar mass and environment}}.
\bibinfo{journal}{{\em \mnras}} \bibinfo{volume}{373} (\bibinfo{number}{2}):
  \bibinfo{pages}{469--483}.
  \bibinfo{doi}{\doi{10.1111/j.1365-2966.2006.11081.x}}.
\eprint{astro-ph/0607648}.

\bibtype{Article}%
\bibitem[{Balogh} et al.(2004)]{Balogh_etal_2004}
\bibinfo{author}{{Balogh} ML}, \bibinfo{author}{{Baldry} IK},
  \bibinfo{author}{{Nichol} R}, \bibinfo{author}{{Miller} C},
  \bibinfo{author}{{Bower} R} and  \bibinfo{author}{{Glazebrook} K}
  (\bibinfo{year}{2004}), \bibinfo{month}{Nov.}
\bibinfo{title}{{The Bimodal Galaxy Color Distribution: Dependence on
  Luminosity and Environment}}.
\bibinfo{journal}{{\em \apjl}} \bibinfo{volume}{615} (\bibinfo{number}{2}):
  \bibinfo{pages}{L101--L104}. \bibinfo{doi}{\doi{10.1086/426079}}.
\eprint{astro-ph/0406266}.

\bibtype{Article}%
\bibitem[{Balogh} et al.(2017)]{Balogh_etal_2017}
\bibinfo{author}{{Balogh} ML}, \bibinfo{author}{{Gilbank} DG},
  \bibinfo{author}{{Muzzin} A}, \bibinfo{author}{{Rudnick} G},
  \bibinfo{author}{{Cooper} MC}, \bibinfo{author}{{Lidman} C},
  \bibinfo{author}{{Biviano} A}, \bibinfo{author}{{Demarco} R},
  \bibinfo{author}{{McGee} SL}, \bibinfo{author}{{Nantais} JB},
  \bibinfo{author}{{Noble} A}, \bibinfo{author}{{Old} L},
  \bibinfo{author}{{Wilson} G}, \bibinfo{author}{{Yee} HKC},
  \bibinfo{author}{{Bellhouse} C}, \bibinfo{author}{{Cerulo} P},
  \bibinfo{author}{{Chan} J}, \bibinfo{author}{{Pintos-Castro} I},
  \bibinfo{author}{{Simpson} R}, \bibinfo{author}{{van der Burg} RFJ},
  \bibinfo{author}{{Zaritsky} D}, \bibinfo{author}{{Ziparo} F},
  \bibinfo{author}{{Alonso} MV}, \bibinfo{author}{{Bower} RG},
  \bibinfo{author}{{De Lucia} G}, \bibinfo{author}{{Finoguenov} A},
  \bibinfo{author}{{Lambas} DG}, \bibinfo{author}{{Muriel} H},
  \bibinfo{author}{{Parker} LC}, \bibinfo{author}{{Rettura} A},
  \bibinfo{author}{{Valotto} C} and  \bibinfo{author}{{Wetzel} A}
  (\bibinfo{year}{2017}), \bibinfo{month}{Oct.}
\bibinfo{title}{{Gemini Observations of Galaxies in Rich Early Environments
  (GOGREEN) I: survey description}}.
\bibinfo{journal}{{\em \mnras}} \bibinfo{volume}{470} (\bibinfo{number}{4}):
  \bibinfo{pages}{4168--4185}. \bibinfo{doi}{\doi{10.1093/mnras/stx1370}}.
\eprint{1705.01606}.

\bibtype{Article}%
\bibitem[{Balogh} et al.(2021)]{Balogh_etal_2021}
\bibinfo{author}{{Balogh} ML}, \bibinfo{author}{{van der Burg} RFJ},
  \bibinfo{author}{{Muzzin} A}, \bibinfo{author}{{Rudnick} G},
  \bibinfo{author}{{Wilson} G}, \bibinfo{author}{{Webb} K},
  \bibinfo{author}{{Biviano} A}, \bibinfo{author}{{Boak} K},
  \bibinfo{author}{{Cerulo} P}, \bibinfo{author}{{Chan} J},
  \bibinfo{author}{{Cooper} MC}, \bibinfo{author}{{Gilbank} DG},
  \bibinfo{author}{{Gwyn} S}, \bibinfo{author}{{Lidman} C},
  \bibinfo{author}{{Matharu} J}, \bibinfo{author}{{McGee} SL},
  \bibinfo{author}{{Old} L}, \bibinfo{author}{{Pintos-Castro} I},
  \bibinfo{author}{{Reeves} AMM}, \bibinfo{author}{{Shipley} H},
  \bibinfo{author}{{Vulcani} B}, \bibinfo{author}{{Yee} HKC},
  \bibinfo{author}{{Alonso} MV}, \bibinfo{author}{{Bellhouse} C},
  \bibinfo{author}{{Cooke} KC}, \bibinfo{author}{{Davidson} A},
  \bibinfo{author}{{De Lucia} G}, \bibinfo{author}{{Demarco} R},
  \bibinfo{author}{{Drakos} N}, \bibinfo{author}{{Fillingham} SP},
  \bibinfo{author}{{Finoguenov} A}, \bibinfo{author}{{Forrest} B},
  \bibinfo{author}{{Golledge} C}, \bibinfo{author}{{Jablonka} P},
  \bibinfo{author}{{Lambas Garcia} D}, \bibinfo{author}{{McNab} K},
  \bibinfo{author}{{Muriel} H}, \bibinfo{author}{{Nantais} JB},
  \bibinfo{author}{{Noble} A}, \bibinfo{author}{{Parker} LC},
  \bibinfo{author}{{Petter} G}, \bibinfo{author}{{Poggianti} BM},
  \bibinfo{author}{{Townsend} M}, \bibinfo{author}{{Valotto} C},
  \bibinfo{author}{{Webb} T} and  \bibinfo{author}{{Zaritsky} D}
  (\bibinfo{year}{2021}), \bibinfo{month}{Jan.}
\bibinfo{title}{{The GOGREEN and GCLASS surveys: first data release}}.
\bibinfo{journal}{{\em \mnras}} \bibinfo{volume}{500} (\bibinfo{number}{1}):
  \bibinfo{pages}{358--387}. \bibinfo{doi}{\doi{10.1093/mnras/staa3008}}.
\eprint{2009.13345}.

\bibtype{Article}%
\bibitem[{Barai} et al.(2018)]{Barai18}
\bibinfo{author}{{Barai} P}, \bibinfo{author}{{Gallerani} S},
  \bibinfo{author}{{Pallottini} A}, \bibinfo{author}{{Ferrara} A},
  \bibinfo{author}{{Marconi} A}, \bibinfo{author}{{Cicone} C},
  \bibinfo{author}{{Maiolino} R} and  \bibinfo{author}{{Carniani} S}
  (\bibinfo{year}{2018}), \bibinfo{month}{Jan.}
\bibinfo{title}{{Quasar outflows at z {\ensuremath{\geq}} 6: the impact on the
  host galaxies}}.
\bibinfo{journal}{{\em \mnras}} \bibinfo{volume}{473} (\bibinfo{number}{3}):
  \bibinfo{pages}{4003--4020}. \bibinfo{doi}{\doi{10.1093/mnras/stx2563}}.
\eprint{1707.03014}.

\bibtype{Article}%
\bibitem[{Bassini} et al.(2020)]{Bassini_etal_2020}
\bibinfo{author}{{Bassini} L}, \bibinfo{author}{{Rasia} E},
  \bibinfo{author}{{Borgani} S}, \bibinfo{author}{{Granato} GL},
  \bibinfo{author}{{Ragone-Figueroa} C}, \bibinfo{author}{{Biffi} V},
  \bibinfo{author}{{Ragagnin} A}, \bibinfo{author}{{Dolag} K},
  \bibinfo{author}{{Lin} W}, \bibinfo{author}{{Murante} G},
  \bibinfo{author}{{Napolitano} NR}, \bibinfo{author}{{Taffoni} G},
  \bibinfo{author}{{Tornatore} L} and  \bibinfo{author}{{Wang} Y}
  (\bibinfo{year}{2020}), \bibinfo{month}{Oct.}
\bibinfo{title}{{The DIANOGA simulations of galaxy clusters: characterising
  star formation in protoclusters}}.
\bibinfo{journal}{{\em \aap}} \bibinfo{volume}{642}, \bibinfo{eid}{A37}.
  \bibinfo{doi}{\doi{10.1051/0004-6361/202038396}}.
\eprint{2006.13951}.

\bibtype{Article}%
\bibitem[{Baugh}(2006)]{Baugh_etal_2006}
\bibinfo{author}{{Baugh} CM} (\bibinfo{year}{2006}), \bibinfo{month}{Dec.}
\bibinfo{title}{{A primer on hierarchical galaxy formation: the semi-analytical
  approach}}.
\bibinfo{journal}{{\em Reports on Progress in Physics}} \bibinfo{volume}{69}
  (\bibinfo{number}{12}): \bibinfo{pages}{3101--3156}.
  \bibinfo{doi}{\doi{10.1088/0034-4885/69/12/R02}}.
\eprint{astro-ph/0610031}.

\bibtype{Article}%
\bibitem[{Bell} et al.(2004)]{Bell_etal_2004}
\bibinfo{author}{{Bell} EF}, \bibinfo{author}{{Wolf} C},
  \bibinfo{author}{{Meisenheimer} K}, \bibinfo{author}{{Rix} HW},
  \bibinfo{author}{{Borch} A}, \bibinfo{author}{{Dye} S},
  \bibinfo{author}{{Kleinheinrich} M}, \bibinfo{author}{{Wisotzki} L} and
  \bibinfo{author}{{McIntosh} DH} (\bibinfo{year}{2004}), \bibinfo{month}{Jun.}
\bibinfo{title}{{Nearly 5000 Distant Early-Type Galaxies in COMBO-17: A Red
  Sequence and Its Evolution since z\raisebox{-0.5ex}\textasciitilde1}}.
\bibinfo{journal}{{\em \apj}} \bibinfo{volume}{608} (\bibinfo{number}{2}):
  \bibinfo{pages}{752--767}. \bibinfo{doi}{\doi{10.1086/420778}}.
\eprint{astro-ph/0303394}.

\bibtype{Article}%
\bibitem[{Benson} and {Bower}(2011)]{Benson_and_Bower_2011}
\bibinfo{author}{{Benson} AJ} and  \bibinfo{author}{{Bower} R}
  (\bibinfo{year}{2011}), \bibinfo{month}{Feb.}
\bibinfo{title}{{Accretion shocks and cold filaments in galaxy formation}}.
\bibinfo{journal}{{\em \mnras}} \bibinfo{volume}{410} (\bibinfo{number}{4}):
  \bibinfo{pages}{2653--2661}.
  \bibinfo{doi}{\doi{10.1111/j.1365-2966.2010.17641.x}}.

\bibtype{Article}%
\bibitem[{Benson} et al.(2003)]{Benson_etal_2003}
\bibinfo{author}{{Benson} AJ}, \bibinfo{author}{{Bower} RG},
  \bibinfo{author}{{Frenk} CS}, \bibinfo{author}{{Lacey} CG},
  \bibinfo{author}{{Baugh} CM} and  \bibinfo{author}{{Cole} S}
  (\bibinfo{year}{2003}), \bibinfo{month}{Dec.}
\bibinfo{title}{{What Shapes the Luminosity Function of Galaxies?}}
\bibinfo{journal}{{\em \apj}} \bibinfo{volume}{599} (\bibinfo{number}{1}):
  \bibinfo{pages}{38--49}. \bibinfo{doi}{\doi{10.1086/379160}}.
\eprint{astro-ph/0302450}.

\bibtype{Article}%
\bibitem[{Best} et al.(2005)]{Best_etal_2005}
\bibinfo{author}{{Best} PN}, \bibinfo{author}{{Kauffmann} G},
  \bibinfo{author}{{Heckman} TM}, \bibinfo{author}{{Brinchmann} J},
  \bibinfo{author}{{Charlot} S}, \bibinfo{author}{{Ivezi{\'c}} {\v{Z}}} and
  \bibinfo{author}{{White} SDM} (\bibinfo{year}{2005}), \bibinfo{month}{Sep.}
\bibinfo{title}{{The host galaxies of radio-loud active galactic nuclei: mass
  dependences, gas cooling and active galactic nuclei feedback}}.
\bibinfo{journal}{{\em \mnras}} \bibinfo{volume}{362} (\bibinfo{number}{1}):
  \bibinfo{pages}{25--40}.
  \bibinfo{doi}{\doi{10.1111/j.1365-2966.2005.09192.x}}.
\eprint{astro-ph/0506269}.

\bibtype{Article}%
\bibitem[{Best} et al.(2007)]{Best_etal_2007}
\bibinfo{author}{{Best} PN}, \bibinfo{author}{{von der Linden} A},
  \bibinfo{author}{{Kauffmann} G}, \bibinfo{author}{{Heckman} TM} and
  \bibinfo{author}{{Kaiser} CR} (\bibinfo{year}{2007}), \bibinfo{month}{Aug.}
\bibinfo{title}{{On the prevalence of radio-loud active galactic nuclei in
  brightest cluster galaxies: implications for AGN heating of cooling flows}}.
\bibinfo{journal}{{\em \mnras}} \bibinfo{volume}{379} (\bibinfo{number}{3}):
  \bibinfo{pages}{894--908}.
  \bibinfo{doi}{\doi{10.1111/j.1365-2966.2007.11937.x}}.
\eprint{astro-ph/0611197}.

\bibtype{Article}%
\bibitem[{Bieri} et al.(2017)]{Bieri17}
\bibinfo{author}{{Bieri} R}, \bibinfo{author}{{Dubois} Y},
  \bibinfo{author}{{Rosdahl} J}, \bibinfo{author}{{Wagner} A},
  \bibinfo{author}{{Silk} J} and  \bibinfo{author}{{Mamon} GA}
  (\bibinfo{year}{2017}), \bibinfo{month}{Jan.}
\bibinfo{title}{{Outflows driven by quasars in high-redshift galaxies with
  radiation hydrodynamics}}.
\bibinfo{journal}{{\em \mnras}} \bibinfo{volume}{464}:
  \bibinfo{pages}{1854--1873}. \bibinfo{doi}{\doi{10.1093/mnras/stw2380}}.
\eprint{1606.06281}.

\bibtype{Article}%
\bibitem[{Binney}(1977)]{Binney_1977}
\bibinfo{author}{{Binney} J} (\bibinfo{year}{1977}), \bibinfo{month}{Jul.}
\bibinfo{title}{{The physics of dissipational galaxy formation.}}
\bibinfo{journal}{{\em \apj}} \bibinfo{volume}{215}: \bibinfo{pages}{483--491}.
  \bibinfo{doi}{\doi{10.1086/155378}}.

\bibtype{Article}%
\bibitem[{Bischetti} et al.(2019)]{Bischetti_etal_2019}
\bibinfo{author}{{Bischetti} M}, \bibinfo{author}{{Maiolino} R},
  \bibinfo{author}{{Carniani} S}, \bibinfo{author}{{Fiore} F},
  \bibinfo{author}{{Piconcelli} E} and  \bibinfo{author}{{Fluetsch} A}
  (\bibinfo{year}{2019}), \bibinfo{month}{Oct.}
\bibinfo{title}{{Widespread QSO-driven outflows in the early Universe}}.
\bibinfo{journal}{{\em \aap}} \bibinfo{volume}{630}, \bibinfo{eid}{A59}.
  \bibinfo{doi}{\doi{10.1051/0004-6361/201833557}}.
\eprint{1806.00786}.

\bibtype{Article}%
\bibitem[{Bisigello} et al.(2020)]{Bisigello_etal_2020}
\bibinfo{author}{{Bisigello} L}, \bibinfo{author}{{Kuchner} U},
  \bibinfo{author}{{Conselice} CJ}, \bibinfo{author}{{Andreon} S},
  \bibinfo{author}{{Bolzonella} M}, \bibinfo{author}{{Duc} PA},
  \bibinfo{author}{{Garilli} B}, \bibinfo{author}{{Humphrey} A},
  \bibinfo{author}{{Maraston} C}, \bibinfo{author}{{Moresco} M},
  \bibinfo{author}{{Pozzetti} L}, \bibinfo{author}{{Tortora} C},
  \bibinfo{author}{{Zamorani} G}, \bibinfo{author}{{Auricchio} N},
  \bibinfo{author}{{Brinchmann} J}, \bibinfo{author}{{Capobianco} V},
  \bibinfo{author}{{Carretero} J}, \bibinfo{author}{{Castander} FJ},
  \bibinfo{author}{{Castellano} M}, \bibinfo{author}{{Cavuoti} S},
  \bibinfo{author}{{Cimatti} A}, \bibinfo{author}{{Cledassou} R},
  \bibinfo{author}{{Congedo} G}, \bibinfo{author}{{Conversi} L},
  \bibinfo{author}{{Corcione} L}, \bibinfo{author}{{Cropper} MS},
  \bibinfo{author}{{Dusini} S}, \bibinfo{author}{{Frailis} M},
  \bibinfo{author}{{Franceschi} E}, \bibinfo{author}{{Franzetti} P},
  \bibinfo{author}{{Fumana} M}, \bibinfo{author}{{Hormuth} F},
  \bibinfo{author}{{Israel} H}, \bibinfo{author}{{Jahnke} K},
  \bibinfo{author}{{Kermiche} S}, \bibinfo{author}{{Kitching} T},
  \bibinfo{author}{{Kohley} R}, \bibinfo{author}{{Kubik} B},
  \bibinfo{author}{{Kunz} M}, \bibinfo{author}{{Le F{\`e}vre} O},
  \bibinfo{author}{{Ligori} S}, \bibinfo{author}{{Lilje} PB},
  \bibinfo{author}{{Lloro} I}, \bibinfo{author}{{Maiorano} E},
  \bibinfo{author}{{Marggraf} O}, \bibinfo{author}{{Massey} R},
  \bibinfo{author}{{Masters} DC}, \bibinfo{author}{{Mei} S},
  \bibinfo{author}{{Mellier} Y}, \bibinfo{author}{{Meylan} G},
  \bibinfo{author}{{Padilla} C}, \bibinfo{author}{{Paltani} S},
  \bibinfo{author}{{Pasian} F}, \bibinfo{author}{{Pettorino} V},
  \bibinfo{author}{{Pires} S}, \bibinfo{author}{{Polenta} G},
  \bibinfo{author}{{Poncet} M}, \bibinfo{author}{{Raison} F},
  \bibinfo{author}{{Rhodes} J}, \bibinfo{author}{{Roncarelli} M},
  \bibinfo{author}{{Rossetti} E}, \bibinfo{author}{{Saglia} R},
  \bibinfo{author}{{Sauvage} M}, \bibinfo{author}{{Schneider} P},
  \bibinfo{author}{{Secroun} A}, \bibinfo{author}{{Serrano} S},
  \bibinfo{author}{{Sureau} F}, \bibinfo{author}{{Taylor} AN},
  \bibinfo{author}{{Tereno} I}, \bibinfo{author}{{Toledo-Moreo} R},
  \bibinfo{author}{{Valenziano} L}, \bibinfo{author}{{Wang} Y},
  \bibinfo{author}{{Wetzstein} M} and  \bibinfo{author}{{Zoubian} J}
  (\bibinfo{year}{2020}), \bibinfo{month}{May}.
\bibinfo{title}{{Euclid: the selection of quiescent and star-forming galaxies
  using observed colours}}.
\bibinfo{journal}{{\em \mnras}} \bibinfo{volume}{494} (\bibinfo{number}{2}):
  \bibinfo{pages}{2337--2354}. \bibinfo{doi}{\doi{10.1093/mnras/staa885}}.
\eprint{2003.07367}.

\bibtype{Article}%
\bibitem[{Blanton} et al.(2003)]{Blanton_etal_2003}
\bibinfo{author}{{Blanton} MR}, \bibinfo{author}{{Hogg} DW},
  \bibinfo{author}{{Bahcall} NA}, \bibinfo{author}{{Baldry} IK},
  \bibinfo{author}{{Brinkmann} J}, \bibinfo{author}{{Csabai} I},
  \bibinfo{author}{{Eisenstein} D}, \bibinfo{author}{{Fukugita} M},
  \bibinfo{author}{{Gunn} JE}, \bibinfo{author}{{Ivezi{\'c}} {\v{Z}}},
  \bibinfo{author}{{Lamb} DQ}, \bibinfo{author}{{Lupton} RH},
  \bibinfo{author}{{Loveday} J}, \bibinfo{author}{{Munn} JA},
  \bibinfo{author}{{Nichol} RC}, \bibinfo{author}{{Okamura} S},
  \bibinfo{author}{{Schlegel} DJ}, \bibinfo{author}{{Shimasaku} K},
  \bibinfo{author}{{Strauss} MA}, \bibinfo{author}{{Vogeley} MS} and
  \bibinfo{author}{{Weinberg} DH} (\bibinfo{year}{2003}), \bibinfo{month}{Sep.}
\bibinfo{title}{{The Broadband Optical Properties of Galaxies with Redshifts
  0.02<z<0.22}}.
\bibinfo{journal}{{\em \apj}} \bibinfo{volume}{594} (\bibinfo{number}{1}):
  \bibinfo{pages}{186--207}. \bibinfo{doi}{\doi{10.1086/375528}}.
\eprint{astro-ph/0209479}.

\bibtype{Article}%
\bibitem[{Bluck} et al.(2014)]{Bluck_etal_2014}
\bibinfo{author}{{Bluck} AFL}, \bibinfo{author}{{Mendel} JT},
  \bibinfo{author}{{Ellison} SL}, \bibinfo{author}{{Moreno} J},
  \bibinfo{author}{{Simard} L}, \bibinfo{author}{{Patton} DR} and
  \bibinfo{author}{{Starkenburg} E} (\bibinfo{year}{2014}),
  \bibinfo{month}{Jun.}
\bibinfo{title}{{Bulge mass is king: the dominant role of the bulge in
  determining the fraction of passive galaxies in the Sloan Digital Sky
  Survey}}.
\bibinfo{journal}{{\em \mnras}} \bibinfo{volume}{441} (\bibinfo{number}{1}):
  \bibinfo{pages}{599--629}. \bibinfo{doi}{\doi{10.1093/mnras/stu594}}.
\eprint{1403.5269}.

\bibtype{Article}%
\bibitem[{Boselli} et al.(2014)]{Boselli_etal_2014}
\bibinfo{author}{{Boselli} A}, \bibinfo{author}{{Cortese} L},
  \bibinfo{author}{{Boquien} M}, \bibinfo{author}{{Boissier} S},
  \bibinfo{author}{{Catinella} B}, \bibinfo{author}{{Gavazzi} G},
  \bibinfo{author}{{Lagos} C} and  \bibinfo{author}{{Saintonge} A}
  (\bibinfo{year}{2014}), \bibinfo{month}{Apr.}
\bibinfo{title}{{Cold gas properties of the Herschel Reference Survey. III.
  Molecular gas stripping in cluster galaxies}}.
\bibinfo{journal}{{\em \aap}} \bibinfo{volume}{564}, \bibinfo{eid}{A67}.
  \bibinfo{doi}{\doi{10.1051/0004-6361/201322313}}.
\eprint{1402.0326}.

\bibtype{Article}%
\bibitem[{Boselli} et al.(2016)]{Boselli_etal_2016a}
\bibinfo{author}{{Boselli} A}, \bibinfo{author}{{Cuillandre} JC},
  \bibinfo{author}{{Fossati} M}, \bibinfo{author}{{Boissier} S},
  \bibinfo{author}{{Bomans} D}, \bibinfo{author}{{Consolandi} G},
  \bibinfo{author}{{Anselmi} G}, \bibinfo{author}{{Cortese} L},
  \bibinfo{author}{{C{\^o}t{\'e}} P}, \bibinfo{author}{{Durrell} P},
  \bibinfo{author}{{Ferrarese} L}, \bibinfo{author}{{Fumagalli} M},
  \bibinfo{author}{{Gavazzi} G}, \bibinfo{author}{{Gwyn} S},
  \bibinfo{author}{{Hensler} G}, \bibinfo{author}{{Sun} M} and
  \bibinfo{author}{{Toloba} E} (\bibinfo{year}{2016}), \bibinfo{month}{Mar.}
\bibinfo{title}{{Spectacular tails of ionized gas in the Virgo cluster galaxy
  NGC 4569}}.
\bibinfo{journal}{{\em \aap}} \bibinfo{volume}{587}, \bibinfo{eid}{A68}.
  \bibinfo{doi}{\doi{10.1051/0004-6361/201527795}}.
\eprint{1601.04978}.

\bibtype{Article}%
\bibitem[{Boselli} et al.(2022)]{Boselli_etal_2022}
\bibinfo{author}{{Boselli} A}, \bibinfo{author}{{Fossati} M} and
  \bibinfo{author}{{Sun} M} (\bibinfo{year}{2022}), \bibinfo{month}{Dec.}
\bibinfo{title}{{Ram pressure stripping in high-density environments}}.
\bibinfo{journal}{{\em \aapr}} \bibinfo{volume}{30} (\bibinfo{number}{1}),
  \bibinfo{eid}{3}. \bibinfo{doi}{\doi{10.1007/s00159-022-00140-3}}.
\eprint{2109.13614}.

\bibtype{Article}%
\bibitem[{Bower} et al.(1998)]{Bower_etal_1998}
\bibinfo{author}{{Bower} RG}, \bibinfo{author}{{Kodama} T} and
  \bibinfo{author}{{Terlevich} A} (\bibinfo{year}{1998}), \bibinfo{month}{Oct.}
\bibinfo{title}{{The colour-magnitude relation as a constraint on the formation
  of rich cluster galaxies}}.
\bibinfo{journal}{{\em \mnras}} \bibinfo{volume}{299} (\bibinfo{number}{4}):
  \bibinfo{pages}{1193--1208}.
  \bibinfo{doi}{\doi{10.1046/j.1365-8711.1998.01868.x}}.
\eprint{astro-ph/9805290}.

\bibtype{Article}%
\bibitem[{Bower} et al.(2006)]{Bower_etal_2006}
\bibinfo{author}{{Bower} RG}, \bibinfo{author}{{Benson} AJ},
  \bibinfo{author}{{Malbon} R}, \bibinfo{author}{{Helly} JC},
  \bibinfo{author}{{Frenk} CS}, \bibinfo{author}{{Baugh} CM},
  \bibinfo{author}{{Cole} S} and  \bibinfo{author}{{Lacey} CG}
  (\bibinfo{year}{2006}), \bibinfo{month}{Aug.}
\bibinfo{title}{{Breaking the hierarchy of galaxy formation}}.
\bibinfo{journal}{{\em \mnras}} \bibinfo{volume}{370} (\bibinfo{number}{2}):
  \bibinfo{pages}{645--655}.
  \bibinfo{doi}{\doi{10.1111/j.1365-2966.2006.10519.x}}.
\eprint{astro-ph/0511338}.

\bibtype{Article}%
\bibitem[{Brammer} et al.(2009)]{Brammer_etal_2009}
\bibinfo{author}{{Brammer} GB}, \bibinfo{author}{{Whitaker} KE},
  \bibinfo{author}{{van Dokkum} PG}, \bibinfo{author}{{Marchesini} D},
  \bibinfo{author}{{Labb{\'e}} I}, \bibinfo{author}{{Franx} M},
  \bibinfo{author}{{Kriek} M}, \bibinfo{author}{{Quadri} RF},
  \bibinfo{author}{{Illingworth} G}, \bibinfo{author}{{Lee} KS},
  \bibinfo{author}{{Muzzin} A} and  \bibinfo{author}{{Rudnick} G}
  (\bibinfo{year}{2009}), \bibinfo{month}{Nov.}
\bibinfo{title}{{The Dead Sequence: A Clear Bimodality in Galaxy Colors from z
  = 0 to z = 2.5}}.
\bibinfo{journal}{{\em \apjl}} \bibinfo{volume}{706} (\bibinfo{number}{1}):
  \bibinfo{pages}{L173--L177}.
  \bibinfo{doi}{\doi{10.1088/0004-637X/706/1/L173}}.
\eprint{0910.2227}.

\bibtype{Article}%
\bibitem[{Brammer} et al.(2011)]{Brammer_etal_2011}
\bibinfo{author}{{Brammer} GB}, \bibinfo{author}{{Whitaker} KE},
  \bibinfo{author}{{van Dokkum} PG}, \bibinfo{author}{{Marchesini} D},
  \bibinfo{author}{{Franx} M}, \bibinfo{author}{{Kriek} M},
  \bibinfo{author}{{Labb{\'e}} I}, \bibinfo{author}{{Lee} KS},
  \bibinfo{author}{{Muzzin} A}, \bibinfo{author}{{Quadri} RF},
  \bibinfo{author}{{Rudnick} G} and  \bibinfo{author}{{Williams} R}
  (\bibinfo{year}{2011}), \bibinfo{month}{Sep.}
\bibinfo{title}{{The Number Density and Mass Density of Star-forming and
  Quiescent Galaxies at 0.4 <= z <= 2.2}}.
\bibinfo{journal}{{\em \apj}} \bibinfo{volume}{739} (\bibinfo{number}{1}),
  \bibinfo{eid}{24}. \bibinfo{doi}{\doi{10.1088/0004-637X/739/1/24}}.
\eprint{1104.2595}.

\bibtype{Article}%
\bibitem[{Bremer} et al.(2018)]{Bremer_etal_2018}
\bibinfo{author}{{Bremer} MN}, \bibinfo{author}{{Phillipps} S},
  \bibinfo{author}{{Kelvin} LS}, \bibinfo{author}{{De Propris} R},
  \bibinfo{author}{{Kennedy} R}, \bibinfo{author}{{Moffett} AJ},
  \bibinfo{author}{{Bamford} S}, \bibinfo{author}{{Davies} LJM},
  \bibinfo{author}{{Driver} SP}, \bibinfo{author}{{H{\"a}u{\ss}ler} B},
  \bibinfo{author}{{Holwerda} B}, \bibinfo{author}{{Hopkins} A},
  \bibinfo{author}{{James} PA}, \bibinfo{author}{{Liske} J},
  \bibinfo{author}{{Percival} S} and  \bibinfo{author}{{Taylor} EN}
  (\bibinfo{year}{2018}), \bibinfo{month}{May}.
\bibinfo{title}{{Galaxy and Mass Assembly (GAMA): Morphological transformation
  of galaxies across the green valley}}.
\bibinfo{journal}{{\em \mnras}} \bibinfo{volume}{476} (\bibinfo{number}{1}):
  \bibinfo{pages}{12--26}. \bibinfo{doi}{\doi{10.1093/mnras/sty124}}.
\eprint{1801.04277}.

\bibtype{Article}%
\bibitem[{Brennan} et al.(2018)]{Brennan_etal_2018}
\bibinfo{author}{{Brennan} R}, \bibinfo{author}{{Choi} E},
  \bibinfo{author}{{Somerville} RS}, \bibinfo{author}{{Hirschmann} M},
  \bibinfo{author}{{Naab} T} and  \bibinfo{author}{{Ostriker} JP}
  (\bibinfo{year}{2018}), \bibinfo{month}{Jun.}
\bibinfo{title}{{Momentum-driven Winds from Radiatively Efficient Black Hole
  Accretion and Their Impact on Galaxies}}.
\bibinfo{journal}{{\em \apj}} \bibinfo{volume}{860} (\bibinfo{number}{1}),
  \bibinfo{eid}{14}. \bibinfo{doi}{\doi{10.3847/1538-4357/aac2c4}}.
\eprint{1805.00946}.

\bibtype{Article}%
\bibitem[{Brinchmann} et al.(2004)]{Brinchmann_etal_2004}
\bibinfo{author}{{Brinchmann} J}, \bibinfo{author}{{Charlot} S},
  \bibinfo{author}{{White} SDM}, \bibinfo{author}{{Tremonti} C},
  \bibinfo{author}{{Kauffmann} G}, \bibinfo{author}{{Heckman} T} and
  \bibinfo{author}{{Brinkmann} J} (\bibinfo{year}{2004}), \bibinfo{month}{Jul.}
\bibinfo{title}{{The physical properties of star-forming galaxies in the
  low-redshift Universe}}.
\bibinfo{journal}{{\em \mnras}} \bibinfo{volume}{351} (\bibinfo{number}{4}):
  \bibinfo{pages}{1151--1179}.
  \bibinfo{doi}{\doi{10.1111/j.1365-2966.2004.07881.x}}.
\eprint{astro-ph/0311060}.

\bibtype{Article}%
\bibitem[{Brodwin} et al.(2013)]{Brodwin_etal_2013}
\bibinfo{author}{{Brodwin} M}, \bibinfo{author}{{Stanford} SA},
  \bibinfo{author}{{Gonzalez} AH}, \bibinfo{author}{{Zeimann} GR},
  \bibinfo{author}{{Snyder} GF}, \bibinfo{author}{{Mancone} CL},
  \bibinfo{author}{{Pope} A}, \bibinfo{author}{{Eisenhardt} PR},
  \bibinfo{author}{{Stern} D}, \bibinfo{author}{{Alberts} S},
  \bibinfo{author}{{Ashby} MLN}, \bibinfo{author}{{Brown} MJI},
  \bibinfo{author}{{Chary} RR}, \bibinfo{author}{{Dey} A},
  \bibinfo{author}{{Galametz} A}, \bibinfo{author}{{Gettings} DP},
  \bibinfo{author}{{Jannuzi} BT}, \bibinfo{author}{{Miller} ED},
  \bibinfo{author}{{Moustakas} J} and  \bibinfo{author}{{Moustakas} LA}
  (\bibinfo{year}{2013}), \bibinfo{month}{Dec.}
\bibinfo{title}{{The Era of Star Formation in Galaxy Clusters}}.
\bibinfo{journal}{{\em \apj}} \bibinfo{volume}{779} (\bibinfo{number}{2}),
  \bibinfo{eid}{138}. \bibinfo{doi}{\doi{10.1088/0004-637X/779/2/138}}.
\eprint{1310.6039}.

\bibtype{Article}%
\bibitem[{Brown} et al.(2017)]{Brown_etal_2017}
\bibinfo{author}{{Brown} T}, \bibinfo{author}{{Catinella} B},
  \bibinfo{author}{{Cortese} L}, \bibinfo{author}{{Lagos} CdP},
  \bibinfo{author}{{Dav{\'e}} R}, \bibinfo{author}{{Kilborn} V},
  \bibinfo{author}{{Haynes} MP}, \bibinfo{author}{{Giovanelli} R} and
  \bibinfo{author}{{Rafieferantsoa} M} (\bibinfo{year}{2017}),
  \bibinfo{month}{Apr.}
\bibinfo{title}{{Cold gas stripping in satellite galaxies: from pairs to
  clusters}}.
\bibinfo{journal}{{\em \mnras}} \bibinfo{volume}{466} (\bibinfo{number}{2}):
  \bibinfo{pages}{1275--1289}. \bibinfo{doi}{\doi{10.1093/mnras/stw2991}}.
\eprint{1611.00896}.

\bibtype{Article}%
\bibitem[{Br{\"u}ggen} and {De Lucia}(2008)]{Brueggen_and_DeLucia_2008}
\bibinfo{author}{{Br{\"u}ggen} M} and  \bibinfo{author}{{De Lucia} G}
  (\bibinfo{year}{2008}), \bibinfo{month}{Feb.}
\bibinfo{title}{{Ram-pressure histories of cluster galaxies}}.
\bibinfo{journal}{{\em \mnras}} \bibinfo{volume}{383} (\bibinfo{number}{4}):
  \bibinfo{pages}{1336--1342}.
  \bibinfo{doi}{\doi{10.1111/j.1365-2966.2007.12670.x}}.
\eprint{0710.5580}.

\bibtype{Article}%
\bibitem[{Bustard} et al.(2018)]{Bustard_etal_2018}
\bibinfo{author}{{Bustard} C}, \bibinfo{author}{{Pardy} SA},
  \bibinfo{author}{{D'Onghia} E}, \bibinfo{author}{{Zweibel} EG} and
  \bibinfo{author}{{Gallagher} J.~S. I} (\bibinfo{year}{2018}),
  \bibinfo{month}{Aug.}
\bibinfo{title}{{The Fate of Supernova-heated Gas in Star-forming Regions of
  the LMC: Lessons for Galaxy Formation?}}
\bibinfo{journal}{{\em \apj}} \bibinfo{volume}{863} (\bibinfo{number}{1}),
  \bibinfo{eid}{49}. \bibinfo{doi}{\doi{10.3847/1538-4357/aad08f}}.
\eprint{1802.07263}.

\bibtype{Article}%
\bibitem[{Butcher} and {Oemler}(1984)]{Butcher_and_Oemler_1984}
\bibinfo{author}{{Butcher} H} and  \bibinfo{author}{{Oemler} A. J}
  (\bibinfo{year}{1984}), \bibinfo{month}{Oct.}
\bibinfo{title}{{The evolution of galaxies in clusters. V. A study of
  populations since Z 0.5.}}
\bibinfo{journal}{{\em \apj}} \bibinfo{volume}{285}: \bibinfo{pages}{426--438}.
  \bibinfo{doi}{\doi{10.1086/162519}}.

\bibtype{incollection}%
\bibitem[{Calzetti}(2013)]{Calzetti_etal_2013}
\bibinfo{author}{{Calzetti} D} (\bibinfo{year}{2013}), \bibinfo{title}{{Star
  Formation Rate Indicators}}, \bibinfo{editor}{{Falc{\'o}n-Barroso} J} and
  \bibinfo{editor}{{Knapen} JH}, (Eds.), \bibinfo{booktitle}{Secular Evolution
  of Galaxies}, pp. \bibinfo{pages}{419}.

\bibtype{Article}%
\bibitem[{Carnall} et al.(2018)]{Carnall_etal_2018}
\bibinfo{author}{{Carnall} AC}, \bibinfo{author}{{McLure} RJ},
  \bibinfo{author}{{Dunlop} JS} and  \bibinfo{author}{{Dav{\'e}} R}
  (\bibinfo{year}{2018}), \bibinfo{month}{Nov.}
\bibinfo{title}{{Inferring the star formation histories of massive quiescent
  galaxies with BAGPIPES: evidence for multiple quenching mechanisms}}.
\bibinfo{journal}{{\em \mnras}} \bibinfo{volume}{480} (\bibinfo{number}{4}):
  \bibinfo{pages}{4379--4401}. \bibinfo{doi}{\doi{10.1093/mnras/sty2169}}.
\eprint{1712.04452}.

\bibtype{Article}%
\bibitem[{Carnall} et al.(2023)]{Carnall_etal_2023}
\bibinfo{author}{{Carnall} AC}, \bibinfo{author}{{McLeod} DJ},
  \bibinfo{author}{{McLure} RJ}, \bibinfo{author}{{Dunlop} JS},
  \bibinfo{author}{{Begley} R}, \bibinfo{author}{{Cullen} F},
  \bibinfo{author}{{Donnan} CT}, \bibinfo{author}{{Hamadouche} ML},
  \bibinfo{author}{{Jewell} SM}, \bibinfo{author}{{Jones} EW},
  \bibinfo{author}{{Pollock} CL} and  \bibinfo{author}{{Wild} V}
  (\bibinfo{year}{2023}), \bibinfo{month}{Apr.}
\bibinfo{title}{{A surprising abundance of massive quiescent galaxies at 3 < z
  < 5 in the first data from JWST CEERS}}.
\bibinfo{journal}{{\em \mnras}} \bibinfo{volume}{520} (\bibinfo{number}{3}):
  \bibinfo{pages}{3974--3985}. \bibinfo{doi}{\doi{10.1093/mnras/stad369}}.
\eprint{2208.00986}.

\bibtype{Article}%
\bibitem[{Carnall} et al.(2024)]{Carnall_etal_2024}
\bibinfo{author}{{Carnall} AC}, \bibinfo{author}{{Cullen} F},
  \bibinfo{author}{{McLure} RJ}, \bibinfo{author}{{McLeod} DJ},
  \bibinfo{author}{{Begley} R}, \bibinfo{author}{{Donnan} CT},
  \bibinfo{author}{{Dunlop} JS}, \bibinfo{author}{{Shapley} AE},
  \bibinfo{author}{{Rowlands} K}, \bibinfo{author}{{Almaini} O},
  \bibinfo{author}{{Arellano-C{\'o}rdova} KZ}, \bibinfo{author}{{Barrufet} L},
  \bibinfo{author}{{Cimatti} A}, \bibinfo{author}{{Ellis} RS},
  \bibinfo{author}{{Grogin} NA}, \bibinfo{author}{{Hamadouche} ML},
  \bibinfo{author}{{Illingworth} GD}, \bibinfo{author}{{Koekemoer} AM},
  \bibinfo{author}{{Leung} HH}, \bibinfo{author}{{Lovell} CC},
  \bibinfo{author}{{P{\'e}rez-Gonz{\'a}lez} PG}, \bibinfo{author}{{Santini} P},
  \bibinfo{author}{{Stanton} TM} and  \bibinfo{author}{{Wild} V}
  (\bibinfo{year}{2024}), \bibinfo{month}{Oct.}
\bibinfo{title}{{The JWST EXCELS survey: too much, too young, too fast?
  Ultra-massive quiescent galaxies at 3 < z < 5}}.
\bibinfo{journal}{{\em \mnras}} \bibinfo{volume}{534} (\bibinfo{number}{1}):
  \bibinfo{pages}{325--348}. \bibinfo{doi}{\doi{10.1093/mnras/stae2092}}.
\eprint{2405.02242}.

\bibtype{Article}%
\bibitem[{Catinella} et al.(2013)]{Catinella_etal_2013}
\bibinfo{author}{{Catinella} B}, \bibinfo{author}{{Schiminovich} D},
  \bibinfo{author}{{Cortese} L}, \bibinfo{author}{{Fabello} S},
  \bibinfo{author}{{Hummels} CB}, \bibinfo{author}{{Moran} SM},
  \bibinfo{author}{{Lemonias} JJ}, \bibinfo{author}{{Cooper} AP},
  \bibinfo{author}{{Wu} R}, \bibinfo{author}{{Heckman} TM} and
  \bibinfo{author}{{Wang} J} (\bibinfo{year}{2013}), \bibinfo{month}{Nov.}
\bibinfo{title}{{The GALEX Arecibo SDSS Survey - VIII. Final data release. The
  effect of group environment on the gas content of massive galaxies}}.
\bibinfo{journal}{{\em \mnras}} \bibinfo{volume}{436} (\bibinfo{number}{1}):
  \bibinfo{pages}{34--70}. \bibinfo{doi}{\doi{10.1093/mnras/stt1417}}.
\eprint{1308.1676}.

\bibtype{Article}%
\bibitem[{Cayatte} et al.(1990)]{Cayatte_etal_1990}
\bibinfo{author}{{Cayatte} V}, \bibinfo{author}{{van Gorkom} JH},
  \bibinfo{author}{{Balkowski} C} and  \bibinfo{author}{{Kotanyi} C}
  (\bibinfo{year}{1990}), \bibinfo{month}{Sep.}
\bibinfo{title}{{VLA Observations of Neutral Hydrogen in Virgo Cluster
  Galaxies. I. The Atlas}}.
\bibinfo{journal}{{\em \aj}} \bibinfo{volume}{100}: \bibinfo{pages}{604}.
  \bibinfo{doi}{\doi{10.1086/115545}}.

\bibtype{Article}%
\bibitem[{Chen} et al.(2024)]{Chen_etal_2024}
\bibinfo{author}{{Chen} H}, \bibinfo{author}{{Xie} L}, \bibinfo{author}{{Wang}
  J}, \bibinfo{author}{{Hu} W}, \bibinfo{author}{{De Lucia} G},
  \bibinfo{author}{{Fontanot} F} and  \bibinfo{author}{{Hirschamnn} M}
  (\bibinfo{year}{2024}), \bibinfo{month}{Feb.}
\bibinfo{title}{{Environmental effects on satellite galaxies from the
  perspective of cold gas}}.
\bibinfo{journal}{{\em \mnras}} \bibinfo{volume}{528} (\bibinfo{number}{2}):
  \bibinfo{pages}{2451--2463}. \bibinfo{doi}{\doi{10.1093/mnras/stae162}}.
\eprint{2401.07158}.

\bibtype{Article}%
\bibitem[{Chevallard} and {Charlot}(2016)]{Chevallard_and_Charlot_2016}
\bibinfo{author}{{Chevallard} J} and  \bibinfo{author}{{Charlot} S}
  (\bibinfo{year}{2016}), \bibinfo{month}{Oct.}
\bibinfo{title}{{Modelling and interpreting spectral energy distributions of
  galaxies with BEAGLE}}.
\bibinfo{journal}{{\em \mnras}} \bibinfo{volume}{462} (\bibinfo{number}{2}):
  \bibinfo{pages}{1415--1443}. \bibinfo{doi}{\doi{10.1093/mnras/stw1756}}.
\eprint{1603.03037}.

\bibtype{Article}%
\bibitem[{Choi} et al.(2015)]{Choi15}
\bibinfo{author}{{Choi} E}, \bibinfo{author}{{Ostriker} JP},
  \bibinfo{author}{{Naab} T}, \bibinfo{author}{{Oser} L} and
  \bibinfo{author}{{Moster} BP} (\bibinfo{year}{2015}), \bibinfo{month}{Jun.}
\bibinfo{title}{{The impact of mechanical AGN feedback on the formation of
  massive early-type galaxies}}.
\bibinfo{journal}{{\em \mnras}} \bibinfo{volume}{449}:
  \bibinfo{pages}{4105--4116}. \bibinfo{doi}{\doi{10.1093/mnras/stv575}}.
\eprint{1403.1257}.

\bibtype{Article}%
\bibitem[{Choi} et al.(2017)]{Choi17}
\bibinfo{author}{{Choi} E}, \bibinfo{author}{{Ostriker} JP},
  \bibinfo{author}{{Naab} T}, \bibinfo{author}{{Somerville} RS},
  \bibinfo{author}{{Hirschmann} M}, \bibinfo{author}{{N{\'u}{\~n}ez} A},
  \bibinfo{author}{{Hu} CY} and  \bibinfo{author}{{Oser} L}
  (\bibinfo{year}{2017}), \bibinfo{month}{Jul.}
\bibinfo{title}{{Physics of Galactic Metals: Evolutionary Effects due to
  Production, Distribution, Feedback, and Interaction with Black Holes}}.
\bibinfo{journal}{{\em \apj}} \bibinfo{volume}{844} (\bibinfo{number}{1}),
  \bibinfo{eid}{31}. \bibinfo{doi}{\doi{10.3847/1538-4357/aa7849}}.
\eprint{1610.09389}.

\bibtype{Article}%
\bibitem[{Cielo} et al.(2018)]{Cielo18}
\bibinfo{author}{{Cielo} S}, \bibinfo{author}{{Babul} A},
  \bibinfo{author}{{Antonuccio-Delogu} V}, \bibinfo{author}{{Silk} J} and
  \bibinfo{author}{{Volonteri} M} (\bibinfo{year}{2018}), \bibinfo{month}{Sep.}
\bibinfo{title}{{Feedback from reorienting AGN jets. I. Jet-ICM coupling,
  cavity properties and global energetics}}.
\bibinfo{journal}{{\em \aap}} \bibinfo{volume}{617}, \bibinfo{eid}{A58}.
  \bibinfo{doi}{\doi{10.1051/0004-6361/201832582}}.

\bibtype{Article}%
\bibitem[{Coil} et al.(2017)]{Coil_etal_2017}
\bibinfo{author}{{Coil} AL}, \bibinfo{author}{{Mendez} AJ},
  \bibinfo{author}{{Eisenstein} DJ} and  \bibinfo{author}{{Moustakas} J}
  (\bibinfo{year}{2017}), \bibinfo{month}{Apr.}
\bibinfo{title}{{PRIMUS+DEEP2: The Dependence of Galaxy Clustering on Stellar
  Mass and Specific Star Formation Rate at 0.2 < z < 1.2}}.
\bibinfo{journal}{{\em \apj}} \bibinfo{volume}{838} (\bibinfo{number}{2}),
  \bibinfo{eid}{87}. \bibinfo{doi}{\doi{10.3847/1538-4357/aa63ec}}.
\eprint{1609.09090}.

\bibtype{Article}%
\bibitem[{Conroy}(2013)]{Conroy_etal_2013}
\bibinfo{author}{{Conroy} C} (\bibinfo{year}{2013}), \bibinfo{month}{Aug.}
\bibinfo{title}{{Modeling the Panchromatic Spectral Energy Distributions of
  Galaxies}}.
\bibinfo{journal}{{\em \araa}} \bibinfo{volume}{51} (\bibinfo{number}{1}):
  \bibinfo{pages}{393--455}.
  \bibinfo{doi}{\doi{10.1146/annurev-astro-082812-141017}}.
\eprint{1301.7095}.

\bibtype{Article}%
\bibitem[{Cora} et al.(2018)]{Cora_etal_2018}
\bibinfo{author}{{Cora} SA}, \bibinfo{author}{{Vega-Mart{\'\i}nez} CA},
  \bibinfo{author}{{Hough} T}, \bibinfo{author}{{Ruiz} AN},
  \bibinfo{author}{{Orsi} {\'A}A}, \bibinfo{author}{{Mu{\~n}oz Arancibia} AM},
  \bibinfo{author}{{Gargiulo} ID}, \bibinfo{author}{{Collacchioni} F},
  \bibinfo{author}{{Padilla} ND}, \bibinfo{author}{{Gottl{\"o}ber} S} and
  \bibinfo{author}{{Yepes} G} (\bibinfo{year}{2018}), \bibinfo{month}{Sep.}
\bibinfo{title}{{Semi-analytic galaxies - I. Synthesis of environmental and
  star-forming regulation mechanisms}}.
\bibinfo{journal}{{\em \mnras}} \bibinfo{volume}{479} (\bibinfo{number}{1}):
  \bibinfo{pages}{2--24}. \bibinfo{doi}{\doi{10.1093/mnras/sty1131}}.
\eprint{1801.03883}.

\bibtype{Article}%
\bibitem[{Costa} et al.(2017)]{Costa17}
\bibinfo{author}{{Costa} T}, \bibinfo{author}{{Rosdahl} J},
  \bibinfo{author}{{Sijacki} D} and  \bibinfo{author}{{Haehnelt} M}
  (\bibinfo{year}{2017}), \bibinfo{month}{Mar.}
\bibinfo{title}{{Driving gas shells with radiation pressure on dust in
  radiation-hydrodynamic simulations}}.
\bibinfo{journal}{{\em ArXiv e-prints}} \eprint{1703.05766}.

\bibtype{Article}%
\bibitem[{Cowie} and {Songaila}(1977)]{Cowie_and_Songaila_1977}
\bibinfo{author}{{Cowie} LL} and  \bibinfo{author}{{Songaila} A}
  (\bibinfo{year}{1977}), \bibinfo{month}{Apr.}
\bibinfo{title}{{Thermal evaporation of gas within galaxies by a hot
  intergalactic medium}}.
\bibinfo{journal}{{\em \nat}} \bibinfo{volume}{266}: \bibinfo{pages}{501--503}.
  \bibinfo{doi}{\doi{10.1038/266501a0}}.

\bibtype{Article}%
\bibitem[{Cox} et al.(2008)]{Cox_etal_2008}
\bibinfo{author}{{Cox} TJ}, \bibinfo{author}{{Jonsson} P},
  \bibinfo{author}{{Somerville} RS}, \bibinfo{author}{{Primack} JR} and
  \bibinfo{author}{{Dekel} A} (\bibinfo{year}{2008}), \bibinfo{month}{Feb.}
\bibinfo{title}{{The effect of galaxy mass ratio on merger-driven starbursts}}.
\bibinfo{journal}{{\em \mnras}} \bibinfo{volume}{384} (\bibinfo{number}{1}):
  \bibinfo{pages}{386--409}.
  \bibinfo{doi}{\doi{10.1111/j.1365-2966.2007.12730.x}}.
\eprint{0709.3511}.

\bibtype{Article}%
\bibitem[{Cresci} et al.(2015)]{Cresci_etal_2015}
\bibinfo{author}{{Cresci} G}, \bibinfo{author}{{Marconi} A},
  \bibinfo{author}{{Zibetti} S}, \bibinfo{author}{{Risaliti} G},
  \bibinfo{author}{{Carniani} S}, \bibinfo{author}{{Mannucci} F},
  \bibinfo{author}{{Gallazzi} A}, \bibinfo{author}{{Maiolino} R},
  \bibinfo{author}{{Balmaverde} B}, \bibinfo{author}{{Brusa} M},
  \bibinfo{author}{{Capetti} A}, \bibinfo{author}{{Cicone} C},
  \bibinfo{author}{{Feruglio} C}, \bibinfo{author}{{Bland-Hawthorn} J},
  \bibinfo{author}{{Nagao} T}, \bibinfo{author}{{Oliva} E},
  \bibinfo{author}{{Salvato} M}, \bibinfo{author}{{Sani} E},
  \bibinfo{author}{{Tozzi} P}, \bibinfo{author}{{Urrutia} T} and
  \bibinfo{author}{{Venturi} G} (\bibinfo{year}{2015}), \bibinfo{month}{Oct.}
\bibinfo{title}{{The MAGNUM survey: positive feedback in the nuclear region of
  NGC 5643 suggested by MUSE}}.
\bibinfo{journal}{{\em \aap}} \bibinfo{volume}{582}, \bibinfo{eid}{A63}.
  \bibinfo{doi}{\doi{10.1051/0004-6361/201526581}}.
\eprint{1508.04464}.

\bibtype{Article}%
\bibitem[{Croton} et al.(2006)]{Croton_etal_2006}
\bibinfo{author}{{Croton} DJ}, \bibinfo{author}{{Springel} V},
  \bibinfo{author}{{White} SDM}, \bibinfo{author}{{De Lucia} G},
  \bibinfo{author}{{Frenk} CS}, \bibinfo{author}{{Gao} L},
  \bibinfo{author}{{Jenkins} A}, \bibinfo{author}{{Kauffmann} G},
  \bibinfo{author}{{Navarro} JF} and  \bibinfo{author}{{Yoshida} N}
  (\bibinfo{year}{2006}), \bibinfo{month}{Jan.}
\bibinfo{title}{{The many lives of active galactic nuclei: cooling flows, black
  holes and the luminosities and colours of galaxies}}.
\bibinfo{journal}{{\em \mnras}} \bibinfo{volume}{365} (\bibinfo{number}{1}):
  \bibinfo{pages}{11--28}.
  \bibinfo{doi}{\doi{10.1111/j.1365-2966.2005.09675.x}}.
\eprint{astro-ph/0508046}.

\bibtype{Article}%
\bibitem[{Darvish} et al.(2016)]{Darvish_etal_2016}
\bibinfo{author}{{Darvish} B}, \bibinfo{author}{{Mobasher} B},
  \bibinfo{author}{{Sobral} D}, \bibinfo{author}{{Rettura} A},
  \bibinfo{author}{{Scoville} N}, \bibinfo{author}{{Faisst} A} and
  \bibinfo{author}{{Capak} P} (\bibinfo{year}{2016}), \bibinfo{month}{Jul.}
\bibinfo{title}{{The Effects of the Local Environment and Stellar Mass on
  Galaxy Quenching to z {\ensuremath{\sim}} 3}}.
\bibinfo{journal}{{\em \apj}} \bibinfo{volume}{825} (\bibinfo{number}{2}),
  \bibinfo{eid}{113}. \bibinfo{doi}{\doi{10.3847/0004-637X/825/2/113}}.
\eprint{1605.03182}.

\bibtype{Article}%
\bibitem[{Dav{\'e}} et al.(2019)]{Dave_etal_2019}
\bibinfo{author}{{Dav{\'e}} R}, \bibinfo{author}{{Angl{\'e}s-Alc{\'a}zar} D},
  \bibinfo{author}{{Narayanan} D}, \bibinfo{author}{{Li} Q},
  \bibinfo{author}{{Rafieferantsoa} MH} and  \bibinfo{author}{{Appleby} S}
  (\bibinfo{year}{2019}), \bibinfo{month}{Jun.}
\bibinfo{title}{{SIMBA: Cosmological simulations with black hole growth and
  feedback}}.
\bibinfo{journal}{{\em \mnras}} \bibinfo{volume}{486} (\bibinfo{number}{2}):
  \bibinfo{pages}{2827--2849}. \bibinfo{doi}{\doi{10.1093/mnras/stz937}}.
\eprint{1901.10203}.

\bibtype{Article}%
\bibitem[{de Graaff} et al.(2024)]{deGraaff_etal_2024}
\bibinfo{author}{{de Graaff} A}, \bibinfo{author}{{Setton} DJ},
  \bibinfo{author}{{Brammer} G}, \bibinfo{author}{{Cutler} S},
  \bibinfo{author}{{Suess} KA}, \bibinfo{author}{{Labbe} I},
  \bibinfo{author}{{Leja} J}, \bibinfo{author}{{Weibel} A},
  \bibinfo{author}{{Maseda} MV}, \bibinfo{author}{{Whitaker} KE},
  \bibinfo{author}{{Bezanson} R}, \bibinfo{author}{{Boogaard} LA},
  \bibinfo{author}{{Cleri} NJ}, \bibinfo{author}{{De Lucia} G},
  \bibinfo{author}{{Franx} M}, \bibinfo{author}{{Greene} JE},
  \bibinfo{author}{{Hirschmann} M}, \bibinfo{author}{{Matthee} J},
  \bibinfo{author}{{McConachie} I}, \bibinfo{author}{{Naidu} RP},
  \bibinfo{author}{{Oesch} PA}, \bibinfo{author}{{Price} SH},
  \bibinfo{author}{{Rix} HW}, \bibinfo{author}{{Valentino} F},
  \bibinfo{author}{{Wang} B} and  \bibinfo{author}{{Williams} CC}
  (\bibinfo{year}{2024}), \bibinfo{month}{Apr.}
\bibinfo{title}{{Efficient formation of a massive quiescent galaxy at redshift
  4.9}}.
\bibinfo{journal}{{\em arXiv e-prints}} ,
  \bibinfo{eid}{arXiv:2404.05683}\bibinfo{doi}{\doi{10.48550/arXiv.2404.05683}}.
\eprint{2404.05683}.

\bibtype{Inproceedings}%
\bibitem[{De Lucia}(2011)]{DeLucia_2011}
\bibinfo{author}{{De Lucia} G} (\bibinfo{year}{2011}), \bibinfo{month}{Jan.},
  \bibinfo{title}{{Modelling the Evolution of Galaxies as a Function of
  Environment}}, \bibinfo{editor}{{Ferreras} I} and
  \bibinfo{editor}{{Pasquali} A}, (Eds.), \bibinfo{booktitle}{Environment and
  the Formation of Galaxies: 30 years later}, \bibinfo{series}{Astrophysics and
  Space Science Proceedings}, \bibinfo{volume}{27}, pp. \bibinfo{pages}{203},
  \eprint{1012.3326}.

\bibtype{Article}%
\bibitem[{De Lucia}(2019)]{DeLucia_etal_2019}
\bibinfo{author}{{De Lucia} G} (\bibinfo{year}{2019}), \bibinfo{month}{May}.
\bibinfo{title}{{Lighting Up Dark Matter Haloes}}.
\bibinfo{journal}{{\em Galaxies}} \bibinfo{volume}{7} (\bibinfo{number}{2}),
  \bibinfo{eid}{56}. \bibinfo{doi}{\doi{10.3390/galaxies7020056}}.

\bibtype{Article}%
\bibitem[{De Lucia} et al.(2004)]{DeLucia_etal_2004}
\bibinfo{author}{{De Lucia} G}, \bibinfo{author}{{Kauffmann} G} and
  \bibinfo{author}{{White} SDM} (\bibinfo{year}{2004}), \bibinfo{month}{Apr.}
\bibinfo{title}{{Chemical enrichment of the intracluster and intergalactic
  medium in a hierarchical galaxy formation model}}.
\bibinfo{journal}{{\em \mnras}} \bibinfo{volume}{349} (\bibinfo{number}{3}):
  \bibinfo{pages}{1101--1116}.
  \bibinfo{doi}{\doi{10.1111/j.1365-2966.2004.07584.x}}.
\eprint{astro-ph/0310268}.

\bibtype{Article}%
\bibitem[{De Lucia} et al.(2006)]{DeLucia_etal_2006}
\bibinfo{author}{{De Lucia} G}, \bibinfo{author}{{Springel} V},
  \bibinfo{author}{{White} SDM}, \bibinfo{author}{{Croton} D} and
  \bibinfo{author}{{Kauffmann} G} (\bibinfo{year}{2006}), \bibinfo{month}{Feb.}
\bibinfo{title}{{The formation history of elliptical galaxies}}.
\bibinfo{journal}{{\em \mnras}} \bibinfo{volume}{366} (\bibinfo{number}{2}):
  \bibinfo{pages}{499--509}.
  \bibinfo{doi}{\doi{10.1111/j.1365-2966.2005.09879.x}}.
\eprint{astro-ph/0509725}.

\bibtype{Article}%
\bibitem[{De Lucia} et al.(2012)]{DeLucia_etal_2012}
\bibinfo{author}{{De Lucia} G}, \bibinfo{author}{{Weinmann} S},
  \bibinfo{author}{{Poggianti} BM}, \bibinfo{author}{{Arag{\'o}n-Salamanca} A}
  and  \bibinfo{author}{{Zaritsky} D} (\bibinfo{year}{2012}),
  \bibinfo{month}{Jun.}
\bibinfo{title}{{The environmental history of group and cluster galaxies in a
  {\ensuremath{\Lambda}} cold dark matter universe}}.
\bibinfo{journal}{{\em \mnras}} \bibinfo{volume}{423} (\bibinfo{number}{2}):
  \bibinfo{pages}{1277--1292}.
  \bibinfo{doi}{\doi{10.1111/j.1365-2966.2012.20983.x}}.
\eprint{1111.6590}.

\bibtype{Article}%
\bibitem[{De Lucia} et al.(2014)]{DeLucia_etal_2014}
\bibinfo{author}{{De Lucia} G}, \bibinfo{author}{{Muzzin} A} and
  \bibinfo{author}{{Weinmann} S} (\bibinfo{year}{2014}), \bibinfo{month}{Oct.}
\bibinfo{title}{{What Regulates Galaxy Evolution? Open questions in our
  understanding of galaxy formation and evolution}}.
\bibinfo{journal}{{\em New Astronomy Reviews}} \bibinfo{volume}{62}:
  \bibinfo{pages}{1--14}. \bibinfo{doi}{\doi{10.1016/j.newar.2014.08.001}}.
\eprint{1407.8366}.

\bibtype{Article}%
\bibitem[{De Lucia} et al.(2017)]{DeLucia_etal_2017}
\bibinfo{author}{{De Lucia} G}, \bibinfo{author}{{Fontanot} F} and
  \bibinfo{author}{{Hirschmann} M} (\bibinfo{year}{2017}),
  \bibinfo{month}{Mar.}
\bibinfo{title}{{AGN feedback and the origin of the {\ensuremath{\alpha}}
  enhancement in early-type galaxies - insights from the GAEA model}}.
\bibinfo{journal}{{\em \mnras}} \bibinfo{volume}{466} (\bibinfo{number}{1}):
  \bibinfo{pages}{L88--L92}. \bibinfo{doi}{\doi{10.1093/mnrasl/slw242}}.
\eprint{1611.04597}.

\bibtype{Article}%
\bibitem[{De Lucia} et al.(2024)]{DeLucia_etal_2024}
\bibinfo{author}{{De Lucia} G}, \bibinfo{author}{{Fontanot} F},
  \bibinfo{author}{{Xie} L} and  \bibinfo{author}{{Hirschmann} M}
  (\bibinfo{year}{2024}), \bibinfo{month}{Jul.}
\bibinfo{title}{{Tracing the quenching journey across cosmic time}}.
\bibinfo{journal}{{\em \aap}} \bibinfo{volume}{687}, \bibinfo{eid}{A68}.
  \bibinfo{doi}{\doi{10.1051/0004-6361/202349045}}.
\eprint{2401.06211}.

\bibtype{Article}%
\bibitem[{Dekel} et al.(2023)]{Dekel_etal_2023}
\bibinfo{author}{{Dekel} A}, \bibinfo{author}{{Sarkar} KC},
  \bibinfo{author}{{Birnboim} Y}, \bibinfo{author}{{Mandelker} N} and
  \bibinfo{author}{{Li} Z} (\bibinfo{year}{2023}), \bibinfo{month}{Aug.}
\bibinfo{title}{{Efficient formation of massive galaxies at cosmic dawn by
  feedback-free starbursts}}.
\bibinfo{journal}{{\em \mnras}} \bibinfo{volume}{523} (\bibinfo{number}{3}):
  \bibinfo{pages}{3201--3218}. \bibinfo{doi}{\doi{10.1093/mnras/stad1557}}.
\eprint{2303.04827}.

\bibtype{Article}%
\bibitem[{Di Mascolo} et al.(2023)]{DiMascolo_etal_2023}
\bibinfo{author}{{Di Mascolo} L}, \bibinfo{author}{{Saro} A},
  \bibinfo{author}{{Mroczkowski} T}, \bibinfo{author}{{Borgani} S},
  \bibinfo{author}{{Churazov} E}, \bibinfo{author}{{Rasia} E},
  \bibinfo{author}{{Tozzi} P}, \bibinfo{author}{{Dannerbauer} H},
  \bibinfo{author}{{Basu} K}, \bibinfo{author}{{Carilli} CL},
  \bibinfo{author}{{Ginolfi} M}, \bibinfo{author}{{Miley} G},
  \bibinfo{author}{{Nonino} M}, \bibinfo{author}{{Pannella} M},
  \bibinfo{author}{{Pentericci} L} and  \bibinfo{author}{{Rizzo} F}
  (\bibinfo{year}{2023}), \bibinfo{month}{Mar.}
\bibinfo{title}{{Forming intracluster gas in a galaxy protocluster at a
  redshift of 2.16}}.
\bibinfo{journal}{{\em \nat}} \bibinfo{volume}{615} (\bibinfo{number}{7954}):
  \bibinfo{pages}{809--812}. \bibinfo{doi}{\doi{10.1038/s41586-023-05761-x}}.
\eprint{2303.16226}.

\bibtype{Article}%
\bibitem[{Di Matteo} et al.(2003)]{DiMatteo_etal_2003}
\bibinfo{author}{{Di Matteo} T}, \bibinfo{author}{{Croft} RAC},
  \bibinfo{author}{{Springel} V} and  \bibinfo{author}{{Hernquist} L}
  (\bibinfo{year}{2003}), \bibinfo{month}{Aug.}
\bibinfo{title}{{Black Hole Growth and Activity in a {$\Lambda$} Cold Dark
  Matter Universe}}.
\bibinfo{journal}{{\em \apj}} \bibinfo{volume}{593}: \bibinfo{pages}{56--68}.
  \bibinfo{doi}{\doi{10.1086/376501}}.
\eprint{astro-ph/0301586}.

\bibtype{Article}%
\bibitem[{Di Matteo} et al.(2008)]{DiMatteo_etal_2008}
\bibinfo{author}{{Di Matteo} P}, \bibinfo{author}{{Bournaud} F},
  \bibinfo{author}{{Martig} M}, \bibinfo{author}{{Combes} F},
  \bibinfo{author}{{Melchior} AL} and  \bibinfo{author}{{Semelin} B}
  (\bibinfo{year}{2008}), \bibinfo{month}{Dec.}
\bibinfo{title}{{On the frequency, intensity, and duration of starburst
  episodes triggered by galaxy interactions and mergers}}.
\bibinfo{journal}{{\em \aap}} \bibinfo{volume}{492} (\bibinfo{number}{1}):
  \bibinfo{pages}{31--49}. \bibinfo{doi}{\doi{10.1051/0004-6361:200809480}}.
\eprint{0809.2592}.

\bibtype{Article}%
\bibitem[{Diemer} et al.(2019)]{Diemer_etal_2019}
\bibinfo{author}{{Diemer} B}, \bibinfo{author}{{Stevens} ARH},
  \bibinfo{author}{{Lagos} CdP}, \bibinfo{author}{{Calette} AR},
  \bibinfo{author}{{Tacchella} S}, \bibinfo{author}{{Hernquist} L},
  \bibinfo{author}{{Marinacci} F}, \bibinfo{author}{{Nelson} D},
  \bibinfo{author}{{Pillepich} A}, \bibinfo{author}{{Rodriguez-Gomez} V},
  \bibinfo{author}{{Villaescusa-Navarro} F} and
  \bibinfo{author}{{Vogelsberger} M} (\bibinfo{year}{2019}),
  \bibinfo{month}{Aug.}
\bibinfo{title}{{Atomic and molecular gas in IllustrisTNG galaxies at low
  redshift}}.
\bibinfo{journal}{{\em \mnras}} \bibinfo{volume}{487} (\bibinfo{number}{2}):
  \bibinfo{pages}{1529--1550}. \bibinfo{doi}{\doi{10.1093/mnras/stz1323}}.
\eprint{1902.10714}.

\bibtype{Article}%
\bibitem[{Donnari} et al.(2021)]{Donnari_etal_2021}
\bibinfo{author}{{Donnari} M}, \bibinfo{author}{{Pillepich} A},
  \bibinfo{author}{{Nelson} D}, \bibinfo{author}{{Marinacci} F},
  \bibinfo{author}{{Vogelsberger} M} and  \bibinfo{author}{{Hernquist} L}
  (\bibinfo{year}{2021}), \bibinfo{month}{Oct.}
\bibinfo{title}{{Quenched fractions in the IllustrisTNG simulations: comparison
  with observations and other theoretical models}}.
\bibinfo{journal}{{\em \mnras}} \bibinfo{volume}{506} (\bibinfo{number}{4}):
  \bibinfo{pages}{4760--4780}. \bibinfo{doi}{\doi{10.1093/mnras/stab1950}}.
\eprint{2008.00004}.

\bibtype{Article}%
\bibitem[{Dressler}(1980)]{Dressler_1980}
\bibinfo{author}{{Dressler} A} (\bibinfo{year}{1980}), \bibinfo{month}{Mar.}
\bibinfo{title}{{Galaxy morphology in rich clusters: implications for the
  formation and evolution of galaxies.}}
\bibinfo{journal}{{\em \apj}} \bibinfo{volume}{236}: \bibinfo{pages}{351--365}.
  \bibinfo{doi}{\doi{10.1086/157753}}.

\bibtype{Article}%
\bibitem[{Dressler} et al.(1997)]{Dressler_etal_1997}
\bibinfo{author}{{Dressler} A}, \bibinfo{author}{{Oemler} Augustus J},
  \bibinfo{author}{{Couch} WJ}, \bibinfo{author}{{Smail} I},
  \bibinfo{author}{{Ellis} RS}, \bibinfo{author}{{Barger} A},
  \bibinfo{author}{{Butcher} H}, \bibinfo{author}{{Poggianti} BM} and
  \bibinfo{author}{{Sharples} RM} (\bibinfo{year}{1997}), \bibinfo{month}{Dec.}
\bibinfo{title}{{Evolution since z = 0.5 of the Morphology-Density Relation for
  Clusters of Galaxies}}.
\bibinfo{journal}{{\em \apj}} \bibinfo{volume}{490} (\bibinfo{number}{2}):
  \bibinfo{pages}{577--591}. \bibinfo{doi}{\doi{10.1086/304890}}.
\eprint{astro-ph/9707232}.

\bibtype{Article}%
\bibitem[{Driver} et al.(2006)]{Driver_etal_2006}
\bibinfo{author}{{Driver} SP}, \bibinfo{author}{{Allen} PD},
  \bibinfo{author}{{Graham} AW}, \bibinfo{author}{{Cameron} E},
  \bibinfo{author}{{Liske} J}, \bibinfo{author}{{Ellis} SC},
  \bibinfo{author}{{Cross} NJG}, \bibinfo{author}{{De Propris} R},
  \bibinfo{author}{{Phillipps} S} and  \bibinfo{author}{{Couch} WJ}
  (\bibinfo{year}{2006}), \bibinfo{month}{May}.
\bibinfo{title}{{The Millennium Galaxy Catalogue: morphological classification
  and bimodality in the colour-concentration plane}}.
\bibinfo{journal}{{\em \mnras}} \bibinfo{volume}{368} (\bibinfo{number}{1}):
  \bibinfo{pages}{414--434}.
  \bibinfo{doi}{\doi{10.1111/j.1365-2966.2006.10126.x}}.
\eprint{astro-ph/0602240}.

\bibtype{Article}%
\bibitem[{Driver} et al.(2011)]{Driver_etal_2011}
\bibinfo{author}{{Driver} SP}, \bibinfo{author}{{Hill} DT},
  \bibinfo{author}{{Kelvin} LS}, \bibinfo{author}{{Robotham} ASG},
  \bibinfo{author}{{Liske} J}, \bibinfo{author}{{Norberg} P},
  \bibinfo{author}{{Baldry} IK}, \bibinfo{author}{{Bamford} SP},
  \bibinfo{author}{{Hopkins} AM}, \bibinfo{author}{{Loveday} J},
  \bibinfo{author}{{Peacock} JA}, \bibinfo{author}{{Andrae} E},
  \bibinfo{author}{{Bland-Hawthorn} J}, \bibinfo{author}{{Brough} S},
  \bibinfo{author}{{Brown} MJI}, \bibinfo{author}{{Cameron} E},
  \bibinfo{author}{{Ching} JHY}, \bibinfo{author}{{Colless} M},
  \bibinfo{author}{{Conselice} CJ}, \bibinfo{author}{{Croom} SM},
  \bibinfo{author}{{Cross} NJG}, \bibinfo{author}{{de Propris} R},
  \bibinfo{author}{{Dye} S}, \bibinfo{author}{{Drinkwater} MJ},
  \bibinfo{author}{{Ellis} S}, \bibinfo{author}{{Graham} AW},
  \bibinfo{author}{{Grootes} MW}, \bibinfo{author}{{Gunawardhana} M},
  \bibinfo{author}{{Jones} DH}, \bibinfo{author}{{van Kampen} E},
  \bibinfo{author}{{Maraston} C}, \bibinfo{author}{{Nichol} RC},
  \bibinfo{author}{{Parkinson} HR}, \bibinfo{author}{{Phillipps} S},
  \bibinfo{author}{{Pimbblet} K}, \bibinfo{author}{{Popescu} CC},
  \bibinfo{author}{{Prescott} M}, \bibinfo{author}{{Roseboom} IG},
  \bibinfo{author}{{Sadler} EM}, \bibinfo{author}{{Sansom} AE},
  \bibinfo{author}{{Sharp} RG}, \bibinfo{author}{{Smith} DJB},
  \bibinfo{author}{{Taylor} E}, \bibinfo{author}{{Thomas} D},
  \bibinfo{author}{{Tuffs} RJ}, \bibinfo{author}{{Wijesinghe} D},
  \bibinfo{author}{{Dunne} L}, \bibinfo{author}{{Frenk} CS},
  \bibinfo{author}{{Jarvis} MJ}, \bibinfo{author}{{Madore} BF},
  \bibinfo{author}{{Meyer} MJ}, \bibinfo{author}{{Seibert} M},
  \bibinfo{author}{{Staveley-Smith} L}, \bibinfo{author}{{Sutherland} WJ} and
  \bibinfo{author}{{Warren} SJ} (\bibinfo{year}{2011}), \bibinfo{month}{May}.
\bibinfo{title}{{Galaxy and Mass Assembly (GAMA): survey diagnostics and core
  data release}}.
\bibinfo{journal}{{\em \mnras}} \bibinfo{volume}{413} (\bibinfo{number}{2}):
  \bibinfo{pages}{971--995}.
  \bibinfo{doi}{\doi{10.1111/j.1365-2966.2010.18188.x}}.
\eprint{1009.0614}.

\bibtype{Article}%
\bibitem[{Dubois} et al.(2016)]{Dubois_etal_2016}
\bibinfo{author}{{Dubois} Y}, \bibinfo{author}{{Peirani} S},
  \bibinfo{author}{{Pichon} C}, \bibinfo{author}{{Devriendt} J},
  \bibinfo{author}{{Gavazzi} R}, \bibinfo{author}{{Welker} C} and
  \bibinfo{author}{{Volonteri} M} (\bibinfo{year}{2016}), \bibinfo{month}{Dec.}
\bibinfo{title}{{The HORIZON-AGN simulation: morphological diversity of
  galaxies promoted by AGN feedback}}.
\bibinfo{journal}{{\em \mnras}} \bibinfo{volume}{463} (\bibinfo{number}{4}):
  \bibinfo{pages}{3948--3964}. \bibinfo{doi}{\doi{10.1093/mnras/stw2265}}.
\eprint{1606.03086}.

\bibtype{Article}%
\bibitem[{Erfanianfar} et al.(2016)]{Erfanianfar_etal_2016}
\bibinfo{author}{{Erfanianfar} G}, \bibinfo{author}{{Popesso} P},
  \bibinfo{author}{{Finoguenov} A}, \bibinfo{author}{{Wilman} D},
  \bibinfo{author}{{Wuyts} S}, \bibinfo{author}{{Biviano} A},
  \bibinfo{author}{{Salvato} M}, \bibinfo{author}{{Mirkazemi} M},
  \bibinfo{author}{{Morselli} L}, \bibinfo{author}{{Ziparo} F},
  \bibinfo{author}{{Nandra} K}, \bibinfo{author}{{Lutz} D},
  \bibinfo{author}{{Elbaz} D}, \bibinfo{author}{{Dickinson} M},
  \bibinfo{author}{{Tanaka} M}, \bibinfo{author}{{Altieri} MB},
  \bibinfo{author}{{Aussel} H}, \bibinfo{author}{{Bauer} F},
  \bibinfo{author}{{Berta} S}, \bibinfo{author}{{Bielby} RM},
  \bibinfo{author}{{Brandt} N}, \bibinfo{author}{{Cappelluti} N},
  \bibinfo{author}{{Cimatti} A}, \bibinfo{author}{{Cooper} MC},
  \bibinfo{author}{{Fadda} D}, \bibinfo{author}{{Ilbert} O},
  \bibinfo{author}{{Le Floch} E}, \bibinfo{author}{{Magnelli} B},
  \bibinfo{author}{{Mulchaey} JS}, \bibinfo{author}{{Nordon} R},
  \bibinfo{author}{{Newman} JA}, \bibinfo{author}{{Poglitsch} A} and
  \bibinfo{author}{{Pozzi} F} (\bibinfo{year}{2016}), \bibinfo{month}{Jan.}
\bibinfo{title}{{Non-linearity and environmental dependence of the star-forming
  galaxies main sequence}}.
\bibinfo{journal}{{\em \mnras}} \bibinfo{volume}{455} (\bibinfo{number}{3}):
  \bibinfo{pages}{2839--2851}. \bibinfo{doi}{\doi{10.1093/mnras/stv2485}}.
\eprint{1511.01899}.

\bibtype{Article}%
\bibitem[{Euclid Collaboration} et al.(2022)]{Scaramella_etal_2022}
\bibinfo{author}{{Euclid Collaboration}}, \bibinfo{author}{{Scaramella} R} and
  \bibinfo{author}{{et al.}} (\bibinfo{year}{2022}), \bibinfo{month}{Jun.}
\bibinfo{title}{{Euclid preparation. I. The Euclid Wide Survey}}.
\bibinfo{journal}{{\em \aap}} \bibinfo{volume}{662}, \bibinfo{eid}{A112}.
  \bibinfo{doi}{\doi{10.1051/0004-6361/202141938}}.
\eprint{2108.01201}.

\bibtype{Article}%
\bibitem[{Euclid Collaboration} et al.(2023)]{Humphrey_etal_2023}
\bibinfo{author}{{Euclid Collaboration}}, \bibinfo{author}{{Humphrey} A} and
  \bibinfo{author}{{et al.}} (\bibinfo{year}{2023}), \bibinfo{month}{Mar.}
\bibinfo{title}{{Euclid preparation. XXII. Selection of quiescent galaxies from
  mock photometry using machine learning}}.
\bibinfo{journal}{{\em \aap}} \bibinfo{volume}{671}, \bibinfo{eid}{A99}.
  \bibinfo{doi}{\doi{10.1051/0004-6361/202244307}}.
\eprint{2209.13074}.

\bibtype{Article}%
\bibitem[{Euclid Collaboration} et al.(2024)]{Mellier_etal_2024}
\bibinfo{author}{{Euclid Collaboration}}, \bibinfo{author}{{Mellier} Y} and
  \bibinfo{author}{{et al.}} (\bibinfo{year}{2024}), \bibinfo{month}{May}.
\bibinfo{title}{{Euclid. I. Overview of the Euclid mission}}.
\bibinfo{journal}{{\em arXiv e-prints}} ,
  \bibinfo{eid}{arXiv:2405.13491}\bibinfo{doi}{\doi{10.48550/arXiv.2405.13491}}.
\eprint{2405.13491}.

\bibtype{Article}%
\bibitem[{Fabian}(2012)]{Fabian_2012}
\bibinfo{author}{{Fabian} AC} (\bibinfo{year}{2012}), \bibinfo{month}{Sep.}
\bibinfo{title}{{Observational Evidence of Active Galactic Nuclei Feedback}}.
\bibinfo{journal}{{\em \araa}} \bibinfo{volume}{50}: \bibinfo{pages}{455--489}.
  \bibinfo{doi}{\doi{10.1146/annurev-astro-081811-125521}}.
\eprint{1204.4114}.

\bibtype{Article}%
\bibitem[{Fang} et al.(2013)]{Fang_etal_2013}
\bibinfo{author}{{Fang} JJ}, \bibinfo{author}{{Faber} SM},
  \bibinfo{author}{{Koo} DC} and  \bibinfo{author}{{Dekel} A}
  (\bibinfo{year}{2013}), \bibinfo{month}{Oct.}
\bibinfo{title}{{A Link between Star Formation Quenching and Inner Stellar Mass
  Density in Sloan Digital Sky Survey Central Galaxies}}.
\bibinfo{journal}{{\em \apj}} \bibinfo{volume}{776} (\bibinfo{number}{1}),
  \bibinfo{eid}{63}. \bibinfo{doi}{\doi{10.1088/0004-637X/776/1/63}}.
\eprint{1308.5224}.

\bibtype{Article}%
\bibitem[{Feruglio} et al.(2010)]{Feruglio_etal_2010}
\bibinfo{author}{{Feruglio} C}, \bibinfo{author}{{Maiolino} R},
  \bibinfo{author}{{Piconcelli} E}, \bibinfo{author}{{Menci} N},
  \bibinfo{author}{{Aussel} H}, \bibinfo{author}{{Lamastra} A} and
  \bibinfo{author}{{Fiore} F} (\bibinfo{year}{2010}), \bibinfo{month}{Jul.}
\bibinfo{title}{{Quasar feedback revealed by giant molecular outflows}}.
\bibinfo{journal}{{\em \aap}} \bibinfo{volume}{518}, \bibinfo{eid}{L155}.
  \bibinfo{doi}{\doi{10.1051/0004-6361/201015164}}.
\eprint{1006.1655}.

\bibtype{Article}%
\bibitem[{Feruglio} et al.(2020)]{Feruglio20}
\bibinfo{author}{{Feruglio} C}, \bibinfo{author}{{Fabbiano} G},
  \bibinfo{author}{{Bischetti} M}, \bibinfo{author}{{Elvis} M},
  \bibinfo{author}{{Travascio} A} and  \bibinfo{author}{{Fiore} F}
  (\bibinfo{year}{2020}), \bibinfo{month}{Feb.}
\bibinfo{title}{{Multiphase Gas Flows in the Nearby Seyfert Galaxy ESO428-G014.
  Paper I}}.
\bibinfo{journal}{{\em \apj}} \bibinfo{volume}{890} (\bibinfo{number}{1}),
  \bibinfo{eid}{29}. \bibinfo{doi}{\doi{10.3847/1538-4357/ab67bd}}.
\eprint{1904.01483}.

\bibtype{Article}%
\bibitem[{Fiore} et al.(2017)]{Fiore_etal_2017}
\bibinfo{author}{{Fiore} F}, \bibinfo{author}{{Feruglio} C},
  \bibinfo{author}{{Shankar} F}, \bibinfo{author}{{Bischetti} M},
  \bibinfo{author}{{Bongiorno} A}, \bibinfo{author}{{Brusa} M},
  \bibinfo{author}{{Carniani} S}, \bibinfo{author}{{Cicone} C},
  \bibinfo{author}{{Duras} F}, \bibinfo{author}{{Lamastra} A},
  \bibinfo{author}{{Mainieri} V}, \bibinfo{author}{{Marconi} A},
  \bibinfo{author}{{Menci} N}, \bibinfo{author}{{Maiolino} R},
  \bibinfo{author}{{Piconcelli} E}, \bibinfo{author}{{Vietri} G} and
  \bibinfo{author}{{Zappacosta} L} (\bibinfo{year}{2017}),
  \bibinfo{month}{May}.
\bibinfo{title}{{AGN wind scaling relations and the co-evolution of black holes
  and galaxies}}.
\bibinfo{journal}{{\em \aap}} \bibinfo{volume}{601}, \bibinfo{eid}{A143}.
  \bibinfo{doi}{\doi{10.1051/0004-6361/201629478}}.
\eprint{1702.04507}.

\bibtype{Article}%
\bibitem[{Font} et al.(2008)]{Font_etal_2008}
\bibinfo{author}{{Font} AS}, \bibinfo{author}{{Bower} RG},
  \bibinfo{author}{{McCarthy} IG}, \bibinfo{author}{{Benson} AJ},
  \bibinfo{author}{{Frenk} CS}, \bibinfo{author}{{Helly} JC},
  \bibinfo{author}{{Lacey} CG}, \bibinfo{author}{{Baugh} CM} and
  \bibinfo{author}{{Cole} S} (\bibinfo{year}{2008}), \bibinfo{month}{Oct.}
\bibinfo{title}{{The colours of satellite galaxies in groups and clusters}}.
\bibinfo{journal}{{\em \mnras}} \bibinfo{volume}{389} (\bibinfo{number}{4}):
  \bibinfo{pages}{1619--1629}.
  \bibinfo{doi}{\doi{10.1111/j.1365-2966.2008.13698.x}}.
\eprint{0807.0001}.

\bibtype{Article}%
\bibitem[{Fontanot} et al.(2009)]{Fontanot_etal_2009}
\bibinfo{author}{{Fontanot} F}, \bibinfo{author}{{De Lucia} G},
  \bibinfo{author}{{Monaco} P}, \bibinfo{author}{{Somerville} RS} and
  \bibinfo{author}{{Santini} P} (\bibinfo{year}{2009}), \bibinfo{month}{Aug.}
\bibinfo{title}{{The many manifestations of downsizing: hierarchical galaxy
  formation models confront observations}}.
\bibinfo{journal}{{\em \mnras}} \bibinfo{volume}{397} (\bibinfo{number}{4}):
  \bibinfo{pages}{1776--1790}.
  \bibinfo{doi}{\doi{10.1111/j.1365-2966.2009.15058.x}}.
\eprint{0901.1130}.

\bibtype{Article}%
\bibitem[{Fontanot} et al.(2017)]{Fontanot_etal_2017}
\bibinfo{author}{{Fontanot} F}, \bibinfo{author}{{De Lucia} G},
  \bibinfo{author}{{Hirschmann} M}, \bibinfo{author}{{Bruzual} G},
  \bibinfo{author}{{Charlot} S} and  \bibinfo{author}{{Zibetti} S}
  (\bibinfo{year}{2017}), \bibinfo{month}{Feb.}
\bibinfo{title}{{Variations of the stellar initial mass function in
  semi-analytical models: implications for the mass assembly and the chemical
  enrichment of galaxies in the GAEA model}}.
\bibinfo{journal}{{\em \mnras}} \bibinfo{volume}{464} (\bibinfo{number}{4}):
  \bibinfo{pages}{3812--3824}. \bibinfo{doi}{\doi{10.1093/mnras/stw2612}}.
\eprint{1606.01908}.

\bibtype{Article}%
\bibitem[{Fontanot} et al.(2020)]{Fontanot_etal_2020}
\bibinfo{author}{{Fontanot} F}, \bibinfo{author}{{De Lucia} G},
  \bibinfo{author}{{Hirschmann} M}, \bibinfo{author}{{Xie} L},
  \bibinfo{author}{{Monaco} P}, \bibinfo{author}{{Menci} N},
  \bibinfo{author}{{Fiore} F}, \bibinfo{author}{{Feruglio} C},
  \bibinfo{author}{{Cristiani} S} and  \bibinfo{author}{{Shankar} F}
  (\bibinfo{year}{2020}), \bibinfo{month}{Aug.}
\bibinfo{title}{{The rise of active galactic nuclei in the galaxy evolution and
  assembly semi-analytic model}}.
\bibinfo{journal}{{\em \mnras}} \bibinfo{volume}{496} (\bibinfo{number}{3}):
  \bibinfo{pages}{3943--3960}. \bibinfo{doi}{\doi{10.1093/mnras/staa1716}}.
\eprint{2002.10576}.

\bibtype{Article}%
\bibitem[{Fumagalli} et al.(2009)]{Fumagalli_etal_2009}
\bibinfo{author}{{Fumagalli} M}, \bibinfo{author}{{Krumholz} MR},
  \bibinfo{author}{{Prochaska} JX}, \bibinfo{author}{{Gavazzi} G} and
  \bibinfo{author}{{Boselli} A} (\bibinfo{year}{2009}), \bibinfo{month}{Jun.}
\bibinfo{title}{{Molecular Hydrogen Deficiency in H I-poor Galaxies and its
  Implications for Star Formation}}.
\bibinfo{journal}{{\em \apj}} \bibinfo{volume}{697} (\bibinfo{number}{2}):
  \bibinfo{pages}{1811--1821}.
  \bibinfo{doi}{\doi{10.1088/0004-637X/697/2/1811}}.
\eprint{0903.3950}.

\bibtype{Article}%
\bibitem[{Gallazzi} et al.(2014)]{Gallazzi14}
\bibinfo{author}{{Gallazzi} A}, \bibinfo{author}{{Bell} EF},
  \bibinfo{author}{{Zibetti} S}, \bibinfo{author}{{Brinchmann} J} and
  \bibinfo{author}{{Kelson} DD} (\bibinfo{year}{2014}), \bibinfo{month}{Jun.}
\bibinfo{title}{{Charting the Evolution of the Ages and Metallicities of
  Massive Galaxies since z = 0.7}}.
\bibinfo{journal}{{\em \apj}} \bibinfo{volume}{788} (\bibinfo{number}{1}),
  \bibinfo{eid}{72}. \bibinfo{doi}{\doi{10.1088/0004-637X/788/1/72}}.
\eprint{1404.5624}.

\bibtype{Article}%
\bibitem[{Gaspari} and {S{\c a}dowski}(2017)]{Gaspari17}
\bibinfo{author}{{Gaspari} M} and  \bibinfo{author}{{S{\c a}dowski} A}
  (\bibinfo{year}{2017}), \bibinfo{month}{Mar.}
\bibinfo{title}{{Unifying the Micro and Macro Properties of AGN Feeding and
  Feedback}}.
\bibinfo{journal}{{\em \apj}} \bibinfo{volume}{837}, \bibinfo{eid}{149}.
  \bibinfo{doi}{\doi{10.3847/1538-4357/aa61a3}}.
\eprint{1701.07030}.

\bibtype{Article}%
\bibitem[{Gatto} et al.(2017)]{Gatto_etal_2017}
\bibinfo{author}{{Gatto} A}, \bibinfo{author}{{Walch} S},
  \bibinfo{author}{{Naab} T}, \bibinfo{author}{{Girichidis} P},
  \bibinfo{author}{{W{\"u}nsch} R}, \bibinfo{author}{{Glover} SCO},
  \bibinfo{author}{{Klessen} RS}, \bibinfo{author}{{Clark} PC},
  \bibinfo{author}{{Peters} T}, \bibinfo{author}{{Derigs} D},
  \bibinfo{author}{{Baczynski} C} and  \bibinfo{author}{{Puls} J}
  (\bibinfo{year}{2017}), \bibinfo{month}{Apr.}
\bibinfo{title}{{The SILCC project - III. Regulation of star formation and
  outflows by stellar winds and supernovae}}.
\bibinfo{journal}{{\em \mnras}} \bibinfo{volume}{466} (\bibinfo{number}{2}):
  \bibinfo{pages}{1903--1924}. \bibinfo{doi}{\doi{10.1093/mnras/stw3209}}.
\eprint{1606.05346}.

\bibtype{Article}%
\bibitem[{Giavalisco} et al.(2004)]{Giavalisco_etal_2004}
\bibinfo{author}{{Giavalisco} M}, \bibinfo{author}{{Ferguson} HC},
  \bibinfo{author}{{Koekemoer} AM}, \bibinfo{author}{{Dickinson} M},
  \bibinfo{author}{{Alexander} DM}, \bibinfo{author}{{Bauer} FE},
  \bibinfo{author}{{Bergeron} J}, \bibinfo{author}{{Biagetti} C},
  \bibinfo{author}{{Brandt} WN}, \bibinfo{author}{{Casertano} S},
  \bibinfo{author}{{Cesarsky} C}, \bibinfo{author}{{Chatzichristou} E},
  \bibinfo{author}{{Conselice} C}, \bibinfo{author}{{Cristiani} S},
  \bibinfo{author}{{Da Costa} L}, \bibinfo{author}{{Dahlen} T},
  \bibinfo{author}{{de Mello} D}, \bibinfo{author}{{Eisenhardt} P},
  \bibinfo{author}{{Erben} T}, \bibinfo{author}{{Fall} SM},
  \bibinfo{author}{{Fassnacht} C}, \bibinfo{author}{{Fosbury} R},
  \bibinfo{author}{{Fruchter} A}, \bibinfo{author}{{Gardner} JP},
  \bibinfo{author}{{Grogin} N}, \bibinfo{author}{{Hook} RN},
  \bibinfo{author}{{Hornschemeier} AE}, \bibinfo{author}{{Idzi} R},
  \bibinfo{author}{{Jogee} S}, \bibinfo{author}{{Kretchmer} C},
  \bibinfo{author}{{Laidler} V}, \bibinfo{author}{{Lee} KS},
  \bibinfo{author}{{Livio} M}, \bibinfo{author}{{Lucas} R},
  \bibinfo{author}{{Madau} P}, \bibinfo{author}{{Mobasher} B},
  \bibinfo{author}{{Moustakas} LA}, \bibinfo{author}{{Nonino} M},
  \bibinfo{author}{{Padovani} P}, \bibinfo{author}{{Papovich} C},
  \bibinfo{author}{{Park} Y}, \bibinfo{author}{{Ravindranath} S},
  \bibinfo{author}{{Renzini} A}, \bibinfo{author}{{Richardson} M},
  \bibinfo{author}{{Riess} A}, \bibinfo{author}{{Rosati} P},
  \bibinfo{author}{{Schirmer} M}, \bibinfo{author}{{Schreier} E},
  \bibinfo{author}{{Somerville} RS}, \bibinfo{author}{{Spinrad} H},
  \bibinfo{author}{{Stern} D}, \bibinfo{author}{{Stiavelli} M},
  \bibinfo{author}{{Strolger} L}, \bibinfo{author}{{Urry} CM},
  \bibinfo{author}{{Vandame} B}, \bibinfo{author}{{Williams} R} and
  \bibinfo{author}{{Wolf} C} (\bibinfo{year}{2004}), \bibinfo{month}{Jan.}
\bibinfo{title}{{The Great Observatories Origins Deep Survey: Initial Results
  from Optical and Near-Infrared Imaging}}.
\bibinfo{journal}{{\em \apjl}} \bibinfo{volume}{600} (\bibinfo{number}{2}):
  \bibinfo{pages}{L93--L98}. \bibinfo{doi}{\doi{10.1086/379232}}.
\eprint{astro-ph/0309105}.

\bibtype{Article}%
\bibitem[{Greene} et al.(2022)]{Greene_et_al_2022}
\bibinfo{author}{{Greene} J}, \bibinfo{author}{{Bezanson} R},
  \bibinfo{author}{{Ouchi} M}, \bibinfo{author}{{Silverman} J} and
  \bibinfo{author}{{the PFS Galaxy Evolution Working Group}}
  (\bibinfo{year}{2022}), \bibinfo{month}{Jun.}
\bibinfo{title}{{The Prime Focus Spectrograph Galaxy Evolution Survey}}.
\bibinfo{journal}{{\em arXiv e-prints}} ,
  \bibinfo{eid}{arXiv:2206.14908}\bibinfo{doi}{\doi{10.48550/arXiv.2206.14908}}.
\eprint{2206.14908}.

\bibtype{Article}%
\bibitem[{Grudi{\'c}} et al.(2021)]{Grudic21}
\bibinfo{author}{{Grudi{\'c}} MY}, \bibinfo{author}{{Guszejnov} D},
  \bibinfo{author}{{Hopkins} PF}, \bibinfo{author}{{Offner} SSR} and
  \bibinfo{author}{{Faucher-Gigu{\`e}re} CA} (\bibinfo{year}{2021}),
  \bibinfo{month}{Sep.}
\bibinfo{title}{{STARFORGE: Towards a comprehensive numerical model of star
  cluster formation and feedback}}.
\bibinfo{journal}{{\em \mnras}} \bibinfo{volume}{506} (\bibinfo{number}{2}):
  \bibinfo{pages}{2199--2231}. \bibinfo{doi}{\doi{10.1093/mnras/stab1347}}.
\eprint{2010.11254}.

\bibtype{Article}%
\bibitem[{Grudi{\'c}} et al.(2022)]{Grudic_etal_2022}
\bibinfo{author}{{Grudi{\'c}} MY}, \bibinfo{author}{{Guszejnov} D},
  \bibinfo{author}{{Offner} SSR}, \bibinfo{author}{{Rosen} AL},
  \bibinfo{author}{{Raju} AN}, \bibinfo{author}{{Faucher-Gigu{\`e}re} CA} and
  \bibinfo{author}{{Hopkins} PF} (\bibinfo{year}{2022}), \bibinfo{month}{May}.
\bibinfo{title}{{The dynamics and outcome of star formation with jets,
  radiation, winds, and supernovae in concert}}.
\bibinfo{journal}{{\em \mnras}} \bibinfo{volume}{512} (\bibinfo{number}{1}):
  \bibinfo{pages}{216--232}. \bibinfo{doi}{\doi{10.1093/mnras/stac526}}.
\eprint{2201.00882}.

\bibtype{Article}%
\bibitem[{Gunn} and {Gott}(1972)]{Gunn_and_Gott_1972}
\bibinfo{author}{{Gunn} JE} and  \bibinfo{author}{{Gott} J.~Richard I}
  (\bibinfo{year}{1972}), \bibinfo{month}{Aug.}
\bibinfo{title}{{On the Infall of Matter Into Clusters of Galaxies and Some
  Effects on Their Evolution}}.
\bibinfo{journal}{{\em \apj}} \bibinfo{volume}{176}: \bibinfo{pages}{1}.
  \bibinfo{doi}{\doi{10.1086/151605}}.

\bibtype{Article}%
\bibitem[{Guo} et al.(2011)]{Guo_etal_2011}
\bibinfo{author}{{Guo} Q}, \bibinfo{author}{{White} S},
  \bibinfo{author}{{Boylan-Kolchin} M}, \bibinfo{author}{{De Lucia} G},
  \bibinfo{author}{{Kauffmann} G}, \bibinfo{author}{{Lemson} G},
  \bibinfo{author}{{Li} C}, \bibinfo{author}{{Springel} V} and
  \bibinfo{author}{{Weinmann} S} (\bibinfo{year}{2011}), \bibinfo{month}{May}.
\bibinfo{title}{{From dwarf spheroidals to cD galaxies: simulating the galaxy
  population in a {\ensuremath{\Lambda}}CDM cosmology}}.
\bibinfo{journal}{{\em \mnras}} \bibinfo{volume}{413} (\bibinfo{number}{1}):
  \bibinfo{pages}{101--131}.
  \bibinfo{doi}{\doi{10.1111/j.1365-2966.2010.18114.x}}.
\eprint{1006.0106}.

\bibtype{Article}%
\bibitem[{Gutcke} et al.(2022)]{Gutcke22}
\bibinfo{author}{{Gutcke} TA}, \bibinfo{author}{{Pakmor} R},
  \bibinfo{author}{{Naab} T} and  \bibinfo{author}{{Springel} V}
  (\bibinfo{year}{2022}), \bibinfo{month}{Jun.}
\bibinfo{title}{{LYRA - II. Cosmological dwarf galaxy formation with
  inhomogeneous Population III enrichment}}.
\bibinfo{journal}{{\em \mnras}} \bibinfo{volume}{513} (\bibinfo{number}{1}):
  \bibinfo{pages}{1372--1385}. \bibinfo{doi}{\doi{10.1093/mnras/stac867}}.
\eprint{2110.06233}.

\bibtype{Article}%
\bibitem[{Hartley} et al.(2010)]{Hartley_etal_2010}
\bibinfo{author}{{Hartley} WG}, \bibinfo{author}{{Almaini} O},
  \bibinfo{author}{{Cirasuolo} M}, \bibinfo{author}{{Foucaud} S},
  \bibinfo{author}{{Simpson} C}, \bibinfo{author}{{Conselice} CJ},
  \bibinfo{author}{{Smail} I}, \bibinfo{author}{{McLure} RJ},
  \bibinfo{author}{{Dunlop} JS}, \bibinfo{author}{{Chuter} RW},
  \bibinfo{author}{{Maddox} S}, \bibinfo{author}{{Lane} KP} and
  \bibinfo{author}{{Bradshaw} EJ} (\bibinfo{year}{2010}), \bibinfo{month}{Sep.}
\bibinfo{title}{{The evolution of galaxy clustering since z = 3 using the
  UKIDSS Ultra Deep Survey: the divergence of passive and star-forming
  galaxies}}.
\bibinfo{journal}{{\em \mnras}} \bibinfo{volume}{407} (\bibinfo{number}{2}):
  \bibinfo{pages}{1212--1222}.
  \bibinfo{doi}{\doi{10.1111/j.1365-2966.2010.16972.x}}.
\eprint{1005.1180}.

\bibtype{Article}%
\bibitem[{Hayward} et al.(2014)]{Hayward_etal_2014}
\bibinfo{author}{{Hayward} CC}, \bibinfo{author}{{Torrey} P},
  \bibinfo{author}{{Springel} V}, \bibinfo{author}{{Hernquist} L} and
  \bibinfo{author}{{Vogelsberger} M} (\bibinfo{year}{2014}),
  \bibinfo{month}{Aug.}
\bibinfo{title}{{Galaxy mergers on a moving mesh: a comparison with smoothed
  particle hydrodynamics}}.
\bibinfo{journal}{{\em \mnras}} \bibinfo{volume}{442} (\bibinfo{number}{3}):
  \bibinfo{pages}{1992--2016}. \bibinfo{doi}{\doi{10.1093/mnras/stu957}}.
\eprint{1309.2942}.

\bibtype{Article}%
\bibitem[{Hirschmann} et al.(2014)]{Hirschmann_etal_2014}
\bibinfo{author}{{Hirschmann} M}, \bibinfo{author}{{De Lucia} G},
  \bibinfo{author}{{Wilman} D}, \bibinfo{author}{{Weinmann} S},
  \bibinfo{author}{{Iovino} A}, \bibinfo{author}{{Cucciati} O},
  \bibinfo{author}{{Zibetti} S} and  \bibinfo{author}{{Villalobos} {\'A}}
  (\bibinfo{year}{2014}), \bibinfo{month}{Nov.}
\bibinfo{title}{{The influence of the environmental history on quenching star
  formation in a {\ensuremath{\Lambda}} cold dark matter universe}}.
\bibinfo{journal}{{\em \mnras}} \bibinfo{volume}{444} (\bibinfo{number}{3}):
  \bibinfo{pages}{2938--2959}. \bibinfo{doi}{\doi{10.1093/mnras/stu1609}}.
\eprint{1407.5621}.

\bibtype{Article}%
\bibitem[{Hirschmann} et al.(2016)]{Hirschmann_etal_2016}
\bibinfo{author}{{Hirschmann} M}, \bibinfo{author}{{De Lucia} G} and
  \bibinfo{author}{{Fontanot} F} (\bibinfo{year}{2016}), \bibinfo{month}{Sep.}
\bibinfo{title}{{Galaxy assembly, stellar feedback and metal enrichment: the
  view from the GAEA model}}.
\bibinfo{journal}{{\em \mnras}} \bibinfo{volume}{461} (\bibinfo{number}{2}):
  \bibinfo{pages}{1760--1785}. \bibinfo{doi}{\doi{10.1093/mnras/stw1318}}.
\eprint{1512.04531}.

\bibtype{Article}%
\bibitem[{Hu} et al.(2016)]{Hu_etal_2016}
\bibinfo{author}{{Hu} CY}, \bibinfo{author}{{Naab} T}, \bibinfo{author}{{Walch}
  S}, \bibinfo{author}{{Glover} SCO} and  \bibinfo{author}{{Clark} PC}
  (\bibinfo{year}{2016}), \bibinfo{month}{Jun.}
\bibinfo{title}{{Star formation and molecular hydrogen in dwarf galaxies: a
  non-equilibrium view}}.
\bibinfo{journal}{{\em \mnras}} \bibinfo{volume}{458} (\bibinfo{number}{4}):
  \bibinfo{pages}{3528--3553}. \bibinfo{doi}{\doi{10.1093/mnras/stw544}}.
\eprint{1510.05644}.

\bibtype{Article}%
\bibitem[{Ilbert} et al.(2013)]{Ilbert_etal_2013}
\bibinfo{author}{{Ilbert} O}, \bibinfo{author}{{McCracken} HJ},
  \bibinfo{author}{{Le F{\`e}vre} O}, \bibinfo{author}{{Capak} P},
  \bibinfo{author}{{Dunlop} J}, \bibinfo{author}{{Karim} A},
  \bibinfo{author}{{Renzini} MA}, \bibinfo{author}{{Caputi} K},
  \bibinfo{author}{{Boissier} S}, \bibinfo{author}{{Arnouts} S},
  \bibinfo{author}{{Aussel} H}, \bibinfo{author}{{Comparat} J},
  \bibinfo{author}{{Guo} Q}, \bibinfo{author}{{Hudelot} P},
  \bibinfo{author}{{Kartaltepe} J}, \bibinfo{author}{{Kneib} JP},
  \bibinfo{author}{{Krogager} JK}, \bibinfo{author}{{Le Floc'h} E},
  \bibinfo{author}{{Lilly} S}, \bibinfo{author}{{Mellier} Y},
  \bibinfo{author}{{Milvang-Jensen} B}, \bibinfo{author}{{Moutard} T},
  \bibinfo{author}{{Onodera} M}, \bibinfo{author}{{Richard} J},
  \bibinfo{author}{{Salvato} M}, \bibinfo{author}{{Sanders} DB},
  \bibinfo{author}{{Scoville} N}, \bibinfo{author}{{Silverman} JD},
  \bibinfo{author}{{Taniguchi} Y}, \bibinfo{author}{{Tasca} L},
  \bibinfo{author}{{Thomas} R}, \bibinfo{author}{{Toft} S},
  \bibinfo{author}{{Tresse} L}, \bibinfo{author}{{Vergani} D},
  \bibinfo{author}{{Wolk} M} and  \bibinfo{author}{{Zirm} A}
  (\bibinfo{year}{2013}), \bibinfo{month}{Aug.}
\bibinfo{title}{{Mass assembly in quiescent and star-forming galaxies since $z \simeq 4$ from UltraVISTA}}.
\bibinfo{journal}{{\em \aap}} \bibinfo{volume}{556}, \bibinfo{eid}{A55}.
  \bibinfo{doi}{\doi{10.1051/0004-6361/201321100}}.
\eprint{1301.3157}.

\bibtype{Article}%
\bibitem[{J{\'a}chym} et al.(2019)]{Jachym_etal_2019}
\bibinfo{author}{{J{\'a}chym} P}, \bibinfo{author}{{Kenney} JDP},
  \bibinfo{author}{{Sun} M}, \bibinfo{author}{{Combes} F},
  \bibinfo{author}{{Cortese} L}, \bibinfo{author}{{Scott} TC},
  \bibinfo{author}{{Sivanandam} S}, \bibinfo{author}{{Brinks} E},
  \bibinfo{author}{{Roediger} E}, \bibinfo{author}{{Palou{\v{s}}} J} and
  \bibinfo{author}{{Fumagalli} M} (\bibinfo{year}{2019}), \bibinfo{month}{Oct.}
\bibinfo{title}{{ALMA Unveils Widespread Molecular Gas Clumps in the Ram
  Pressure Stripped Tail of the Norma Jellyfish Galaxy}}.
\bibinfo{journal}{{\em \apj}} \bibinfo{volume}{883} (\bibinfo{number}{2}),
  \bibinfo{eid}{145}. \bibinfo{doi}{\doi{10.3847/1538-4357/ab3e6c}}.
\eprint{1905.13249}.

\bibtype{Article}%
\bibitem[{Jin} et al.(2024)]{Jin_etal_2024}
\bibinfo{author}{{Jin} S}, \bibinfo{author}{{Sillassen} NB},
  \bibinfo{author}{{Magdis} GE}, \bibinfo{author}{{Brinch} M},
  \bibinfo{author}{{Shuntov} M}, \bibinfo{author}{{Brammer} G},
  \bibinfo{author}{{Gobat} R}, \bibinfo{author}{{Valentino} F},
  \bibinfo{author}{{Carnall} AC}, \bibinfo{author}{{Lee} M},
  \bibinfo{author}{{Vijayan} AP}, \bibinfo{author}{{Gillman} S},
  \bibinfo{author}{{Kokorev} V}, \bibinfo{author}{{Le Bail} A},
  \bibinfo{author}{{Greve} TR}, \bibinfo{author}{{Gullberg} B},
  \bibinfo{author}{{Gould} KML} and  \bibinfo{author}{{Toft} S}
  (\bibinfo{year}{2024}), \bibinfo{month}{Mar.}
\bibinfo{title}{{Cosmic Vine: A z = 3.44 large-scale structure hosting massive
  quiescent galaxies}}.
\bibinfo{journal}{{\em \aap}} \bibinfo{volume}{683}, \bibinfo{eid}{L4}.
  \bibinfo{doi}{\doi{10.1051/0004-6361/202348540}}.
\eprint{2311.04867}.

\bibtype{Article}%
\bibitem[{Johansson} et al.(2009)]{Johannson_etal_2009}
\bibinfo{author}{{Johansson} PH}, \bibinfo{author}{{Naab} T} and
  \bibinfo{author}{{Ostriker} JP} (\bibinfo{year}{2009}), \bibinfo{month}{May}.
\bibinfo{title}{{Gravitational Heating Helps Make Massive Galaxies Red and
  Dead}}.
\bibinfo{journal}{{\em \apjl}} \bibinfo{volume}{697} (\bibinfo{number}{1}):
  \bibinfo{pages}{L38--L43}. \bibinfo{doi}{\doi{10.1088/0004-637X/697/1/L38}}.
\eprint{0903.2840}.

\bibtype{Article}%
\bibitem[{Kakimoto} et al.(2024)]{Kakimoto_etal_2024}
\bibinfo{author}{{Kakimoto} T}, \bibinfo{author}{{Tanaka} M},
  \bibinfo{author}{{Onodera} M}, \bibinfo{author}{{Shimakawa} R},
  \bibinfo{author}{{Wu} PF}, \bibinfo{author}{{Gould} KML},
  \bibinfo{author}{{Ito} K}, \bibinfo{author}{{Jin} S}, \bibinfo{author}{{Kubo}
  M}, \bibinfo{author}{{Suzuki} TL}, \bibinfo{author}{{Toft} S},
  \bibinfo{author}{{Valentino} F} and  \bibinfo{author}{{Yabe} K}
  (\bibinfo{year}{2024}), \bibinfo{month}{Mar.}
\bibinfo{title}{{A Massive Quiescent Galaxy in a Group Environment at z =
  4.53}}.
\bibinfo{journal}{{\em \apj}} \bibinfo{volume}{963} (\bibinfo{number}{1}),
  \bibinfo{eid}{49}. \bibinfo{doi}{\doi{10.3847/1538-4357/ad1ff1}}.
\eprint{2308.15011}.

\bibtype{Article}%
\bibitem[{Kang} and {van den Bosch}(2008)]{Kang_and_vandenBosch_2008}
\bibinfo{author}{{Kang} X} and  \bibinfo{author}{{van den Bosch} FC}
  (\bibinfo{year}{2008}), \bibinfo{month}{Apr.}
\bibinfo{title}{{New Constraints on the Efficiencies of Ram Pressure Stripping
  and the Tidal Disruption of Satellite Galaxies}}.
\bibinfo{journal}{{\em \apjl}} \bibinfo{volume}{676} (\bibinfo{number}{2}):
  \bibinfo{pages}{L101}. \bibinfo{doi}{\doi{10.1086/587620}}.
\eprint{0801.1843}.

\bibtype{Article}%
\bibitem[{Kasen} and {Woosley}(2009)]{Kasen_Woosley_2009}
\bibinfo{author}{{Kasen} D} and  \bibinfo{author}{{Woosley} SE}
  (\bibinfo{year}{2009}), \bibinfo{month}{Oct.}
\bibinfo{title}{{Type II Supernovae: Model Light Curves and Standard Candle
  Relationships}}.
\bibinfo{journal}{{\em \apj}} \bibinfo{volume}{703} (\bibinfo{number}{2}):
  \bibinfo{pages}{2205--2216}.
  \bibinfo{doi}{\doi{10.1088/0004-637X/703/2/2205}}.
\eprint{0910.1590}.

\bibtype{Article}%
\bibitem[{Kauffmann} et al.(1999)]{Kauffmann_etal_1999}
\bibinfo{author}{{Kauffmann} G}, \bibinfo{author}{{Colberg} JM},
  \bibinfo{author}{{Diaferio} A} and  \bibinfo{author}{{White} SDM}
  (\bibinfo{year}{1999}), \bibinfo{month}{Feb.}
\bibinfo{title}{{Clustering of galaxies in a hierarchical universe - I. Methods
  and results at z=0}}.
\bibinfo{journal}{{\em \mnras}} \bibinfo{volume}{303} (\bibinfo{number}{1}):
  \bibinfo{pages}{188--206}.
  \bibinfo{doi}{\doi{10.1046/j.1365-8711.1999.02202.x}}.
\eprint{astro-ph/9805283}.

\bibtype{Article}%
\bibitem[{Kauffmann} et al.(2003)]{Kauffmann_etal_2003}
\bibinfo{author}{{Kauffmann} G}, \bibinfo{author}{{Heckman} TM},
  \bibinfo{author}{{White} SDM}, \bibinfo{author}{{Charlot} S},
  \bibinfo{author}{{Tremonti} C}, \bibinfo{author}{{Brinchmann} J},
  \bibinfo{author}{{Bruzual} G}, \bibinfo{author}{{Peng} EW},
  \bibinfo{author}{{Seibert} M}, \bibinfo{author}{{Bernardi} M},
  \bibinfo{author}{{Blanton} M}, \bibinfo{author}{{Brinkmann} J},
  \bibinfo{author}{{Castander} F}, \bibinfo{author}{{Cs{\'a}bai} I},
  \bibinfo{author}{{Fukugita} M}, \bibinfo{author}{{Ivezic} Z},
  \bibinfo{author}{{Munn} JA}, \bibinfo{author}{{Nichol} RC},
  \bibinfo{author}{{Padmanabhan} N}, \bibinfo{author}{{Thakar} AR},
  \bibinfo{author}{{Weinberg} DH} and  \bibinfo{author}{{York} D}
  (\bibinfo{year}{2003}), \bibinfo{month}{May}.
\bibinfo{title}{{Stellar masses and star formation histories for {}10$^{5}$
  galaxies from the Sloan Digital Sky Survey}}.
\bibinfo{journal}{{\em \mnras}} \bibinfo{volume}{341} (\bibinfo{number}{1}):
  \bibinfo{pages}{33--53}.
  \bibinfo{doi}{\doi{10.1046/j.1365-8711.2003.06291.x}}.
\eprint{astro-ph/0204055}.

\bibtype{Article}%
\bibitem[{Kauffmann} et al.(2004)]{Kauffmann_etal_2004}
\bibinfo{author}{{Kauffmann} G}, \bibinfo{author}{{White} SDM},
  \bibinfo{author}{{Heckman} TM}, \bibinfo{author}{{M{\'e}nard} B},
  \bibinfo{author}{{Brinchmann} J}, \bibinfo{author}{{Charlot} S},
  \bibinfo{author}{{Tremonti} C} and  \bibinfo{author}{{Brinkmann} J}
  (\bibinfo{year}{2004}), \bibinfo{month}{Sep.}
\bibinfo{title}{{The environmental dependence of the relations between stellar
  mass, structure, star formation and nuclear activity in galaxies}}.
\bibinfo{journal}{{\em \mnras}} \bibinfo{volume}{353} (\bibinfo{number}{3}):
  \bibinfo{pages}{713--731}.
  \bibinfo{doi}{\doi{10.1111/j.1365-2966.2004.08117.x}}.
\eprint{astro-ph/0402030}.

\bibtype{Article}%
\bibitem[{Kennicutt} and {Evans}(2012)]{Kennicutt_and_Evans_2012}
\bibinfo{author}{{Kennicutt} RC} and  \bibinfo{author}{{Evans} NJ}
  (\bibinfo{year}{2012}), \bibinfo{month}{Sep.}
\bibinfo{title}{{Star Formation in the Milky Way and Nearby Galaxies}}.
\bibinfo{journal}{{\em \araa}} \bibinfo{volume}{50}: \bibinfo{pages}{531--608}.
  \bibinfo{doi}{\doi{10.1146/annurev-astro-081811-125610}}.
\eprint{1204.3552}.

\bibtype{Article}%
\bibitem[{Kimmig} et al.(2023)]{Kimmig_etal_2023}
\bibinfo{author}{{Kimmig} LC}, \bibinfo{author}{{Remus} RS},
  \bibinfo{author}{{Seidel} B}, \bibinfo{author}{{Valenzuela} LM},
  \bibinfo{author}{{Dolag} K} and  \bibinfo{author}{{Burkert} A}
  (\bibinfo{year}{2023}), \bibinfo{month}{Oct.}
\bibinfo{title}{{Blowing out the Candle: How to Quench Galaxies at High
  Redshift -- an Ensemble of Rapid Starbursts, AGN Feedback and Environment}}.
\bibinfo{journal}{{\em arXiv e-prints}} ,
  \bibinfo{eid}{arXiv:2310.16085}\bibinfo{doi}{\doi{10.48550/arXiv.2310.16085}}.
\eprint{2310.16085}.

\bibtype{Article}%
\bibitem[{Klutse} et al.(2024)]{Klutse_etal_2024}
\bibinfo{author}{{Klutse} DY}, \bibinfo{author}{{Hilton} M},
  \bibinfo{author}{{Heywood} I}, \bibinfo{author}{{Smail} I},
  \bibinfo{author}{{Swinbank} AM}, \bibinfo{author}{{Knowles} K} and
  \bibinfo{author}{{Sikhosana} SP} (\bibinfo{year}{2024}),
  \bibinfo{month}{Aug.}
\bibinfo{title}{{MeerKAT observations of starburst galaxies and AGNs within the
  core of XMMXCS J2215.9-1738 at z = 1.46}}.
\bibinfo{journal}{{\em \mnras}} \bibinfo{volume}{532} (\bibinfo{number}{2}):
  \bibinfo{pages}{2842--2859}. \bibinfo{doi}{\doi{10.1093/mnras/stae1640}}.
\eprint{2407.03667}.

\bibtype{Article}%
\bibitem[{Kocevski} et al.(2023)]{Kocevski_etal_2023}
\bibinfo{author}{{Kocevski} DD}, \bibinfo{author}{{Barro} G},
  \bibinfo{author}{{McGrath} EJ}, \bibinfo{author}{{Finkelstein} SL},
  \bibinfo{author}{{Bagley} MB}, \bibinfo{author}{{Ferguson} HC},
  \bibinfo{author}{{Jogee} S}, \bibinfo{author}{{Yang} G},
  \bibinfo{author}{{Dickinson} M}, \bibinfo{author}{{Hathi} NP},
  \bibinfo{author}{{Backhaus} BE}, \bibinfo{author}{{Bell} EF},
  \bibinfo{author}{{Bisigello} L}, \bibinfo{author}{{Buat} V},
  \bibinfo{author}{{Burgarella} D}, \bibinfo{author}{{Casey} CM},
  \bibinfo{author}{{Cleri} NJ}, \bibinfo{author}{{Cooper} MC},
  \bibinfo{author}{{Costantin} L}, \bibinfo{author}{{Croton} D},
  \bibinfo{author}{{Daddi} E}, \bibinfo{author}{{Fontana} A},
  \bibinfo{author}{{Fujimoto} S}, \bibinfo{author}{{Gardner} JP},
  \bibinfo{author}{{Gawiser} E}, \bibinfo{author}{{Giavalisco} M},
  \bibinfo{author}{{Grazian} A}, \bibinfo{author}{{Grogin} NA},
  \bibinfo{author}{{Guo} Y}, \bibinfo{author}{{Arrabal Haro} P},
  \bibinfo{author}{{Hirschmann} M}, \bibinfo{author}{{Holwerda} BW},
  \bibinfo{author}{{Huertas-Company} M}, \bibinfo{author}{{Hutchison} TA},
  \bibinfo{author}{{Iyer} KG}, \bibinfo{author}{{Jones} B},
  \bibinfo{author}{{Juneau} S}, \bibinfo{author}{{Kartaltepe} JS},
  \bibinfo{author}{{Kewley} LJ}, \bibinfo{author}{{Kirkpatrick} A},
  \bibinfo{author}{{Koekemoer} AM}, \bibinfo{author}{{Kurczynski} P},
  \bibinfo{author}{{Le Bail} A}, \bibinfo{author}{{Long} AS},
  \bibinfo{author}{{Lotz} JM}, \bibinfo{author}{{Lucas} RA},
  \bibinfo{author}{{Papovich} C}, \bibinfo{author}{{Pentericci} L},
  \bibinfo{author}{{P{\'e}rez-Gonz{\'a}lez} PG}, \bibinfo{author}{{Pirzkal} N},
  \bibinfo{author}{{Rafelski} M}, \bibinfo{author}{{Ravindranath} S},
  \bibinfo{author}{{Somerville} RS}, \bibinfo{author}{{Straughn} AN},
  \bibinfo{author}{{Tacchella} S}, \bibinfo{author}{{Trump} JR},
  \bibinfo{author}{{Wilkins} SM}, \bibinfo{author}{{Wuyts} S},
  \bibinfo{author}{{Yung} LYA} and  \bibinfo{author}{{Zavala} JA}
  (\bibinfo{year}{2023}), \bibinfo{month}{Mar.}
\bibinfo{title}{{CEERS Key Paper. II. A First Look at the Resolved Host
  Properties of AGN at 3 < z < 5 with JWST}}.
\bibinfo{journal}{{\em \apjl}} \bibinfo{volume}{946} (\bibinfo{number}{1}),
  \bibinfo{eid}{L14}. \bibinfo{doi}{\doi{10.3847/2041-8213/acad00}}.
\eprint{2208.14480}.

\bibtype{Article}%
\bibitem[{Kormendy} and {Ho}(2013)]{Kormendy_Ho_2013}
\bibinfo{author}{{Kormendy} J} and  \bibinfo{author}{{Ho} LC}
  (\bibinfo{year}{2013}), \bibinfo{month}{Aug.}
\bibinfo{title}{{Coevolution (Or Not) of Supermassive Black Holes and Host
  Galaxies}}.
\bibinfo{journal}{{\em \araa}} \bibinfo{volume}{51} (\bibinfo{number}{1}):
  \bibinfo{pages}{511--653}.
  \bibinfo{doi}{\doi{10.1146/annurev-astro-082708-101811}}.
\eprint{1304.7762}.

\bibtype{Article}%
\bibitem[{Kriek} et al.(2024)]{Kriek_etal_2024}
\bibinfo{author}{{Kriek} M}, \bibinfo{author}{{Beverage} AG},
  \bibinfo{author}{{Price} SH}, \bibinfo{author}{{Suess} KA},
  \bibinfo{author}{{Barro} G}, \bibinfo{author}{{Bezanson} RS},
  \bibinfo{author}{{Conroy} C}, \bibinfo{author}{{Cutler} SE},
  \bibinfo{author}{{Franx} M}, \bibinfo{author}{{Lin} J},
  \bibinfo{author}{{Lorenz} B}, \bibinfo{author}{{Ma} Y},
  \bibinfo{author}{{Momcheva} IG}, \bibinfo{author}{{Mowla} LA},
  \bibinfo{author}{{Pasha} I}, \bibinfo{author}{{van Dokkum} P} and
  \bibinfo{author}{{Whitaker} KE} (\bibinfo{year}{2024}), \bibinfo{month}{May}.
\bibinfo{title}{{The Heavy Metal Survey: Star Formation Constraints and
  Dynamical Masses of 21 Massive Quiescent Galaxies at z =
  1.3{\textendash}2.3}}.
\bibinfo{journal}{{\em \apj}} \bibinfo{volume}{966} (\bibinfo{number}{1}),
  \bibinfo{eid}{36}. \bibinfo{doi}{\doi{10.3847/1538-4357/ad2df9}}.
\eprint{2311.16232}.

\bibtype{Article}%
\bibitem[{Krumholz} et al.(2019)]{Krumholz_etal_2019}
\bibinfo{author}{{Krumholz} MR}, \bibinfo{author}{{McKee} CF} and
  \bibinfo{author}{{Bland-Hawthorn} J} (\bibinfo{year}{2019}),
  \bibinfo{month}{Aug.}
\bibinfo{title}{{Star Clusters Across Cosmic Time}}.
\bibinfo{journal}{{\em \araa}} \bibinfo{volume}{57}: \bibinfo{pages}{227--303}.
  \bibinfo{doi}{\doi{10.1146/annurev-astro-091918-104430}}.
\eprint{1812.01615}.

\bibtype{Article}%
\bibitem[{Kukstas} et al.(2023)]{Kukstas_etal_2023}
\bibinfo{author}{{Kukstas} E}, \bibinfo{author}{{Balogh} ML},
  \bibinfo{author}{{McCarthy} IG}, \bibinfo{author}{{Bah{\'e}} YM},
  \bibinfo{author}{{De Lucia} G}, \bibinfo{author}{{Jablonka} P},
  \bibinfo{author}{{Vulcani} B}, \bibinfo{author}{{Baxter} DC},
  \bibinfo{author}{{Biviano} A}, \bibinfo{author}{{Cerulo} P},
  \bibinfo{author}{{Chan} JC}, \bibinfo{author}{{Cooper} MC},
  \bibinfo{author}{{Demarco} R}, \bibinfo{author}{{Finoguenov} A},
  \bibinfo{author}{{Font} AS}, \bibinfo{author}{{Lidman} C},
  \bibinfo{author}{{Marchioni} J}, \bibinfo{author}{{McGee} S},
  \bibinfo{author}{{Muzzin} A}, \bibinfo{author}{{Nantais} J},
  \bibinfo{author}{{Old} L}, \bibinfo{author}{{Pintos-Castro} I},
  \bibinfo{author}{{Poggianti} B}, \bibinfo{author}{{Reeves} AMM},
  \bibinfo{author}{{Rudnick} G}, \bibinfo{author}{{Sarron} F},
  \bibinfo{author}{{van der Burg} R}, \bibinfo{author}{{Webb} K},
  \bibinfo{author}{{Wilson} G}, \bibinfo{author}{{Yee} HKC} and
  \bibinfo{author}{{Zaritsky} D} (\bibinfo{year}{2023}), \bibinfo{month}{Jan.}
\bibinfo{title}{{GOGREEN: A critical assessment of environmental trends in
  cosmological hydrodynamical simulations at z {\ensuremath{\approx}} 1}}.
\bibinfo{journal}{{\em \mnras}} \bibinfo{volume}{518} (\bibinfo{number}{3}):
  \bibinfo{pages}{4782--4800}. \bibinfo{doi}{\doi{10.1093/mnras/stac3438}}.
\eprint{2210.10803}.

\bibtype{Article}%
\bibitem[{Kulier} et al.(2023)]{Kulier_etal_2023}
\bibinfo{author}{{Kulier} A}, \bibinfo{author}{{Poggianti} B},
  \bibinfo{author}{{Tonnesen} S}, \bibinfo{author}{{Smith} R},
  \bibinfo{author}{{Ignesti} A}, \bibinfo{author}{{Akerman} N},
  \bibinfo{author}{{Marasco} A}, \bibinfo{author}{{Vulcani} B},
  \bibinfo{author}{{Moretti} A} and  \bibinfo{author}{{Wolter} A}
  (\bibinfo{year}{2023}), \bibinfo{month}{Sep.}
\bibinfo{title}{{Ram Pressure Stripping in the EAGLE Simulation}}.
\bibinfo{journal}{{\em \apj}} \bibinfo{volume}{954} (\bibinfo{number}{2}),
  \bibinfo{eid}{177}. \bibinfo{doi}{\doi{10.3847/1538-4357/aceda3}}.
\eprint{2305.03758}.

\bibtype{Article}%
\bibitem[{Kurinchi-Vendhan} et al.(2024)]{Kurinchi-Vendhan_etal_2024}
\bibinfo{author}{{Kurinchi-Vendhan} S}, \bibinfo{author}{{Farcy} M},
  \bibinfo{author}{{Hirschmann} M} and  \bibinfo{author}{{Valentino} F}
  (\bibinfo{year}{2024}), \bibinfo{month}{Nov.}
\bibinfo{title}{{On the origin of star formation quenching in massive galaxies
  at {\ensuremath{\gtrsim}} in the cosmological simulations IllustrisTNG}}.
\bibinfo{journal}{{\em \mnras}} \bibinfo{volume}{534} (\bibinfo{number}{4}):
  \bibinfo{pages}{3974--3988}. \bibinfo{doi}{\doi{10.1093/mnras/stae2297}}.
\eprint{2310.03083}.

\bibtype{Article}%
\bibitem[{Labb{\'e}} et al.(2005)]{Labbe_etal_2005}
\bibinfo{author}{{Labb{\'e}} I}, \bibinfo{author}{{Huang} J},
  \bibinfo{author}{{Franx} M}, \bibinfo{author}{{Rudnick} G},
  \bibinfo{author}{{Barmby} P}, \bibinfo{author}{{Daddi} E},
  \bibinfo{author}{{van Dokkum} PG}, \bibinfo{author}{{Fazio} GG},
  \bibinfo{author}{{F{\"o}rster Schreiber} NM}, \bibinfo{author}{{Moorwood}
  AFM}, \bibinfo{author}{{Rix} HW}, \bibinfo{author}{{R{\"o}ttgering} H},
  \bibinfo{author}{{Trujillo} I} and  \bibinfo{author}{{van der Werf} P}
  (\bibinfo{year}{2005}), \bibinfo{month}{May}.
\bibinfo{title}{{IRAC Mid-Infrared Imaging of the Hubble Deep Field-South: Star
  Formation Histories and Stellar Masses of Red Galaxies at z>2}}.
\bibinfo{journal}{{\em \apjl}} \bibinfo{volume}{624} (\bibinfo{number}{2}):
  \bibinfo{pages}{L81--L84}. \bibinfo{doi}{\doi{10.1086/430700}}.
\eprint{astro-ph/0504219}.

\bibtype{Article}%
\bibitem[{Lagos} et al.(2011)]{Lagos_etal_2011}
\bibinfo{author}{{Lagos} CDP}, \bibinfo{author}{{Lacey} CG},
  \bibinfo{author}{{Baugh} CM}, \bibinfo{author}{{Bower} RG} and
  \bibinfo{author}{{Benson} AJ} (\bibinfo{year}{2011}), \bibinfo{month}{Sep.}
\bibinfo{title}{{On the impact of empirical and theoretical star formation laws
  on galaxy formation}}.
\bibinfo{journal}{{\em \mnras}} \bibinfo{volume}{416} (\bibinfo{number}{2}):
  \bibinfo{pages}{1566--1584}.
  \bibinfo{doi}{\doi{10.1111/j.1365-2966.2011.19160.x}}.
\eprint{1011.5506}.

\bibtype{Article}%
\bibitem[{Lagos} et al.(2024{\natexlab{a}})]{Lagos_etal_2024a}
\bibinfo{author}{{Lagos} CdP}, \bibinfo{author}{{Bravo} M},
  \bibinfo{author}{{Tobar} R}, \bibinfo{author}{{Obreschkow} D},
  \bibinfo{author}{{Power} C}, \bibinfo{author}{{Robotham} ASG},
  \bibinfo{author}{{Proctor} KL}, \bibinfo{author}{{Hansen} S},
  \bibinfo{author}{{Chandro-G{\'o}mez} {\'A}} and  \bibinfo{author}{{Carrivick}
  J} (\bibinfo{year}{2024}{\natexlab{a}}), \bibinfo{month}{Jul.}
\bibinfo{title}{{Quenching massive galaxies across cosmic time with the
  semi-analytic model SHARK V2.0}}.
\bibinfo{journal}{{\em \mnras}} \bibinfo{volume}{531} (\bibinfo{number}{3}):
  \bibinfo{pages}{3551--3578}. \bibinfo{doi}{\doi{10.1093/mnras/stae1024}}.
\eprint{2309.02310}.

\bibtype{Article}%
\bibitem[{Lagos} et al.(2024{\natexlab{b}})]{Lagos_etal_2024b}
\bibinfo{author}{{Lagos} CdP}, \bibinfo{author}{{Valentino} F},
  \bibinfo{author}{{Wright} RJ}, \bibinfo{author}{{de Graaff} A},
  \bibinfo{author}{{Glazebrook} K}, \bibinfo{author}{{De Lucia} G},
  \bibinfo{author}{{Robotham} ASG}, \bibinfo{author}{{Nanayakkara} T},
  \bibinfo{author}{{Chandro-Gomez} A}, \bibinfo{author}{{Bravo} M},
  \bibinfo{author}{{Baugh} CM}, \bibinfo{author}{{Harborne} KE},
  \bibinfo{author}{{Hirschmann} M}, \bibinfo{author}{{Fontanot} F},
  \bibinfo{author}{{Xie} L} and  \bibinfo{author}{{Chittenden} H}
  (\bibinfo{year}{2024}{\natexlab{b}}), \bibinfo{month}{Sep.}
\bibinfo{title}{{The diverse star formation histories of early massive,
  quenched galaxies in modern galaxy formation simulations}}.
\bibinfo{journal}{{\em arXiv e-prints}} ,
  \bibinfo{eid}{arXiv:2409.16916}\bibinfo{doi}{\doi{10.48550/arXiv.2409.16916}}.
\eprint{2409.16916}.

\bibtype{Article}%
\bibitem[{Larson} et al.(1980)]{Larson_etal_1980}
\bibinfo{author}{{Larson} RB}, \bibinfo{author}{{Tinsley} BM} and
  \bibinfo{author}{{Caldwell} CN} (\bibinfo{year}{1980}), \bibinfo{month}{May}.
\bibinfo{title}{{The evolution of disk galaxies and the origin of S0
  galaxies}}.
\bibinfo{journal}{{\em \apj}} \bibinfo{volume}{237}: \bibinfo{pages}{692--707}.
  \bibinfo{doi}{\doi{10.1086/157917}}.

\bibtype{Article}%
\bibitem[{Le F{\`e}vre} et al.(2005)]{LeFevre_etal_2005}
\bibinfo{author}{{Le F{\`e}vre} O}, \bibinfo{author}{{Vettolani} G},
  \bibinfo{author}{{Garilli} B}, \bibinfo{author}{{Tresse} L},
  \bibinfo{author}{{Bottini} D}, \bibinfo{author}{{Le Brun} V},
  \bibinfo{author}{{Maccagni} D}, \bibinfo{author}{{Picat} JP},
  \bibinfo{author}{{Scaramella} R}, \bibinfo{author}{{Scodeggio} M},
  \bibinfo{author}{{Zanichelli} A}, \bibinfo{author}{{Adami} C},
  \bibinfo{author}{{Arnaboldi} M}, \bibinfo{author}{{Arnouts} S},
  \bibinfo{author}{{Bardelli} S}, \bibinfo{author}{{Bolzonella} M},
  \bibinfo{author}{{Cappi} A}, \bibinfo{author}{{Charlot} S},
  \bibinfo{author}{{Ciliegi} P}, \bibinfo{author}{{Contini} T},
  \bibinfo{author}{{Foucaud} S}, \bibinfo{author}{{Franzetti} P},
  \bibinfo{author}{{Gavignaud} I}, \bibinfo{author}{{Guzzo} L},
  \bibinfo{author}{{Ilbert} O}, \bibinfo{author}{{Iovino} A},
  \bibinfo{author}{{McCracken} HJ}, \bibinfo{author}{{Marano} B},
  \bibinfo{author}{{Marinoni} C}, \bibinfo{author}{{Mathez} G},
  \bibinfo{author}{{Mazure} A}, \bibinfo{author}{{Meneux} B},
  \bibinfo{author}{{Merighi} R}, \bibinfo{author}{{Paltani} S},
  \bibinfo{author}{{Pell{\`o}} R}, \bibinfo{author}{{Pollo} A},
  \bibinfo{author}{{Pozzetti} L}, \bibinfo{author}{{Radovich} M},
  \bibinfo{author}{{Zamorani} G}, \bibinfo{author}{{Zucca} E},
  \bibinfo{author}{{Bondi} M}, \bibinfo{author}{{Bongiorno} A},
  \bibinfo{author}{{Busarello} G}, \bibinfo{author}{{Lamareille} F},
  \bibinfo{author}{{Mellier} Y}, \bibinfo{author}{{Merluzzi} P},
  \bibinfo{author}{{Ripepi} V} and  \bibinfo{author}{{Rizzo} D}
  (\bibinfo{year}{2005}), \bibinfo{month}{Sep.}
\bibinfo{title}{{The VIMOS VLT deep survey. First epoch VVDS-deep survey: 11
  564 spectra with 17.5 {\ensuremath{\leq}} IAB {\ensuremath{\leq}} 24, and the
  redshift distribution over 0 {\ensuremath{\leq}} z {\ensuremath{\leq}} 5}}.
\bibinfo{journal}{{\em \aap}} \bibinfo{volume}{439} (\bibinfo{number}{3}):
  \bibinfo{pages}{845--862}. \bibinfo{doi}{\doi{10.1051/0004-6361:20041960}}.
\eprint{astro-ph/0409133}.

\bibtype{Article}%
\bibitem[{Lee-Brown} et al.(2017)]{Lee-Brown_etal_2017}
\bibinfo{author}{{Lee-Brown} DB}, \bibinfo{author}{{Rudnick} GH},
  \bibinfo{author}{{Momcheva} IG}, \bibinfo{author}{{Papovich} C},
  \bibinfo{author}{{Lotz} JM}, \bibinfo{author}{{Tran} KVH},
  \bibinfo{author}{{Henke} B}, \bibinfo{author}{{Willmer} CNA},
  \bibinfo{author}{{Brammer} GB}, \bibinfo{author}{{Brodwin} M},
  \bibinfo{author}{{Dunlop} J} and  \bibinfo{author}{{Farrah} D}
  (\bibinfo{year}{2017}), \bibinfo{month}{Jul.}
\bibinfo{title}{{The Ages of Passive Galaxies in a z = 1.62 Protocluster}}.
\bibinfo{journal}{{\em \apj}} \bibinfo{volume}{844} (\bibinfo{number}{1}),
  \bibinfo{eid}{43}. \bibinfo{doi}{\doi{10.3847/1538-4357/aa7948}}.
\eprint{1706.05017}.

\bibtype{Article}%
\bibitem[{Lee} et al.(2022)]{Lee_Kimm_etal_2022}
\bibinfo{author}{{Lee} J}, \bibinfo{author}{{Kimm} T},
  \bibinfo{author}{{Blaizot} J}, \bibinfo{author}{{Katz} H},
  \bibinfo{author}{{Lee} W}, \bibinfo{author}{{Sheen} YK},
  \bibinfo{author}{{Devriendt} J} and  \bibinfo{author}{{Slyz} A}
  (\bibinfo{year}{2022}), \bibinfo{month}{Apr.}
\bibinfo{title}{{Simulating Jellyfish Galaxies: A Case Study for a Gas-rich
  Dwarf Galaxy}}.
\bibinfo{journal}{{\em \apj}} \bibinfo{volume}{928} (\bibinfo{number}{2}),
  \bibinfo{eid}{144}. \bibinfo{doi}{\doi{10.3847/1538-4357/ac5595}}.
\eprint{2201.01316}.

\bibtype{Article}%
\bibitem[{Leja} et al.(2019)]{Leja_etal_2019}
\bibinfo{author}{{Leja} J}, \bibinfo{author}{{Carnall} AC},
  \bibinfo{author}{{Johnson} BD}, \bibinfo{author}{{Conroy} C} and
  \bibinfo{author}{{Speagle} JS} (\bibinfo{year}{2019}), \bibinfo{month}{May}.
\bibinfo{title}{{How to Measure Galaxy Star Formation Histories. II.
  Nonparametric Models}}.
\bibinfo{journal}{{\em \apj}} \bibinfo{volume}{876} (\bibinfo{number}{1}),
  \bibinfo{eid}{3}. \bibinfo{doi}{\doi{10.3847/1538-4357/ab133c}}.
\eprint{1811.03637}.

\bibtype{Article}%
\bibitem[{Li} et al.(2006)]{Li_etal_2006}
\bibinfo{author}{{Li} C}, \bibinfo{author}{{Kauffmann} G},
  \bibinfo{author}{{Jing} YP}, \bibinfo{author}{{White} SDM},
  \bibinfo{author}{{B{\"o}rner} G} and  \bibinfo{author}{{Cheng} FZ}
  (\bibinfo{year}{2006}), \bibinfo{month}{May}.
\bibinfo{title}{{The dependence of clustering on galaxy properties}}.
\bibinfo{journal}{{\em \mnras}} \bibinfo{volume}{368} (\bibinfo{number}{1}):
  \bibinfo{pages}{21--36}.
  \bibinfo{doi}{\doi{10.1111/j.1365-2966.2006.10066.x}}.
\eprint{astro-ph/0509873}.

\bibtype{Article}%
\bibitem[{Lilly} et al.(2007)]{Lilly_etal_2007}
\bibinfo{author}{{Lilly} SJ}, \bibinfo{author}{{Le F{\`e}vre} O},
  \bibinfo{author}{{Renzini} A}, \bibinfo{author}{{Zamorani} G},
  \bibinfo{author}{{Scodeggio} M}, \bibinfo{author}{{Contini} T},
  \bibinfo{author}{{Carollo} CM}, \bibinfo{author}{{Hasinger} G},
  \bibinfo{author}{{Kneib} JP}, \bibinfo{author}{{Iovino} A},
  \bibinfo{author}{{Le Brun} V}, \bibinfo{author}{{Maier} C},
  \bibinfo{author}{{Mainieri} V}, \bibinfo{author}{{Mignoli} M},
  \bibinfo{author}{{Silverman} J}, \bibinfo{author}{{Tasca} LAM},
  \bibinfo{author}{{Bolzonella} M}, \bibinfo{author}{{Bongiorno} A},
  \bibinfo{author}{{Bottini} D}, \bibinfo{author}{{Capak} P},
  \bibinfo{author}{{Caputi} K}, \bibinfo{author}{{Cimatti} A},
  \bibinfo{author}{{Cucciati} O}, \bibinfo{author}{{Daddi} E},
  \bibinfo{author}{{Feldmann} R}, \bibinfo{author}{{Franzetti} P},
  \bibinfo{author}{{Garilli} B}, \bibinfo{author}{{Guzzo} L},
  \bibinfo{author}{{Ilbert} O}, \bibinfo{author}{{Kampczyk} P},
  \bibinfo{author}{{Kovac} K}, \bibinfo{author}{{Lamareille} F},
  \bibinfo{author}{{Leauthaud} A}, \bibinfo{author}{{Le Borgne} JF},
  \bibinfo{author}{{McCracken} HJ}, \bibinfo{author}{{Marinoni} C},
  \bibinfo{author}{{Pello} R}, \bibinfo{author}{{Ricciardelli} E},
  \bibinfo{author}{{Scarlata} C}, \bibinfo{author}{{Vergani} D},
  \bibinfo{author}{{Sanders} DB}, \bibinfo{author}{{Schinnerer} E},
  \bibinfo{author}{{Scoville} N}, \bibinfo{author}{{Taniguchi} Y},
  \bibinfo{author}{{Arnouts} S}, \bibinfo{author}{{Aussel} H},
  \bibinfo{author}{{Bardelli} S}, \bibinfo{author}{{Brusa} M},
  \bibinfo{author}{{Cappi} A}, \bibinfo{author}{{Ciliegi} P},
  \bibinfo{author}{{Finoguenov} A}, \bibinfo{author}{{Foucaud} S},
  \bibinfo{author}{{Franceschini} A}, \bibinfo{author}{{Halliday} C},
  \bibinfo{author}{{Impey} C}, \bibinfo{author}{{Knobel} C},
  \bibinfo{author}{{Koekemoer} A}, \bibinfo{author}{{Kurk} J},
  \bibinfo{author}{{Maccagni} D}, \bibinfo{author}{{Maddox} S},
  \bibinfo{author}{{Marano} B}, \bibinfo{author}{{Marconi} G},
  \bibinfo{author}{{Meneux} B}, \bibinfo{author}{{Mobasher} B},
  \bibinfo{author}{{Moreau} C}, \bibinfo{author}{{Peacock} JA},
  \bibinfo{author}{{Porciani} C}, \bibinfo{author}{{Pozzetti} L},
  \bibinfo{author}{{Scaramella} R}, \bibinfo{author}{{Schiminovich} D},
  \bibinfo{author}{{Shopbell} P}, \bibinfo{author}{{Smail} I},
  \bibinfo{author}{{Thompson} D}, \bibinfo{author}{{Tresse} L},
  \bibinfo{author}{{Vettolani} G}, \bibinfo{author}{{Zanichelli} A} and
  \bibinfo{author}{{Zucca} E} (\bibinfo{year}{2007}), \bibinfo{month}{Sep.}
\bibinfo{title}{{zCOSMOS: A Large VLT/VIMOS Redshift Survey Covering 0 < z < 3
  in the COSMOS Field}}.
\bibinfo{journal}{{\em \apjs}} \bibinfo{volume}{172} (\bibinfo{number}{1}):
  \bibinfo{pages}{70--85}. \bibinfo{doi}{\doi{10.1086/516589}}.
\eprint{astro-ph/0612291}.

\bibtype{Article}%
\bibitem[{Lim} et al.(2024)]{Lim_etal_2024}
\bibinfo{author}{{Lim} S}, \bibinfo{author}{{Tacchella} S},
  \bibinfo{author}{{Schaye} J}, \bibinfo{author}{{Schaller} M},
  \bibinfo{author}{{Helton} JM}, \bibinfo{author}{{Kugel} R} and
  \bibinfo{author}{{Maiolino} R} (\bibinfo{year}{2024}), \bibinfo{month}{Aug.}
\bibinfo{title}{{The FLAMINGO simulation view of cluster progenitors observed
  in the epoch of reionization with JWST}}.
\bibinfo{journal}{{\em \mnras}} \bibinfo{volume}{532} (\bibinfo{number}{4}):
  \bibinfo{pages}{4551--4569}. \bibinfo{doi}{\doi{10.1093/mnras/stae1790}}.
\eprint{2402.17819}.

\bibtype{Article}%
\bibitem[{Liske} et al.(2015)]{Liske_etal_2015}
\bibinfo{author}{{Liske} J}, \bibinfo{author}{{Baldry} IK},
  \bibinfo{author}{{Driver} SP}, \bibinfo{author}{{Tuffs} RJ},
  \bibinfo{author}{{Alpaslan} M}, \bibinfo{author}{{Andrae} E},
  \bibinfo{author}{{Brough} S}, \bibinfo{author}{{Cluver} ME},
  \bibinfo{author}{{Grootes} MW}, \bibinfo{author}{{Gunawardhana} MLP},
  \bibinfo{author}{{Kelvin} LS}, \bibinfo{author}{{Loveday} J},
  \bibinfo{author}{{Robotham} ASG}, \bibinfo{author}{{Taylor} EN},
  \bibinfo{author}{{Bamford} SP}, \bibinfo{author}{{Bland-Hawthorn} J},
  \bibinfo{author}{{Brown} MJI}, \bibinfo{author}{{Drinkwater} MJ},
  \bibinfo{author}{{Hopkins} AM}, \bibinfo{author}{{Meyer} MJ},
  \bibinfo{author}{{Norberg} P}, \bibinfo{author}{{Peacock} JA},
  \bibinfo{author}{{Agius} NK}, \bibinfo{author}{{Andrews} SK},
  \bibinfo{author}{{Bauer} AE}, \bibinfo{author}{{Ching} JHY},
  \bibinfo{author}{{Colless} M}, \bibinfo{author}{{Conselice} CJ},
  \bibinfo{author}{{Croom} SM}, \bibinfo{author}{{Davies} LJM},
  \bibinfo{author}{{De Propris} R}, \bibinfo{author}{{Dunne} L},
  \bibinfo{author}{{Eardley} EM}, \bibinfo{author}{{Ellis} S},
  \bibinfo{author}{{Foster} C}, \bibinfo{author}{{Frenk} CS},
  \bibinfo{author}{{H{\"a}u{\ss}ler} B}, \bibinfo{author}{{Holwerda} BW},
  \bibinfo{author}{{Howlett} C}, \bibinfo{author}{{Ibarra} H},
  \bibinfo{author}{{Jarvis} MJ}, \bibinfo{author}{{Jones} DH},
  \bibinfo{author}{{Kafle} PR}, \bibinfo{author}{{Lacey} CG},
  \bibinfo{author}{{Lange} R}, \bibinfo{author}{{Lara-L{\'o}pez} MA},
  \bibinfo{author}{{L{\'o}pez-S{\'a}nchez} {\'A}R}, \bibinfo{author}{{Maddox}
  S}, \bibinfo{author}{{Madore} BF}, \bibinfo{author}{{McNaught-Roberts} T},
  \bibinfo{author}{{Moffett} AJ}, \bibinfo{author}{{Nichol} RC},
  \bibinfo{author}{{Owers} MS}, \bibinfo{author}{{Palamara} D},
  \bibinfo{author}{{Penny} SJ}, \bibinfo{author}{{Phillipps} S},
  \bibinfo{author}{{Pimbblet} KA}, \bibinfo{author}{{Popescu} CC},
  \bibinfo{author}{{Prescott} M}, \bibinfo{author}{{Proctor} R},
  \bibinfo{author}{{Sadler} EM}, \bibinfo{author}{{Sansom} AE},
  \bibinfo{author}{{Seibert} M}, \bibinfo{author}{{Sharp} R},
  \bibinfo{author}{{Sutherland} W}, \bibinfo{author}{{V{\'a}zquez-Mata} JA},
  \bibinfo{author}{{van Kampen} E}, \bibinfo{author}{{Wilkins} SM},
  \bibinfo{author}{{Williams} R} and  \bibinfo{author}{{Wright} AH}
  (\bibinfo{year}{2015}), \bibinfo{month}{Sep.}
\bibinfo{title}{{Galaxy And Mass Assembly (GAMA): end of survey report and data
  release 2}}.
\bibinfo{journal}{{\em \mnras}} \bibinfo{volume}{452} (\bibinfo{number}{2}):
  \bibinfo{pages}{2087--2126}. \bibinfo{doi}{\doi{10.1093/mnras/stv1436}}.
\eprint{1506.08222}.

\bibtype{Article}%
\bibitem[{Loh} et al.(2008)]{Loh_etal_2008}
\bibinfo{author}{{Loh} YS}, \bibinfo{author}{{Ellingson} E},
  \bibinfo{author}{{Yee} HKC}, \bibinfo{author}{{Gilbank} DG},
  \bibinfo{author}{{Gladders} MD} and  \bibinfo{author}{{Barrientos} LF}
  (\bibinfo{year}{2008}), \bibinfo{month}{Jun.}
\bibinfo{title}{{The Color Bimodality in Galaxy Clusters since z
  \raisebox{-0.5ex}\textasciitilde 0.9}}.
\bibinfo{journal}{{\em \apj}} \bibinfo{volume}{680} (\bibinfo{number}{1}):
  \bibinfo{pages}{214--223}. \bibinfo{doi}{\doi{10.1086/587830}}.
\eprint{0802.3726}.

\bibtype{Article}%
\bibitem[{Lotz} et al.(2019)]{Lotz_etal_2019}
\bibinfo{author}{{Lotz} M}, \bibinfo{author}{{Remus} RS},
  \bibinfo{author}{{Dolag} K}, \bibinfo{author}{{Biviano} A} and
  \bibinfo{author}{{Burkert} A} (\bibinfo{year}{2019}), \bibinfo{month}{Oct.}
\bibinfo{title}{{Gone after one orbit: How cluster environments quench
  galaxies}}.
\bibinfo{journal}{{\em \mnras}} \bibinfo{volume}{488} (\bibinfo{number}{4}):
  \bibinfo{pages}{5370--5389}. \bibinfo{doi}{\doi{10.1093/mnras/stz2070}}.
\eprint{1810.02382}.

\bibtype{Article}%
\bibitem[{Lustig} et al.(2023)]{Lustig_etal_2023}
\bibinfo{author}{{Lustig} P}, \bibinfo{author}{{Strazzullo} V},
  \bibinfo{author}{{Remus} RS}, \bibinfo{author}{{D'Eugenio} C},
  \bibinfo{author}{{Daddi} E}, \bibinfo{author}{{Burkert} A},
  \bibinfo{author}{{De Lucia} G}, \bibinfo{author}{{Delvecchio} I},
  \bibinfo{author}{{Dolag} K}, \bibinfo{author}{{Fontanot} F},
  \bibinfo{author}{{Gobat} R}, \bibinfo{author}{{Mohr} JJ},
  \bibinfo{author}{{Onodera} M}, \bibinfo{author}{{Pannella} M} and
  \bibinfo{author}{{Pillepich} A} (\bibinfo{year}{2023}), \bibinfo{month}{Feb.}
\bibinfo{title}{{Massive quiescent galaxies at z 3: A comparison of selection,
  stellar population, and structural properties with simulation predictions}}.
\bibinfo{journal}{{\em \mnras}} \bibinfo{volume}{518} (\bibinfo{number}{4}):
  \bibinfo{pages}{5953--5975}. \bibinfo{doi}{\doi{10.1093/mnras/stac3450}}.
\eprint{2201.09068}.

\bibtype{Article}%
\bibitem[{Mac Low} and {Klessen}(2004)]{MacLow_and_Klessen_2004}
\bibinfo{author}{{Mac Low} MM} and  \bibinfo{author}{{Klessen} RS}
  (\bibinfo{year}{2004}), \bibinfo{month}{Jan.}
\bibinfo{title}{{Control of star formation by supersonic turbulence}}.
\bibinfo{journal}{{\em Reviews of Modern Physics}} \bibinfo{volume}{76}
  (\bibinfo{number}{1}): \bibinfo{pages}{125--194}.
  \bibinfo{doi}{\doi{10.1103/RevModPhys.76.125}}.
\eprint{astro-ph/0301093}.

\bibtype{Article}%
\bibitem[{Mainieri} and {et al.}(2024)]{Mainieri_etal_2024}
\bibinfo{author}{{Mainieri} V} and  \bibinfo{author}{{et al.}}
  (\bibinfo{year}{2024}), \bibinfo{month}{Mar.}
\bibinfo{title}{{The Wide-field Spectroscopic Telescope (WST) Science White
  Paper}}.
\bibinfo{journal}{{\em arXiv e-prints}} ,
  \bibinfo{eid}{arXiv:2403.05398}\bibinfo{doi}{\doi{10.48550/arXiv.2403.05398}}.
\eprint{2403.05398}.

\bibtype{Article}%
\bibitem[{Maiolino} et al.(2020)]{Maiolino_etal_2020}
\bibinfo{author}{{Maiolino} R}, \bibinfo{author}{{Cirasuolo} M},
  \bibinfo{author}{{Afonso} J}, \bibinfo{author}{{Bauer} FE},
  \bibinfo{author}{{Bowler} R}, \bibinfo{author}{{Cucciati} O},
  \bibinfo{author}{{Daddi} E}, \bibinfo{author}{{De Lucia} G},
  \bibinfo{author}{{Evans} C}, \bibinfo{author}{{Flores} H},
  \bibinfo{author}{{Gargiulo} A}, \bibinfo{author}{{Garilli} B},
  \bibinfo{author}{{Jablonka} P}, \bibinfo{author}{{Jarvis} M},
  \bibinfo{author}{{Kneib} JP}, \bibinfo{author}{{Lilly} S},
  \bibinfo{author}{{Looser} T}, \bibinfo{author}{{Magliocchetti} M},
  \bibinfo{author}{{Man} Z}, \bibinfo{author}{{Mannucci} F},
  \bibinfo{author}{{Maurogordato} S}, \bibinfo{author}{{McLure} RJ},
  \bibinfo{author}{{Norberg} P}, \bibinfo{author}{{Oesch} P},
  \bibinfo{author}{{Oliva} E}, \bibinfo{author}{{Paltani} S},
  \bibinfo{author}{{Pappalardo} C}, \bibinfo{author}{{Peng} Y},
  \bibinfo{author}{{Pentericci} L}, \bibinfo{author}{{Pozzetti} L},
  \bibinfo{author}{{Renzini} A}, \bibinfo{author}{{Rodrigues} M},
  \bibinfo{author}{{Royer} F}, \bibinfo{author}{{Serjeant} S},
  \bibinfo{author}{{Vanzi} L}, \bibinfo{author}{{Wild} V} and
  \bibinfo{author}{{Zamorani} G} (\bibinfo{year}{2020}), \bibinfo{month}{Jun.}
\bibinfo{title}{{MOONRISE: The Main MOONS GTO Extragalactic Survey}}.
\bibinfo{journal}{{\em The Messenger}} \bibinfo{volume}{180}:
  \bibinfo{pages}{24--29}. \bibinfo{doi}{\doi{10.18727/0722-6691/5197}}.
\eprint{2009.00644}.

\bibtype{Article}%
\bibitem[{Man} and {Belli}(2018)]{Man_and_Belli_2018}
\bibinfo{author}{{Man} A} and  \bibinfo{author}{{Belli} S}
  (\bibinfo{year}{2018}), \bibinfo{month}{Sep.}
\bibinfo{title}{{Star formation quenching in massive galaxies}}.
\bibinfo{journal}{{\em Nature Astronomy}} \bibinfo{volume}{2}:
  \bibinfo{pages}{695--697}. \bibinfo{doi}{\doi{10.1038/s41550-018-0558-1}}.
\eprint{1809.00722}.

\bibtype{Article}%
\bibitem[{Martig} et al.(2009)]{Martig_etal_2009}
\bibinfo{author}{{Martig} M}, \bibinfo{author}{{Bournaud} F},
  \bibinfo{author}{{Teyssier} R} and  \bibinfo{author}{{Dekel} A}
  (\bibinfo{year}{2009}), \bibinfo{month}{Dec.}
\bibinfo{title}{{Morphological Quenching of Star Formation: Making Early-Type
  Galaxies Red}}.
\bibinfo{journal}{{\em \apj}} \bibinfo{volume}{707} (\bibinfo{number}{1}):
  \bibinfo{pages}{250--267}. \bibinfo{doi}{\doi{10.1088/0004-637X/707/1/250}}.
\eprint{0905.4669}.

\bibtype{Article}%
\bibitem[{Martis} et al.(2016)]{Martis_etal_2016}
\bibinfo{author}{{Martis} NS}, \bibinfo{author}{{Marchesini} D},
  \bibinfo{author}{{Brammer} GB}, \bibinfo{author}{{Muzzin} A},
  \bibinfo{author}{{Labb{\'e}} I}, \bibinfo{author}{{Momcheva} IG},
  \bibinfo{author}{{Skelton} RE}, \bibinfo{author}{{Stefanon} M},
  \bibinfo{author}{{van Dokkum} PG} and  \bibinfo{author}{{Whitaker} KE}
  (\bibinfo{year}{2016}), \bibinfo{month}{Aug.}
\bibinfo{title}{{The Evolution of the Fractions of Quiescent and Star-forming
  Galaxies as a Function of Stellar Mass Since z = 3: Increasing Importance of
  Massive, Dusty Star-forming Galaxies in the Early Universe}}.
\bibinfo{journal}{{\em \apjl}} \bibinfo{volume}{827} (\bibinfo{number}{2}),
  \bibinfo{eid}{L25}. \bibinfo{doi}{\doi{10.3847/2041-8205/827/2/L25}}.
\eprint{1606.04090}.

\bibtype{Article}%
\bibitem[{McCarthy} et al.(2008)]{McCarthy_etal_2008}
\bibinfo{author}{{McCarthy} IG}, \bibinfo{author}{{Frenk} CS},
  \bibinfo{author}{{Font} AS}, \bibinfo{author}{{Lacey} CG},
  \bibinfo{author}{{Bower} RG}, \bibinfo{author}{{Mitchell} NL},
  \bibinfo{author}{{Balogh} ML} and  \bibinfo{author}{{Theuns} T}
  (\bibinfo{year}{2008}), \bibinfo{month}{Jan.}
\bibinfo{title}{{Ram pressure stripping the hot gaseous haloes of galaxies in
  groups and clusters}}.
\bibinfo{journal}{{\em \mnras}} \bibinfo{volume}{383} (\bibinfo{number}{2}):
  \bibinfo{pages}{593--605}.
  \bibinfo{doi}{\doi{10.1111/j.1365-2966.2007.12577.x}}.
\eprint{0710.0964}.

\bibtype{Article}%
\bibitem[{McNamara} and {Nulsen}(2007)]{McNamara_and_Nulsen_2007}
\bibinfo{author}{{McNamara} BR} and  \bibinfo{author}{{Nulsen} PEJ}
  (\bibinfo{year}{2007}), \bibinfo{month}{Sep.}
\bibinfo{title}{{Heating Hot Atmospheres with Active Galactic Nuclei}}.
\bibinfo{journal}{{\em \araa}} \bibinfo{volume}{45} (\bibinfo{number}{1}):
  \bibinfo{pages}{117--175}.
  \bibinfo{doi}{\doi{10.1146/annurev.astro.45.051806.110625}}.
\eprint{0709.2152}.

\bibtype{Article}%
\bibitem[{Mei} et al.(2009)]{Mei_etal_2009}
\bibinfo{author}{{Mei} S}, \bibinfo{author}{{Holden} BP},
  \bibinfo{author}{{Blakeslee} JP}, \bibinfo{author}{{Ford} HC},
  \bibinfo{author}{{Franx} M}, \bibinfo{author}{{Homeier} NL},
  \bibinfo{author}{{Illingworth} GD}, \bibinfo{author}{{Jee} MJ},
  \bibinfo{author}{{Overzier} R}, \bibinfo{author}{{Postman} M},
  \bibinfo{author}{{Rosati} P}, \bibinfo{author}{{Van der Wel} A} and
  \bibinfo{author}{{Bartlett} JG} (\bibinfo{year}{2009}), \bibinfo{month}{Jan.}
\bibinfo{title}{{Evolution of the Color-Magnitude Relation in Galaxy Clusters
  at z \raisebox{-0.5ex}\textasciitilde 1 from the ACS Intermediate Redshift
  Cluster Survey}}.
\bibinfo{journal}{{\em \apj}} \bibinfo{volume}{690} (\bibinfo{number}{1}):
  \bibinfo{pages}{42--68}. \bibinfo{doi}{\doi{10.1088/0004-637X/690/1/42}}.
\eprint{0810.1917}.

\bibtype{Article}%
\bibitem[{Merlin} et al.(2018)]{Merlin_etal_2018}
\bibinfo{author}{{Merlin} E}, \bibinfo{author}{{Fontana} A},
  \bibinfo{author}{{Castellano} M}, \bibinfo{author}{{Santini} P},
  \bibinfo{author}{{Torelli} M}, \bibinfo{author}{{Boutsia} K},
  \bibinfo{author}{{Wang} T}, \bibinfo{author}{{Grazian} A},
  \bibinfo{author}{{Pentericci} L}, \bibinfo{author}{{Schreiber} C},
  \bibinfo{author}{{Ciesla} L}, \bibinfo{author}{{McLure} R},
  \bibinfo{author}{{Derriere} S}, \bibinfo{author}{{Dunlop} JS} and
  \bibinfo{author}{{Elbaz} D} (\bibinfo{year}{2018}), \bibinfo{month}{Jan.}
\bibinfo{title}{{Chasing passive galaxies in the early Universe: a critical
  analysis in CANDELS GOODS-South}}.
\bibinfo{journal}{{\em \mnras}} \bibinfo{volume}{473} (\bibinfo{number}{2}):
  \bibinfo{pages}{2098--2123}. \bibinfo{doi}{\doi{10.1093/mnras/stx2385}}.
\eprint{1709.00429}.

\bibtype{Article}%
\bibitem[{Monaco} and {Fontanot}(2005)]{Monaco_and_Fontanot_2005}
\bibinfo{author}{{Monaco} P} and  \bibinfo{author}{{Fontanot} F}
  (\bibinfo{year}{2005}), \bibinfo{month}{May}.
\bibinfo{title}{{Feedback from quasars in star-forming galaxies and the
  triggering of massive galactic winds}}.
\bibinfo{journal}{{\em \mnras}} \bibinfo{volume}{359} (\bibinfo{number}{1}):
  \bibinfo{pages}{283--294}.
  \bibinfo{doi}{\doi{10.1111/j.1365-2966.2005.08884.x}}.
\eprint{astro-ph/0502145}.

\bibtype{Article}%
\bibitem[{Moretti} et al.(2018)]{Moretti_etal_2018}
\bibinfo{author}{{Moretti} A}, \bibinfo{author}{{Paladino} R},
  \bibinfo{author}{{Poggianti} BM}, \bibinfo{author}{{D'Onofrio} M},
  \bibinfo{author}{{Bettoni} D}, \bibinfo{author}{{Gullieuszik} M},
  \bibinfo{author}{{Jaff{\'e}} YL}, \bibinfo{author}{{Vulcani} B},
  \bibinfo{author}{{Fasano} G}, \bibinfo{author}{{Fritz} J} and
  \bibinfo{author}{{Torstensson} K} (\bibinfo{year}{2018}),
  \bibinfo{month}{Oct.}
\bibinfo{title}{{GASP - X. APEX observations of molecular gas in the discs and
  in the tails of ram-pressure stripped galaxies}}.
\bibinfo{journal}{{\em \mnras}} \bibinfo{volume}{480} (\bibinfo{number}{2}):
  \bibinfo{pages}{2508--2520}. \bibinfo{doi}{\doi{10.1093/mnras/sty2021}}.
\eprint{1803.06183}.

\bibtype{Article}%
\bibitem[{Mostek} et al.(2013)]{Mostek_etal_2013}
\bibinfo{author}{{Mostek} N}, \bibinfo{author}{{Coil} AL},
  \bibinfo{author}{{Cooper} M}, \bibinfo{author}{{Davis} M},
  \bibinfo{author}{{Newman} JA} and  \bibinfo{author}{{Weiner} BJ}
  (\bibinfo{year}{2013}), \bibinfo{month}{Apr.}
\bibinfo{title}{{The DEEP2 Galaxy Redshift Survey: Clustering Dependence on
  Galaxy Stellar Mass and Star Formation Rate at z
  \raisebox{-0.5ex}\textasciitilde 1}}.
\bibinfo{journal}{{\em \apj}} \bibinfo{volume}{767} (\bibinfo{number}{1}),
  \bibinfo{eid}{89}. \bibinfo{doi}{\doi{10.1088/0004-637X/767/1/89}}.
\eprint{1210.6694}.

\bibtype{Article}%
\bibitem[{Muzzin} et al.(2013)]{Muzzin_etal_2013}
\bibinfo{author}{{Muzzin} A}, \bibinfo{author}{{Marchesini} D},
  \bibinfo{author}{{Stefanon} M}, \bibinfo{author}{{Franx} M},
  \bibinfo{author}{{McCracken} HJ}, \bibinfo{author}{{Milvang-Jensen} B},
  \bibinfo{author}{{Dunlop} JS}, \bibinfo{author}{{Fynbo} JPU},
  \bibinfo{author}{{Brammer} G}, \bibinfo{author}{{Labb{\'e}} I} and
  \bibinfo{author}{{van Dokkum} PG} (\bibinfo{year}{2013}),
  \bibinfo{month}{Nov.}
\bibinfo{title}{{The Evolution of the Stellar Mass Functions of Star-forming
  and Quiescent Galaxies to z = 4 from the COSMOS/UltraVISTA Survey}}.
\bibinfo{journal}{{\em \apj}} \bibinfo{volume}{777} (\bibinfo{number}{1}),
  \bibinfo{eid}{18}. \bibinfo{doi}{\doi{10.1088/0004-637X/777/1/18}}.
\eprint{1303.4409}.

\bibtype{Article}%
\bibitem[{Naab} and {Ostriker}(2017)]{Naab_and_Ostriker_2017}
\bibinfo{author}{{Naab} T} and  \bibinfo{author}{{Ostriker} JP}
  (\bibinfo{year}{2017}), \bibinfo{month}{Aug.}
\bibinfo{title}{{Theoretical Challenges in Galaxy Formation}}.
\bibinfo{journal}{{\em \araa}} \bibinfo{volume}{55} (\bibinfo{number}{1}):
  \bibinfo{pages}{59--109}.
  \bibinfo{doi}{\doi{10.1146/annurev-astro-081913-040019}}.
\eprint{1612.06891}.

\bibtype{Article}%
\bibitem[{Nanayakkara} et al.(2024)]{Nanayakkara_etal_2024}
\bibinfo{author}{{Nanayakkara} T}, \bibinfo{author}{{Glazebrook} K},
  \bibinfo{author}{{Jacobs} C}, \bibinfo{author}{{Kawinwanichakij} L},
  \bibinfo{author}{{Schreiber} C}, \bibinfo{author}{{Brammer} G},
  \bibinfo{author}{{Esdaile} J}, \bibinfo{author}{{Kacprzak} GG},
  \bibinfo{author}{{Labbe} I}, \bibinfo{author}{{Lagos} C},
  \bibinfo{author}{{Marchesini} D}, \bibinfo{author}{{Marsan} ZC},
  \bibinfo{author}{{Oesch} PA}, \bibinfo{author}{{Papovich} C},
  \bibinfo{author}{{Remus} RS} and  \bibinfo{author}{{Tran} KVH}
  (\bibinfo{year}{2024}), \bibinfo{month}{Feb.}
\bibinfo{title}{{A population of faint, old, and massive quiescent galaxies at
  3 <z <4 revealed by JWST NIRSpec Spectroscopy}}.
\bibinfo{journal}{{\em Scientific Reports}} \bibinfo{volume}{14},
  \bibinfo{eid}{3724}. \bibinfo{doi}{\doi{10.1038/s41598-024-52585-4}}.
\eprint{2212.11638}.

\bibtype{Article}%
\bibitem[{Nantais} et al.(2017)]{Nantais_etal_2017}
\bibinfo{author}{{Nantais} JB}, \bibinfo{author}{{Muzzin} A},
  \bibinfo{author}{{van der Burg} RFJ}, \bibinfo{author}{{Wilson} G},
  \bibinfo{author}{{Lidman} C}, \bibinfo{author}{{Foltz} R},
  \bibinfo{author}{{DeGroot} A}, \bibinfo{author}{{Noble} A},
  \bibinfo{author}{{Cooper} MC} and  \bibinfo{author}{{Demarco} R}
  (\bibinfo{year}{2017}), \bibinfo{month}{Feb.}
\bibinfo{title}{{Evidence for strong evolution in galaxy environmental
  quenching efficiency between z = 1.6 and z = 0.9}}.
\bibinfo{journal}{{\em \mnras}} \bibinfo{volume}{465} (\bibinfo{number}{1}):
  \bibinfo{pages}{L104--L108}. \bibinfo{doi}{\doi{10.1093/mnrasl/slw224}}.
\eprint{1610.08058}.

\bibtype{Article}%
\bibitem[{Nesvadba} et al.(2020)]{Nesvadba_etal_2020}
\bibinfo{author}{{Nesvadba} NPH}, \bibinfo{author}{{Bicknell} GV},
  \bibinfo{author}{{Mukherjee} D} and  \bibinfo{author}{{Wagner} AY}
  (\bibinfo{year}{2020}), \bibinfo{month}{Jul.}
\bibinfo{title}{{Gas, dust, and star formation in the positive AGN feedback
  candidate 4C 41.17 at z = 3.8}}.
\bibinfo{journal}{{\em \aap}} \bibinfo{volume}{639}, \bibinfo{eid}{L13}.
  \bibinfo{doi}{\doi{10.1051/0004-6361/202038269}}.
\eprint{2006.10572}.

\bibtype{Article}%
\bibitem[{Nulsen}(1982)]{Nulsen_etal_1982}
\bibinfo{author}{{Nulsen} PEJ} (\bibinfo{year}{1982}), \bibinfo{month}{Mar.}
\bibinfo{title}{{Transport processes and the stripping of cluster galaxies.}}
\bibinfo{journal}{{\em \mnras}} \bibinfo{volume}{198}:
  \bibinfo{pages}{1007--1016}. \bibinfo{doi}{\doi{10.1093/mnras/198.4.1007}}.

\bibtype{Article}%
\bibitem[{Overzier}(2016)]{Overzier_2016}
\bibinfo{author}{{Overzier} RA} (\bibinfo{year}{2016}), \bibinfo{month}{Nov.}
\bibinfo{title}{{The realm of the galaxy protoclusters. A review}}.
\bibinfo{journal}{{\em \aapr}} \bibinfo{volume}{24} (\bibinfo{number}{1}),
  \bibinfo{eid}{14}. \bibinfo{doi}{\doi{10.1007/s00159-016-0100-3}}.
\eprint{1610.05201}.

\bibtype{Article}%
\bibitem[{Paccagnella} et al.(2016)]{Paccagnella_etal_2016}
\bibinfo{author}{{Paccagnella} A}, \bibinfo{author}{{Vulcani} B},
  \bibinfo{author}{{Poggianti} BM}, \bibinfo{author}{{Moretti} A},
  \bibinfo{author}{{Fritz} J}, \bibinfo{author}{{Gullieuszik} M},
  \bibinfo{author}{{Couch} W}, \bibinfo{author}{{Bettoni} D},
  \bibinfo{author}{{Cava} A}, \bibinfo{author}{{D'Onofrio} M} and
  \bibinfo{author}{{Fasano} G} (\bibinfo{year}{2016}), \bibinfo{month}{Jan.}
\bibinfo{title}{{Slow Quenching of Star Formation in OMEGAWINGS Clusters:
  Galaxies in Transition in the Local Universe}}.
\bibinfo{journal}{{\em \apjl}} \bibinfo{volume}{816} (\bibinfo{number}{2}),
  \bibinfo{eid}{L25}. \bibinfo{doi}{\doi{10.3847/2041-8205/816/2/L25}}.
\eprint{1512.04549}.

\bibtype{Article}%
\bibitem[{Pacifici} et al.(2015)]{Pacifici_etal_2015}
\bibinfo{author}{{Pacifici} C}, \bibinfo{author}{{da Cunha} E},
  \bibinfo{author}{{Charlot} S}, \bibinfo{author}{{Rix} HW},
  \bibinfo{author}{{Fumagalli} M}, \bibinfo{author}{{Wel} Avd},
  \bibinfo{author}{{Franx} M}, \bibinfo{author}{{Maseda} MV},
  \bibinfo{author}{{van Dokkum} PG}, \bibinfo{author}{{Brammer} GB},
  \bibinfo{author}{{Momcheva} I}, \bibinfo{author}{{Skelton} RE},
  \bibinfo{author}{{Whitaker} K}, \bibinfo{author}{{Leja} J},
  \bibinfo{author}{{Lundgren} B}, \bibinfo{author}{{Kassin} SA} and
  \bibinfo{author}{{Yi} SK} (\bibinfo{year}{2015}), \bibinfo{month}{Feb.}
\bibinfo{title}{{On the importance of using appropriate spectral models to
  derive physical properties of galaxies at 0.7 < z < 2.8}}.
\bibinfo{journal}{{\em \mnras}} \bibinfo{volume}{447} (\bibinfo{number}{1}):
  \bibinfo{pages}{786--805}. \bibinfo{doi}{\doi{10.1093/mnras/stu2447}}.
\eprint{1411.5689}.

\bibtype{Article}%
\bibitem[{Parente} et al.(2023)]{Parente_etal_2023}
\bibinfo{author}{{Parente} M}, \bibinfo{author}{{Ragone-Figueroa} C},
  \bibinfo{author}{{Granato} GL} and  \bibinfo{author}{{Lapi} A}
  (\bibinfo{year}{2023}), \bibinfo{month}{Jun.}
\bibinfo{title}{{The z {\ensuremath{\lesssim}} 1 drop of cosmic dust abundance
  in a semi-analytic framework}}.
\bibinfo{journal}{{\em \mnras}} \bibinfo{volume}{521} (\bibinfo{number}{4}):
  \bibinfo{pages}{6105--6123}. \bibinfo{doi}{\doi{10.1093/mnras/stad907}}.
\eprint{2302.03058}.

\bibtype{Article}%
\bibitem[{Park} et al.(2024)]{Park_etal_2024}
\bibinfo{author}{{Park} M}, \bibinfo{author}{{Belli} S},
  \bibinfo{author}{{Conroy} C}, \bibinfo{author}{{Johnson} BD},
  \bibinfo{author}{{Davies} RL}, \bibinfo{author}{{Leja} J},
  \bibinfo{author}{{Tacchella} S}, \bibinfo{author}{{Mendel} JT},
  \bibinfo{author}{{Benton} C}, \bibinfo{author}{{Bugiani} L},
  \bibinfo{author}{{Emami} R}, \bibinfo{author}{{Khoram} AH},
  \bibinfo{author}{{Li} Y}, \bibinfo{author}{{Maheson} G},
  \bibinfo{author}{{Mathews} EP}, \bibinfo{author}{{Naidu} RP},
  \bibinfo{author}{{Nelson} EJ}, \bibinfo{author}{{Terrazas} BA} and
  \bibinfo{author}{{Weinberger} R} (\bibinfo{year}{2024}),
  \bibinfo{month}{Nov.}
\bibinfo{title}{{Widespread Rapid Quenching at Cosmic Noon Revealed by JWST
  Deep Spectroscopy}}.
\bibinfo{journal}{{\em \apj}} \bibinfo{volume}{976} (\bibinfo{number}{1}),
  \bibinfo{eid}{72}. \bibinfo{doi}{\doi{10.3847/1538-4357/ad7e15}}.
\eprint{2404.17945}.

\bibtype{Article}%
\bibitem[{Pawlik} et al.(2019)]{Pawlik_etal_2019}
\bibinfo{author}{{Pawlik} MM}, \bibinfo{author}{{McAlpine} S},
  \bibinfo{author}{{Trayford} JW}, \bibinfo{author}{{Wild} V},
  \bibinfo{author}{{Bower} R}, \bibinfo{author}{{Crain} RA},
  \bibinfo{author}{{Schaller} M} and  \bibinfo{author}{{Schaye} J}
  (\bibinfo{year}{2019}), \bibinfo{month}{Mar.}
\bibinfo{title}{{The diverse evolutionary pathways of post-starburst
  galaxies}}.
\bibinfo{journal}{{\em Nature Astronomy}} \bibinfo{volume}{3}:
  \bibinfo{pages}{440--446}. \bibinfo{doi}{\doi{10.1038/s41550-019-0725-z}}.
\eprint{1903.11050}.

\bibtype{Article}%
\bibitem[{Peng} et al.(2010)]{Peng_etal_2010}
\bibinfo{author}{{Peng} Yj}, \bibinfo{author}{{Lilly} SJ},
  \bibinfo{author}{{Kova{\v{c}}} K}, \bibinfo{author}{{Bolzonella} M},
  \bibinfo{author}{{Pozzetti} L}, \bibinfo{author}{{Renzini} A},
  \bibinfo{author}{{Zamorani} G}, \bibinfo{author}{{Ilbert} O},
  \bibinfo{author}{{Knobel} C}, \bibinfo{author}{{Iovino} A},
  \bibinfo{author}{{Maier} C}, \bibinfo{author}{{Cucciati} O},
  \bibinfo{author}{{Tasca} L}, \bibinfo{author}{{Carollo} CM},
  \bibinfo{author}{{Silverman} J}, \bibinfo{author}{{Kampczyk} P},
  \bibinfo{author}{{de Ravel} L}, \bibinfo{author}{{Sanders} D},
  \bibinfo{author}{{Scoville} N}, \bibinfo{author}{{Contini} T},
  \bibinfo{author}{{Mainieri} V}, \bibinfo{author}{{Scodeggio} M},
  \bibinfo{author}{{Kneib} JP}, \bibinfo{author}{{Le F{\`e}vre} O},
  \bibinfo{author}{{Bardelli} S}, \bibinfo{author}{{Bongiorno} A},
  \bibinfo{author}{{Caputi} K}, \bibinfo{author}{{Coppa} G},
  \bibinfo{author}{{de la Torre} S}, \bibinfo{author}{{Franzetti} P},
  \bibinfo{author}{{Garilli} B}, \bibinfo{author}{{Lamareille} F},
  \bibinfo{author}{{Le Borgne} JF}, \bibinfo{author}{{Le Brun} V},
  \bibinfo{author}{{Mignoli} M}, \bibinfo{author}{{Perez Montero} E},
  \bibinfo{author}{{Pello} R}, \bibinfo{author}{{Ricciardelli} E},
  \bibinfo{author}{{Tanaka} M}, \bibinfo{author}{{Tresse} L},
  \bibinfo{author}{{Vergani} D}, \bibinfo{author}{{Welikala} N},
  \bibinfo{author}{{Zucca} E}, \bibinfo{author}{{Oesch} P},
  \bibinfo{author}{{Abbas} U}, \bibinfo{author}{{Barnes} L},
  \bibinfo{author}{{Bordoloi} R}, \bibinfo{author}{{Bottini} D},
  \bibinfo{author}{{Cappi} A}, \bibinfo{author}{{Cassata} P},
  \bibinfo{author}{{Cimatti} A}, \bibinfo{author}{{Fumana} M},
  \bibinfo{author}{{Hasinger} G}, \bibinfo{author}{{Koekemoer} A},
  \bibinfo{author}{{Leauthaud} A}, \bibinfo{author}{{Maccagni} D},
  \bibinfo{author}{{Marinoni} C}, \bibinfo{author}{{McCracken} H},
  \bibinfo{author}{{Memeo} P}, \bibinfo{author}{{Meneux} B},
  \bibinfo{author}{{Nair} P}, \bibinfo{author}{{Porciani} C},
  \bibinfo{author}{{Presotto} V} and  \bibinfo{author}{{Scaramella} R}
  (\bibinfo{year}{2010}), \bibinfo{month}{Sep.}
\bibinfo{title}{{Mass and Environment as Drivers of Galaxy Evolution in SDSS
  and zCOSMOS and the Origin of the Schechter Function}}.
\bibinfo{journal}{{\em \apj}} \bibinfo{volume}{721} (\bibinfo{number}{1}):
  \bibinfo{pages}{193--221}. \bibinfo{doi}{\doi{10.1088/0004-637X/721/1/193}}.
\eprint{1003.4747}.

\bibtype{Article}%
\bibitem[{Pillepich} et al.(2018)]{Pillepich_etal_2018}
\bibinfo{author}{{Pillepich} A}, \bibinfo{author}{{Springel} V},
  \bibinfo{author}{{Nelson} D}, \bibinfo{author}{{Genel} S},
  \bibinfo{author}{{Naiman} J}, \bibinfo{author}{{Pakmor} R},
  \bibinfo{author}{{Hernquist} L}, \bibinfo{author}{{Torrey} P},
  \bibinfo{author}{{Vogelsberger} M}, \bibinfo{author}{{Weinberger} R} and
  \bibinfo{author}{{Marinacci} F} (\bibinfo{year}{2018}), \bibinfo{month}{Jan.}
\bibinfo{title}{{Simulating galaxy formation with the IllustrisTNG model}}.
\bibinfo{journal}{{\em \mnras}} \bibinfo{volume}{473} (\bibinfo{number}{3}):
  \bibinfo{pages}{4077--4106}. \bibinfo{doi}{\doi{10.1093/mnras/stx2656}}.
\eprint{1703.02970}.

\bibtype{Article}%
\bibitem[{Pintos-Castro} et al.(2019)]{Pintos-Castro_etal_2019}
\bibinfo{author}{{Pintos-Castro} I}, \bibinfo{author}{{Yee} HKC},
  \bibinfo{author}{{Muzzin} A}, \bibinfo{author}{{Old} L} and
  \bibinfo{author}{{Wilson} G} (\bibinfo{year}{2019}), \bibinfo{month}{May}.
\bibinfo{title}{{The Evolution of the Quenching of Star Formation in Cluster
  Galaxies since z {\ensuremath{\sim}} 1}}.
\bibinfo{journal}{{\em \apj}} \bibinfo{volume}{876} (\bibinfo{number}{1}),
  \bibinfo{eid}{40}. \bibinfo{doi}{\doi{10.3847/1538-4357/ab14ee}}.
\eprint{1904.00023}.

\bibtype{Article}%
\bibitem[{Poggianti} et al.(1999)]{Poggianti_etal_1999}
\bibinfo{author}{{Poggianti} BM}, \bibinfo{author}{{Smail} I},
  \bibinfo{author}{{Dressler} A}, \bibinfo{author}{{Couch} WJ},
  \bibinfo{author}{{Barger} AJ}, \bibinfo{author}{{Butcher} H},
  \bibinfo{author}{{Ellis} RS} and  \bibinfo{author}{{Oemler} Augustus J}
  (\bibinfo{year}{1999}), \bibinfo{month}{Jun.}
\bibinfo{title}{{The Star Formation Histories of Galaxies in Distant
  Clusters}}.
\bibinfo{journal}{{\em \apj}} \bibinfo{volume}{518} (\bibinfo{number}{2}):
  \bibinfo{pages}{576--593}. \bibinfo{doi}{\doi{10.1086/307322}}.
\eprint{astro-ph/9901264}.

\bibtype{Article}%
\bibitem[{Poggianti} et al.(2006)]{Poggianti_etal_2006}
\bibinfo{author}{{Poggianti} BM}, \bibinfo{author}{{von der Linden} A},
  \bibinfo{author}{{De Lucia} G}, \bibinfo{author}{{Desai} V},
  \bibinfo{author}{{Simard} L}, \bibinfo{author}{{Halliday} C},
  \bibinfo{author}{{Arag{\'o}n-Salamanca} A}, \bibinfo{author}{{Bower} R},
  \bibinfo{author}{{Varela} J}, \bibinfo{author}{{Best} P},
  \bibinfo{author}{{Clowe} DI}, \bibinfo{author}{{Dalcanton} J},
  \bibinfo{author}{{Jablonka} P}, \bibinfo{author}{{Milvang-Jensen} B},
  \bibinfo{author}{{Pello} R}, \bibinfo{author}{{Rudnick} G},
  \bibinfo{author}{{Saglia} R}, \bibinfo{author}{{White} SDM} and
  \bibinfo{author}{{Zaritsky} D} (\bibinfo{year}{2006}), \bibinfo{month}{May}.
\bibinfo{title}{{The Evolution of the Star Formation Activity in Galaxies and
  Its Dependence on Environment}}.
\bibinfo{journal}{{\em \apj}} \bibinfo{volume}{642} (\bibinfo{number}{1}):
  \bibinfo{pages}{188--215}. \bibinfo{doi}{\doi{10.1086/500666}}.
\eprint{astro-ph/0512391}.

\bibtype{Article}%
\bibitem[{Poggianti} et al.(2017)]{Poggianti17}
\bibinfo{author}{{Poggianti} BM}, \bibinfo{author}{{Moretti} A},
  \bibinfo{author}{{Gullieuszik} M}, \bibinfo{author}{{Fritz} J},
  \bibinfo{author}{{Jaff{\'e}} Y}, \bibinfo{author}{{Bettoni} D},
  \bibinfo{author}{{Fasano} G}, \bibinfo{author}{{Bellhouse} C},
  \bibinfo{author}{{Hau} G}, \bibinfo{author}{{Vulcani} B},
  \bibinfo{author}{{Biviano} A}, \bibinfo{author}{{Omizzolo} A},
  \bibinfo{author}{{Paccagnella} A}, \bibinfo{author}{{D'Onofrio} M},
  \bibinfo{author}{{Cava} A}, \bibinfo{author}{{Sheen} YK},
  \bibinfo{author}{{Couch} W} and  \bibinfo{author}{{Owers} M}
  (\bibinfo{year}{2017}), \bibinfo{month}{Jul.}
\bibinfo{title}{{GASP. I. Gas Stripping Phenomena in Galaxies with MUSE}}.
\bibinfo{journal}{{\em \apj}} \bibinfo{volume}{844} (\bibinfo{number}{1}),
  \bibinfo{eid}{48}. \bibinfo{doi}{\doi{10.3847/1538-4357/aa78ed}}.
\eprint{1704.05086}.

\bibtype{Article}%
\bibitem[{Popesso} et al.(2023)]{Popesso_etal_2023}
\bibinfo{author}{{Popesso} P}, \bibinfo{author}{{Concas} A},
  \bibinfo{author}{{Cresci} G}, \bibinfo{author}{{Belli} S},
  \bibinfo{author}{{Rodighiero} G}, \bibinfo{author}{{Inami} H},
  \bibinfo{author}{{Dickinson} M}, \bibinfo{author}{{Ilbert} O},
  \bibinfo{author}{{Pannella} M} and  \bibinfo{author}{{Elbaz} D}
  (\bibinfo{year}{2023}), \bibinfo{month}{Feb.}
\bibinfo{title}{{The main sequence of star-forming galaxies across cosmic
  times}}.
\bibinfo{journal}{{\em \mnras}} \bibinfo{volume}{519} (\bibinfo{number}{1}):
  \bibinfo{pages}{1526--1544}. \bibinfo{doi}{\doi{10.1093/mnras/stac3214}}.
\eprint{2203.10487}.

\bibtype{Article}%
\bibitem[{Popping} et al.(2017)]{Popping_etal_2017}
\bibinfo{author}{{Popping} G}, \bibinfo{author}{{Somerville} RS} and
  \bibinfo{author}{{Galametz} M} (\bibinfo{year}{2017}), \bibinfo{month}{Nov.}
\bibinfo{title}{{The dust content of galaxies from z = 0 to z = 9}}.
\bibinfo{journal}{{\em \mnras}} \bibinfo{volume}{471} (\bibinfo{number}{3}):
  \bibinfo{pages}{3152--3185}. \bibinfo{doi}{\doi{10.1093/mnras/stx1545}}.
\eprint{1609.08622}.

\bibtype{Article}%
\bibitem[{Postman} et al.(2005)]{Postman_etal_2005}
\bibinfo{author}{{Postman} M}, \bibinfo{author}{{Franx} M},
  \bibinfo{author}{{Cross} NJG}, \bibinfo{author}{{Holden} B},
  \bibinfo{author}{{Ford} HC}, \bibinfo{author}{{Illingworth} GD},
  \bibinfo{author}{{Goto} T}, \bibinfo{author}{{Demarco} R},
  \bibinfo{author}{{Rosati} P}, \bibinfo{author}{{Blakeslee} JP},
  \bibinfo{author}{{Tran} KV}, \bibinfo{author}{{Ben{\'\i}tez} N},
  \bibinfo{author}{{Clampin} M}, \bibinfo{author}{{Hartig} GF},
  \bibinfo{author}{{Homeier} N}, \bibinfo{author}{{Ardila} DR},
  \bibinfo{author}{{Bartko} F}, \bibinfo{author}{{Bouwens} RJ},
  \bibinfo{author}{{Bradley} LD}, \bibinfo{author}{{Broadhurst} TJ},
  \bibinfo{author}{{Brown} RA}, \bibinfo{author}{{Burrows} CJ},
  \bibinfo{author}{{Cheng} ES}, \bibinfo{author}{{Feldman} PD},
  \bibinfo{author}{{Golimowski} DA}, \bibinfo{author}{{Gronwall} C},
  \bibinfo{author}{{Infante} L}, \bibinfo{author}{{Kimble} RA},
  \bibinfo{author}{{Krist} JE}, \bibinfo{author}{{Lesser} MP},
  \bibinfo{author}{{Martel} AR}, \bibinfo{author}{{Mei} S},
  \bibinfo{author}{{Menanteau} F}, \bibinfo{author}{{Meurer} GR},
  \bibinfo{author}{{Miley} GK}, \bibinfo{author}{{Motta} V},
  \bibinfo{author}{{Sirianni} M}, \bibinfo{author}{{Sparks} WB},
  \bibinfo{author}{{Tran} HD}, \bibinfo{author}{{Tsvetanov} ZI},
  \bibinfo{author}{{White} RL} and  \bibinfo{author}{{Zheng} W}
  (\bibinfo{year}{2005}), \bibinfo{month}{Apr.}
\bibinfo{title}{{The Morphology-Density Relation in z
  \raisebox{-0.5ex}\textasciitilde 1 Clusters}}.
\bibinfo{journal}{{\em \apj}} \bibinfo{volume}{623} (\bibinfo{number}{2}):
  \bibinfo{pages}{721--741}. \bibinfo{doi}{\doi{10.1086/428881}}.
\eprint{astro-ph/0501224}.

\bibtype{Article}%
\bibitem[{Rees} and {Ostriker}(1977)]{Rees_and_Ostriker_1977}
\bibinfo{author}{{Rees} MJ} and  \bibinfo{author}{{Ostriker} JP}
  (\bibinfo{year}{1977}), \bibinfo{month}{Jun.}
\bibinfo{title}{{Cooling, dynamics and fragmentation of massive gas clouds:
  clues to the masses and radii of galaxies and clusters.}}
\bibinfo{journal}{{\em \mnras}} \bibinfo{volume}{179}:
  \bibinfo{pages}{541--559}. \bibinfo{doi}{\doi{10.1093/mnras/179.4.541}}.

\bibtype{Article}%
\bibitem[{Roediger} and {Br{\"u}ggen}(2006)]{Roediger_Bruggen_2006}
\bibinfo{author}{{Roediger} E} and  \bibinfo{author}{{Br{\"u}ggen} M}
  (\bibinfo{year}{2006}), \bibinfo{month}{Jun.}
\bibinfo{title}{{Ram pressure stripping of disc galaxies: the role of the
  inclination angle}}.
\bibinfo{journal}{{\em \mnras}} \bibinfo{volume}{369} (\bibinfo{number}{2}):
  \bibinfo{pages}{567--580}.
  \bibinfo{doi}{\doi{10.1111/j.1365-2966.2006.10335.x}}.
\eprint{astro-ph/0512365}.

\bibtype{Article}%
\bibitem[{Saintonge} and {Catinella}(2022)]{Saintonge_Catinella_2022}
\bibinfo{author}{{Saintonge} A} and  \bibinfo{author}{{Catinella} B}
  (\bibinfo{year}{2022}), \bibinfo{month}{Aug.}
\bibinfo{title}{{The Cold Interstellar Medium of Galaxies in the Local
  Universe}}.
\bibinfo{journal}{{\em \araa}} \bibinfo{volume}{60}: \bibinfo{pages}{319--361}.
  \bibinfo{doi}{\doi{10.1146/annurev-astro-021022-043545}}.
\eprint{2202.00690}.

\bibtype{Article}%
\bibitem[{Schaye} et al.(2015)]{Schaye_etal_2015}
\bibinfo{author}{{Schaye} J}, \bibinfo{author}{{Crain} RA},
  \bibinfo{author}{{Bower} RG}, \bibinfo{author}{{Furlong} M},
  \bibinfo{author}{{Schaller} M}, \bibinfo{author}{{Theuns} T},
  \bibinfo{author}{{Dalla Vecchia} C}, \bibinfo{author}{{Frenk} CS},
  \bibinfo{author}{{McCarthy} IG}, \bibinfo{author}{{Helly} JC},
  \bibinfo{author}{{Jenkins} A}, \bibinfo{author}{{Rosas-Guevara} YM},
  \bibinfo{author}{{White} SDM}, \bibinfo{author}{{Baes} M},
  \bibinfo{author}{{Booth} CM}, \bibinfo{author}{{Camps} P},
  \bibinfo{author}{{Navarro} JF}, \bibinfo{author}{{Qu} Y},
  \bibinfo{author}{{Rahmati} A}, \bibinfo{author}{{Sawala} T},
  \bibinfo{author}{{Thomas} PA} and  \bibinfo{author}{{Trayford} J}
  (\bibinfo{year}{2015}), \bibinfo{month}{Jan.}
\bibinfo{title}{{The EAGLE project: simulating the evolution and assembly of
  galaxies and their environments}}.
\bibinfo{journal}{{\em \mnras}} \bibinfo{volume}{446} (\bibinfo{number}{1}):
  \bibinfo{pages}{521--554}. \bibinfo{doi}{\doi{10.1093/mnras/stu2058}}.
\eprint{1407.7040}.

\bibtype{Article}%
\bibitem[{Schaye} et al.(2023)]{Schaye_etal_2023}
\bibinfo{author}{{Schaye} J}, \bibinfo{author}{{Kugel} R},
  \bibinfo{author}{{Schaller} M}, \bibinfo{author}{{Helly} JC},
  \bibinfo{author}{{Braspenning} J}, \bibinfo{author}{{Elbers} W},
  \bibinfo{author}{{McCarthy} IG}, \bibinfo{author}{{van Daalen} MP},
  \bibinfo{author}{{Vandenbroucke} B}, \bibinfo{author}{{Frenk} CS},
  \bibinfo{author}{{Kwan} J}, \bibinfo{author}{{Salcido} J},
  \bibinfo{author}{{Bah{\'e}} YM}, \bibinfo{author}{{Borrow} J},
  \bibinfo{author}{{Chaikin} E}, \bibinfo{author}{{Hahn} O},
  \bibinfo{author}{{Hu{\v{s}}ko} F}, \bibinfo{author}{{Jenkins} A},
  \bibinfo{author}{{Lacey} CG} and  \bibinfo{author}{{Nobels} FSJ}
  (\bibinfo{year}{2023}), \bibinfo{month}{Dec.}
\bibinfo{title}{{The FLAMINGO project: cosmological hydrodynamical simulations
  for large-scale structure and galaxy cluster surveys}}.
\bibinfo{journal}{{\em \mnras}} \bibinfo{volume}{526} (\bibinfo{number}{4}):
  \bibinfo{pages}{4978--5020}. \bibinfo{doi}{\doi{10.1093/mnras/stad2419}}.
\eprint{2306.04024}.

\bibtype{Article}%
\bibitem[{Schreiber} et al.(2018)]{Schreiber_etal_2018}
\bibinfo{author}{{Schreiber} C}, \bibinfo{author}{{Glazebrook} K},
  \bibinfo{author}{{Nanayakkara} T}, \bibinfo{author}{{Kacprzak} GG},
  \bibinfo{author}{{Labb{\'e}} I}, \bibinfo{author}{{Oesch} P},
  \bibinfo{author}{{Yuan} T}, \bibinfo{author}{{Tran} KV},
  \bibinfo{author}{{Papovich} C}, \bibinfo{author}{{Spitler} L} and
  \bibinfo{author}{{Straatman} C} (\bibinfo{year}{2018}), \bibinfo{month}{Oct.}
\bibinfo{title}{{Near infrared spectroscopy and star-formation histories of 3
  {\ensuremath{\leq}} z {\ensuremath{\leq}} 4 quiescent galaxies}}.
\bibinfo{journal}{{\em \aap}} \bibinfo{volume}{618}, \bibinfo{eid}{A85}.
  \bibinfo{doi}{\doi{10.1051/0004-6361/201833070}}.
\eprint{1807.02523}.

\bibtype{Article}%
\bibitem[{Scoville} et al.(2007)]{Scoville_etal_2007}
\bibinfo{author}{{Scoville} N}, \bibinfo{author}{{Abraham} RG},
  \bibinfo{author}{{Aussel} H}, \bibinfo{author}{{Barnes} JE},
  \bibinfo{author}{{Benson} A}, \bibinfo{author}{{Blain} AW},
  \bibinfo{author}{{Calzetti} D}, \bibinfo{author}{{Comastri} A},
  \bibinfo{author}{{Capak} P}, \bibinfo{author}{{Carilli} C},
  \bibinfo{author}{{Carlstrom} JE}, \bibinfo{author}{{Carollo} CM},
  \bibinfo{author}{{Colbert} J}, \bibinfo{author}{{Daddi} E},
  \bibinfo{author}{{Ellis} RS}, \bibinfo{author}{{Elvis} M},
  \bibinfo{author}{{Ewald} SP}, \bibinfo{author}{{Fall} M},
  \bibinfo{author}{{Franceschini} A}, \bibinfo{author}{{Giavalisco} M},
  \bibinfo{author}{{Green} W}, \bibinfo{author}{{Griffiths} RE},
  \bibinfo{author}{{Guzzo} L}, \bibinfo{author}{{Hasinger} G},
  \bibinfo{author}{{Impey} C}, \bibinfo{author}{{Kneib} JP},
  \bibinfo{author}{{Koda} J}, \bibinfo{author}{{Koekemoer} A},
  \bibinfo{author}{{Lefevre} O}, \bibinfo{author}{{Lilly} S},
  \bibinfo{author}{{Liu} CT}, \bibinfo{author}{{McCracken} HJ},
  \bibinfo{author}{{Massey} R}, \bibinfo{author}{{Mellier} Y},
  \bibinfo{author}{{Miyazaki} S}, \bibinfo{author}{{Mobasher} B},
  \bibinfo{author}{{Mould} J}, \bibinfo{author}{{Norman} C},
  \bibinfo{author}{{Refregier} A}, \bibinfo{author}{{Renzini} A},
  \bibinfo{author}{{Rhodes} J}, \bibinfo{author}{{Rich} M},
  \bibinfo{author}{{Sanders} DB}, \bibinfo{author}{{Schiminovich} D},
  \bibinfo{author}{{Schinnerer} E}, \bibinfo{author}{{Scodeggio} M},
  \bibinfo{author}{{Sheth} K}, \bibinfo{author}{{Shopbell} PL},
  \bibinfo{author}{{Taniguchi} Y}, \bibinfo{author}{{Tyson} ND},
  \bibinfo{author}{{Urry} CM}, \bibinfo{author}{{Van Waerbeke} L},
  \bibinfo{author}{{Vettolani} P}, \bibinfo{author}{{White} SDM} and
  \bibinfo{author}{{Yan} L} (\bibinfo{year}{2007}), \bibinfo{month}{Sep.}
\bibinfo{title}{{COSMOS: Hubble Space Telescope Observations}}.
\bibinfo{journal}{{\em \apjs}} \bibinfo{volume}{172} (\bibinfo{number}{1}):
  \bibinfo{pages}{38--45}. \bibinfo{doi}{\doi{10.1086/516580}}.
\eprint{astro-ph/0612306}.

\bibtype{Article}%
\bibitem[{Setton} et al.(2024)]{Setton_etal_2024}
\bibinfo{author}{{Setton} DJ}, \bibinfo{author}{{Khullar} G},
  \bibinfo{author}{{Miller} TB}, \bibinfo{author}{{Bezanson} R},
  \bibinfo{author}{{Greene} JE}, \bibinfo{author}{{Suess} KA},
  \bibinfo{author}{{Whitaker} KE}, \bibinfo{author}{{Antwi-Danso} J},
  \bibinfo{author}{{Atek} H}, \bibinfo{author}{{Brammer} G},
  \bibinfo{author}{{Cutler} SE}, \bibinfo{author}{{Dayal} P},
  \bibinfo{author}{{Feldmann} R}, \bibinfo{author}{{Furtak} LJ},
  \bibinfo{author}{{Fujimoto} S}, \bibinfo{author}{{Glazebrook} K},
  \bibinfo{author}{{Goulding} AD}, \bibinfo{author}{{Kokorev} V},
  \bibinfo{author}{{Labbe} I}, \bibinfo{author}{{Leja} J},
  \bibinfo{author}{{Ma} Y}, \bibinfo{author}{{Marchesini} D},
  \bibinfo{author}{{Nanayakkara} T}, \bibinfo{author}{{Pan} R},
  \bibinfo{author}{{Price} SH}, \bibinfo{author}{{Siegel} JC},
  \bibinfo{author}{{Shipley} H}, \bibinfo{author}{{Weaver} JR},
  \bibinfo{author}{{van Dokkum} P}, \bibinfo{author}{{Wang} B} and
  \bibinfo{author}{{Williams} CC} (\bibinfo{year}{2024}), \bibinfo{month}{Feb.}
\bibinfo{title}{{UNCOVER NIRSpec/PRISM Spectroscopy Unveils Evidence of Early
  Core Formation in a Massive, Centrally Dusty Quiescent Galaxy at
  $z_{spec}=3.97$}}.
\bibinfo{journal}{{\em arXiv e-prints}} ,
  \bibinfo{eid}{arXiv:2402.05664}\bibinfo{doi}{\doi{10.48550/arXiv.2402.05664}}.
\eprint{2402.05664}.

\bibtype{Article}%
\bibitem[{Shankar} et al.(2020)]{Shankar_etal_2020}
\bibinfo{author}{{Shankar} F}, \bibinfo{author}{{Weinberg} DH},
  \bibinfo{author}{{Marsden} C}, \bibinfo{author}{{Grylls} PJ},
  \bibinfo{author}{{Bernardi} M}, \bibinfo{author}{{Yang} G},
  \bibinfo{author}{{Moster} B}, \bibinfo{author}{{Fu} H},
  \bibinfo{author}{{Carraro} R}, \bibinfo{author}{{Alexander} DM},
  \bibinfo{author}{{Allevato} V}, \bibinfo{author}{{Ananna} TT},
  \bibinfo{author}{{Bongiorno} A}, \bibinfo{author}{{Calderone} G},
  \bibinfo{author}{{Civano} F}, \bibinfo{author}{{Daddi} E},
  \bibinfo{author}{{Delvecchio} I}, \bibinfo{author}{{Duras} F},
  \bibinfo{author}{{La Franca} F}, \bibinfo{author}{{Lapi} A},
  \bibinfo{author}{{Lu} Y}, \bibinfo{author}{{Menci} N},
  \bibinfo{author}{{Mezcua} M}, \bibinfo{author}{{Ricci} F},
  \bibinfo{author}{{Rodighiero} G}, \bibinfo{author}{{Sheth} RK},
  \bibinfo{author}{{Suh} H}, \bibinfo{author}{{Villforth} C} and
  \bibinfo{author}{{Zanisi} L} (\bibinfo{year}{2020}), \bibinfo{month}{Mar.}
\bibinfo{title}{{Probing black hole accretion tracks, scaling relations, and
  radiative efficiencies from stacked X-ray active galactic nuclei}}.
\bibinfo{journal}{{\em \mnras}} \bibinfo{volume}{493} (\bibinfo{number}{1}):
  \bibinfo{pages}{1500--1511}. \bibinfo{doi}{\doi{10.1093/mnras/stz3522}}.
\eprint{1912.06153}.

\bibtype{Article}%
\bibitem[{Shin} et al.(2019)]{Shin_etal_2019}
\bibinfo{author}{{Shin} J}, \bibinfo{author}{{Woo} JH},
  \bibinfo{author}{{Chung} A}, \bibinfo{author}{{Baek} J},
  \bibinfo{author}{{Cho} K}, \bibinfo{author}{{Kang} D} and
  \bibinfo{author}{{Bae} HJ} (\bibinfo{year}{2019}), \bibinfo{month}{Aug.}
\bibinfo{title}{{Positive and Negative Feedback of AGN Outflows in NGC 5728}}.
\bibinfo{journal}{{\em \apj}} \bibinfo{volume}{881} (\bibinfo{number}{2}),
  \bibinfo{eid}{147}. \bibinfo{doi}{\doi{10.3847/1538-4357/ab2e72}}.
\eprint{1907.00982}.

\bibtype{Article}%
\bibitem[{Silk} and {Rees}(1998)]{Silk_and_Rees_1998}
\bibinfo{author}{{Silk} J} and  \bibinfo{author}{{Rees} MJ}
  (\bibinfo{year}{1998}), \bibinfo{month}{Mar}.
\bibinfo{title}{{Quasars and galaxy formation}}.
\bibinfo{journal}{{\em \aap}} \bibinfo{volume}{331}: \bibinfo{pages}{L1--L4}.
\eprint{astro-ph/9801013}.

\bibtype{Article}%
\bibitem[{Snyder} et al.(2011)]{Snyder_etal_2011}
\bibinfo{author}{{Snyder} GF}, \bibinfo{author}{{Cox} TJ},
  \bibinfo{author}{{Hayward} CC}, \bibinfo{author}{{Hernquist} L} and
  \bibinfo{author}{{Jonsson} P} (\bibinfo{year}{2011}), \bibinfo{month}{Nov.}
\bibinfo{title}{{K+A Galaxies as the Aftermath of Gas-rich Mergers: Simulating
  the Evolution of Galaxies as Seen by Spectroscopic Surveys}}.
\bibinfo{journal}{{\em \apj}} \bibinfo{volume}{741} (\bibinfo{number}{2}),
  \bibinfo{eid}{77}. \bibinfo{doi}{\doi{10.1088/0004-637X/741/2/77}}.
\eprint{1102.3689}.

\bibtype{Article}%
\bibitem[{Somerville} and {Dav{\'e}}(2015)]{Somerville_and_Dave_2015}
\bibinfo{author}{{Somerville} RS} and  \bibinfo{author}{{Dav{\'e}} R}
  (\bibinfo{year}{2015}), \bibinfo{month}{Aug.}
\bibinfo{title}{{Physical Models of Galaxy Formation in a Cosmological
  Framework}}.
\bibinfo{journal}{{\em \araa}} \bibinfo{volume}{53}: \bibinfo{pages}{51--113}.
  \bibinfo{doi}{\doi{10.1146/annurev-astro-082812-140951}}.
\eprint{1412.2712}.

\bibtype{Article}%
\bibitem[{Somerville} et al.(2015)]{Somerville_etal_2015}
\bibinfo{author}{{Somerville} RS}, \bibinfo{author}{{Popping} G} and
  \bibinfo{author}{{Trager} SC} (\bibinfo{year}{2015}), \bibinfo{month}{Nov.}
\bibinfo{title}{{Star formation in semi-analytic galaxy formation models with
  multiphase gas}}.
\bibinfo{journal}{{\em \mnras}} \bibinfo{volume}{453} (\bibinfo{number}{4}):
  \bibinfo{pages}{4337--4367}. \bibinfo{doi}{\doi{10.1093/mnras/stv1877}}.
\eprint{1503.00755}.

\bibtype{Article}%
\bibitem[{Springel} et al.(2005)]{Springel_etal_2005}
\bibinfo{author}{{Springel} V}, \bibinfo{author}{{Di Matteo} T} and
  \bibinfo{author}{{Hernquist} L} (\bibinfo{year}{2005}), \bibinfo{month}{Aug.}
\bibinfo{title}{{Modelling feedback from stars and black holes in galaxy
  mergers}}.
\bibinfo{journal}{{\em \mnras}} \bibinfo{volume}{361}:
  \bibinfo{pages}{776--794}.
  \bibinfo{doi}{\doi{10.1111/j.1365-2966.2005.09238.x}}.
\eprint{arXiv:astro-ph/0411108}.

\bibtype{Article}%
\bibitem[{Steinhauser} et al.(2016)]{Steinhauser_etal_2016}
\bibinfo{author}{{Steinhauser} D}, \bibinfo{author}{{Schindler} S} and
  \bibinfo{author}{{Springel} V} (\bibinfo{year}{2016}), \bibinfo{month}{Jun.}
\bibinfo{title}{{Simulations of ram-pressure stripping in galaxy-cluster
  interactions}}.
\bibinfo{journal}{{\em \aap}} \bibinfo{volume}{591}, \bibinfo{eid}{A51}.
  \bibinfo{doi}{\doi{10.1051/0004-6361/201527705}}.
\eprint{1604.05193}.

\bibtype{Article}%
\bibitem[{Stevens} and {Brown}(2017)]{Stevens_etal_2017}
\bibinfo{author}{{Stevens} ARH} and  \bibinfo{author}{{Brown} T}
  (\bibinfo{year}{2017}), \bibinfo{month}{Oct.}
\bibinfo{title}{{Physical drivers of galaxies' cold-gas content: exploring
  environmental and evolutionary effects with Dark Sage}}.
\bibinfo{journal}{{\em \mnras}} \bibinfo{volume}{471} (\bibinfo{number}{1}):
  \bibinfo{pages}{447--462}. \bibinfo{doi}{\doi{10.1093/mnras/stx1596}}.
\eprint{1706.07434}.

\bibtype{Article}%
\bibitem[{Stevens} et al.(2016)]{Stevens_etal_2016}
\bibinfo{author}{{Stevens} ARH}, \bibinfo{author}{{Croton} DJ} and
  \bibinfo{author}{{Mutch} SJ} (\bibinfo{year}{2016}), \bibinfo{month}{Sep.}
\bibinfo{title}{{Building disc structure and galaxy properties through angular
  momentum: the DARK SAGE semi-analytic model}}.
\bibinfo{journal}{{\em \mnras}} \bibinfo{volume}{461} (\bibinfo{number}{1}):
  \bibinfo{pages}{859--876}. \bibinfo{doi}{\doi{10.1093/mnras/stw1332}}.
\eprint{1605.00647}.

\bibtype{Article}%
\bibitem[{Strazzullo} et al.(2010)]{Strazzullo_etal_2010}
\bibinfo{author}{{Strazzullo} V}, \bibinfo{author}{{Rosati} P},
  \bibinfo{author}{{Pannella} M}, \bibinfo{author}{{Gobat} R},
  \bibinfo{author}{{Santos} JS}, \bibinfo{author}{{Nonino} M},
  \bibinfo{author}{{Demarco} R}, \bibinfo{author}{{Lidman} C},
  \bibinfo{author}{{Tanaka} M}, \bibinfo{author}{{Mullis} CR},
  \bibinfo{author}{{Nu{\~n}ez} C}, \bibinfo{author}{{Rettura} A},
  \bibinfo{author}{{Jee} MJ}, \bibinfo{author}{{B{\"o}hringer} H},
  \bibinfo{author}{{Bender} R}, \bibinfo{author}{{Bouwens} RJ},
  \bibinfo{author}{{Dawson} K}, \bibinfo{author}{{Fassbender} R},
  \bibinfo{author}{{Franx} M}, \bibinfo{author}{{Perlmutter} S} and
  \bibinfo{author}{{Postman} M} (\bibinfo{year}{2010}), \bibinfo{month}{Dec.}
\bibinfo{title}{{Cluster galaxies in XMMU J2235-2557: galaxy population
  properties in most massive environments at z \raisebox{-0.5ex}\textasciitilde
  1.4}}.
\bibinfo{journal}{{\em \aap}} \bibinfo{volume}{524}, \bibinfo{eid}{A17}.
  \bibinfo{doi}{\doi{10.1051/0004-6361/201015251}}.
\eprint{1009.1423}.

\bibtype{Article}%
\bibitem[{Strazzullo} et al.(2019)]{Strazzullo_etal_2019}
\bibinfo{author}{{Strazzullo} V}, \bibinfo{author}{{Pannella} M},
  \bibinfo{author}{{Mohr} JJ}, \bibinfo{author}{{Saro} A},
  \bibinfo{author}{{Ashby} MLN}, \bibinfo{author}{{Bayliss} MB},
  \bibinfo{author}{{Bocquet} S}, \bibinfo{author}{{Bulbul} E},
  \bibinfo{author}{{Khullar} G}, \bibinfo{author}{{Mantz} AB},
  \bibinfo{author}{{Stanford} SA}, \bibinfo{author}{{Benson} BA},
  \bibinfo{author}{{Bleem} LE}, \bibinfo{author}{{Brodwin} M},
  \bibinfo{author}{{Canning} REA}, \bibinfo{author}{{Capasso} R},
  \bibinfo{author}{{Chiu} I}, \bibinfo{author}{{Gonzalez} AH},
  \bibinfo{author}{{Gupta} N}, \bibinfo{author}{{Hlavacek-Larrondo} J},
  \bibinfo{author}{{Klein} M}, \bibinfo{author}{{McDonald} M},
  \bibinfo{author}{{Noordeh} E}, \bibinfo{author}{{Rapetti} D},
  \bibinfo{author}{{Reichardt} CL}, \bibinfo{author}{{Schrabback} T},
  \bibinfo{author}{{Sharon} K} and  \bibinfo{author}{{Stalder} B}
  (\bibinfo{year}{2019}), \bibinfo{month}{Feb.}
\bibinfo{title}{{Galaxy populations in the most distant SPT-SZ clusters. I.
  Environmental quenching in massive clusters at 1.4 {\ensuremath{\lesssim}} z
  {\ensuremath{\lesssim}} 1.7}}.
\bibinfo{journal}{{\em \aap}} \bibinfo{volume}{622}, \bibinfo{eid}{A117}.
  \bibinfo{doi}{\doi{10.1051/0004-6361/201833944}}.
\eprint{1807.09768}.

\bibtype{Article}%
\bibitem[{Su} et al.(2019)]{Su_etal_2019}
\bibinfo{author}{{Su} KY}, \bibinfo{author}{{Hopkins} PF},
  \bibinfo{author}{{Hayward} CC}, \bibinfo{author}{{Ma} X},
  \bibinfo{author}{{Faucher-Gigu{\`e}re} CA}, \bibinfo{author}{{Kere{\v{s}}}
  D}, \bibinfo{author}{{Orr} ME}, \bibinfo{author}{{Chan} TK} and
  \bibinfo{author}{{Robles} VH} (\bibinfo{year}{2019}), \bibinfo{month}{Aug.}
\bibinfo{title}{{The failure of stellar feedback, magnetic fields, conduction,
  and morphological quenching in maintaining red galaxies}}.
\bibinfo{journal}{{\em \mnras}} \bibinfo{volume}{487} (\bibinfo{number}{3}):
  \bibinfo{pages}{4393--4408}. \bibinfo{doi}{\doi{10.1093/mnras/stz1494}}.
\eprint{1809.09120}.

\bibtype{Article}%
\bibitem[{Szpila} et al.(2024)]{Szpila_etal_2024}
\bibinfo{author}{{Szpila} J}, \bibinfo{author}{{Dav{\'e}} R},
  \bibinfo{author}{{Rennehan} D}, \bibinfo{author}{{Cui} W} and
  \bibinfo{author}{{Hough} R} (\bibinfo{year}{2024}), \bibinfo{month}{Feb.}
\bibinfo{title}{{The Nature and Evolution of Early Massive Quenched Galaxies in
  the Simba-C Simulation}}.
\bibinfo{journal}{{\em arXiv e-prints}} ,
  \bibinfo{eid}{arXiv:2402.08729}\bibinfo{doi}{\doi{10.48550/arXiv.2402.08729}}.
\eprint{2402.08729}.

\bibtype{Article}%
\bibitem[{Talbot} et al.(2022)]{Talbot22}
\bibinfo{author}{{Talbot} RY}, \bibinfo{author}{{Sijacki} D} and
  \bibinfo{author}{{Bourne} MA} (\bibinfo{year}{2022}), \bibinfo{month}{Aug.}
\bibinfo{title}{{Blandford-Znajek jets in galaxy formation simulations:
  exploring the diversity of outflows produced by spin-driven AGN jets in
  Seyfert galaxies}}.
\bibinfo{journal}{{\em \mnras}} \bibinfo{volume}{514} (\bibinfo{number}{3}):
  \bibinfo{pages}{4535--4559}. \bibinfo{doi}{\doi{10.1093/mnras/stac1566}}.
\eprint{2111.01801}.

\bibtype{Article}%
\bibitem[{The MSE Science Team}(2019)]{MSE_2019}
\bibinfo{author}{{The MSE Science Team}} (\bibinfo{year}{2019}),
  \bibinfo{month}{Apr.}
\bibinfo{title}{{The Detailed Science Case for the Maunakea Spectroscopic
  Explorer, 2019 edition}}.
\bibinfo{journal}{{\em arXiv e-prints}} ,
  \bibinfo{eid}{arXiv:1904.04907}\bibinfo{doi}{\doi{10.48550/arXiv.1904.04907}}.
\eprint{1904.04907}.

\bibtype{Article}%
\bibitem[{Tonnesen}(2019)]{Tonnesen_2019}
\bibinfo{author}{{Tonnesen} S} (\bibinfo{year}{2019}), \bibinfo{month}{Apr.}
\bibinfo{title}{{The Journey Counts: The Importance of Including Orbits when
  Simulating Ram Pressure Stripping}}.
\bibinfo{journal}{{\em \apj}} \bibinfo{volume}{874} (\bibinfo{number}{2}),
  \bibinfo{eid}{161}. \bibinfo{doi}{\doi{10.3847/1538-4357/ab0960}}.
\eprint{1903.08178}.

\bibtype{Article}%
\bibitem[{Tran} et al.(2015)]{Tran_etal_2015}
\bibinfo{author}{{Tran} KVH}, \bibinfo{author}{{Nanayakkara} T},
  \bibinfo{author}{{Yuan} T}, \bibinfo{author}{{Kacprzak} GG},
  \bibinfo{author}{{Glazebrook} K}, \bibinfo{author}{{Kewley} LJ},
  \bibinfo{author}{{Momcheva} I}, \bibinfo{author}{{Papovich} CJ},
  \bibinfo{author}{{Quadri} R}, \bibinfo{author}{{Rudnick} G},
  \bibinfo{author}{{Saintonge} A}, \bibinfo{author}{{Spitler} LR},
  \bibinfo{author}{{Straatman} C} and  \bibinfo{author}{{Tomczak} A}
  (\bibinfo{year}{2015}), \bibinfo{month}{Sep.}
\bibinfo{title}{{ZFIRE: Galaxy Cluster Kinematics, H alpha Star Formation
  Rates, and Gas Phase Metallicities of XMM-LSS J02182-05102 at zcl = 1.6232}}.
\bibinfo{journal}{{\em \apj}} \bibinfo{volume}{811} (\bibinfo{number}{1}),
  \bibinfo{eid}{28}. \bibinfo{doi}{\doi{10.1088/0004-637X/811/1/28}}.
\eprint{1508.03057}.

\bibtype{Article}%
\bibitem[{Triani} et al.(2021)]{Triani_etal_2021}
\bibinfo{author}{{Triani} DP}, \bibinfo{author}{{Sinha} M},
  \bibinfo{author}{{Croton} DJ}, \bibinfo{author}{{Dwek} E} and
  \bibinfo{author}{{Pacifici} C} (\bibinfo{year}{2021}), \bibinfo{month}{May}.
\bibinfo{title}{{Exploring the relation between dust mass and galaxy properties
  using Dusty SAGE}}.
\bibinfo{journal}{{\em \mnras}} \bibinfo{volume}{503} (\bibinfo{number}{1}):
  \bibinfo{pages}{1005--1016}. \bibinfo{doi}{\doi{10.1093/mnras/stab558}}.
\eprint{2102.12652}.

\bibtype{Article}%
\bibitem[{T{\"u}llmann} et al.(2008)]{Tuellemann_etal_2008}
\bibinfo{author}{{T{\"u}llmann} R}, \bibinfo{author}{{Gaetz} TJ},
  \bibinfo{author}{{Plucinsky} PP}, \bibinfo{author}{{Long} KS},
  \bibinfo{author}{{Hughes} JP}, \bibinfo{author}{{Blair} WP},
  \bibinfo{author}{{Winkler} PF}, \bibinfo{author}{{Pannuti} TG},
  \bibinfo{author}{{Breitschwerdt} D} and  \bibinfo{author}{{Ghavamian} P}
  (\bibinfo{year}{2008}), \bibinfo{month}{Oct.}
\bibinfo{title}{{The Chandra ACIS Survey of M33 (ChASeM33): Investigating the
  Hot Ionized Medium in NGC 604}}.
\bibinfo{journal}{{\em \apj}} \bibinfo{volume}{685} (\bibinfo{number}{2}):
  \bibinfo{pages}{919--932}. \bibinfo{doi}{\doi{10.1086/591019}}.
\eprint{0806.1527}.

\bibtype{Article}%
\bibitem[{Valentini} et al.(2017)]{Valentini_etal_2017}
\bibinfo{author}{{Valentini} M}, \bibinfo{author}{{Murante} G},
  \bibinfo{author}{{Borgani} S}, \bibinfo{author}{{Monaco} P},
  \bibinfo{author}{{Bressan} A} and  \bibinfo{author}{{Beck} AM}
  (\bibinfo{year}{2017}), \bibinfo{month}{Sep.}
\bibinfo{title}{{On the effect of galactic outflows in cosmological simulations
  of disc galaxies}}.
\bibinfo{journal}{{\em \mnras}} \bibinfo{volume}{470} (\bibinfo{number}{3}):
  \bibinfo{pages}{3167--3193}. \bibinfo{doi}{\doi{10.1093/mnras/stx1352}}.
\eprint{1705.10325}.

\bibtype{Article}%
\bibitem[{Valentino} et al.(2023)]{Valentino_etal_2023}
\bibinfo{author}{{Valentino} F}, \bibinfo{author}{{Brammer} G},
  \bibinfo{author}{{Gould} KML}, \bibinfo{author}{{Kokorev} V},
  \bibinfo{author}{{Fujimoto} S}, \bibinfo{author}{{Jespersen} CK},
  \bibinfo{author}{{Vijayan} AP}, \bibinfo{author}{{Weaver} JR},
  \bibinfo{author}{{Ito} K}, \bibinfo{author}{{Tanaka} M},
  \bibinfo{author}{{Ilbert} O}, \bibinfo{author}{{Magdis} GE},
  \bibinfo{author}{{Whitaker} KE}, \bibinfo{author}{{Faisst} AL},
  \bibinfo{author}{{Gallazzi} A}, \bibinfo{author}{{Gillman} S},
  \bibinfo{author}{{Gim{\'e}nez-Arteaga} C},
  \bibinfo{author}{{G{\'o}mez-Guijarro} C}, \bibinfo{author}{{Kubo} M},
  \bibinfo{author}{{Heintz} KE}, \bibinfo{author}{{Hirschmann} M},
  \bibinfo{author}{{Oesch} P}, \bibinfo{author}{{Onodera} M},
  \bibinfo{author}{{Rizzo} F}, \bibinfo{author}{{Lee} M},
  \bibinfo{author}{{Strait} V} and  \bibinfo{author}{{Toft} S}
  (\bibinfo{year}{2023}), \bibinfo{month}{Apr.}
\bibinfo{title}{{An Atlas of Color-selected Quiescent Galaxies at z > 3 in
  Public JWST Fields}}.
\bibinfo{journal}{{\em \apj}} \bibinfo{volume}{947} (\bibinfo{number}{1}),
  \bibinfo{eid}{20}. \bibinfo{doi}{\doi{10.3847/1538-4357/acbefa}}.
\eprint{2302.10936}.

\bibtype{Article}%
\bibitem[{van der Burg} et al.(2020)]{vanderBurg_etal_2020}
\bibinfo{author}{{van der Burg} RFJ}, \bibinfo{author}{{Rudnick} G},
  \bibinfo{author}{{Balogh} ML}, \bibinfo{author}{{Muzzin} A},
  \bibinfo{author}{{Lidman} C}, \bibinfo{author}{{Old} LJ},
  \bibinfo{author}{{Shipley} H}, \bibinfo{author}{{Gilbank} D},
  \bibinfo{author}{{McGee} S}, \bibinfo{author}{{Biviano} A},
  \bibinfo{author}{{Cerulo} P}, \bibinfo{author}{{Chan} JCC},
  \bibinfo{author}{{Cooper} M}, \bibinfo{author}{{De Lucia} G},
  \bibinfo{author}{{Demarco} R}, \bibinfo{author}{{Forrest} B},
  \bibinfo{author}{{Gwyn} S}, \bibinfo{author}{{Jablonka} P},
  \bibinfo{author}{{Kukstas} E}, \bibinfo{author}{{Marchesini} D},
  \bibinfo{author}{{Nantais} J}, \bibinfo{author}{{Noble} A},
  \bibinfo{author}{{Pintos-Castro} I}, \bibinfo{author}{{Poggianti} B},
  \bibinfo{author}{{Reeves} AMM}, \bibinfo{author}{{Stefanon} M},
  \bibinfo{author}{{Vulcani} B}, \bibinfo{author}{{Webb} K},
  \bibinfo{author}{{Wilson} G}, \bibinfo{author}{{Yee} H} and
  \bibinfo{author}{{Zaritsky} D} (\bibinfo{year}{2020}), \bibinfo{month}{Jun.}
\bibinfo{title}{{The GOGREEN Survey: A deep stellar mass function of cluster
  galaxies at 1.0 < z < 1.4 and the complex nature of satellite quenching}}.
\bibinfo{journal}{{\em \aap}} \bibinfo{volume}{638}, \bibinfo{eid}{A112}.
  \bibinfo{doi}{\doi{10.1051/0004-6361/202037754}}.
\eprint{2004.10757}.

\bibtype{Article}%
\bibitem[{Vijayan} et al.(2019)]{Vijayan_etal_2019}
\bibinfo{author}{{Vijayan} AP}, \bibinfo{author}{{Clay} SJ},
  \bibinfo{author}{{Thomas} PA}, \bibinfo{author}{{Yates} RM},
  \bibinfo{author}{{Wilkins} SM} and  \bibinfo{author}{{Henriques} BM}
  (\bibinfo{year}{2019}), \bibinfo{month}{Nov.}
\bibinfo{title}{{Detailed dust modelling in the L-GALAXIES semi-analytic model
  of galaxy formation}}.
\bibinfo{journal}{{\em \mnras}} \bibinfo{volume}{489} (\bibinfo{number}{3}):
  \bibinfo{pages}{4072--4089}. \bibinfo{doi}{\doi{10.1093/mnras/stz1948}}.
\eprint{1904.02196}.

\bibtype{Article}%
\bibitem[{Villalobos} et al.(2012)]{Villalobos_etal_2012}
\bibinfo{author}{{Villalobos} {\'A}}, \bibinfo{author}{{De Lucia} G},
  \bibinfo{author}{{Borgani} S} and  \bibinfo{author}{{Murante} G}
  (\bibinfo{year}{2012}), \bibinfo{month}{Aug.}
\bibinfo{title}{{Simulating the evolution of disc galaxies in a group
  environment - I. The influence of the global tidal field}}.
\bibinfo{journal}{{\em \mnras}} \bibinfo{volume}{424} (\bibinfo{number}{4}):
  \bibinfo{pages}{2401--2428}.
  \bibinfo{doi}{\doi{10.1111/j.1365-2966.2012.20667.x}}.
\eprint{1202.0550}.

\bibtype{Article}%
\bibitem[{Villalobos} et al.(2014)]{Villalobos_etal_2014}
\bibinfo{author}{{Villalobos} {\'A}}, \bibinfo{author}{{De Lucia} G} and
  \bibinfo{author}{{Murante} G} (\bibinfo{year}{2014}), \bibinfo{month}{Oct.}
\bibinfo{title}{{Simulating the evolution of disc galaxies in a group
  environment - II. The influence of close encounters between galaxies}}.
\bibinfo{journal}{{\em \mnras}} \bibinfo{volume}{444} (\bibinfo{number}{1}):
  \bibinfo{pages}{313--326}. \bibinfo{doi}{\doi{10.1093/mnras/stu1278}}.
\eprint{1404.2942}.

\bibtype{Article}%
\bibitem[{Vogelsberger} et al.(2020)]{Vogelsberger_etal_2020}
\bibinfo{author}{{Vogelsberger} M}, \bibinfo{author}{{Marinacci} F},
  \bibinfo{author}{{Torrey} P} and  \bibinfo{author}{{Puchwein} E}
  (\bibinfo{year}{2020}), \bibinfo{month}{Jan.}
\bibinfo{title}{{Cosmological simulations of galaxy formation}}.
\bibinfo{journal}{{\em Nature Reviews Physics}} \bibinfo{volume}{2}
  (\bibinfo{number}{1}): \bibinfo{pages}{42--66}.
  \bibinfo{doi}{\doi{10.1038/s42254-019-0127-2}}.
\eprint{1909.07976}.

\bibtype{Article}%
\bibitem[{von der Linden} et al.(2010)]{vonderLinden_etal_2010}
\bibinfo{author}{{von der Linden} A}, \bibinfo{author}{{Wild} V},
  \bibinfo{author}{{Kauffmann} G}, \bibinfo{author}{{White} SDM} and
  \bibinfo{author}{{Weinmann} S} (\bibinfo{year}{2010}), \bibinfo{month}{May}.
\bibinfo{title}{{Star formation and AGN activity in SDSS cluster galaxies}}.
\bibinfo{journal}{{\em \mnras}} \bibinfo{volume}{404} (\bibinfo{number}{3}):
  \bibinfo{pages}{1231--1246}.
  \bibinfo{doi}{\doi{10.1111/j.1365-2966.2010.16375.x}}.
\eprint{0909.3522}.

\bibtype{Article}%
\bibitem[{Vulcani} et al.(2010)]{Vulcani_etal_2010}
\bibinfo{author}{{Vulcani} B}, \bibinfo{author}{{Poggianti} BM},
  \bibinfo{author}{{Finn} RA}, \bibinfo{author}{{Rudnick} G},
  \bibinfo{author}{{Desai} V} and  \bibinfo{author}{{Bamford} S}
  (\bibinfo{year}{2010}), \bibinfo{month}{Feb.}
\bibinfo{title}{{Comparing the Relation Between Star Formation and Galaxy Mass
  in Different Environments}}.
\bibinfo{journal}{{\em \apjl}} \bibinfo{volume}{710} (\bibinfo{number}{1}):
  \bibinfo{pages}{L1--L6}. \bibinfo{doi}{\doi{10.1088/2041-8205/710/1/L1}}.
\eprint{0912.1180}.

\bibtype{Article}%
\bibitem[{Vulcani} et al.(2021)]{Vulcani21}
\bibinfo{author}{{Vulcani} B}, \bibinfo{author}{{Poggianti} BM},
  \bibinfo{author}{{Moretti} A}, \bibinfo{author}{{Franchetto} A},
  \bibinfo{author}{{Bacchini} C}, \bibinfo{author}{{McGee} S},
  \bibinfo{author}{{Jaff{\'e}} YL}, \bibinfo{author}{{Mingozzi} M},
  \bibinfo{author}{{Werle} A}, \bibinfo{author}{{Tomi{\v{c}}i{\'c}} N},
  \bibinfo{author}{{Fritz} J}, \bibinfo{author}{{Bettoni} D},
  \bibinfo{author}{{Wolter} A} and  \bibinfo{author}{{Gullieuszik} M}
  (\bibinfo{year}{2021}), \bibinfo{month}{Jun.}
\bibinfo{title}{{GASP. XXXIII. The Ability of Spatially Resolved Data to
  Distinguish among the Different Physical Mechanisms Affecting Galaxies in
  Low-density Environments}}.
\bibinfo{journal}{{\em \apj}} \bibinfo{volume}{914} (\bibinfo{number}{1}),
  \bibinfo{eid}{27}. \bibinfo{doi}{\doi{10.3847/1538-4357/abf655}}.
\eprint{2104.02089}.

\bibtype{Article}%
\bibitem[{Wang} et al.(2007)]{Wang_etal_2007}
\bibinfo{author}{{Wang} L}, \bibinfo{author}{{Li} C},
  \bibinfo{author}{{Kauffmann} G} and  \bibinfo{author}{{De Lucia} G}
  (\bibinfo{year}{2007}), \bibinfo{month}{Jun.}
\bibinfo{title}{{Modelling and interpreting the dependence of clustering on the
  spectral energy distributions of galaxies}}.
\bibinfo{journal}{{\em \mnras}} \bibinfo{volume}{377} (\bibinfo{number}{4}):
  \bibinfo{pages}{1419--1430}.
  \bibinfo{doi}{\doi{10.1111/j.1365-2966.2007.11737.x}}.
\eprint{astro-ph/0701682}.

\bibtype{Article}%
\bibitem[{Weaver} et al.(2023)]{Weaver_etal_2023}
\bibinfo{author}{{Weaver} JR}, \bibinfo{author}{{Davidzon} I},
  \bibinfo{author}{{Toft} S}, \bibinfo{author}{{Ilbert} O},
  \bibinfo{author}{{McCracken} HJ}, \bibinfo{author}{{Gould} KML},
  \bibinfo{author}{{Jespersen} CK}, \bibinfo{author}{{Steinhardt} C},
  \bibinfo{author}{{Lagos} CDP}, \bibinfo{author}{{Capak} PL},
  \bibinfo{author}{{Casey} CM}, \bibinfo{author}{{Chartab} N},
  \bibinfo{author}{{Faisst} AL}, \bibinfo{author}{{Hayward} CC},
  \bibinfo{author}{{Kartaltepe} JS}, \bibinfo{author}{{Kauffmann} OB},
  \bibinfo{author}{{Koekemoer} AM}, \bibinfo{author}{{Kokorev} V},
  \bibinfo{author}{{Laigle} C}, \bibinfo{author}{{Liu} D},
  \bibinfo{author}{{Long} A}, \bibinfo{author}{{Magdis} GE},
  \bibinfo{author}{{McPartland} CJR}, \bibinfo{author}{{Milvang-Jensen} B},
  \bibinfo{author}{{Mobasher} B}, \bibinfo{author}{{Moneti} A},
  \bibinfo{author}{{Peng} Y}, \bibinfo{author}{{Sanders} DB},
  \bibinfo{author}{{Shuntov} M}, \bibinfo{author}{{Sneppen} A},
  \bibinfo{author}{{Valentino} F}, \bibinfo{author}{{Zalesky} L} and
  \bibinfo{author}{{Zamorani} G} (\bibinfo{year}{2023}), \bibinfo{month}{Sep.}
\bibinfo{title}{{COSMOS2020: The galaxy stellar mass function. The assembly and
  star formation cessation of galaxies at 0.2< z {\ensuremath{\leq}} 7.5}}.
\bibinfo{journal}{{\em \aap}} \bibinfo{volume}{677}, \bibinfo{eid}{A184}.
  \bibinfo{doi}{\doi{10.1051/0004-6361/202245581}}.
\eprint{2212.02512}.

\bibtype{Article}%
\bibitem[{Weibel} et al.(2024)]{Weibel_etal_2024}
\bibinfo{author}{{Weibel} A}, \bibinfo{author}{{de Graaff} A},
  \bibinfo{author}{{Setton} DJ}, \bibinfo{author}{{Miller} TB},
  \bibinfo{author}{{Oesch} PA}, \bibinfo{author}{{Brammer} G},
  \bibinfo{author}{{Lagos} CDP}, \bibinfo{author}{{Whitaker} KE},
  \bibinfo{author}{{Williams} CC}, \bibinfo{author}{{Baggen} JFW},
  \bibinfo{author}{{Bezanson} R}, \bibinfo{author}{{Boogaard} LA},
  \bibinfo{author}{{Cleri} NJ}, \bibinfo{author}{{Greene} JE},
  \bibinfo{author}{{Hirschmann} M}, \bibinfo{author}{{Hviding} RE},
  \bibinfo{author}{{Kuruvanthodi} A}, \bibinfo{author}{{Labb{\'e}} I},
  \bibinfo{author}{{Leja} J}, \bibinfo{author}{{Maseda} MV},
  \bibinfo{author}{{Matthee} J}, \bibinfo{author}{{McConachie} I},
  \bibinfo{author}{{Naidu} RP}, \bibinfo{author}{{Roberts-Borsani} G},
  \bibinfo{author}{{Schaerer} D}, \bibinfo{author}{{Suess} KA},
  \bibinfo{author}{{Valentino} F}, \bibinfo{author}{{van Dokkum} P} and
  \bibinfo{author}{{Wang} B} (\bibinfo{year}{2024}), \bibinfo{month}{Sep.}
\bibinfo{title}{{RUBIES Reveals a Massive Quiescent Galaxy at z=7.3}}.
\bibinfo{journal}{{\em arXiv e-prints}} ,
  \bibinfo{eid}{arXiv:2409.03829}\bibinfo{doi}{\doi{10.48550/arXiv.2409.03829}}.
\eprint{2409.03829}.

\bibtype{Article}%
\bibitem[{Weiner} et al.(2005)]{Weiner_etal_2005}
\bibinfo{author}{{Weiner} BJ}, \bibinfo{author}{{Phillips} AC},
  \bibinfo{author}{{Faber} SM}, \bibinfo{author}{{Willmer} CNA},
  \bibinfo{author}{{Vogt} NP}, \bibinfo{author}{{Simard} L},
  \bibinfo{author}{{Gebhardt} K}, \bibinfo{author}{{Im} M},
  \bibinfo{author}{{Koo} DC}, \bibinfo{author}{{Sarajedini} VL},
  \bibinfo{author}{{Wu} KL}, \bibinfo{author}{{Forbes} DA},
  \bibinfo{author}{{Gronwall} C}, \bibinfo{author}{{Groth} EJ},
  \bibinfo{author}{{Illingworth} GD}, \bibinfo{author}{{Kron} RG},
  \bibinfo{author}{{Rhodes} J}, \bibinfo{author}{{Szalay} AS} and
  \bibinfo{author}{{Takamiya} M} (\bibinfo{year}{2005}), \bibinfo{month}{Feb.}
\bibinfo{title}{{The DEEP Groth Strip Galaxy Redshift Survey. III. Redshift
  Catalog and Properties of Galaxies}}.
\bibinfo{journal}{{\em \apj}} \bibinfo{volume}{620} (\bibinfo{number}{2}):
  \bibinfo{pages}{595--617}. \bibinfo{doi}{\doi{10.1086/427256}}.
\eprint{astro-ph/0411128}.

\bibtype{Article}%
\bibitem[{Weinmann} et al.(2006)]{Weinmann_etal_2006}
\bibinfo{author}{{Weinmann} SM}, \bibinfo{author}{{van den Bosch} FC},
  \bibinfo{author}{{Yang} X}, \bibinfo{author}{{Mo} HJ},
  \bibinfo{author}{{Croton} DJ} and  \bibinfo{author}{{Moore} B}
  (\bibinfo{year}{2006}), \bibinfo{month}{Nov.}
\bibinfo{title}{{Properties of galaxy groups in the Sloan Digital Sky Survey -
  II. Active galactic nucleus feedback and star formation truncation}}.
\bibinfo{journal}{{\em \mnras}} \bibinfo{volume}{372} (\bibinfo{number}{3}):
  \bibinfo{pages}{1161--1174}.
  \bibinfo{doi}{\doi{10.1111/j.1365-2966.2006.10932.x}}.
\eprint{astro-ph/0606458}.

\bibtype{Article}%
\bibitem[{Weinmann} et al.(2010)]{Weinmann_etal_2010}
\bibinfo{author}{{Weinmann} SM}, \bibinfo{author}{{Kauffmann} G},
  \bibinfo{author}{{von der Linden} A} and  \bibinfo{author}{{De Lucia} G}
  (\bibinfo{year}{2010}), \bibinfo{month}{Aug.}
\bibinfo{title}{{Cluster galaxies die hard}}.
\bibinfo{journal}{{\em \mnras}} \bibinfo{volume}{406} (\bibinfo{number}{4}):
  \bibinfo{pages}{2249--2266}.
  \bibinfo{doi}{\doi{10.1111/j.1365-2966.2010.16855.x}}.
\eprint{0912.2741}.

\bibtype{Article}%
\bibitem[{Wetzel} et al.(2012)]{Wetzel_etal_2012}
\bibinfo{author}{{Wetzel} AR}, \bibinfo{author}{{Tinker} JL} and
  \bibinfo{author}{{Conroy} C} (\bibinfo{year}{2012}), \bibinfo{month}{Jul.}
\bibinfo{title}{{Galaxy evolution in groups and clusters: star formation rates,
  red sequence fractions and the persistent bimodality}}.
\bibinfo{journal}{{\em \mnras}} \bibinfo{volume}{424} (\bibinfo{number}{1}):
  \bibinfo{pages}{232--243}.
  \bibinfo{doi}{\doi{10.1111/j.1365-2966.2012.21188.x}}.
\eprint{1107.5311}.

\bibtype{Article}%
\bibitem[{Wetzel} et al.(2013)]{Wetzel_etal_2013}
\bibinfo{author}{{Wetzel} AR}, \bibinfo{author}{{Tinker} JL},
  \bibinfo{author}{{Conroy} C} and  \bibinfo{author}{{van den Bosch} FC}
  (\bibinfo{year}{2013}), \bibinfo{month}{Jun.}
\bibinfo{title}{{Galaxy evolution in groups and clusters: satellite star
  formation histories and quenching time-scales in a hierarchical Universe}}.
\bibinfo{journal}{{\em \mnras}} \bibinfo{volume}{432} (\bibinfo{number}{1}):
  \bibinfo{pages}{336--358}. \bibinfo{doi}{\doi{10.1093/mnras/stt469}}.
\eprint{1206.3571}.

\bibtype{Article}%
\bibitem[{Whitaker} et al.(2011)]{Whitaker_etal_2011}
\bibinfo{author}{{Whitaker} KE}, \bibinfo{author}{{Labb{\'e}} I},
  \bibinfo{author}{{van Dokkum} PG}, \bibinfo{author}{{Brammer} G},
  \bibinfo{author}{{Kriek} M}, \bibinfo{author}{{Marchesini} D},
  \bibinfo{author}{{Quadri} RF}, \bibinfo{author}{{Franx} M},
  \bibinfo{author}{{Muzzin} A}, \bibinfo{author}{{Williams} RJ},
  \bibinfo{author}{{Bezanson} R}, \bibinfo{author}{{Illingworth} GD},
  \bibinfo{author}{{Lee} KS}, \bibinfo{author}{{Lundgren} B},
  \bibinfo{author}{{Nelson} EJ}, \bibinfo{author}{{Rudnick} G},
  \bibinfo{author}{{Tal} T} and  \bibinfo{author}{{Wake} DA}
  (\bibinfo{year}{2011}), \bibinfo{month}{Jul.}
\bibinfo{title}{{The NEWFIRM Medium-band Survey: Photometric Catalogs,
  Redshifts, and the Bimodal Color Distribution of Galaxies out to z
  \raisebox{-0.5ex}\textasciitilde 3}}.
\bibinfo{journal}{{\em \apj}} \bibinfo{volume}{735} (\bibinfo{number}{2}),
  \bibinfo{eid}{86}. \bibinfo{doi}{\doi{10.1088/0004-637X/735/2/86}}.
\eprint{1105.4609}.

\bibtype{Article}%
\bibitem[{Williams} et al.(2009)]{Williams_etal_2009}
\bibinfo{author}{{Williams} RJ}, \bibinfo{author}{{Quadri} RF},
  \bibinfo{author}{{Franx} M}, \bibinfo{author}{{van Dokkum} P} and
  \bibinfo{author}{{Labb{\'e}} I} (\bibinfo{year}{2009}), \bibinfo{month}{Feb.}
\bibinfo{title}{{Detection of Quiescent Galaxies in a Bicolor Sequence from Z =
  0-2}}.
\bibinfo{journal}{{\em \apj}} \bibinfo{volume}{691} (\bibinfo{number}{2}):
  \bibinfo{pages}{1879--1895}.
  \bibinfo{doi}{\doi{10.1088/0004-637X/691/2/1879}}.
\eprint{0806.0625}.

\bibtype{Article}%
\bibitem[{Willis} et al.(2020)]{Willis_etal_2020}
\bibinfo{author}{{Willis} JP}, \bibinfo{author}{{Canning} REA},
  \bibinfo{author}{{Noordeh} ES}, \bibinfo{author}{{Allen} SW},
  \bibinfo{author}{{King} AL}, \bibinfo{author}{{Mantz} A},
  \bibinfo{author}{{Morris} RG}, \bibinfo{author}{{Stanford} SA} and
  \bibinfo{author}{{Brammer} G} (\bibinfo{year}{2020}), \bibinfo{month}{Jan.}
\bibinfo{title}{{Spectroscopic confirmation of a mature galaxy cluster at a
  redshift of 2}}.
\bibinfo{journal}{{\em \nat}} \bibinfo{volume}{577} (\bibinfo{number}{7788}):
  \bibinfo{pages}{39--41}. \bibinfo{doi}{\doi{10.1038/s41586-019-1829-4}}.
\eprint{2001.00549}.

\bibtype{Article}%
\bibitem[{Wolf} et al.(2003)]{Wolf_etal_2003}
\bibinfo{author}{{Wolf} C}, \bibinfo{author}{{Meisenheimer} K},
  \bibinfo{author}{{Rix} HW}, \bibinfo{author}{{Borch} A},
  \bibinfo{author}{{Dye} S} and  \bibinfo{author}{{Kleinheinrich} M}
  (\bibinfo{year}{2003}), \bibinfo{month}{Apr.}
\bibinfo{title}{{The COMBO-17 survey: Evolution of the galaxy luminosity
  function from 25 000 galaxies with 0.2< z <1.2}}.
\bibinfo{journal}{{\em \aap}} \bibinfo{volume}{401}: \bibinfo{pages}{73--98}.
  \bibinfo{doi}{\doi{10.1051/0004-6361:20021513}}.
\eprint{astro-ph/0208345}.

\bibtype{Article}%
\bibitem[{Xie} et al.(2017)]{Xie_etal_2017}
\bibinfo{author}{{Xie} L}, \bibinfo{author}{{De Lucia} G},
  \bibinfo{author}{{Hirschmann} M}, \bibinfo{author}{{Fontanot} F} and
  \bibinfo{author}{{Zoldan} A} (\bibinfo{year}{2017}), \bibinfo{month}{Jul.}
\bibinfo{title}{{H$_{2}$-based star formation laws in hierarchical models of
  galaxy formation}}.
\bibinfo{journal}{{\em \mnras}} \bibinfo{volume}{469} (\bibinfo{number}{1}):
  \bibinfo{pages}{968--993}. \bibinfo{doi}{\doi{10.1093/mnras/stx889}}.
\eprint{1611.09372}.

\bibtype{Article}%
\bibitem[{Xie} et al.(2020)]{Xie_etal_2020}
\bibinfo{author}{{Xie} L}, \bibinfo{author}{{De Lucia} G},
  \bibinfo{author}{{Hirschmann} M} and  \bibinfo{author}{{Fontanot} F}
  (\bibinfo{year}{2020}), \bibinfo{month}{Nov.}
\bibinfo{title}{{The influence of environment on satellite galaxies in the GAEA
  semi-analytic model}}.
\bibinfo{journal}{{\em \mnras}} \bibinfo{volume}{498} (\bibinfo{number}{3}):
  \bibinfo{pages}{4327--4344}. \bibinfo{doi}{\doi{10.1093/mnras/staa2370}}.
\eprint{2003.12757}.

\bibtype{Article}%
\bibitem[{Xie} et al.(2024)]{Xie_etal_2024}
\bibinfo{author}{{Xie} L}, \bibinfo{author}{{De Lucia} G},
  \bibinfo{author}{{Fontanot} F}, \bibinfo{author}{{Hirschmann} M},
  \bibinfo{author}{{Bah{\'e}} YM}, \bibinfo{author}{{Balogh} ML},
  \bibinfo{author}{{Muzzin} A}, \bibinfo{author}{{Vulcani} B},
  \bibinfo{author}{{Baxter} DC}, \bibinfo{author}{{Forrest} B},
  \bibinfo{author}{{Wilson} G}, \bibinfo{author}{{Rudnick} GH},
  \bibinfo{author}{{Cooper} MC} and  \bibinfo{author}{{Rescigno} U}
  (\bibinfo{year}{2024}), \bibinfo{month}{May}.
\bibinfo{title}{{The First Quenched Galaxies: When and How?}}
\bibinfo{journal}{{\em \apjl}} \bibinfo{volume}{966} (\bibinfo{number}{1}),
  \bibinfo{eid}{L2}. \bibinfo{doi}{\doi{10.3847/2041-8213/ad380a}}.
\eprint{2402.01314}.

\bibtype{Article}%
\bibitem[{Yang} et al.(2007)]{Yang_etal_2007}
\bibinfo{author}{{Yang} X}, \bibinfo{author}{{Mo} HJ}, \bibinfo{author}{{van
  den Bosch} FC}, \bibinfo{author}{{Pasquali} A}, \bibinfo{author}{{Li} C} and
  \bibinfo{author}{{Barden} M} (\bibinfo{year}{2007}), \bibinfo{month}{Dec.}
\bibinfo{title}{{Galaxy Groups in the SDSS DR4. I. The Catalog and Basic
  Properties}}.
\bibinfo{journal}{{\em \apj}} \bibinfo{volume}{671} (\bibinfo{number}{1}):
  \bibinfo{pages}{153--170}. \bibinfo{doi}{\doi{10.1086/522027}}.
\eprint{0707.4640}.

\bibtype{Article}%
\bibitem[{Yang} et al.(2008)]{Yang_etal_2008}
\bibinfo{author}{{Yang} X}, \bibinfo{author}{{Mo} HJ} and
  \bibinfo{author}{{van den Bosch} FC} (\bibinfo{year}{2008}),
  \bibinfo{month}{Mar.}
\bibinfo{title}{{Galaxy Groups in the SDSS DR4. II. Halo Occupation
  Statistics}}.
\bibinfo{journal}{{\em \apj}} \bibinfo{volume}{676} (\bibinfo{number}{1}):
  \bibinfo{pages}{248--261}. \bibinfo{doi}{\doi{10.1086/528954}}.
\eprint{0710.5096}.

\bibtype{Article}%
\bibitem[{Yang} et al.(2009)]{Yang_etal_2009}
\bibinfo{author}{{Yang} X}, \bibinfo{author}{{Mo} HJ} and
  \bibinfo{author}{{van den Bosch} FC} (\bibinfo{year}{2009}),
  \bibinfo{month}{Apr.}
\bibinfo{title}{{Galaxy Groups in the SDSS DR4. III. The Luminosity and Stellar
  Mass Functions}}.
\bibinfo{journal}{{\em \apj}} \bibinfo{volume}{695} (\bibinfo{number}{2}):
  \bibinfo{pages}{900--916}. \bibinfo{doi}{\doi{10.1088/0004-637X/695/2/900}}.
\eprint{0808.0539}.

\bibtype{Article}%
\bibitem[{Yang} et al.(2012)]{Yang_etal_2012}
\bibinfo{author}{{Yang} X}, \bibinfo{author}{{Mo} HJ}, \bibinfo{author}{{van
  den Bosch} FC}, \bibinfo{author}{{Zhang} Y} and  \bibinfo{author}{{Han} J}
  (\bibinfo{year}{2012}), \bibinfo{month}{Jun.}
\bibinfo{title}{{Evolution of the Galaxy-Dark Matter Connection and the
  Assembly of Galaxies in Dark Matter Halos}}.
\bibinfo{journal}{{\em \apj}} \bibinfo{volume}{752} (\bibinfo{number}{1}),
  \bibinfo{eid}{41}. \bibinfo{doi}{\doi{10.1088/0004-637X/752/1/41}}.
\eprint{1110.1420}.

\bibtype{Article}%
\bibitem[{Yang} et al.(2024)]{Hang_etal_2024}
\bibinfo{author}{{Yang} H}, \bibinfo{author}{{Liao} S},
  \bibinfo{author}{{Fattahi} A}, \bibinfo{author}{{Frenk} CS},
  \bibinfo{author}{{Gao} L}, \bibinfo{author}{{Guo} Q}, \bibinfo{author}{{Shao}
  S}, \bibinfo{author}{{Wang} L}, \bibinfo{author}{{Wright} RJ} and
  \bibinfo{author}{{Zeng} G} (\bibinfo{year}{2024}), \bibinfo{month}{Dec.}
\bibinfo{title}{{APOSTLE-AURIGA: effects of stellar feedback subgrid models on
  the evolution of angular momentum in disc galaxies}}.
\bibinfo{journal}{{\em \mnras}} \bibinfo{volume}{535} (\bibinfo{number}{2}):
  \bibinfo{pages}{1394--1405}. \bibinfo{doi}{\doi{10.1093/mnras/stae2411}}.
\eprint{2408.09784}.

\bibtype{Article}%
\bibitem[{Yates} et al.(2021)]{Yates_etal_2021}
\bibinfo{author}{{Yates} RM}, \bibinfo{author}{{Henriques} BMB},
  \bibinfo{author}{{Fu} J}, \bibinfo{author}{{Kauffmann} G},
  \bibinfo{author}{{Thomas} PA}, \bibinfo{author}{{Guo} Q},
  \bibinfo{author}{{White} SDM} and  \bibinfo{author}{{Schady} P}
  (\bibinfo{year}{2021}), \bibinfo{month}{May}.
\bibinfo{title}{{L-GALAXIES 2020: The evolution of radial metallicity profiles
  and global metallicities in disc galaxies}}.
\bibinfo{journal}{{\em \mnras}} \bibinfo{volume}{503} (\bibinfo{number}{3}):
  \bibinfo{pages}{4474--4495}. \bibinfo{doi}{\doi{10.1093/mnras/stab741}}.
\eprint{2011.04670}.

\bibtype{Article}%
\bibitem[{Yi} et al.(2005)]{Yi_etal_2005}
\bibinfo{author}{{Yi} SK}, \bibinfo{author}{{Yoon} SJ},
  \bibinfo{author}{{Kaviraj} S}, \bibinfo{author}{{Deharveng} JM},
  \bibinfo{author}{{Rich} RM}, \bibinfo{author}{{Salim} S},
  \bibinfo{author}{{Boselli} A}, \bibinfo{author}{{Lee} YW},
  \bibinfo{author}{{Ree} CH}, \bibinfo{author}{{Sohn} YJ},
  \bibinfo{author}{{Rey} SC}, \bibinfo{author}{{Lee} JW},
  \bibinfo{author}{{Rhee} J}, \bibinfo{author}{{Bianchi} L},
  \bibinfo{author}{{Byun} YI}, \bibinfo{author}{{Donas} J},
  \bibinfo{author}{{Friedman} PG}, \bibinfo{author}{{Heckman} TM},
  \bibinfo{author}{{Jelinsky} P}, \bibinfo{author}{{Madore} BF},
  \bibinfo{author}{{Malina} R}, \bibinfo{author}{{Martin} DC},
  \bibinfo{author}{{Milliard} B}, \bibinfo{author}{{Morrissey} P},
  \bibinfo{author}{{Neff} S}, \bibinfo{author}{{Schiminovich} D},
  \bibinfo{author}{{Siegmund} O}, \bibinfo{author}{{Small} T},
  \bibinfo{author}{{Szalay} AS}, \bibinfo{author}{{Jee} MJ},
  \bibinfo{author}{{Kim} SW}, \bibinfo{author}{{Barlow} T},
  \bibinfo{author}{{Forster} K}, \bibinfo{author}{{Welsh} B} and
  \bibinfo{author}{{Wyder} TK} (\bibinfo{year}{2005}), \bibinfo{month}{Jan.}
\bibinfo{title}{{Galaxy Evolution Explorer Ultraviolet Color-Magnitude
  Relations and Evidence of Recent Star Formation in Early-Type Galaxies}}.
\bibinfo{journal}{{\em \apjl}} \bibinfo{volume}{619} (\bibinfo{number}{1}):
  \bibinfo{pages}{L111--L114}. \bibinfo{doi}{\doi{10.1086/422811}}.
\eprint{astro-ph/0411327}.

\bibtype{Article}%
\bibitem[{York} et al.(2000)]{York_etal_2000}
\bibinfo{author}{{York} DG}, \bibinfo{author}{{Adelman} J},
  \bibinfo{author}{{Anderson} John~E. J}, \bibinfo{author}{{Anderson} SF},
  \bibinfo{author}{{Annis} J}, \bibinfo{author}{{Bahcall} NA},
  \bibinfo{author}{{Bakken} JA}, \bibinfo{author}{{Barkhouser} R},
  \bibinfo{author}{{Bastian} S}, \bibinfo{author}{{Berman} E},
  \bibinfo{author}{{Boroski} WN}, \bibinfo{author}{{Bracker} S},
  \bibinfo{author}{{Briegel} C}, \bibinfo{author}{{Briggs} JW},
  \bibinfo{author}{{Brinkmann} J}, \bibinfo{author}{{Brunner} R},
  \bibinfo{author}{{Burles} S}, \bibinfo{author}{{Carey} L},
  \bibinfo{author}{{Carr} MA}, \bibinfo{author}{{Castander} FJ},
  \bibinfo{author}{{Chen} B}, \bibinfo{author}{{Colestock} PL},
  \bibinfo{author}{{Connolly} AJ}, \bibinfo{author}{{Crocker} JH},
  \bibinfo{author}{{Csabai} I}, \bibinfo{author}{{Czarapata} PC},
  \bibinfo{author}{{Davis} JE}, \bibinfo{author}{{Doi} M},
  \bibinfo{author}{{Dombeck} T}, \bibinfo{author}{{Eisenstein} D},
  \bibinfo{author}{{Ellman} N}, \bibinfo{author}{{Elms} BR},
  \bibinfo{author}{{Evans} ML}, \bibinfo{author}{{Fan} X},
  \bibinfo{author}{{Federwitz} GR}, \bibinfo{author}{{Fiscelli} L},
  \bibinfo{author}{{Friedman} S}, \bibinfo{author}{{Frieman} JA},
  \bibinfo{author}{{Fukugita} M}, \bibinfo{author}{{Gillespie} B},
  \bibinfo{author}{{Gunn} JE}, \bibinfo{author}{{Gurbani} VK},
  \bibinfo{author}{{de Haas} E}, \bibinfo{author}{{Haldeman} M},
  \bibinfo{author}{{Harris} FH}, \bibinfo{author}{{Hayes} J},
  \bibinfo{author}{{Heckman} TM}, \bibinfo{author}{{Hennessy} GS},
  \bibinfo{author}{{Hindsley} RB}, \bibinfo{author}{{Holm} S},
  \bibinfo{author}{{Holmgren} DJ}, \bibinfo{author}{{Huang} Ch},
  \bibinfo{author}{{Hull} C}, \bibinfo{author}{{Husby} D},
  \bibinfo{author}{{Ichikawa} SI}, \bibinfo{author}{{Ichikawa} T},
  \bibinfo{author}{{Ivezi{\'c}} {\v{Z}}}, \bibinfo{author}{{Kent} S},
  \bibinfo{author}{{Kim} RSJ}, \bibinfo{author}{{Kinney} E},
  \bibinfo{author}{{Klaene} M}, \bibinfo{author}{{Kleinman} AN},
  \bibinfo{author}{{Kleinman} S}, \bibinfo{author}{{Knapp} GR},
  \bibinfo{author}{{Korienek} J}, \bibinfo{author}{{Kron} RG},
  \bibinfo{author}{{Kunszt} PZ}, \bibinfo{author}{{Lamb} DQ},
  \bibinfo{author}{{Lee} B}, \bibinfo{author}{{Leger} RF},
  \bibinfo{author}{{Limmongkol} S}, \bibinfo{author}{{Lindenmeyer} C},
  \bibinfo{author}{{Long} DC}, \bibinfo{author}{{Loomis} C},
  \bibinfo{author}{{Loveday} J}, \bibinfo{author}{{Lucinio} R},
  \bibinfo{author}{{Lupton} RH}, \bibinfo{author}{{MacKinnon} B},
  \bibinfo{author}{{Mannery} EJ}, \bibinfo{author}{{Mantsch} PM},
  \bibinfo{author}{{Margon} B}, \bibinfo{author}{{McGehee} P},
  \bibinfo{author}{{McKay} TA}, \bibinfo{author}{{Meiksin} A},
  \bibinfo{author}{{Merelli} A}, \bibinfo{author}{{Monet} DG},
  \bibinfo{author}{{Munn} JA}, \bibinfo{author}{{Narayanan} VK},
  \bibinfo{author}{{Nash} T}, \bibinfo{author}{{Neilsen} E},
  \bibinfo{author}{{Neswold} R}, \bibinfo{author}{{Newberg} HJ},
  \bibinfo{author}{{Nichol} RC}, \bibinfo{author}{{Nicinski} T},
  \bibinfo{author}{{Nonino} M}, \bibinfo{author}{{Okada} N},
  \bibinfo{author}{{Okamura} S}, \bibinfo{author}{{Ostriker} JP},
  \bibinfo{author}{{Owen} R}, \bibinfo{author}{{Pauls} AG},
  \bibinfo{author}{{Peoples} J}, \bibinfo{author}{{Peterson} RL},
  \bibinfo{author}{{Petravick} D}, \bibinfo{author}{{Pier} JR},
  \bibinfo{author}{{Pope} A}, \bibinfo{author}{{Pordes} R},
  \bibinfo{author}{{Prosapio} A}, \bibinfo{author}{{Rechenmacher} R},
  \bibinfo{author}{{Quinn} TR}, \bibinfo{author}{{Richards} GT},
  \bibinfo{author}{{Richmond} MW}, \bibinfo{author}{{Rivetta} CH},
  \bibinfo{author}{{Rockosi} CM}, \bibinfo{author}{{Ruthmansdorfer} K},
  \bibinfo{author}{{Sandford} D}, \bibinfo{author}{{Schlegel} DJ},
  \bibinfo{author}{{Schneider} DP}, \bibinfo{author}{{Sekiguchi} M},
  \bibinfo{author}{{Sergey} G}, \bibinfo{author}{{Shimasaku} K},
  \bibinfo{author}{{Siegmund} WA}, \bibinfo{author}{{Smee} S},
  \bibinfo{author}{{Smith} JA}, \bibinfo{author}{{Snedden} S},
  \bibinfo{author}{{Stone} R}, \bibinfo{author}{{Stoughton} C},
  \bibinfo{author}{{Strauss} MA}, \bibinfo{author}{{Stubbs} C},
  \bibinfo{author}{{SubbaRao} M}, \bibinfo{author}{{Szalay} AS},
  \bibinfo{author}{{Szapudi} I}, \bibinfo{author}{{Szokoly} GP},
  \bibinfo{author}{{Thakar} AR}, \bibinfo{author}{{Tremonti} C},
  \bibinfo{author}{{Tucker} DL}, \bibinfo{author}{{Uomoto} A},
  \bibinfo{author}{{Vanden Berk} D}, \bibinfo{author}{{Vogeley} MS},
  \bibinfo{author}{{Waddell} P}, \bibinfo{author}{{Wang} Si},
  \bibinfo{author}{{Watanabe} M}, \bibinfo{author}{{Weinberg} DH},
  \bibinfo{author}{{Yanny} B}, \bibinfo{author}{{Yasuda} N} and
  \bibinfo{author}{{SDSS Collaboration}} (\bibinfo{year}{2000}),
  \bibinfo{month}{Sep.}
\bibinfo{title}{{The Sloan Digital Sky Survey: Technical Summary}}.
\bibinfo{journal}{{\em \aj}} \bibinfo{volume}{120} (\bibinfo{number}{3}):
  \bibinfo{pages}{1579--1587}. \bibinfo{doi}{\doi{10.1086/301513}}.
\eprint{astro-ph/0006396}.

\end{thebibliography*}

\end{document}